%% file: review.tex
\documentclass[3p]{elsarticle}

\usepackage{amsmath, amssymb, bm, graphicx}
\usepackage[breaklinks]{hyperref}
\usepackage{xcolor}

\journal{Physics Reports}

\biboptions{sort&compress}

\begin{document}

\begin{frontmatter}

\title{Three-dimensional Abelian and non-Abelian gauge Higgs theories}

\author{Claudio Bonati}
\ead{claudio.bonati@unipi.it}

\address{Dipartimento di Fisica dell'Universit\`a di Pisa and 
  INFN Sezione di Pisa, Largo Pontecorvo 3, I-56127 Pisa, Italy}

\author{Andrea Pelissetto}
\ead{andrea.pelissetto@roma1.infn.it}

\address{Dipartimento di Fisica dell'Universit\`a di Roma Sapienza and
  INFN Sezione di Roma I, I-00185 Roma, Italy}
  
\author{Ettore Vicari}
\ead{ettore.vicari@unipi.it}

\address{Dipartimento di Fisica dell'Universit\`a di Pisa,
  Largo Pontecorvo 3, I-56127 Pisa, Italy}

\input{abstract}

\end{frontmatter}

\tableofcontents

\vspace*{5mm}

\hrulefill

\input{introduction}

\input{LGWapproach}

\input{FTapproachestoLGT}

\input{AHSFT}

\input{ncLAHM}

\input{cLAHM}

\input{discretegauge}

\input{NAHSFT}

\input{NAHL}

\input{twomodels}

\input{gaugebreaking}

\input{concluding}


\appendix

\input{appa}

\input{appb}

\input{appc}


\end{document}

%% file: abstract.tex
\begin{abstract}
Gauge symmetries and Higgs mechanisms are key features of theories
describing high-energy particle physics and collective phenomena in
statistical and condensed-matter physics. In this review we address
the collective behavior of systems of multicomponent scalar fields
interacting with gauge fields, which can be already present in the
underlying microscopic system or emerge only at criticality.  The
interplay between local gauge and global symmetries determines the
phase diagram, the nature of the Higgs phases, and the nature of phase
transitions between the high-temperature disordered and the
low-temperature Higgs phases.  However, additional crucial features
determine the universal properties of the critical behavior at
continuous transitions. Specifically, their nature also depends on the
role played by the gauge modes at criticality. Effective (Abelian or
non-Abelian) gauge Higgs field theories emerge when gauge modes
develop critical correlations. On the other hand, a more standard
critical behavior, which admits an effective description in terms of
Landau-Ginzburg-Wilson $\Phi^4$ theories, occurs when gauge-field
modes are short ranged at the transition.  In the latter case, gauge
fields only prevent non-gauge invariant correlation functions from
becoming critical.  This review covers the recent progress made in the
study of Higgs systems with Abelian and non-Abelian gauge fields.  We
discuss the equilibrium thermodynamic properties of systems with a
classical partition function, focusing mainly on three-dimensional
systems, and only briefly discussing two-dimensional models.  However,
by using the quantum-to-classical mapping, the results on the critical
behavior for classical systems in $D=d+1$ dimensions can be extended
to quantum transitions in $d$ dimensions.
\end{abstract}


%% file: introduction.tex
\section{Plan of the review}
\label{firstsec}

\subsection{Introduction}
\label{intro}

Gauge symmetries and Higgs mechanisms are key features of theories
describing high-energy particle
physics~\cite{Weinberg-book,ZJ-book,MM-book,Rothe-book} and collective
phenomena in statistical and condensed-matter
physics~\cite{Landau-book,Wilson-83,Anderson-book,ZJ-book,ID-book1,ID-book2,
  Parisi-book,AM-book,Wen-book,Fradkin-book,
  Sachdev-book,ZCZW-book,Sachdev-book2}.  The large-scale properties
of three-dimensional (3D) gauge models and the nature of their phase
transitions are of interest in several physical contexts. For
instance, they are relevant for
superconductivity~\cite{GL-50,HLM-74,Herbut-book}, for topological
order and unconventional quantum
transitions~\cite{Sachdev-19,Senthil-23,Sachdev-book2}, and also in
high-energy physics, to address some aspects of the finite-temperature
electroweak and strong-interaction transitions (or, most likely, sharp
crossovers) occurring in the early universe~\cite{KT-book,BVS-06}, and
the quark-gluon and chiral transitions in hadronic
matter~\cite{GPY-81,PW-84}, presently studied in heavy-ion
collisions~\cite{STAR-05}.  In both high-energy and condensed-matter
contexts, it is crucial to have a solid understanding of the interplay
between global and gauge symmetries, and, in particular, of the role
that local gauge symmetries play in determining the phase structure of
a model, the nature of its different phases and of its thermal and
quantum transitions.

Many collective phenomena in condensed-matter physics are modelled by using
effective Abelian Higgs (AH) theories, which describe a
$d$-dimensional system of degenerate $N$-component scalar fields
minimally coupled with an Abelian U(1) gauge field.  We mention
transitions in superconductors, see, e.g.,
Refs.~\cite{HLM-74,Herbut-book}, and in quantum SU($N$)
antiferromagnets, see, e.g., Refs.~\cite{RS-90, Kaul-12, KS-12,
  BMK-13, NCSOS-15, WNMXS-17,Sachdev-19}, and some unconventional
quantum transitions characterized by the so-called deconfined quantum
criticality, see, e.g., Refs.~\cite{SBSVF-04,Sandvik-07, MK-08,
  JNCW-08, Sandvik-10, HSOMLWTK-13, CHDKPS-13, PDA-13}. The phase
structure and universal features of AH models have been extensively
studied, see, e.g., Refs.~\cite{HLM-74,Herbut-book,FS-79,DH-81,FMS-81,
  BF-83,FM-83,KK-85,KK-86,BN-86,BN-86b,BN-87,KKS-94,BFLLW-96,
  HT-96,FH-96,IKK-96,KKLP-98,OT-98, CN-99, HS-00, KNS-02,MHS-02,
  SSSNH-02,SSNHS-03,MZ-03,NRR-03, MV-04, NSSS-04, SSS-04, HW-05,
  CFIS-05, TIM-05, TIM-06, CIS-06, KPST-06, WBJS-08, MK-08, KS-08,
  ODHIM-09, CGTAB-09, BMK-13, BS-13, NCSOS-15, SP-15, WNMXS-17, FH-17,
  PV-19-CP, IZMHS-19, PV-19-AH3d, Sachdev-19,PV-20-mfcp,
  BPV-20-hcAH,BPV-21-ncAH, WB-21, BPV-22-mpf, BPV-22, BPV-23-chgf,
  Senthil-23, BPV-24-cAH, BPV-24-ncAH,RS-24}, paying particular attention to
the role of the gauge fields and of the related topological features,
like monopoles and Berry phases, which cannot be captured by effective
Landau-Ginzburg-Wilson (LGW) 
theories~\cite{Landau-book,WK-74,Wilson-83,ZJ-book,PV-02}
with gauge-invariant scalar order parameters,
see, e.g., Refs.~\cite{SBSVF-04, Sachdev-19, Senthil-23}.

Various lattice AH models have been considered, using both compact and
noncompact gauge fields, with the purpose of identifying the possible
universality classes of the continuous transitions occurring in the
presence of gauge invariance.  These models provide examples of
topological transitions, which are driven by extended charged
excitations with no local order parameter, and of transitions
characterized by a nontrivial interplay between long-range scalar
fluctuations and nonlocal topological gauge modes.  Most of the
literature focused so far on Abelian gauge models, but, recently,
these studies have been extended to non-Abelian Higgs (NAH) theories,
in which multicomponent scalar fields are coupled with non-Abelian
gauge fields, see, e.g.,
Refs.~\cite{Sachdev-19,SSST-19,BPV-19,BFPV-21-adj,BFPV-21-mpsun},
formally similar to those used in the Standard Model of electroweak
interactions~\cite{Weinberg-book,ZJ-book}, in which the interplay
between gauge and global symmetries becomes even more complex.

From the theoretical point of view, the existence of continuous phase
transitions that cannot be described by the standard gauge-invariant
LGW paradigm is related to the nonperturbative infrared behavior of
strongly coupled gauge theories. In particular, the existence
of infrared stable charged fixed points (FPs) of the
renormalization-group (RG) flow implies that  gauge-field
correlations can also be critical at some continuous 
transition. At a fundamental level, the existence of these
nontrivial transitions also provides a way to define gauge field
theories in a nonperturbative framework.

Over the past decade, researchers with different backgrounds, working
in high-energy, statistical, and condensed-matter physics, have made
important steps forward in the theoretical understanding of the
nonperturbative physics of Abelian and non-Abelian gauge Higgs models
in less than four dimensions.  In light of this steady progress, a
survey of the field seems to be useful for two different reasons:
First, one can present the state of the art in the field; second,
there is the possibility to collect and discuss in a unified fashion
results that appeared in different contexts. This will also promote a
more efficient communication and interdisciplinary interactions among
researchers working in different areas.

To the best of our knowledge, a review on lower-dimensional, i.e.,
three-dimensional (3D) and two-dimensional (2D), AH and NAH theories
is still missing, and we would like to fill such gap. Of course, some
of the aspects we plan to cover have already been reviewed--- emerging
gauge fields and deconfined criticality are obvious examples, see,
e.g., Refs.~\cite{Sachdev-19,Senthil-23}. However, the existing
reviews focus on specific issues and do not report a general
discussion of lower-dimensional lattice AH and NAH theories.  In this
review we thoroughly examine the phenomenology of AH and NAH systems,
in particular their phase diagram and the nature of the critical
transitions observed when varying the gauge and global symmetries.

\subsection{Plan}
\label{plan}

This review focuses on lower-dimensional AH and NAH theories, in which
multicomponent scalar fields are coupled with Abelian and non-Abelian
gauge fields. We mostly discuss these models in a statistical lattice
gauge
framework~\cite{Wilson-74,ZJ-book,Parisi-book,MM-book,Rothe-book}, in
which the partition function is defined as a sum over classical
lattice configurations, thus providing a regularization of the
functional path integral and, possibly, paving the way for a
nonperturbative definition of the corresponding statistical field
theory. The quantum-to-classical mapping allows us to apply the
classical results for the critical behavior (equivalently, continuum
limit) in $(d+1)$ dimensions to quantum transitions of many-body
systems in $d$ dimensions, see, e.g.,
Refs.~\cite{Sachdev-book,RV-21}. In particular, since we consider
isotropic $(d+1)$-dimensional statistical models, the results can be
applied to the large class of continuous quantum transitions
characterized by a dynamic exponent $z=1$, i.e., to transitions in
which the critical gap decays as the inverse of the spatial
size~\cite{Sachdev-book,RV-21}. We mainly consider 3D classical
systems, but we also briefly discuss 2D systems.

The review covers the recent developments in the studies of lattice AH
and NAH systems (LAH and LNAH, respectively). We discuss their phase
diagram, the main features of their Higgs ordered
phases~\cite{Anderson-62,SSBgauge1,SSBgauge2,SSBgauge3}, where gauge
correlations are gapped and nonlocal charged operators may condense,
and the nature of their phase transitions, paying particular attention
to transitions characterized by a nontrivial interplay of gauge and
scalar modes.  We will show that the standard effective LGW $\Phi^4$
field theory approach~\cite{WK-74,ZJ-book,PV-02}, which is commonly
applied to critical phenomena in statistical systems without gauge
symmetries, is not always able to describe transitions in gauge
models. In some cases different effective approaches are needed to
correctly capture the nature of the critical behavior.  We pay
particular attention to the comparison with the predictions of the
corresponding AH and NAH field theories (AHFTs and NAHFTs,
respectively), to understand when, and how, the critical behavior of
lattice systems is effectively described by a corresponding
field-theory model.  If this is the case, the lattice model provides a
way, the so-called continuum limit, to define nonperturbatively the
field theory.

In Secs.~\ref{LGWfra} and \ref{gaugesymtr} we begin by discussing the
possible types of critical behavior that emerge at continuous
transitions in the presence of gauge symmetries.  This requires an
extension of the standard LGW paradigm, in particular when gauge modes
play an active role at criticality.  Secs.~\ref{AHSFT}-\ref{LAHMc}
focus on systems characterized by an Abelian U(1) gauge invariance,
i.e., the AHFTs and the corresponding LAH models. We consider models
with both noncompact and compact formulations of the gauge fields,
which turn out to present significantly different features. Lattice
models with discrete Abelian gauge symmetries are addressed in
Sec.~\ref{discgauge}.  Field theories and lattice systems with
non-Abelian gauge symmetries are discussed in Secs.~\ref{NAHSFT} and
\ref{NAHLM}, respectively. Two-dimensional models, with Abelian and
non-Abelian gauge symmetries, are discussed in
Sec.~\ref{2dmod}. Finally, in Sec.~\ref{gaugebreaking} we discuss the
effects of perturbations explicitly breaking the gauge symmetry, an
issue which is particularly relevant for systems in which the gauge
symmetry effectively emerges at the transition but which are not
gauge invariant at the microscopic level. A more detailed description
of the content of the review follows.

\begin{itemize}

\item[$\bullet$] In Sec.~\ref{LGWfra} we introduce the main ideas at
  the basis of the modern understanding of phase transitions and
  critical phenomena in statistical systems. In particular, we discuss
  the LGW approach in which transitions have an effective description
  in terms of $\Phi^4$ field theories. Their RG flows, and in
  particular the corresponding stable FPs, characterize the universal
  features of the critical behavior of systems without gauge
  symmetries.  We also discuss some notable examples, in which the LGW
  field-theoretical approach is applied.

\item[$\bullet$] In Sec.~\ref{gaugesymtr}, we discuss the key features
  of the critical behavior of statistical systems with local gauge
  symmetries.  We argue that different types of critical behaviors may
  be realized.  There are transitions driven by scalar fields whose
  nature depends on the role played by the gauge modes, in particular
  whether they develop or not critical correlations, and transitions
  driven by topological gauge modes without local order parameters.
  The effective description of these different critical behaviors
  requires different effective approaches, such as the standard
  gauge-invariant LGW approach when gauge fields are not critical, or
  the more complex gauge field theory (GFT) approach with explicit
  gauge fields in the presence of critical gauge correlations.  This
  distinction is crucial to understand which statistical field theory
  (SFT) is actually realized by the various continuous phase
  transitions occurring in the presence of gauge symmetries.

\item[$\bullet$] In Sec.~\ref{AHSFT} we focus on AHFTs, or scalar
  quantum electrodynamics (QED), which describe $d$-dimensional
  systems of degenerate $N$-component scalar fields minimally coupled
  with Abelian U(1) gauge fields. In particular, we consider systems
  with global SU($N$) and SO($N$) symmetry. We 
  review the main features of their RG flow, whose stable charged FP
  is expected to describe the critical behavior at {\em charged}
  continuous transitions, when both scalar and gauge fields are
  critical.

\item[$\bullet$] In Sec.~\ref{LAHMnc} we analyze lattice models
  corresponding to the AHFTs discussed in Sec.~\ref{AHSFT}, which are
  characterized by the same global and local symmetries.  We consider
  3D LAH models in which a noncompact Abelian gauge field is coupled
  with an $N$-component scalar field, providing a straightforward
  lattice discretization of the AHFT.  We review results for their
  phase diagrams, which present various phases, including a Higgs
  phase, and for their critical behavior along the various transition
  lines.  Different critical behaviors occur, which can be described
  by using the different approaches outlined in
  Sec.~\ref{gaugesymtr}. In particular, the transitions separating the
  Coulomb and Higgs phases realize the critical continuum limit
  associated with the RG flow of the AHFTs for a sufficiently large
  number of scalar components.
  
\item[$\bullet$] In Sec.~\ref{LAHMc} we consider an alternative
  lattice discretization of the AHFT, based on a compact formulation
  of the gauge variables. Its phase diagram and transitions show
  notable differences with respect to those of the noncompact
  formulation considered in Sec.~\ref{LAHMnc}.  In particular, the
  phase diagram also depends on the charge of the scalar field. These
  LAH systems present transition lines in which the gauge modes play
  different roles and require different effective descriptions.  The
  relation with the AHFT turns out to be less straightforward. The
  differences with the noncompact LAH models, discussed in
  Sec.~\ref{LAHMnc}, may be related to the existence of topological
  objects. We close this section by discussing the nontrivial relation
  between the LAH models with noncompact gauge fields and the compact
  LAH models with higher-charge scalar fields.
  
\item[$\bullet$] In Sec.~\ref{discgauge} we discuss lattice Higgs
  systems characterized by discrete Abelian gauge symmetries.  We
  mainly discuss the 3D lattice ${\mathbb Z}_2$-gauge $N$-vector
  models, obtained by minimally coupling $N$-component real variables
  with ${\mathbb Z}_2$-gauge variables.  They are paradigmatic models
  with different phases characterized by the spontaneous breaking of
  the global O($N$) symmetry and by the different topological
  properties of the ${\mathbb Z}_2$-gauge correlations.  In
  particular, the $N=1$ model corresponds to the ${\mathbb Z}_2$-gauge
  Higgs model, which is an interesting model showing (self-)duality
  and a nontrivial multicritical behavior. Other models with discrete
  gauge symmetries are also briefly discussed.

\item[$\bullet$] In Sec.~\ref{NAHSFT} we review NAHFTs, in which
  multiflavor scalar fields are coupled with non-Abelian gauge
  fields. We discuss models with SU($N_c$) and SO($N_c$) gauge
  symmetries, with scalar fields transforming in different
  representations of the gauge group, such as the fundamental and
  adjoint representations. We outline the known features of their RG
  flow, whose stable infrared FPs are expected to describe the
  critical behavior at some continuous phase transitions in the 
  corresponding lattice statistical models.

\item[$\bullet$] In Sec.~\ref{NAHLM} we consider LNAH models, which
  are lattice non-Abelian gauge theories with multicomponent
  degenerate scalar fields. We present results for their phase
  diagram, paying particular attention to their low-temperature Higgs
  phases.  We discuss the interplay between non-Abelian gauge
  symmetries and global symmetries, which crucially determines the
  properties of the various Higgs phases, and therefore the nature of
  the phase transitions between the disordered and the Higgs
  phases. In particular we consider SU($N_c$) gauge theories with
  multicomponent scalar fields in the fundamental and adjoint
  representations of the gauge group, two choices which lead to
  different Higgs phases. Some of the transitions between the
  disordered and the Higgs phases are continuous and can be associated
  with the charged FP of the corresponding NAHFT.  We close this
  section by briefly discussing SO($N_c$) gauge theories with
  multiflavor scalar matter.
  
\item[$\bullet$] In Sec.~\ref{2dmod} we review results for the
  critical continuum limit of 2D LAH and LNAH models with continuous
  global symmetries. They show that also the critical continuum limit
  of 2D LAH and LNAH models in the zero-temperature limit arises from
  a nontrivial interplay between gauge and global symmetries. The
  emerging results lead us to conjecture that the zero-temperature
  critical behavior, and therefore the continuum limit, of 2D lattice
  gauge models with scalar fields belongs to the zero-temperature
  universality classes associated with 2D field theories ($\sigma$
  models) defined on symmetric spaces.

\item[$\bullet$] In Sec.~\ref{gaugebreaking} we discuss the effects of
  perturbations that give rise to an explicit gauge-symmetry breaking
  (GSB) in lattice gauge theories. This issue is discussed in LAH
  models with noncompact and compact gauge fields in the presence of
  photon-mass terms. We argue that their effect is irrelevant at the
  continuous transitions where gauge correlations are not critical,
  and the critical behavior can be described by effective LGW theories
  with gauge-invariant order-parameter fields. On the other hand,
  photon-mass terms are expected to be relevant at charged transitions
  where the gauge modes develop critical correlations.
    
\end{itemize}  
  
Finally, in Sec.~\ref{conclu} we report some concluding remarks and
outlook.  To make this review self-contained some appendices have been
added, with the aim of summarizing a few related topics required to
fully appreciate some details of the methodology adopted in the main
text.

\begin{itemize}
   
\item[$\bullet$] In \ref{rgtheory} we report an overview of the
  homogeneous scaling laws expected at generic continuous phase
  transitions, as inferred from the RG theory of critical phenomena.
  We report the RG scaling relations valid in the thermodynamic limit
  and in the finite-size scaling (FSS) limit, at critical and
  multicritical points. We also briefly discuss FSS at first-order
  transitions.
    
\item[$\bullet$] For reference, in \ref{univclass} we report the
  presently most accurate estimates of the critical exponents of the
  3D Ising, XY, and O(3) vector (Heisenberg) universality classes,
  which are often referred to in this review.

\item[$\bullet$] In \ref{meanfieldHiggs} we present some details on
  the mean-field analyses of the Higgs phases of LNAH models with
  SU($N_c$) gauge symmetry and $N_f$ flavors in the fundamental
  representation.
    
 \end{itemize}

We limit our review to low-dimensional Abelian and non-Abelian gauge
Higgs models, in particular to three-dimensional models.  We do not
discuss four-dimensional (4D) Abelian and non-Abelian gauge Higgs
models, which are particularly relevant for high-energy physics and
are generally investigated by using perturbative approaches.  Beside
the extensive studies of the Standard Model of the electroweak
interactions~\cite{Weinberg-book}, based on a SU(2)$\otimes$U(1) gauge
model with a doublet of scalar fields, gauge theories with scalar
fields have been investigated as possible extensions beyond the
Standard Model, for example in the context of the so-called
technicolor or composite Higgs models, see, e.g., Refs.~\cite{FS-81,
  Kaul-81, Lane-00, Contino-10, CPS-20}. Note also that 4D gauge
theories with scalar fields are believed to be generally affected by
the so-called triviality problem, like the simpler 4D $\Phi^4$ field
theories~\cite{Callaway-88,ZJ-book,PV-02,MZ-03}. This implies the
impossibility of (nonperturbatively) constructing a corresponding 4D
field theory that is consistent (in the sense of satisfying all usual
physical requirements) on all energy scales for nonzero coupling (the
renormalized coupling goes to zero logarithmically when the
ultraviolet cutoff is removed, so the 4D field theory turns out to
eventually describe free fields). Note, however, that in the framework
of effective field theories~\cite{Weinberg-book} triviality is not a
relevant issue.

Another important issue that is not addressed in this review concerns
the new features that may arise when fermionic fields are added to
lattice gauge Higgs models.  This topic is quite interesting, since
models with emerging gauge symmetries in condensed-matter physics
often involve fermionic excitations beside scalar modes.  We believe
that this issue may provide further interesting developments, that
would extend our present understanding of the phenomenology of the
transitions in the presence of gauge symmetries.  Recent developments,
mostly in field-theoretical frameworks, have been already reported
in the literature; e.g., in  the review~\cite{Sachdev-19} and, in
particular, in the recent book~\cite{Sachdev-book2}, which also 
discusses boson-fermion and fermion-fermion dualities.  However, 3D
lattice gauge Higgs models including fermionic fields have been much
less studied so far. Therefore, we prefer to restrict our review to
lattice gauge models with scalar matter only, for which there is
already a large amount of results that have not been surveyed yet.

To conclude, we may summarize the content of this review as follows.
It is meant to provide a comprehensive presentation of the state of the art of
the research on the nonperturbative behavior of low-dimensional (2D and 3D)
Abelian and non-Abelian Higgs theories.  For this purpose we present a survey
of results for gauge Higgs theories, focusing mainly on 2D and 3D lattice
realizations, in which the partition function is defined as a sum over
classical lattice configurations.  These models are particularly effective to
understand the nonperturbative properties of gauge-Higgs theories and the
universal features of their transitions.  We emphasize that their critical
behaviors allow one to obtain nonperturbative definitions of the statistical
field theories sharing the same global and local gauge symmetries, which can
also be investigated by standard field-theoretical approaches.

In the following we report a list of the abbreviations and symbols used
throughout the review (in alphabetic order):
\begin{table}[h]
\centering
\begin{tabular}{ll|ll}
  \hline\hline
2D & Two-dimensional                  & LGW & Landau-Ginzburg-Wilson \\
3D & Three-dimensional                & LAH & Lattice Abelian Higgs\\
4D & Four-dimensional                 & LNAH & Lattice non-Abelian Higgs \\
AH & Abelian Higgs                    & MS & Minimal subtraction \\
AHFT & Abelian Higgs field theory     & MC & Monte Carlo \\
BKT & Berezinskii-Kosterlitz-Thouless & MCP & Multicritical point \\
CFT & Conformal field theory          & NAH & Non-Abelian Higgs \\
$d$ & Space dimensionality        & NAHFT & Non-Abelian Higgs field theory \\
FSS & Finite-size scaling             & QCD   & Quantum chromodynamics \\
FP  & Fixed Point                     & QED   & Quantum electrodynamics \\
GFT & Gauge fied theory               & QFT   & Quantum field theory \\
GSB & Gauge symmetry breaking         & RG    & Renormalization-group \\
IXY & Inverted XY                     & SFT   & Statistical field theory \\
$L$ & Lattice size                    & $V$   & Volume \\
  \hline\hline
\end{tabular}
\end{table}

%% file: LGWapproach.tex
\section{Phase transitions, critical phenomena,
  and the Landau-Ginzburg-Wilson framework}
\label{LGWfra}

\subsection{Phase transitions and critical phenomena}
\label{phatrcriphe}

Classical and quantum phase transitions are phenomena of great interest in
modern physics, both theoretically and experimentally. Classical phase
transitions are generally driven by thermal
fluctuations~\cite{Landau-book,Fisher-65,WK-74,Fisher-74,Ma-book,
Wilson-83,Binder-87,ZJ-book,PV-02,Fisher-16,PV-24}, while their quantum
counterparts arise from quantum fluctuations in the zero-temperature
limit~\cite{SGCS-97,Sachdev-book,RV-21,PV-24}.  Phase transitions separate
different phases characterized by distinctive properties.  In many cases they
are associated with the spontaneous breaking of a global symmetry, due to the
condensation of an order parameter and the emergence of a corresponding
ordering. Notable exceptions occur in the presence of gauge symmetries. In this
case the transition may be driven by topological
excitations~\cite{Wegner-71,Wen-book,Sachdev-16,Sachdev-19}, without a local
order parameter and without breaking any global symmetry.

Phase transitions occur when the free-energy density, or the
ground-state properties at quantum transitions, are singular in the
thermodynamic
limit~\cite{Ruelle-63-a,Ruelle-63-b,Fisher-64,Fisher-65,WK-74,
  Fisher-74, Ma-book, BGZ-76, Wegner-76, Wilson-83, FM-86, Binder-87,
  Fisher-98,ZJ-book,PV-02,SGCS-97,Sachdev-book,Fisher-16,RV-21,PV-24}.
Depending on the nature of the singularity, phase transitions are
generally distinguished as first-order or continuous phase
transitions.  They are continuous when the bulk properties change
continuously at the transition point and the correlation functions are
characterized by a length scale that diverges at criticality. They are
of first order when the thermodynamic or ground-state properties in
the infinite-volume (thermodynamic) limit are discontinuous across the
transition point.

The basic concepts underlying our understanding of phase transitions were
developed in a classical setting, in several seminal works by Kadanoff, Fisher,
Wilson, among others (see, e.g., Refs.~\cite{Kadanoff-66,Fisher-71, WK-74,
Fisher-74, Ma-book, BGZ-76, Wegner-76, Aharony-76, Wilson-83}).  Prototypical
$d$-dimensional continuous transitions are characterized by one relevant
parameter $r$, which is associated with a perturbation of the critical theory
that does not break the global symmetry.  It is defined so that the critical
point corresponds to $r=0$, the disordered phase to $r > 0$, and the ordered
phase to $r < 0$. At thermal transitions $r$ corresponds to the reduced
temperature, i.e., $r\sim T/T_c-1$, where $T_c$ is the critical temperature.
When approaching the critical point from the disordered phase, the length scale
$\xi$ of the critical modes diverges as $\xi\sim r^{-\nu}$, where $\nu$ is a
universal length-scale critical exponent that is related to the RG dimension
$y_r=1/\nu$ of the parameter $r$.  The description of the critical behavior at
a phase transition is generally supplemented with a second relevant parameter
$h$---e.g., an external homogeneous magnetic field in magnetic
systems---coupled with the order parameter whose condensation drives the
spontaneous breaking of the symmetry.  The parameter $h$ corresponds to a RG
perturbation that explicitly breaks the global symmetry.  This perturbation is
relevant and therefore the correlation length $\xi$ does not diverge when
$h\neq 0$. On the other hand, for $r=0$ the length scale diverges as $\xi \sim
|h|^{-1/y_h}$ when decreasing $|h|$, where $y_h$ is the RG dimension of the
parameter $h$.  Therefore, the critical behavior occurs only for $r=0$ and
$h=0$. This critical behavior is crucial to define the continuum limit of
statistical models, providing a nonperturbative realization of a corresponding
continuum SFT~\cite{Wilson-74,ZJ-book,Creutz-book,MM-book, Rothe-book}.

Paradigmatic examples of models undergoing finite-temperature
continuous transitions are the $N$-vector models, defined by the
Hamiltonian
\begin{equation}
H = - J \sum_{{\bm x},\mu} {\bm s}_{\bm x} \cdot {\bm s}_{{\bm
      x}+\hat{\mu}} - \sum_{\bm x} {\bm h} \cdot {\bm s}_{\bm x},\qquad {\bm
    s}_{\bm x}\cdot {\bm s}_{\bm x} = 1,
  \label{Nvectormod}
\end{equation}
on a cubic lattice, with the partition function $Z = \sum_{\{{\bm
    s}\}} e^{-H/T}$. The one-component ($N=1$) case corresponds to the
Ising model.

A summary of the most important scaling aspects of the RG theory of
critical phenomena at classical finite-temperature transitions, see,
e.g., Refs.~\cite{Fisher-71, WK-74, Fisher-74, Ma-book, BGZ-76,
  Wegner-76, Aharony-76, Wilson-83, Abraham-86, CJ-86, Cardy-editor,
  Privman-90, PHA-91, Fisher-98, ZJ-book, ID-book1, ID-book2,
  Cardy-book,PV-02}, is reported in \ref{rgtheory}.  The extension to
quantum systems is based on the quantum-to-classical mapping, which
allows us to map a quantum system defined in a spatial volume $V_s$
onto a classical one defined in a box of volume $V_c = V_s \times
L_T$, with $L_T=1/T$ (in the appropriate units)~\cite{SGCS-97,
  Sachdev-book, CPV-14, RV-21} and periodic or antiperiodic boundary
conditions for bosonic and fermionic excitations, respectively. Under
the quantum-to-classical mapping, the inverse temperature corresponds
to the system size in the imaginary-time direction.  Thus, the
universal zero-temperature critical behavior observed at a quantum
transition in $d$ dimensions is analogous to the behavior of a
corresponding $D$-dimensional classical system with
$D=d+1$.\footnote{It is important to remark that the
quantum-to-classical mapping does not generally lead to standard
classical isotropic systems. In some instances one obtains
complex-valued Boltzmann weights.  Moreover, the corresponding
classical systems are generally anisotropic.  If the dynamic exponent
$z$ that controls the scaling behavior of the energy
gap~\cite{Sachdev-book,RV-21} is equal to 1, as for Ising-like
quantum transitions, the anisotropy is weak, as in the classical Ising
model with direction-dependent couplings. In these cases, a
straightforward rescaling of the imaginary time allows one to recover
space-time rotationally invariant (relativistic) statistical field
theories. Therefore, the critical behavior in isotropic
$(d+1)$-dimensional classical statistical systems is directly related
with that in $d$-dimensional quantum systems only if the dynamic
exponent satisfies $z=1$. However, there are also interesting quantum
transitions with $z\not=1$, such as the superfluid-to-vacuum and Mott
transitions in lattice particle systems described by the Hubbard and
Bose-Hubbard models \cite{FWGF-89,Sachdev-book} (in this case, $z=2$
when the transitions are driven by the chemical potential).  For
continuous quantum transitions with $z \neq 1$, the anisotropy is
strong, i.e., correlations have different exponents in the spatial and
thermal directions.  Indeed, in the case of quantum systems of size
$L$, under a RG rescaling by a factor $b$ such that $\xi\to\xi/b$ and
$L\to L/b$, the additional length $L_T$ scales differently, as $L_T
\to L_T/b^z$. However, RG theory also applies to classical strongly
anisotropic systems~\cite{SZ-95}. An even stronger anisotropy arises
in first-order quantum transitions, which may actually qualitatively
depend on the spatial boundary conditions~\cite{PV-24}.}

In this review we discuss statistical systems undergoing classical
transitions. However, many results on the general features of the
transitions and the corresponding critical behavior can be extended to
quantum transitions in a lower dimension by means of the
quantum-to-classical mapping.

\subsection{The field-theoretical Landau-Ginzburg-Wilson 
  approach to continuous phase transitions}
\label{LGWappro}

The Landau paradigm~\cite{Landau-37a, Landau-37b} for continuous phase
transitions is based on the idea that a phase transition is driven by
the condensation of an order parameter, which characterizes the two
phases separated by the transition: it vanishes in the disordered
phase, up to the critical point, then it takes a nonzero value (in the
limit $h\to 0$), giving rise to the spontaneous breaking of the
symmetry.\footnote{Note, however, that the symmetry of the long-range
behavior of the correlation functions at criticality does not
necessarily coincide with the actual global symmetry of the model. In
some cases the critical modes show an effective enlargement of the
symmetry. If $G$ is the symmetry group of the model, the critical
behavior is invariant under a larger symmetry group $G'$, if the
operators responsible for the breaking $G'\to G$ are irrelevant at the
transition. For instance, this occurs in the 3D ${\mathbb Z}_Q$ clock
models~\cite{Savit-80}, which are invariant under ${\mathbb Z}_Q$
transformations. For $Q>4$ the critical behavior belongs to the XY
universality class with an enlarged U(1) global symmetry (see
Sec.~\ref{sec.ZQ}).}

In the Landau description of continuous phase transitions, extended to
incorporate the scaling and universality
hypotheses~\cite{Kadanoff-66,Kadanoff-66-b,Fisher-71,
  Wilson-71a,WK-74,Fisher-74}, the most important assumptions are the
following:

\begin{itemize}

\item[(i)] The existence of an order parameter, which effectively
  describes the critical modes and signals the spontaneous breaking of
  the global symmetry.

\item[(ii)] The scaling hypothesis, which assumes that 
  the singular part of the thermodynamic observables and 
  of the correlation functions of the order parameter 
  are scale invariant at criticality. Close to the critical point,
  they satisfy scaling relations
  when expressed in terms of appropriately defined scaling fields.

\item[(iii)] The universality of the critical behavior, which is
  only determined by a few global properties of the model, such as the space
  dimensionality, the nature and the symmetry of the order parameter,
  the symmetry-breaking pattern, and the range of the effective interactions.

\end{itemize}
  
The RG theory of critical phenomena~\cite{Wilson-71a, Wilson-71b,
  WK-74, Fisher-74} provides a general framework, in which these
properties naturally arise. It considers the RG flow in the
Hamiltonian-parameter space.  The critical behavior is associated with
a stable FP of the RG flow, where only a few RG perturbations are
relevant.  The corresponding positive eigenvalues of the linearized
theory around the FP are related to the critical exponents $\nu$,
$\eta$, etc.  In \ref{rgtheory} we briefly review the key points of
the RG theory, such as the homogeneous scaling laws for the
thermodynamic functions and for the correlation functions of the
order-parameter field.

In the RG framework, a quantitative description of many continuous
phase transitions can be obtained by considering effective LGW
$\Phi^4$ field theories. The fundamental field is a multicomponent
field ${\bm \Phi}({\bm x})$, which can be thought as obtained by
coarse-graining the order parameter.  The dynamics of ${\bm \Phi}({\bm
  x})$ is specified by the LGW Lagrangian, which is the most general
fourth-order polynomial in ${\bm \Phi}({\bm x})$, which is invariant
under the global symmetry transformations of the system.  Higher
powers of ${\bm \Phi}({\bm x})$ are supposed to be irrelevant (this is
an exact result sufficiently close to four dimensions, which is
assumed to be valid in three dimensions, too). The LGW theory predicts
the possible symmetry-breaking patterns at the transition.  The
corresponding critical behaviors are controlled by the FPs of the RG
flow in the space of the parameters of the quartic scalar potential,
which is determined by the corresponding $\beta$
functions~\cite{ZJ-book,PV-02,ID-book1}.

As already mentioned, the quantum-to-classical mapping allows us to
extend the classical results to the corresponding quantum transitions,
since $d$-dimensional quantum transitions are described by
$(d+1)$-dimensional statistical (quantum) field theories.

We now present more details for some specific LGW theories that 
will be particularly relevant in the following.

\subsubsection{O($N$)-symmetric $\Phi^4$ theory} \label{sec:ON}

The simplest LGW model is the O($N$)-symmetric $\Phi^4$ theory, defined
by the Lagrangian density
\begin{equation}
  {\cal L}_{{\rm O}(N)} = \partial_\mu {\bm \Phi} \cdot
  \partial_\mu {\bm \Phi} + r \, {\bm \Phi}
  \cdot {\bm \Phi} + u \big({\bm \Phi}\cdot
  {\bm \Phi}\big)^2,
\label{ONLGW}
\end{equation}
where ${\bm \Phi}({\bm x})$ is an $N$-component real field. It is
associated with the so-called $N$-vector universality class,
characterized by an $N$-component vector order parameter and by the
symmetry-breaking pattern O($N$)$\rightarrow$O($N-1$).  Several
continuous transitions belong to these universality classes, such as
the finite-temperature transitions in the $N$-vector models
with Hamiltonian (\ref{Nvectormod}). The Ising universality class corresponds to
$N=1$. It is relevant for the liquid-vapor transition in simple
fluids, for the Curie transition in uniaxial magnetic systems,
etc. The XY universality class ($N=2$) describes the superfluid
transition in $^4$He, the formation of Bose-Einstein condensates in
interacting bosonic gases, transitions in magnets with easy-plane
anisotropy and in superconductors.  The Heisenberg universality class
($N=3$) describes the Curie transition in isotropic magnets. The
O(4)-symmetric model is believed to be the effective critical theory
for the hadronic finite-temperature transition with two light quarks
in the chiral limit.  Moreover, the $N$-vector model for $N\rightarrow
0$ describes the behavior of dilute homopolymers in a good solvent, in
the limit of large polymerization. See, e.g., Refs.~\cite{ZJ-book,
  PV-02} for reviews on the O($N$)-symmetric $\Phi^4$ field
theories. In \ref{univclass} we report accurate estimates of
the critical exponents for the 3D Ising, XY, and Heisenberg 
universality classes.
O($N$) models have been studied with a variety of methods.
Critical properties have been determined by using
perturbative and nonperturbative methods applied to the field-theoretical
$\Phi^4$ model and numerical methods applied to a variety of lattice
O($N$)-invariant scalar models; see, e.g.,
Refs.~\cite{BGZ-76,BNGM-76,GZ-98,CPRV-02,PV-02,DBN-05,HPV-05,
BMPS-06,CHPV-06,Clisby-10,Hasenbusch-10,HV-11,CD-16,KPSV-16,
Clisby-17,KP-17,FXL-18,
Hasenbusch-19,DBTW-20,CLLPSSV-20,CLLPSSV-20-o3,
Hasenbusch-20,Shalaby-21,Hasenbusch-21,Hasenbusch-23,ZHHHH-23,HHZ-23,
Schnetz-23,A-etal-24, LHWR-25}. 

\subsubsection{U($N$)-symmetric $\Phi^4$ theory} \label{sec:UN}

We now discuss a straightforward generalization of the O($N$) $\Phi^4$
theory, appropriate to describe transitions in U($N$)-symmetric
models driven by a complex $N$-dimensional vector order parameter.
The corresponding LGW theory is defined in terms of a complex
$N$-component vector field ${\bm \Psi}({\bm x})$.  The most general
U($N$)-symmetric LGW Lagrangian is
\begin{equation}
  {\cal L}_{{\rm U}(N)} = \partial_\mu \bar{\bm \Psi} \cdot
  \partial_\mu {\bm \Psi} + r \, \bar{\bm \Psi} \cdot {\bm \Psi} + u
  \big(\bar{\bm \Psi}\cdot {\bm \Psi}\big)^2.
  \label{UNLGW}
\end{equation}
A remarkable property of the U($N$)-invariant LGW Lagrangian is that
it is symmetric under a larger O($2N$) symmetry group.  This can be
easily seen by noting that the Lagrangian (\ref{UNLGW}) is equivalent
to that of the O($2N$)-vector models, reported in Eq.~(\ref{ONLGW}), if the
complex field component $\Psi^a$ is rewritten as $\Psi^a = \Phi^a + i
\Phi^{a+N}$, where $\Phi^a$ is a real $(2N)$-component vector.  As one
can easily check, U($N$)-symmetric terms that break the O($2N$)
symmetry require higher powers of the field ${\bm \Psi}$.  Indeed, the
lowest-dimensional scalar operator breaking the O($2N$) symmetry is
the dimension-six operator $\sum_\mu |\bar{\bm \Psi} \cdot
\partial_\mu {\bm \Psi}|^2$.

Therefore, the LGW theory predicts that the critical behavior of
U($N$)-symmetric systems, at phase transitions driven by the
condensation of a complex $N$-component order parameter, belong to the
O($2N$) vector universality class, which implies an effective
enlargement of the symmetry of the critical modes, from U($N$) to
O($2N$). The U($N$)-symmetric RG perturbations breaking the O($2N$)
symmetry are irrelevant in the critical limit, giving only rise to
suppressed scaling corrections.

Note however that, although the critical modes show an effective
enlarged O($2N$) symmetry, and therefore they are associated with a
O($2N$)$\to$O($2N-1$) symmetry-breaking pattern, there is no
enlargement of the symmetry in the low-temperature phase, which is
only invariant under U(1)$\otimes$U($N-1$) transformations.  This is
crucial to determine the properties of the gapless Goldstone modes
characterizing the low-temperature phase. In particular, their number
is significantly smaller than that associated with the spontaneous
symmetry breaking O($2N$)$\to$O($2N-1$), see, e.g.,
Ref.~\cite{ZJ-book}.

\subsubsection{Generic Landau-Ginzburg-Wilson $\Phi^4$
  theory with a single quadratic invariant}
\label{gensinqua}

Beside O($N$) and U($N$) invariant models, there are physically
interesting transitions that are described by LGW $\Phi^4$ field
theories characterized by more complex symmetries and
symmetry-breaking patterns, see, e.g.,
Refs.~\cite{BLZ-74,Aharony-76,ZJ-book,PV-02,VZ-06,Vicari-07}.  The
most general LGW Lagrangian for an $N$-component order-parameter field
$\Phi_i({\bm x})$ (with $i=1,2,...N$) can be written as
\begin{equation}
{\cal L}_{\rm LGW} =  \sum_i (\partial_\mu \Phi_{i})^2 + 
\sum_i r_i \Phi_{i}^2  + 
\sum_{ijkl} u_{ijkl} \; \Phi_i\Phi_j\Phi_k\Phi_l.
\label{genH}
\end{equation}
The number of independent quadratic $r_i$ and quartic $u_{ijkl}$
parameters crucially depends on the symmetry group of the theory. An
interesting class of models are those in which $\sum_i \Phi^2_i$ is
the unique quadratic term that is invariant under the symmetry group
of the theory~\cite{BLZ-74,ZJ-book}. In this case, all $r_i$ are
equal, $r_i = r$, and $u_{ijkl}$ must satisfy the trace
condition~\cite{BLZ-74} $\sum_i u_{iikl} \propto \delta_{kl}$, to
avoid the generation of other quadratic invariant terms under RG
transformations.  In these models, criticality is driven by tuning the
single parameter $r$, which may correspond to the reduced temperature.
Therefore, these cases correspond to the prototypical transitions
outlined in Sec.~\ref{phatrcriphe}.  All field components become
critical simultaneously and the two-point function in the disordered
phase is diagonal, that is, $G_{ij}({\bm x},{\bm y}) \equiv \langle
\Phi_i({\bm x}) \Phi_j({\bm y})\rangle = \delta_{ij} G({\bm x},{\bm
  y})$.  LGW theories with multiparameter quartic terms provide
effective descriptions of complex phase transitions in several
physical contexts~\cite{PV-02}. We mention anisotropic magnets (see,
e.g., Refs.~\cite{Aharony-76, CPV-00}), disordered systems (see, e.g.,
Refs.~\cite{Harris-74,PV-00,Aharony-75,CPV-04}), frustrated systems
(see, e.g., Refs.~\cite{Kawamura-98, PRV-01-a,PRV-01, CPS-02, DMT-04,
  CPPV-04}), spin and density wave models (see, e.g.,
Refs.~\cite{ZDS-02, Sachdev-03, DPV-04, DPV-06, PSV-08, Kim-etal-08}),
LAH models in the strong gauge-coupling limit (see, e.g.,
Refs.~\cite{PTV-17,PTV-18,PV-19-AH3d}), multiflavor scalar
chromodynamics~\cite{BPV-19,BPV-20-sun}, and also the
finite-temperature chiral transition in hadronic matter with $N_f$
light flavors~\cite{PW-84,BPV-03,PV-13}.

The RG flow in $\Phi^4$ theories, such as those with Lagrangian
density (\ref{genH}), has generally several FPs. Among them, the
stable FP controls the critical behavior of the corresponding
continuous transitions, and therefore can be associated with their
universality class.  If there is no stable FP, the transitions in the
corresponding class of models are generally expected to be of first
order, unless the considered field-theoretical model misses some
relevant features of the transitions under investigation.

\subsubsection{Multicritical O($N_1$)$\oplus$O($N_2$) $\Phi^4$ theories}
\label{sec:multi}

\begin{figure}
\begin{center}
    \includegraphics[width=0.3\columnwidth]{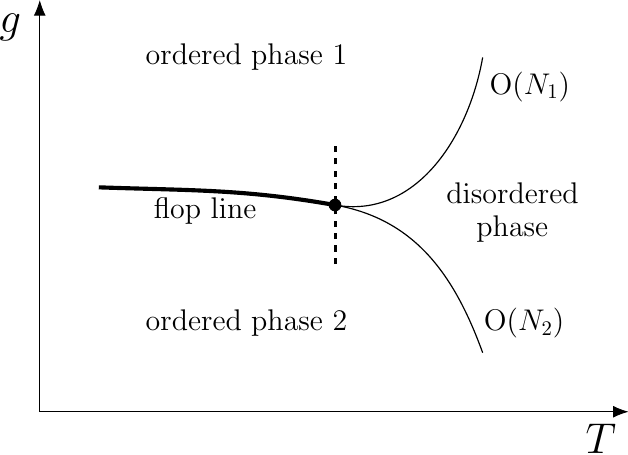}
    \hspace*{3mm}
    \includegraphics[width=0.3\columnwidth]{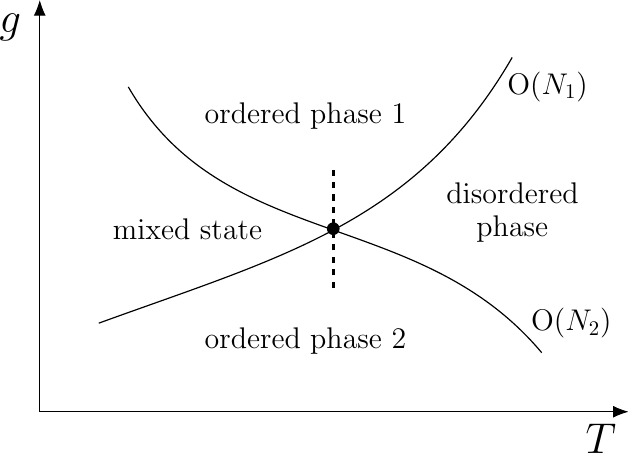}
        \hspace*{3mm}
    \includegraphics[width=0.3\columnwidth]{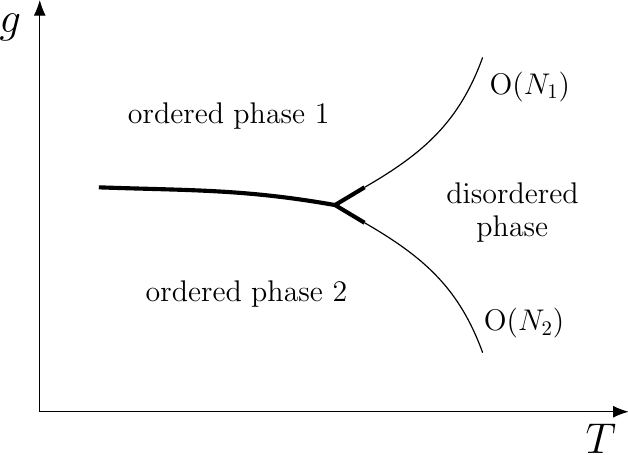}
    \caption{ Possible phase diagrams in the $T$-$g$ plane for the
      O$(N_1$)$\oplus$O$(N_2$) LGW theory. The critical lines
      intersect at a bicritical point (left), a tetracritical point
      (center), and at a first-order transition point (right), located
      at $(T_{\rm mc},g_{\rm mc})$.  Here, $T$ is the temperature and
      $g$ is a second relevant parameter.  The thick line (``flop
      line") represents first-order transitions.  A bicritical
      behavior (left) is characterized by the presence of a
      first-order line that starts at the MCP and separates the two
      different ordered low-temperature phases.  In the tetracritical
      case (center), there exists a mixed low-temperature phase in
      which both types of ordering coexist and which is bounded by two
      critical lines meeting at the tetracritical point. The dashed
      lines in the two leftmost figures show trajectories at fixed
      $T=T_{\rm mc}$, going from one ordered phase to the other one
      across a continuous transition at the MCP.  It is also possible
      that the transition at the meeting point is of first order
      (right).  In this case the two first-order lines, which start at
      the intersection point and separate the disordered phase from
      the ordered phases, end in tricritical points and then continue
      as critical lines. See
      Refs.~\cite{FN-74,NKF-74,KNF-76,PV-02,CPV-03,BPV-22-z2g} for
      studies of the RG flow and scaling behavior at MCPs, and also
      \ref{fssmcp} for a brief overview.  }
\label{multicri}
\end{center}
\end{figure}

LGW theories (\ref{genH}) in which the symmetry allows several
quadratic terms are also relevant phenomenologically. In particular,
they provide effective field-theoretical models describing the
multicritical behaviors that occur in the presence of competing
orderings~\cite{FN-74, NKF-74, KNF-76}. The main features of the
critical behavior at multicritical points (MCPs) are discussed in
\ref{fssmcp}.

As a paradigmatic example, one may consider multicritical LGW theories
that describe the competition of order parameters associated with two
different global O($N_1$) and O($N_2$) symmetries.  They can be
obtained by constructing the most general
O($N_1$)$\oplus$O($N_2$)-symmetric $\Phi^4$ theory with two
real vector fields ${\bm \phi}_1({\bm x})$ and
  ${\bm \phi}_2({\bm x})$,
with $N_1$ and $N_2$ components, respectively.  Their Lagrangian
density reads~\cite{FN-74,NKF-74,KNF-76,PV-02,CPV-03}
\begin{eqnarray}
  {\cal L}_{\rm mc} = \sum_{a=1}^2 \Bigl[\partial_\mu {\bm \phi}_a\cdot
    \partial_\mu {\bm \phi}_a + r_a \, {\bm \phi}_a\cdot{\bm \phi}_a
   + v_a\,({\bm \phi}_a\cdot{\bm \phi}_a)^2\Bigr]
+ w \,{\bm \phi}_1\cdot{\bm \phi}_1 \, {\bm \phi}_2\cdot{\bm \phi}_2.
\label{on1on2} 
\end{eqnarray}
The multicritical theory is characterized by two relevant quadratic
operators, those proportional to $r_1$ and $r_2$ in the Lagrangian
(\ref{on1on2}). Therefore, in the microscopic model and in the absence
of external fields breaking the global O($N_1$)$\oplus$O($N_2$)
symmetry, there are two relevant parameters which should be tuned to
approach the MCP: beside the temperature $T$, there is a second
relevant parameter that is referred to as $g$ in
Fig.~\ref{multicri}. The parameters $r_1$ and $r_2$ will be in general
linear combinations of $T-T_{\rm mc}$ and of $g-g_{\rm mc}$, where
$T_{\rm mc}$ and $g_{\rm mc}$ are the values of $T$ and $g$ at the
MCP.

The phase diagram of the model (\ref{on1on2}) shows different
disordered and ordered phases, separated by transition lines meeting
in one point. A standard mean-field analysis of the model shows that,
by varying the quartic coupligs $v_1$, $v_2$, and $w$, one can obtain
qualitatively different phase diagrams~\cite{FN-74, NKF-74, KNF-76},
as shown in Fig.~\ref{multicri}.  The multicritical behavior of models
associated with a multicritical LGW theory is again determined by the
RG flow in the space of the quartic parameters~\cite{FN-74, NKF-74,
  KNF-76,PV-02,CPV-03,PV-05-mc,HPV-05,BPV-22-z2g,KAE-23,Aharony-24}.
A multicritical universality class is generally associated with
a stable FP of such RG flow.

This issue has been addressed by using perturbative field-theoretical
methods~\cite{FN-74,KNF-76,CPV-03,KAE-23}, for example by computing
and analyzing high-order $\varepsilon$-expansion series~\cite{CPV-03},
and numerical methods, see, e.g., Ref.~\cite{HPV-05,BPV-22-z2g}. The
predicted behavior crucially depends on the number $N_1$ and $N_2$ of
components of the competing order parameters.  The RG flow of the
multicritical O($N_1$)$\oplus$O($N_2$) LGW theories for $N_1+N_2 \ge
4$, see, e.g.,
Refs.~\cite{FN-74,NKF-74,KNF-76,PV-02,CPV-03,HPV-05,KAE-23,Aharony-24},
shows that continuous MCPs are tetracritical (see the 
central panel of Fig.~\ref{multicri}), because the only stable FP is
associated with a decoupled O($N_1$) and O($N_2$) critical
behavior. Therefore, when the phase diagram is characterized by three
transition lines meeting in one point, the competition of the O($N_1$)
and O($N_2$) order parameters gives generally rise to first-order
transitions close to the meeting point, as shown in the right panel of
Fig.~\ref{multicri} (unless an additional tuning of the model parameters
is performed). On the other hand, the competition of two Ising order
parameters---$N_1=N_2=1$ in this case---(see also
Sec.~\ref{mcxybeh}) can give rise to a phase diagram such as that 
shown in the left panel of Fig.~\ref{multicri}. In this case 
the multicritical 
behavior is controlled by the XY FP, thus,
the global symmetry of the multicritical modes at the MCP enlarges 
from ${\mathbb Z}_2\oplus{\mathbb Z}_2$ to O(2).
Finally, for $N_1+N_2=3$
the stable FP of the 3D RG flow is the so-called biconical FP, which
is believed to be associated with a tetracritical
point~\cite{FN-74,NKF-74,KNF-76,CPV-03,KAE-23,Aharony-24}. No
stable FP corresponding to a bicritical point (left panel of
Fig.~\ref{multicri}) exists, although the corresponding bicritical
O(3)-symmetric FP turns out to be very weakly
unstable~\cite{CPV-03,KAE-23}.

We finally remark that a continuous MCP separates different ordered
phases, see Fig.~\ref{multicri}.  For example, assume that $g$ is
increased for fixed $T=T_{\rm mc}$ (see the dashed lines reported in
the leftmost and central panel of Fig.~\ref{multicri}).  The system is
at the beginning in the low-$g$ ordered phase, then undergoes a
continuous transition at the MCP, and finally is in the second ordered
phase.  Therefore, multicritical LGW theories can describe continuous
transitions between phases characterized by different orderings.
Note, however, that the multicritical behavior requires the tuning of
an additional parameter. This observation clarifies when the Landau
paradigm--- there are no continuous transitions between phases with
different symmetry---holds.  Indeed, while it holds for generic
transitions, it may be violated by an appropriate tuning of the
parameters.

%% file: FTapproachestoLGT.tex
\section{Phase transitions and
  critical phenomena in the presence of gauge symmetries}
\label{gaugesymtr}

The description of the main features of phase transitions in the
presence of gauge symmetries is more complex than that outlined in the
previous Section. First of all, while conventional transitions are
always related to the condensation of a local order parameter, in
the presence of gauge fields, we may have transitions without local
order parameter (topological transitions). When an order parameter is
present, we also need to distinguish its behavior under gauge
transformations: the order parameter may be gauge invariant as well as
not gauge invariant. Moreover, the role of gauge fields may differ.
In some transitions, gauge fields are only relevant to define the set
of critical modes (the gauge-invariant ones), while in some other
cases, the phase transition is due to a nontrivial interplay between 
matter and gauge fields, so that gauge fields determine the critical
behavior and should be included in the corresponding effective field
theory (these transitions will be named charged transitions).

The effective description of the different critical behaviors
developed by systems with gauge symmetries requires different
effective approaches, such as the standard gauge-invariant LGW
approach when gauge fields are not critical, or the more complex gauge
field-theory approach with explicit gauge fields in the presence of
critical gauge correlations.  This distinction is crucial to
understand which statistical field theory is actually realized by the
various continuous phase transitions occurring in the presence of
gauge symmetries.

It is worth noting that the study of the critical behavior of
gauge-symmetric models is also relevant to understand the critical
behavior of unconventional transitions in condensed-matter physics, in
which a gauge symmetry is supposed to emerge at criticality, see, e.g.,
Refs.~~\cite{Anderson-book, Wen-book,Fradkin-book,
  ZCZW-book,HK-10,QZ-11,Sachdev-19,Senthil-23}.  Unconventional
quantum transitions~\cite{Xu-12, Sachdev-19, Senthil-23} are observed
between topological phases~\cite{GJWK-16,GHMS-11,DKOSV-11,HW-05}, in
the context of the integer and fractional quantum Hall
effect~\cite{Huckestein-95, HHSV-17, Giesbers-etal-09, OGE-12}. We
also mention transitions between topologically-ordered and
spin-ordered phases, between topologically-ordered and
valence-bond-solid (VBS) phases~\cite{RS-90}, and direct transitions
between differently ordered phases, such as the N\'eel-to-VBS
transition of 2D quantum magnets~\cite{SBSVF-04, SVBSF-04}.
{\em Deconfined} quantum transitions that exhibit fractionalization of
the low-energy excitations, topological order, and long-range
entanglement also admit an effective description in terms of
field theories with emergent gauge fields coupled with matter fields
carrying fractional quantum numbers of the microscopic global
symmetry, see, e.g., Refs.~\cite{Sachdev-19,Senthil-23} for reviews.
Studies of deconfined transitions are reported in, e.g.,
Refs.~\cite{SBSVF-04,SVBSF-04,KPST-06,TIM-06,Sandvik-07,Sandvik-10,
  KMPST-08,BDA-10,Bartosch-13,CHDKPS-13,HSOMLWTK-13,PDA-13,SP-15,NCSOS-15,
  WNMXS-17,IZMHS-19,MW-19,SN-19,SZ-20,ZTS-20,LXY-21,
  Cui-etal-23,SZJSM-23,DES-24,Senthil-23}, and references therein.

\subsection{Different  types of critical scenarios
  in statistical models with
   gauge symmetries}
\label{difftype}

As we have mentioned above, in the presence of gauge symmetries, the
effective LGW approach discussed in Sec.~\ref{LGWfra} cannot be
generically applied.  Other approaches should also be used to obtain
an effective description of the critical behavior. The choice of the
appropriate approach depends on the main features of the mechanism
driving the particular phase transition of the lattice gauge model
under study.

In general, four classes of transitions may be distinguished in
statistical systems with gauge symmetry, each one admitting a
different description:
\begin{itemize}

\item[$\bullet$] {\em LGW transitions}, in which only gauge-invariant
  correlations of the matter field are critical. They admit a
  conventional effective LGW field-theoretical description in terms of
  a gauge-invariant order-parameter field, which orders at the
  transition, driving the spontaneous breaking of the global symmetry.
  Gauge fields do not display critical correlations. The gauge
  symmetry only prevents nongauge-invariant observables from becoming
  critical.

\item[$\bullet$] {\em LGW$^{\,\times}$ transitions}, in which only
  matter fields develop critical correlations, similarly to LGW
  transitions.  They also admit a LGW description, but in this case
  the order-parameter field is not gauge invariant.\footnote{It 
  is important to note that the gauge symmetry can never be 
  broken~\cite{Elitzur-75, DFG-78, FMS-81, ID-book1}, despite
  the common use of referring to the Higgs mechanism as a spontaneous
  gauge symmetry breaking.  The statement that in some cases a
  nongauge invariant order parameter is needed does not refer to the
  description of the thermodynamic phases, but to the definition of
  the effective model which encodes the universal critical properties
  of the continuous phase transition. 
  In LGW$^{\times}$ transitions
  the critical behavior of gauge-invariant quantities is the same 
  as that of composite operators defined in terms of a nongauge invariant
  field, which is the fundamental field in 
  the effective LGW description. As it is not gauge invariant, 
  the order-parameter field can be identified in the gauge model 
  only if an appropriate gauge fixing condition is added. 
  }
  Therefore, to distinguish them from standard LGW transitions, we
  add the superscript $\times$.\footnote{Note that transitions with
  similar properties have been sometimes called LGW$^*$ transitions,
  see, e.g., Ref.~\cite{Senthil-23}.}  In particular, systems with discrete
  gauge groups undergo LGW$^{\times}$ transitions for sufficiently
  weak gauge couplings.
  
\item[$\bullet$] {\em GFT transitions}, in which the critical behavior
  is due to a nontrivial interplay of scalar and gauge critical excitations.
  Therefore, their description requires an
  effective gauge field theory (GFT), with dynamical gauge and matter
  fields.  GFT transitions are observed in the presence of
  continuous Abelian and non-Abelian gauge symmetries. They also occur
  in models that are locally invariant under discrete groups when the
  gauge symmetry enlarges at criticality, so that the critical modes
  are invariant under a continuous group.

\item[$\bullet$] {\em Topological transitions}, driven by topological
  modes, in which one cannot identify a local order-parameter field.
  A typical example is provided by the topological transition of the
  3D lattice ${\mathbb Z}_2$-gauge model~\cite{Wegner-71}.
\end{itemize}

We do not claim that the above classification is exhaustive, i.e.,
that all classical and quantum phase transitions in the presence of
gauge symmetries belong to one of these four classes.  However, the
above classes of transitions cover all classical (thermal) phase
transitions that we will discuss in this review, and the corresponding
quantum transitions related by the quantum-to-classical mapping.  In
the following subsections we discuss the four possibilities in more
detail, mentioning some examples. Other examples will be found in the
following sections.

\subsection{LGW transitions with a gauge-invariant
  order parameter}
\label{ginvLGW}

The first class of transitions, the LGW transitions, is characterized
by the fact that only gauge-invariant matter correlations are
critical. Gauge fields have apparently only the role of hindering some
scalar degrees of freedom, those that are not gauge invariant, from
becoming critical.  It seems thus natural to assume that the
transition is driven by a gauge-invariant order parameter and that
only gauge-invariant matter modes are relevant at
criticality. Although this assumption seems natural, it is important
to stress that it is not strictly necessary.  There are indeed
transitions (the LGW$^{\times}$ transitions that we will discuss in
Sec.~\ref{nonginvLGW}) that are analogous to the LGW transitions as
only matter fields are critical, but that require a scalar nongauge
invariant order parameter.  If the scalar order parameter is gauge
invariant, the approach is the conventional one. Once an order
parameter has been identified, one considers the corresponding
order-parameter field and LGW Lagrangian. The critical behavior is
then determined by studying the field-theoretical model as outlined in
Sec.~\ref{LGWappro}.  Note that the gauge fields do not appear in the
effective description: the gauge symmetry plays only a role in
selecting the order-parameter field, which should be gauge invariant.

In the next subsections we present some examples of LGW transitions,
which will be also useful for the remainder of the review.  Other examples
will be presented later in the review.

\subsubsection{Three-dimensional CP$^{N-1}$ models}
\label{cpnmodels}

The transition in the 3D lattice CP$^{N-1}$ model, and also the
transitions in LAH models (they are discussed in Secs.~\ref{LAHMnc}
and \ref{LAHMc}) in the strong gauge-coupling (small-$\kappa$) regime,
are examples of LGW transitions.  In the CP$^{N-1}$ field theory the
fundamental field is a complex $N$-component unit-length vector ${\bm
  z}({\bm x})$, associated with an element of the complex projective
manifold CP$^{N-1}$~\cite{ZJ-book}. The Lagrangian reads
\begin{eqnarray}
{\cal L}_{{\rm CP}} = {1\over 2 g}
  \overline{D_{\mu}{\bm z}}\cdot D_\mu {\bm z},
\qquad D_\mu =
  \partial_\mu + i A_\mu, \qquad A_\mu = i\bar{{\bm z}}\cdot
  \partial_\mu {\bm z},
\label{contham}
\end{eqnarray}
where $A_\mu$ is a composite real gauge field.  The model is 
characterized by a global U($N$) symmetry ${\bm
  z}({\bm x}) \to U {\bm z}({\bm x})$ with $U\in {\rm U}(N)$, and a
local U(1) gauge symmetry ${\bm z}({\bm x}) \to e^{i\Lambda({\bm x})}
{\bm z}({\bm x})$.\footnote{The CP$^1$ field theory is
equivalent  to the O(3) non-linear $\sigma$ model with the
identification $s^a=\sum_{ij} \bar{z}^i \sigma_{ij}^{a} z^j$, where
$a=1,2,3$ and $\sigma^{a}$ are the Pauli matrices. An analogous
equivalence applies to the corresponding $N=2$ lattice models
(\ref{hcpn}) and (\ref{altcpnmod}).\label{cp1f}} 

The corresponding simplest lattice formulation is
\begin{equation}
  Z = \sum_{\{{\bm z}\}} e^{-H_{\rm CP}/T},\qquad
  H_{{\rm CP}} = - J \sum_{{\bm x},\mu} | \bar{\bm{z}}_{\bm
  x} \cdot {\bm z}_{{\bm x}+\hat\mu} |^2, \qquad
\bar{\bm{z}}_{\bm x} \cdot {\bm z}_{{\bm x}} = 1,
\label{hcpn}
\end{equation}
where the sum is over the sites ${\bm x}$ of a cubic lattice, $\mu$ runs from 1
to 3, and $\hat\mu=\hat{1},\hat{2},\ldots$ are unit vectors along the lattice
directions. An alternative formulation (gauge CP$^{N-1}$ model) is obtained by
introducing link gauge variables $\lambda_{{\bm x},\mu}$ with $|\lambda_{{\bm
x},\mu}|=1$, and defining ~\cite{RS-81,DHMNP-81,BL-81,PV-19-CP}\footnote{Since
there is no kinetic term for the gauge fields, the link variables can be
integrated out, obtaining $Z = \sum_{\{{\bm z},\lambda\}} \exp(-H_{\lambda}) =
\sum_{\{{\bm z}\}} \prod_{{\bm x},\mu} I_0(2 N J|\bar{\bm z}_{\bm x} \cdot {\bm
z}_{{\bm x}+\hat\mu}|)$, where $I_0(x)$ is a modified Bessel function (here we
set $T=1$).}
\begin{eqnarray}
    Z = \sum_{\{{\bm z},\lambda\}} e^{- H_{\lambda}/T},\qquad
  H_\lambda = - 2 N J \sum_{{\bm x}, \mu} {\rm Re} 
\,(\lambda_{{\bm x},\mu} \,\bar{\bm{z}}_{\bm x} \cdot
       {\bm z}_{{\bm x}+\hat\mu}),\qquad
       \bar{\bm{z}}_{\bm x} \cdot {\bm z}_{{\bm x}} = 1.
\label{altcpnmod}
\end{eqnarray}
The lattice CP$^{N-1}$ models have a finite-temperature transition in
3D that admits a LGW description. The order parameter is the
gauge-invariant bilinear perator
\begin{equation}
Q_{{\bm x}}^{ab} = \bar{z}_{\bm x}^a z_{\bm x}^b - {1\over N}
\delta^{ab},
\label{qdefcpn}
\end{equation}
which is a Hermitian and traceless $N\times N$ matrix.  Therefore, to
study the nature of the transition, one may construct an effective LGW
theory in terms of a traceless Hermitian field $\Phi^{ab}({\bm x})$,
which can be formally obtained from the average of
$Q_{\bm x}^{ab}$ over a large but finite lattice domain.  The LGW
field theory is obtained by considering the most general fourth-order
polynomial in $\Phi$ consistent with the global U($N$) symmetry, i.e.,
\begin{eqnarray}
{\cal L} = {\rm Tr} (\partial_\mu \Phi)^2 
+ r \,{\rm Tr} \,\Phi^2 
+  w \,{\rm Tr} \,\Phi^3 
+  \,u\, ({\rm Tr} \,\Phi^2)^2  + v\, {\rm Tr}\, \Phi^4 .
\label{hlgcpn}
\end{eqnarray}
The stable FP of the RG flow of the LGW theory (\ref{hlgcpn}) is
expected to control the critical behavior of the continuous
transitions in lattice CP$^{N-1}$ models.  In the absence of a stable
FP, a first-order transition is expected.\footnote{The same effective
description applies to antiferromagnetic CP$^{N-1}$ models, i.e., to the
model with Hamiltonian (\ref{hcpn}) and $J < 0$ (in the Hamiltonian 
  (\ref{altcpnmod}) the sign of $J$ is irrelevant, and therefore the model is 
always ferromagnetic).
In this case, however, there
is an additional ${\mathbb Z}_2$ symmetry related to translations of
one site~\cite{DPV-15}, which implies that the corresponding effective
LGW theory (the order-parameter field is related to a staggered
gauge-invariant composite operator) must also be invariant under $\Phi
\to - \Phi$. This implies the absence of the cubic term in the LGW
theory (\ref{hlgcpn}), i.e., $w=0$.  For $N=3$ this LGW Hamiltonian is
equivalent to that of the O(8) vector model, thus predicting a
critical behavior belonging to the O(8) vector universality class,
with an effective enlargement of the symmetry group of the critical
modes from SU(3) to O(8).  This prediction has been confirmed
numerically~\cite{DPV-15}.  The critical behavior for larger values of
$N$ is discussed in Refs.~\cite{DPV-15,PTV-17}.
\label{foot:antiferroCPN}
}

For $N=2$, the cubic term in Eq.~(\ref{hlgcpn}) is absent, and the two
quartic terms are equivalent.  Therefore, one recovers the
O(3)-symmetric LGW theory, consistently with the equivalence between
the CP$^1$ and the Heisenberg model, see footnote \ref{cp1f}. For
$N\ge 3$, the cubic term is generically expected to be present. This
is usually considered as the indication that the phase transitions
occurring in this class of systems are of first order, as one can
easily infer using mean-field arguments.\footnote{\label{meanfo} The
mean-field argument that predicts a first-order transition in the
presence of a cubic term in the LGW Lagrangian is strictly valid in
four dimensions. The nature of the transition should not change
sufficiently close to four dimensions, as long as statistical
fluctuations are small. In particular, it is usually assumed that the
four-dimensional mean-field prediction also applies in three
dimensions. }
Continuous transitions may still occur, but they require a fine tuning of the
microscopic parameters leading to the effective cancellation of the cubic
term.\footnote{If this occurs, the critical behavior is controlled by the
effective theory relevant for antiferromagnetic models, see footnote
\ref{foot:antiferroCPN}.  For $N=3$ continuous transitions would belong to the
O(8) vector universality class~\cite{DPV-15}.}

Numerical simulations of the lattice gauge CP$^{N-1}$ model
(\ref{altcpnmod}) confirm the LGW predictions~\cite{PV-19-CP}.  For
$N\ge 3$ the transition is of first order, the first-order nature
becoming stronger and stronger as $N$ increases
\cite{PV-20-largeN}.\footnote{It should be noted that results for
other models that are expected to behave as the CP$^{N-1}$ model are
less clear. In particular, there is no general agreement that all
models undergo a first-order transition for $N=3$, see, e.g.,
Refs.~\cite{MZ-03,KHI-11,NCSOS-11,NCSOS-13,DPV-15,PTV-17,PV-19-CP}. }
In the effective description the gauge group plays no role and thus
one expects the same critical behavior in any SU($N$) invariant model
that has a bilinear Hermitian order parameter, irrespective of the
nature of the gauge group. This conclusion is supported by the results
of Ref.~\cite{BP-23}, which studied a discretized gauge model with
Hamiltonian still given by Eq.~(\ref{altcpnmod}) but with a reduced
${\mathbb Z}_Q$ gauge symmetry, obtained by requiring the gauge fields
to take only the values $\exp(2 \pi i n/Q)$ with $n=0,\ldots Q-1$. For
$N=2$ the model undergoes an O(3) transition for any $Q \ge 3$, as the
CP$^{1}$ model---this can be shown by explicitly performing the
summation over the gauge degrees of freedom---indicating that the
nature of the gauge group is not important.

The LGW prediction is apparently in contradiction with the results
obtained by performing a standard $1/N$ expansion in the lattice model
(\ref{altcpnmod}), as they predict a continuous transition
\cite{RS-81,DHMNP-81,BL-81}.  It turns out that these calculations are
based on the incorrect assumption that the relevant large-$N$
saddle-point solutions are spatially homogeneous and that the gauge
fields are ordered for all values of $J$ \cite{PV-20-largeN}. While
this assumption is correct in the large-$J$ phase, it does not hold
for small values of $J$, because of the presence of topologically
nontrivial configurations, for instance those discussed in
Ref.~\cite{MS-90}, which forbid the ordering of the gauge fields.

Finally, it is interesting to note that an analysis in the
$1/N$-expansion framework indicates that the CP$^{N-1}$ field theory
with Lagrangian (\ref{contham}) is equivalent~\cite{MZ-03,ZJ-book} to
the Abelian Higgs field theory (AHFT), or scalar electrodynamics,
which will be presented in Sec.~\ref{gauFT}, whose RG flow has a
stable charged FP for large values of $N$, see Sec.~\ref{AHSFT}.  This
result however does not apply to the lattice CP$^{N-1}$ models.
Indeed, gauge fields can be integrated out in the lattice gauge
CP$^{N-1}$ models, obtaining a Hamiltonian that can be written as a
sum of local gauge-invariant scalar terms.  Thus, gauge modes cannot
be relevant for the critical behavior, as instead would be the case if
the CP$^{N-1}$ transition were effectively described by a gauge field
theory with dynamical gauge fields, as the AHFT presented in
Sec.~\ref{gauFT} below. A more detailed discussion is presented in
Secs.~\ref{LAHMnc} and \ref{LAHMc}.

\subsubsection{Three-dimensional RP$^{N-1}$ models} 
\label{RPN}

RP$^{N-1}$ models are the real analogues of the CP$^{N-1}$ models.
The fundamental field is a real $N$-vector field ${\bm s}_{\bm x}$
satisfying ${\bm s}_{\bm x}\cdot {\bm s}_{\bm x}=1$.\footnote{
Formally, the model corresponds to fields that are defined on the real
projective space, which is obtained starting from the sphere $S^{N-1}$
in $N$ dimensions and identifying points that differ by a reflection,
i.e., ${\bm s}_{\bm x}$ is identified with $-{\bm s}_{\bm x}$. In the
lattice model, this identification is obtained by requiring the
Hamiltonian to be invariant under local reflections of the spins.}
The standard RP$^{N-1}$ Hamiltonian is
\begin{equation}
H_{\rm RP}=-J\sum_{{\bm x},\mu} ({\bm s}_{\bm x} \cdot {\bm s}_{{\bm
    x}+\hat{\mu}})^2.
\label{hrpn}
\end{equation}
Equivalently, one can introduce a ${\mathbb Z}_2$ gauge field
$\sigma_{{\bm x},\mu}$ associated with the links of the lattice, and
define
\begin{equation}
H_{\sigma}=-J\sum_{{\bm x},\mu} \sigma_{{\bm x},\mu}\,{\bm s}_{\bm x} \cdot {\bm
  s}_{{\bm x}+\hat{\mu}},\qquad \sigma_{{\bm x},\mu}=\pm 1.
\label{hrpn2}
\end{equation}
Both models are invariant under SO($N$) global transformations and
the ${\mathbb Z}_2$ gauge transformations
\begin{equation}
{\bm s}_{\bm x} \to w_{\bm x} {\bm s}_{\bm x}, \qquad 
\sigma_{{\bm x},\mu} \to w_{\bm x} \sigma_{{\bm x},\mu} w_{{\bm x}+\hat{\mu}},
\qquad w_{{\bm x}}=\pm 1.
\label{rpn-gauge}
\end{equation}
RP$^{N-1}$ models are expected to describe the universal features of
the isotropic-nematic transition in liquid
crystals~\cite{deGennes-book}.  They have been mostly investigated in
two dimensions~~\cite{CEPS-93,CPS-94,Hasenbusch-96,NWS-96,CHHR-98,
  DDL-20,BFPV-20-rpn}.

3D RP$^{N-1}$ models undergo a finite-temperature transition. Since
the gauge fields are not dynamical, the transition is a LGW transition
in terms of a gauge-invariant order parameter, which is identified
with the bilinear symmetric tensor
\begin{equation}
   R^{ab}_{\bm x} = s^a_{\bm x} s^b_{\bm x} - {1\over N} \delta^{ab}.
\end{equation}
Therefore, the LGW field is a symmetric and traceless real tensor
$\Phi^{ab}({\bm x})$, obtained by coarse-graining $R^{ab}_{\bm x}$, and the LGW
Lagrangian~\cite{PTV-17,PTV-18} is\footnote{The LGW Lagrangian
(\ref{LGWRPN}) is also relevant for the critical behavior of
antiferromagnetic RP$^{N-1}$ models, such as the lattice models
(\ref{hrpn}) with $J<0$, at least for $N\le 3$.  In this case,
however, the effective theory is also invariant under $\Phi \to -
\Phi$, which implies $w=0$. For $N=3$ the corresponding LGW
Hamiltonian is equivalent to that of the O(5) vector theory, implying
an effective enlargement of the symmetry group, from O(3) to O(5), at
the transition \cite{FMSTV-05}. Numerical results confirm this
prediction.  For $N\ge 4$ there are discrepancies between the
numerical results and the field-theoretical predictions obtained from
high-order perturbative analyses of the RG flow of the LGW theory. The
origin of these differences is still an open problem, see
Refs.~\cite{PTV-18} and \cite{RRV-23}.}
\begin{eqnarray}
{\cal L} =  {\rm Tr} (\partial_\mu \Phi)^2 
+ r \,{\rm Tr} \,\Phi^2 +
w \,{\rm Tr} \,\Phi^3 
+  \,u\, ({\rm Tr} \,\Phi^2)^2  + v\, {\rm Tr}\, \Phi^4.
\label{LGWRPN}
\end{eqnarray}
For $N=2$ the field $\Phi^{ab}$ has only two independent components,
the $\Phi^3$ term is absent, and the Lagrangian (\ref{LGWRPN}) is
equivalent to that of the standard XY vector model.  Thus, continuous
transitions should belong to the XY universality class.  In this case
the field $\Phi^{ab}$ is equivalent to an O(2) vector field.  This
implies that the RG dimension of $R_{\bm x}^{ab}$ coincides with the
RG dimension of the vector field in the standard XY model.

For larger values of $N$, the LGW approach predicts ~\cite{PV-19-AH3d}
first-order transitions due to the presence of the $\Phi^3$ term (see
footnote \ref{meanfo} for a discussion of the relation between cubic
terms in the LGW Lagrangian and first-order transitions).

\subsubsection{Finite-temperature transition of
  hadronic matter in the chiral limit}
\label{hamatter}

As another example where the LGW approach is expected to hold, we
consider the finite-temperature transition of hadronic matter in the
chiral limit, which was analyzed by Pisarski and Wilczek~\cite{PW-84}
using the LGW approach.  The strong hadronic interactions are
described by quantum chromodynamics (QCD), which is a non-Abelian
gauge theory with gauge group SU(3) and quark fields in the
fundamental representation of the gauge group.  The thermodynamics of
the system is characterized by a finite-temperature transition at
$T_c\simeq 200$ MeV, which separates a low-temperature confined
hadronic phase, in which chiral symmetry is broken, from a
high-temperature phase with deconfined quarks and gluons (quark-gluon
plasma), in which chiral symmetry is restored~\cite{Wilczek-00,
  Karsch-02,Sharma-19}. Our present understanding of this finite-$T$
transition is based on the symmetry properties of the system. In the
presence of $N_f$ light quarks, and, in particular, in their massless
limit, the relevant symmetry is the chiral symmetry ${\rm
  SU}(N_f)_L\otimes{\rm SU}(N_f)_R$, where $L,R$ stand for left and
right quark components. Below $T_c$, this symmetry is spontaneously
broken to SU($N_f$)$_V$, the group of vector quark transformations,
with a nonzero quark condensate $\langle {\bar \psi} \psi \rangle$.
The finite-$T$ transition is therefore characterized by the
symmetry-breaking pattern ${\rm SU}(N_f)_L\otimes {\rm SU}(N_f)_R
\rightarrow {\rm SU}(N_f)_V$.\footnote{If the axial U(1)$_A$ symmetry
is effectively restored at $T_c$, the symmetry-breaking pattern
becomes ${\rm U}(N_f)_L\otimes {\rm U}(N_f)_R \rightarrow {\rm
  U}(N_f)_V$. The restoration of this symmetry is not expected in QCD.
However, semiclassical calculations in the high-temperature
phase~\cite{GPY-81} show that instanton effects are exponentially
small for $T\gg T_c$, implying that the effect of the breaking the
axial U(1)$_A$ symmetry becomes small in this limit. This scenario has
been also confirmed by numerical studies, see,
e.g., Refs.~\cite{BDPV-13,BDMMMNSV-15,VK-21}.  RG studies of the LGW
theory appropriate to describe the finite-$T$ transition with
symmetry-breaking pattern ${\rm U}(N_f)_L\otimes {\rm U}(N_f)_R
\rightarrow {\rm U}(N_f)_V$ are reported in Refs.~\cite{PV-13,CP-04}.}

Let us report here the main assumptions of the approach used by
Pisarski and Wilczek to predict the nature of the finite-$T$ chiral
transition ~\cite{PW-84,BPV-03,Vicari-07,PV-13}:

\begin{itemize}

\item[(i)] One assumes that the phase transition is
  continuous for vanishing quark masses.  In this case the length
  scale of the critical modes diverges approaching $T_c$, becoming
  eventually much larger than $1/T_c$, which is the size of the
  euclidean ``temporal'' dimension at $T_c$.  Therefore, the
  asymptotic critical behavior must be associated with a 3D
  universality class with the same symmetry-breaking pattern.

\item[(ii)] One assumes---this is the crucial unproved assumption in the 
  approach---that gauge modes do not become critical at the transition
  and that only gauge-invariant quark correlations are relevant 
  at criticality. In other words, one assumes that the finite-$T$ 
  transition is a LGW transition, according to the classification 
  given in Sec.~\ref{difftype}.

\item[(iii)] 
  To define the LGW theory, an  order parameter should be specified.
  It is identified with the gauge-invariant
  bilinear quark matrix $\bar{\psi}_{Li}\psi_{Rj}$, $i,j=1,\ldots N_f$.
  In the LGW framework one introduces a complex-matrix field $\Psi_{ij}$,
  which is obtained by coarse-graining the order parameter.

\item[(iv)] The LGW Lagrangian is obtained by considering the 
  most general Lagrangian that is symmetric under the global symmetry group
  of the original model~\cite{BPV-03,PV-13}:
\begin{eqnarray}
  &&{\cal L}_{{\rm SU}(N_f)} =  {\cal L}_{{\rm U}(N_f)} + {\cal L}_{\rm det},  
  \label{LSUN}\\
&& {\cal L}_{{\rm U}(N_f)} = {\rm Tr} (\partial_\mu \Psi^\dagger)
  (\partial_\mu \Psi) +r {\rm Tr} \Psi^\dagger \Psi + u \left( {\rm
    Tr} \Psi^\dagger \Psi \right)^2 + v \, {\rm Tr}\left(\Psi^\dagger
  \Psi\right)^2, \nonumber\\
  &&{\cal L}_{\rm det} = w_1 \left( {\rm det}
  \Psi^\dagger + {\rm det} \Psi \right) + w_2 \left( {\rm Tr}
  \Psi^\dagger \Psi \right) \left( {\rm det} \Psi^\dagger + {\rm det}
  \Psi \right) + w_3 \left[ ({\rm det} \Psi^\dagger)^2 + ({\rm det}
    \Psi)^2 \right], \nonumber
\end{eqnarray}
where ${\cal L}_{{\rm U}(N_f)}$ is invariant under ${\rm
  U}(N_f)_L\otimes {\rm U}(N_f)_R$ transformations. 
The Lagrangian ${\cal L}_{\rm
  det}$ contains terms involving the determinant of $\Psi_{ij}$, which
effectively takes into account the explicit breaking of the U(1)$_A$
axial symmetry due to the anomaly (Eq.~(\ref{LSUN}) reports all terms
  that are relevant for $N_f=2$).  Nonvanishing quark masses can be
accounted for by adding an external-field term ${\rm Tr} \left( H \Psi
+ {\rm h.c.}\right)$.
\end{itemize}

The analysis of the RG flow of the model depends on the value of
$N_f$. For $N_f \ge 4$, a continuous finite-$T$ transition is only possible if
there is a stable FP of the RG flow that lies in the parameter region
in which the symmetry-breaking pattern ${\rm SU}(N_f)_L\otimes {\rm
  SU}(N_f)_R \rightarrow {\rm SU}(N_f)_V$ occurs. Since there is no FP
associated with this symmetry-breaking pattern, continuous
transitions cannot occur (thus, contradicting the first 
assumption leading to the LGW approach), and therefore first-order
transitions are generally predicted. For $N_f = 3$, the Lagrangian
contains a cubic term (the term proportional to $w_1$), and therefore
the transition is expected to be of first order, too. For $N_f=1$ the
Lagrangian (\ref{LSUN}) with $w_i=0$ reduces to the O(2) symmetric
$\Phi^4$ theory, corresponding to the 3D XY universality class, and
the term proportional to $w_1$ plays the role of an external field.
Thus, no transitions are expected, but only a crossover.

Given the actual values of the quark masses~\cite{PDG-24}, this
approach is mainly of interest for systems with two flavors, i.e., for
$N_f=2$, in which 
one only considers the lightest quarks $u$ and $d$, that 
are the building blocks of ordinary hadronic matter, e.g., of the
nucleons. The effective LGW theory contains two quadratic terms (those
proportional to $r$ and $w_1$), making the analysis more complex, see
Refs.~\cite{BPV-03,Vicari-07,PV-13}.  A detailed analysis suggests
that, if the transition is continuous in the chiral limit, it must
belong to the 3D O(4) vector universality class, as originally
predicted in Ref.~\cite{PW-84}.  This result can also be understood by
noting that the symmetry-breaking pattern ${\rm SU}(2)_L\otimes {\rm
  SU}(2)_R \rightarrow {\rm SU}(2)_V$ is equivalent to
O(4)$\rightarrow$O(3) (apart from discrete groups), i.e., it is the
same that characterizes transitions in the O(4) vector universality
class.

We stress that these predictions are based on a number of assumptions,
the main one being that the colored gauge modes are effectively
decoupled from the critical modes at the transition. If this
assumption is correct, then the effective LGW theory in terms of a
gauge-invariant bilinear quark operator can provide reliable
predictions. See, e.g., Refs.~\cite{Sharma-19,Aarts-etal-23} for a report on the
lattice numerical results obtained for the chiral transition.

\subsection{LGW$^{\,\times}$ transitions with a gauge-dependent
order parameter}
\label{nonginvLGW}

In Sec.~\ref{ginvLGW} we discussed LGW transitions where only matter
fields show a critical behavior and that admit a conventional LGW
description in terms of a scalar gauge-invariant order parameter.  In
this Section, we discuss a second class of transitions, which we name
LGW$^\times$ transitions, which are also characterized by the fact
  that only scalar modes have a critical dynamics, but that do not
  admit a local gauge-invariant order-parameter field.

3D O($N$)$^{\times}$ (Ising$^{\times}$ and
XY$^{\times}$ for $N=1$ and 2, respectively) transitions are the simplest 
examples of this type of unconventional transitions. They
admit an effective description in terms of the 
O($N$)-symmetric $\Phi^4$ theory with Lagrangian ~(\ref{ONLGW}) 
and  can be observed, e.g., in the lattice $N$-vector model minimally
coupled with ${\mathbb Z}_2$-gauge
variables~\cite{BPV-24-z2Nv,BPV-24-unco}.  On a cubic lattice its
partition function is given by
\begin{eqnarray}
Z=\sum_{\{{\bm s},\sigma\}} e^{-H/T},
\qquad
H  = - J N \sum_{{\bm x},\mu} \sigma_{{\bm x},\mu} \, {\bm s}_{\bm
    x} \cdot {\bm s}_{{\bm x}+\hat{\mu}} - K \sum_{{\bm
      x},\mu>\nu} \sigma_{{\bm
      x},\mu} \,\sigma_{{\bm x}+\hat{\mu},\nu} \,\sigma_{{\bm
    x}+\hat{\nu},\mu} \,\sigma_{{\bm x},\nu},\qquad
   \label{hamz2N}
\end{eqnarray}
where the site variables ${\bm s}_{\bm x}$ are unit-length
$N$-component real vectors (i.e., ${\bm s}_{\bm x}\cdot {\bm s}_{\bm
  x}=1$), and the link variables $\sigma_{{\bm x},\mu}$ 
take the values $\pm 1$. The Hamiltonian
parameter $K$ plays the role of inverse gauge coupling. 
For $K\to\infty$ (small gauge-coupling limit) the variable 
$\sigma_{{\bm x},\mu}$ can be set equal to 1 (modulo gauge transformations)
and the standard O($N$) vector model is recovered. For $N=1$ the spin
variables take the integer values $s_{\bm x}=\pm 1$, and the model
corresponds to the so-called ${\mathbb Z}_2$-gauge Higgs
model~\cite{Wegner-71,FS-79,Kogut-79}.  As we shall discuss in
Sec.~\ref{discgauge}, O($N$)$^{\times}$ transitions are observed along
the transition line that separates the spin disordered phase from the
spin ordered one, for sufficiently small values of the gauge coupling,
i.e., for sufficiently large $K$.

In conventional $N$-vector systems with $N\ge 2$ the spontaneous
breaking of the O($N$) symmetry at the transition is driven by the
condensation of the $N$-component fundamental vector field. However,
in the gauge system with Hamiltonian (\ref{hamz2N}), as a consequence
of the ${\mathbb Z}_2$-gauge symmetry, the local vector correlation
$\langle {\bm s}_{\bm x} \cdot {\bm s}_{\bm y}\rangle$ trivially vanishes for
${\bm x}\neq {\bm y}$.
Therefore, the spontaneous breaking of the O($N$) symmetry can only be
observed by considering correlations of composite gauge-invariant
operators, such as the gauge-invariant bilinear operator $S^{ab}_{\bm
  x} = s^a_{\bm x} s^b_{\bm x} - \delta^{ab}/N$, which transforms as a
spin-two tensor under O($N$) transformations.  A distinctive feature
of the O($N)^{\times}$ transitions is that this bilinear operator, as
well as all gauge-invariant operators, have the same critical behavior
as in the conventional $N$-vector model without gauge invariance.  The
equivalent behavior of gauge-invariant correlations in O($N$) and
O($N)^{\times}$ transitions implies that gauge modes do not drive the
critical behavior. But, more importantly for the characterization of
the effective critical behavior, this result allows us to exclude that
the composite gauge-invariant operator $S^{ab}_{\bm x}$ can be taken
as the fundamental field in a LGW effective theory.  Indeed, if one
identifies the LGW field $\Phi$ with the coarse-grained
gauge-invariant spin-2 order parameter, the corresponding LGW theory
turns out to be that of the RP$^{N-1}$ models, cf. Eq.~(\ref{LGWRPN}),
which would predict a different universality class with a different
symmetry-breaking pattern~\cite{PTV-18}, see also Sec.~\ref{ddo}.  On
the contrary, we conclude that O($N$) and O($N)^{\times}$ transitions
have the same effective description, which in turn implies that also
O($N)^{\times}$ transitions admit a LGW characterization in terms of
a vector field $\Phi$.  The main issue here is the identification of
the correct fundamental vector field ${\bm \Phi}$.  As shown in
Refs.~\cite{BPV-24-z2Nv,BPV-24-unco} and also discussed in 
Sec.~\ref{unconstar}, this
problem can be solved by identifying ${\bm \Phi}$ as the
coarse-grained analogue of the fundamental field $s_{\bm x}$ after an
appropriate gauge-fixing procedure is considered.  This implies that
O($N$)$^{\times}$ transitions can still be described by the
O($N$)-symmetric LGW theory (\ref{ONLGW}), but with a gauge-dependent
order-parameter field.

For $N=1$ there is no global symmetry group, since the global
${\mathbb Z}_2$ symmetry of the ungauged model has been turned into a
local one.  However, the global ${\mathbb Z}_2$ symmetry reemerges in
the effective description of the large-$K$ transitions. Indeed, their critical
features are the same as in the conventional Ising transition. 
Therefore, we will name these
transitions Ising$^{\times}$ transitions.

It is interesting to remark that, most probably, our interpretation of
the transition as due to the condensation of a non-gauge-invariant
local order parameter, is not the only one possible. Indeed, an
alternative interpretation is that LGW$^{\times}$ transitions are also
controlled by a gauge-invariant order parameter. However, the order
parameter is nonlocal, although it admits a local representation when
some specific gauge-fixing condition is considered.  An explicit
realization of this mechanism occurs in noncompact LAH models and is
discussed in Sec.~\ref{nc-orderparameters}.

LGW$^{\times}$ transitions do not only occur in O($N$)-symmetric models, but
are expected in generic spin models with more complex global
symmetries, when they are coupled with discrete gauge variables.  The
discrete nature of the gauge group appears to be a crucial property
for the existence of LGW$^{\times}$ transitions.  Indeed, if the gauge group
is discrete it is possible that the pure spin FP remains stable in the
presence of a small gauge coupling.  If this occurs, in the weak
gauge-coupling limit, continuous transitions should be of LGW$^{\times}$
type.

\subsection{GFT transitions described 
  by effective field theories with gauge fields}
\label{gauFT}

LGW and LGW$^{\times}$ transitions are both characterized by the fact
that the critical behavior is completely determined by the behavior of
the scalar variables. Here we consider transitions in which both
scalar and gauge correlations are critical at the transition. In this
case an appropriate effective field-theory description of the critical
behavior requires a gauge field theory (GFT) with the explicit
inclusion of gauge fields.  Transitions of this type are observed in
LAH models, as we shall discuss in Secs.~\ref{LAHMnc}
and~\ref{LAHMc}. The corresponding field theory description is
provided by the Abelian Higgs field theory (AHFT), or scalar
QED.  In the AHFT an $N$-component complex scalar field
${\bm \Phi}({\bm x})$ is minimally coupled with an electromagnetic
field $A_\mu({\bm x})$. The AHFT Lagrangian reads \cite{GL-50}
\begin{equation}
{\cal L}_{\rm AH} =  
\frac{1}{4 g^2} \,F_{\mu\nu}^2 
+ \overline{D_\mu{\bm\Phi}} \cdot D_\mu{\bm\Phi}
+ r\, \bar{\bm \Phi} \cdot {\bm \Phi} + 
u \,(\bar{\bm \Phi}\cdot{\bm \Phi})^2,
\quad
F_{\mu\nu}\equiv \partial_\mu A_\nu - \partial_\nu A_\mu,
\quad
D_\mu \equiv \partial_\mu + i A_\mu.
\label{AHFTL}
\end{equation}
Beside the Abelian U(1) gauge invariance, the theory has a global
SU($N$) symmetry, ${\bm \Phi} \to V {\bm \Phi}$ with $V\in {\rm
  SU}(N)$.  Its RG flow will be discussed in Sec.~\ref{AHSFT}.  One
expects that the stable charged FP of the RG flow of the AHFT, i.e., a
stable FP with a nonvanishing gauge coupling, is the FP that controls
the universal features of the charged critical transitions of lattice
AH models, when both gauge and scalar fields are critical.

It is interesting to note that the effective description of GFT
transitions requires non-gauge-invariant fundamental fields, like
LGW$^{\times}$ transitions. Thus, the issues we have discussed in
Sec.~\ref{nonginvLGW} occur also here. In particular, vector
correlations and gauge-field correlations are trivial in the gauge
theory, because of gauge invariance. However, in GFTs the solution of
these problems is well known: A gauge fixing should be added to the
theory to be able to observe the critical correlations of the
fundamental fields (this is a necessary step for perturbative
calculations).

As we shall review in the next sections, critical behaviors consistent
with the charged universality classes of the AHFT have been identified
in lattice AH (LAH) models with noncompact gauge variables in the weak
gauge-coupling regime, see Sec.~\ref{LAHMnc}, along the transition
line that separates the Coulomb and Higgs
phases~\cite{BPV-21-ncAH,BPV-23-chgf}, for $N > N^*$, with $N^* =
7(2)$.  On the other hand, in the strong gauge-coupling regime,
transitions between the Coulomb and molecular phase have an effective
LGW description.  Charged universality classes have been identified
also in compact formulations, again in the weak gauge-coupling limit,
see Sec.~\ref{LAHMc}.

It is interesting to note that in lattice gauge models the global
symmetry group and the symmetry breaking pattern do not completely
characterize the nature of the phase transition. The properties of the
gauge modes are crucial to determine the critical behavior.  Systems
with global SU($N$) symmetry, symmetry-breaking pattern SU($N$)$\to$
SU($N-1$), and local U(1) gauge symmetry provide an example of such
behavior. If the gauge modes are not critical the transition is of LGW
type. The corresponding $\Phi^4$ theory is given in
Eq.~(\ref{hlgcpn}), which predicts first-order transitions for $N\ge
3$ and continuous transitions belonging to the O(3) vector
universality class for $N=2$.  On the other hand, if gauge modes are
relevant, the transitions are described by the AHFT (\ref{AHFTL}):
they can be continuous for $N>N^*\approx 7$, and of first order in the
opposite case, and in particular for $N=2$.  Therefore, in the two
cases the expected behavior at the transition is quite different.

\subsection{Topological transitions in lattice gauge models}
\label{toptra}

The first three types of transitions that we have discussed so far are all
characterized by the spontaneous breaking of a global symmetry.  In
gauge models there are however also transitions that are not
associated with a symmetry breaking and thus lie beyond the Landau
paradigm. These transitions separate phases that only differ in the
topological order of the gauge excitations. The scalar degrees of
freedom, if present, play no role.  Thus, they do not admit a LGW or
LGW$^{\times}$ description, nor can they be described using a GFT with scalar
matter, as in all these cases matter fields are critical. They
therefore represent a novel class of transitions that do not obey the
standard paradigm, see, e.g., Refs.~\cite{Xu-12, Sachdev-19, BKV-02,
  BLS-20, BPV-24-ncAH, BPV-24-cAH, Senthil-23}.

Topological phase transitions are characterized by the divergence of
the length scale $\xi$ of the critical modes, as in more conventional
cases.  However, the absence of a local gauge-invariant order
parameter forces one to define the length scale $\xi$ from the
behavior of extended objects, for instance Wilson loops in gauge
theories.  The presence of a divergent length allows one to define a
universal critical exponent $\nu$. Other exponents that are typically
considered in conventional transitions, such as $\eta$ and $\beta$,
are instead not defined in the absence of a local order parameter.
Topological transitions are also defined in the quantum setting, where
one can use $\nu$ and the dynamic exponent $z$ (defined in terms of
the size dependence of the gap at the critical point) to characterize
the critical behavior.\footnote{ We mention that topological
transitions have also been associated with the breaking of
higher-order symmetries, see, e.g.,
Refs.~\cite{McGreevy-23,XRKP-23,GKSW-15,NO-09,BZ-05}.  While
conventional symmetries act on local objects defined on the entire
system, higher-order symmetries act on extended objects associated
with lower-dimensional submanifolds. Topological transitions can be
interpreted as due to the spontaneous breaking of such generalized
symmetries, somehow extending the Landau approach in which transitions
are related to the spontaneous breaking of global symmetries.}

Two important examples of models that undergo topological transitions
are the inverted XY (IXY) model and the lattice ${\mathbb Z}_Q$-gauge
model, which are reviewed below.

\subsubsection{The three-dimensional  inverted XY model} \label{sec.IXY}

The inverted XY model (IXY) is a gauge model with Hamiltonian
\begin{equation}
      H_{\rm IXY}= 
      \frac{\kappa}{2} \sum_{{\bm x},\mu>\nu} 
         (\nabla_\mu A_{{\bm x},\nu} - \nabla_\nu A_{{\bm x},\mu})^2 , 
\end{equation}
where the sum is over all lattice plaquettes, $\nabla_\mu$ is the
forward lattice derivative, $\nabla_\mu f({\bm x}) = f({\bm
  x}+\hat{\mu}) - f({\bm x})$, and the field $A_{{\bm x},\mu}$ takes
only values that are multiples of $2 \pi$, i.e., $A_{{\bm x},\mu} = 2
\pi n_{{\bm x},\mu}$, with $n_{{\bm x},\mu} \in \mathbb Z$.  The
Hamiltonian $H_{\rm IXY}$ is invariant under the gauge transformations
$n_{{\bm x},\mu} \to n_{{\bm x},\mu} + m_{{\bm x}+\hat{\mu}} - m_{\bm
  x}$, with $m_{\bm x} \in \mathbb Z$.

The free energy of the IXY model is related by duality to that of the
XY model with Villain action~\cite{DH-81,NRR-03}, which implies that
the IXY model undergoes a transition (at $\kappa_c =
0.076051(2)$~\cite{NRR-03,BPV-21-ncAH} when setting the temperature
$T=1$) belonging to the XY universality class, but with inverted high-
and low-temperature phases~\cite{DH-81}.

It is important to stress that the above identification of the
universality classes is done using the duality between the gauge and
the spin system, and that duality only maps the free energy and the
related thermal observables. In particular, the magnetic sector
present in the spin universality class has no counterpart in the gauge
universality class. Viceversa, the charged sector that characterizes
the IXY transition, see Secs. \ref{nc-orderparameters} and
\ref{cboncLAHM}, is not present in the spin model.  Thus, the XY FP
that controls the behavior of the gauge model differs from the FP that
is relevant in the XY model.  Indeed, they represent FPs obtained by
performing RG transformations on two different classes of
Hamiltonians---those with local O(2) and global O(2)
symmetry, respectively. Therefore, it is more appropriate to
distinguish the IXY or gauge XY universality class from the spin XY
universality class.\footnote{The IXY or gauge XY universality class
is sometimes also referred to as XY$^*$ universality class in the
literature, see, e.g., Ref.~\cite{Sachdev-19}.}

\subsubsection{Three-dimensional lattice ${\mathbb Z}_Q$ gauge theories}
\label{sec.ZQ}

The lattice ${\mathbb Z}_Q$ gauge theories with $Q\ge 2$ are paradigmatic
models undergoing finite-temperature topological transitions~\cite{Sachdev-19},
separating a high-temperature deconfined phase from a low-temperature confined
phase, that have been investigated in several works, see, e.g.,
Refs.~\cite{Wegner-71,BDI-74, BKKLS-90, CFGHP-97, GPP-02, BCCGPS-14, XCMCS-18,
ACFIP-24}. Their lattice Hamiltonian reads\footnote{It is interesting to
observe that, by means of the Fortuin-Kasteleyn representation \cite{FK-71},
the ${\mathbb Z}_Q$ gauge theory can be extended to any complex value of $Q$,
see, e.g., Ref.~\cite{Gliozzi-07}. Gauge models with real values of $Q$ in the
interval [0,1] have been studied, e.g., in Refs.\cite{Gliozzi-07,GLPN-05}. }
\begin{equation}
  H_{{\mathbb Z}_Q} = - 2 K \, \sum_{{\bm x},\mu>\nu} {\rm Re}
  \left(\lambda_{{\bm x},{\mu}} \,\lambda_{{\bm x}+\hat{\mu},{\nu}}
  \,\bar{\lambda}_{{\bm x}+\hat{\nu},{\mu}} \,\bar{\lambda}_{{\bm
      x},{\nu}}\right), \quad \lambda_{{\bm x},\mu}=\exp\left(i 
    \frac{2\pi n_{{\bm x},\mu}}{Q}\right),\quad n_{{\bm x},\mu}\in 0,1,\ldots,Q-1,
\label{zqgaumodels}
\end{equation}
where $\lambda_{{\bm x},\mu}$ are ${\mathbb Z}_Q$-group variables
associated with the links of a cubic lattice.

The critical behavior can be inferred by using duality.  The
${\mathbb Z}_Q$ lattice gauge model is dual to a specific $Q$-state
clock model~\cite{Savit-80}, with $\mathbb{Z}_Q$ spin variables
$\exp(2\pi i n_{\bm x}/Q)$ (where $n_{\bm x}=1,...,Q$) associated with
the lattice sites, and a Hamiltonian which is symmetric under global
${\mathbb Z}_Q$ transformations. The 2-state clock model is equivalent
to the Ising model, while the 3-state clock model is equivalent to the
3-state Potts model, which undergoes a first-order transition. The
nature of the transitions for $Q\ge 4$ can be inferred by studying the
RG flow of the LGW field theory associated with a
$\mathbb{Z}_Q$-symmetric spin system~\cite{BPV-22-dis}
\begin{eqnarray}
    {\cal L}_{\mathbb{Z}_Q} = |\partial_\mu \varphi|^2 + r\, |\varphi|^2 + u
    \,|\varphi|^4 + v \,(\varphi^Q + \bar\varphi^Q),
\label{lgwzq}
\end{eqnarray}
where $\varphi({\bm x})$ is a complex field.  The $Q$-dependent
potential has dimension $Q$ and is therefore irrelevant for $Q >
4$. In this case we can thus set $v=0$, obtaining the standard
U(1)-symmetric $\Phi^4$ theory for a complex field. This implies an XY
critical behavior and an effective enlargement of the symmetry of the
critical modes at the transition: While the lattice model is ${\mathbb
  Z}_Q$-symmetric, critical modes are U(1)-symmetric.  For $Q=4$, the
$Q$-dependent potential has dimension four and represents a cubic
anisotropy.  The stable FP of its RG flow is again the XY FP
with $v=0$~\cite{Aharony-76, PV-02, CPV-00,
  CPV-03,HV-11,Hasenbusch-23}.\footnote{ For $Q=4$ the anisotropic
interaction with coefficient $v$ appearing in the Hamiltonian
(\ref{lgwzq}) gives only rise to scaling corrections.  However, due to
the small absolute value of the corresponding RG
dimension~\cite{CPV-00,CPV-03,HV-11} $y_v=-0.108(6)$, these scaling
corrections decay slowly. Beside the stable FP with $v=0$, the LGW
model also admits an unstable FP on the line $w=u-6v=0$, where the LGW
Hamiltonian can be written as the sum of two identical LGW
Hamiltonians with a real scalar field, so this FP corresponds to
an Ising critical behavior. The parameter $w$ corresponds to a relevant
perturbation of the {decoupled} Ising fixed point, with RG dimension
$y_w=d - 2/\nu_I=0.17475(2)$, where~\cite{KPSV-16} $\nu_I=0.629971(4)$
is the Ising critical exponent. The standard $\mathbb{Z}_4$ clock
model can be exactly rewritten as a sum of two Ising
models~\cite{Suzuki-67,HS-03}. Thus, it corresponds to the LGW theory
with $w=0$ and hence it undergoes an Ising transition. Generic
four-state clock models, however, undergo XY transitions.
\label{clockfoot}}

Using duality, the results for models with global $\mathbb{Z}_Q$
symmetry allow us to infer the nature of the transitions in the
${\mathbb Z}_Q$ gauge models.  These arguments predict an Ising
transition for $Q=2$, a first-order transition for $Q=3$, and an XY
transition for any $Q> 4$.  For $Q=4$ the ${\mathbb Z}_4$ gauge model
has a transition in the Ising universality class because its partition
function can be written as that of two independent ${\mathbb Z}_2$
gauge models~\cite{Korthals-78, BPV-24-cAH} (this is analogous to what
happens in the standard ${\mathbb Z}_4$ clock model, see
footnote~\ref{clockfoot}). However, this result is specific of the
gauge formulation with Hamiltonian (\ref{zqgaumodels}).  Generic
discrete models with ${\mathbb Z}_4$ gauge symmetry are instead
expected to undergo transitions belonging to the gauge XY (IXY) universality
class.

For $Q\to \infty$ we obtain the compact U(1) gauge theory in which the
fields $\lambda_{{\bm x},\mu}$ are U(1) phases. This model does not
have finite-$K$ transitions because of the presence of topological
excitations (monopoles) that keep the system always disordered
\cite{Polyakov-77}. The limit $K\to \infty$, $Q\to \infty$ at fixed
$K/Q^2$ is more interesting \cite{BPV-22}. In this case one obtains
the IXY model defined in Sec.~\ref{sec.IXY} with $\kappa =
2K/Q^2$.\footnote{ This result implies that the critical value
$K_{c}(Q)$ for the $Q$-state gauge model scales as $2K_c(Q) \simeq
\kappa_{\rm nc}^{(c)} \, Q^2$ for large $Q$, where $\kappa_{\rm
  nc}^{(c)}=0.076051(2)$ is the critical coupling of the IXY model.
Using the existing estimates~\cite{BCCGPS-14,BC-80} of $K_c(Q)$,
Ref.~\cite{BCCGPS-14} obtained $2K_c(Q)\simeq w \,Q^2$ with
$w=0.076053(4)$, in good agreement with the theoretical
prediction, see also Ref.~\cite{BPV-22}.}  The IXY model is the
relevant one for the behavior of the gauge model for $Q\ge 5$ (and
also for $Q=4$ in the generic case): For these values of $Q$,
transitions in the ${\mathbb Z}_Q$ gauge model have the same critical
behavior as the transition in the IXY model. It is interesting to give
a RG interpretation of these results, considering the large-$Q$
${\mathbb Z}_Q$ gauge model as a perturbed IXY model, in which the
inverse charge $1/Q$ plays the role of perturbation parameter. The IXY
critical behavior of all ${\mathbb Z}_Q$ gauge models with $Q\ge 4$
indicates that the $1/Q$ perturbation of the IXY critical behavior is
irrelevant.

As already stressed above for the IXY model, the identification of the
universality classes is done using duality, that only maps the free
energy and the related thermal observables.  Thus, we distinguish the
gauge XY (a representative is the IXY transition) from the spin XY
universality class and analogously, the gauge Ising from the spin
Ising universality class.

Finally, we mention that the Wilson loops
$W_C$, defined as the product of the link variables along a closed
contour $C$ within a plane, provide a nonlocal order parameter for the
topological transition of the 3D lattice ${\mathbb Z}_Q$ gauge
models~\cite{Wegner-71}. Indeed, their asymptotic size dependence for
large contours changes at the transition, varying from the
area law $W_C\sim \exp(- c_a A_C)$, where $A_C$ is the area enclosed by
the contour $C$ and $c_a>0$ is a constant, which is valid for small 
values of $K$,  (this behavior can be easily
checked in the strong-coupling regime for $K\ll 1$
\cite{OS-78, ID-book1}), to the perimeter law $W_c\sim
\exp(- c_p P_C)$, where $P_C$ is the perimeter of the contour $C$ and
$c_p>0$ is a constant, which is valid for large values of $K$. 
An analogous behavior is observed in 2D
quantum formulations of the ${\mathbb Z}_2$ gauge theory, see, e.g.,
Ref.~\cite{Sachdev-19}.

\subsubsection{Topological transitions in low-temperature Higgs phases}
\label{sec.ZQhiggs}

Topological transitions, such as those arising in
3D ${\mathbb Z}_Q$ gauge theories, also occur in theories
with a larger gauge group if the reduced global symmetry of the
low-temperature Higgs phase gives rise to an effective breaking of the
gauge group to a residual smaller subgroup and, in particular, to the
${\mathbb Z}_Q$ group.  Topological transitions of this
kind~\cite{Sachdev-19,SSST-19,SPSS-20,BFPV-21-adj} are realized in
various lattice gauge systems, as we shall see in the following. For
example, we mention LAH models with compact gauge variables and
higher-charge scalar fields, see Sec.~\ref{coLAJHM}, and SU($N_c$)
gauge LNAH model with scalar matter in the adjoint representation, see
Sec.~\ref{modelsuncadj}.

In this context, the relevant gauge symmetry is the residual gauge
symmetry $G_R$ of the Higgs phase, which is defined as the group of
gauge transformations that leave invariant a representative scalar
field in the minimum-potential configuration. The representative
scalar field is supposed to have been fixed to a specific value by
using both the global and the local gauge symmetry.  Note that the
group $G_R$ does not depend on the chosen representative of the scalar
field in the Higgs phase, and it is determined by the global symmetry
of the ordered phase. Therefore, it does not represent an additional
characterization of the Higgs phase, and, more importantly, it has a
gauge-independent relevance.  The gauge modes associated with the
residual gauge group $G_R$ may give rise to gauge-independent
topological transitions, as supported by the numerical results
that we will present in the next sections.  More precisely,
topological transitions are possible if the gauge theory with gauge
group $G_R$ (and no scalar fields) has a topological transition, such
as the ${\mathbb Z}_Q$ gauge model in three dimensions.

To refer to the residual gauge symmetry $G_R$, we will usually speak
of a formal breaking of the gauge symmetry $G$ with gauge-symmetry
breaking pattern $G\to G_R$, consistently with the common use of
referring to the Higgs mechanism as realizing a gauge symmetry
breaking (for example, the Standard Model of the electroweak
  interactions is often associated with the gauge symmetry breaking
  pattern SU(2)$\otimes$ U$_Y(1)\to
  \mathrm{U}_\mathrm{em}$(1)~\cite{Weinberg-book}). However, we should
stress that the gauge symmetry cannot be spontaneously broken in the
standard sense of statistical mechanics (i.e.,
the system cannot be forced in one specific minimum, for instance,
by appropriately fixing the boundary conditions), as this is forbidden
by well-known rigorous arguments~\cite{Elitzur-75, DFG-78, FMS-81,
  ID-book1}.  Thus, to identify uniquely the scalar fields in the
low-temperature phase and the residual gauge symmetry, one can choose
appropriate boundary conditions that break the global symmetry, but one
should also add an appropriate complete gauge fixing to eliminate the
gauge redundancy.

%% file: AHSFT.tex
\section{Abelian Higgs gauge field theories}
\label{AHSFT}

Many collective phenomena in condensed-matter
physics~\cite{Anderson-book,Wen-book} are described by effective 3D
U(1) gauge models in which scalar fields are coupled with an Abelian
gauge field. We mention the transitions in
superconductors~\cite{HLM-74,Herbut-book}, in quantum SU($N$)
antiferromagnets~\cite{RS-90, TIM-05, TIM-06, Kaul-12, KS-12, BMK-13,
  NCSOS-15, WNMXS-17,Sachdev-19}, and the unconventional quantum
transitions between the N\'eel and the valence-bond-solid phases in
two-dimensional antiferromagnetic SU(2) quantum
systems~\cite{Sandvik-07, MK-08, JNCW-08, Sandvik-10, HSOMLWTK-13,
  CHDKPS-13, PDA-13, SGS-16}, which represent the paradigmatic models
for the so-called deconfined quantum criticality~\cite{SBSVF-04}. The
phase structure and the universal features of the transitions in
scalar gauge models have been extensively
studied~\cite{HLM-74,Herbut-book,FS-79,DH-81,FMS-81,BF-81,DHMNP-81,CC-82,BF-83,
  FM-83,KK-85,KK-86,BN-86,BN-86b,BN-87,RS-90,MS-90,KKS-94,BFLLW-96,
  HT-96,FH-96,IKK-96,KKLP-98,OT-98, CN-99, HS-00, KNS-02,MHS-02,
  SSSNH-02,SSNHS-03,MZ-03,NRR-03, MV-04,SBSVF-04, NSSS-04, SSS-04,
  HW-05, WBJSS-05, CFIS-05, TIM-05, TIM-06, CIS-06, KPST-06,
  Sandvik-07, WBJS-08, MK-08, JNCW-08, MV-08, KMPST-08, CAP-08, KS-08,
  ODHIM-09, LSK-09, CGTAB-09, CA-10, BDA-10, Sandvik-10, Kaul-12,
  KS-12, BMK-13, HBBS-13, Bartosch-13, HSOMLWTK-13, CHDKPS-13, PDA-13,
  BS-13, NCSOS-15, NSCOS-15, SP-15, SGS-16,WNMXS-17, FH-17, PV-19-CP,
  IZMHS-19, PV-19-AH3d, SN-19, Sachdev-19, PV-20-largeN, SZ-20,
  PV-20-mfcp, BPV-20-hcAH, HRS-21, BPV-21-bgi, BPV-21-bgi2, BPV-21-ncAH,
  WB-21, BPV-22-mpf, BPV-22, BPV-23-chgf,SZJSM-23, BPV-24-cAH,
  BPV-24-ncAH}, and particular attention has been paid to the role of the
gauge fields and of the related topological features, like monopoles and Berry
phases, which cannot be captured by effective LGW
theories with gauge-invariant scalar order parameters~\cite{ZJ-book, PV-02,
Sachdev-book, SBSVF-04, Sachdev-19}.

In this section we introduce the Abelian Higgs (AH) field theory
(AHFT), or scalar QED, which describes a $d$-dimensional system of
degenerate $N$-component scalar fields minimally coupled with an
Abelian U(1) gauge field. By properly choosing the Lagrangian
potential of the scalar fields, we can define models with different
global symmetry: here we discuss SU($N$) and SO($N$) invariant models.
We outline the known results for the renormalization-group (RG) flow
of the models and the corresponding FPs, which are expected
to describe the critical behavior of transitions in which both scalar
and gauge field are critical (GFT transitions in the classification of
Sec.~\ref{difftype}).

\subsection{SU($N$)-symmetric Abelian Higgs field theory}
\label{suAHFT}

\subsubsection{The model}
\label{suAHFTm}

The SU($N$)-symmetric AHFT is obtained by minimally coupling an
$N$-component complex scalar field ${\bm \Phi}({\bm x})$ with an
electromagnetic U(1) gauge field $A_\mu({\bm x})$.
The corresponding SFT (or Euclidean QFT) is formally defined by the
functional path integral
\begin{equation}
  Z = \int  [d{\bm \Phi}]\,[d{\bm A}]\,
  e^{- S({\bm \Phi},{\bm A})} ,\quad
  S({\bm \Phi},{\bm A}) = \int d^d x \, \left[
    {\cal L}_{\rm AH}({\bm \Phi},{\bm A}) + {\cal L}_{\rm gf}({\bm A})\right],
    \quad
  {\cal L}_{\rm gf}({\bm A}) = {1\over 2\zeta} (\partial_\mu A_\mu)^2,
  \label{partfunc}
\end{equation}
where the gauge-invariant Lagrangian ${\cal L}_{\rm AH}$ is reported
in Eq.~(\ref{AHFTL}). We have added here the Lorenz gauge-fixing term
${\cal L}_{\rm gf}$, which is necessary to obtain a well-defined
theory, see, e.g., Ref.~\cite{ZJ-book}. As it stands here, the AHFT is
only formally defined in perturbation theory. To go beyond
perturbation theory, one must consider a nonperturbative
regularization and show that the properly renormalized theory admits a
finite limit when the regularization is eliminated. If we use the
lattice regularization, this is possible only if the lattice model
admits a continuous
transition~\cite{ZJ-book,Creutz-book,MM-book,Rothe-book,Callaway-88,GJ-book}.

The RG flow of the AHFT has been studied using the perturbative
$\varepsilon\equiv 4-d$
expansion~\cite{HLM-74,Hikami-80,MU-90,FH-96,IZMHS-19} and the
functional RG approach ~\cite{FH-17}. The analysis of the
corresponding FPs allows one to determine the nature of the
transitions in this class of systems.  In particular, if a stable FP
exists, continuous transitions are possible in systems with analogous
features.  The model has also been studied using the large-$N$
expansion, obtaining the expansion of some critical exponents for
large $N$~\cite{HLM-74,DHMNP-81,IKK-96,MZ-03,KS-08}.

For $N\ge 2$, the pattern of the spontaneous breaking of the global
symmetry is
\begin{equation}
\mathrm{SU}(N)\to \mathrm{U}(N-1).
  \label{symbrUN}
\end{equation}
The ordered phase is 
characterized by the condensation of the local gauge-invariant
bilinear operator
\begin{equation}
  Q^{ab}({\bm x}) = \bar\Phi^{a}({\bm x}) \Phi^b({\bm x})-{1\over
    N}\delta^{ab} \bar{\bm \Phi}({\bm x})\cdot{\bm \Phi}({\bm x}).
  \label{qahft}
\end{equation}
Its two-point correlation function
\begin{equation}
    G_Q({\bm x},{\bm y}) = \langle {\rm Tr}\, Q({\bm x}) Q({\bm y}) \rangle,
    \label{GQdef}
\end{equation}
approaches a constant nonvanishing value for large distances (in the presence 
of symmetry-breaking boundary conditions) in the ordered phase, while 
at the critical point it behaves as 
\begin{equation}
  G_Q({\bm x},{\bm y}) 
\sim |{\bm x}-{\bm  y}|^{-2 y_q},\qquad y_q={d-2+\eta_q\over 2},
\label{gqcribeh}
\end{equation}
where $y_q$ is the RG dimension of the operator $Q$ and $\eta_q$ is the
usual critical exponent.  The approach to the critical point is
controlled by the length-scale critical exponent $\nu$, which can be
defined from the power-law divergence of the length scale $\xi$ of the
critical modes, i.e. $\xi \sim |r-r_c|^{-\nu}$, where $r$ is the
coefficient of the quadratic term in Eq.~\eqref{AHFTL} and $r_c$ its
critical value.

\subsubsection{Renormalization-group flow in 
  the perturbative $\varepsilon$-expansion}
\label{epsexpAHFT}

The RG flow of the SU($N$)-symmetric AHFT field theory has been
determined close to four dimensions in the $\varepsilon$-expansion
framework. The RG scaling functions that are related with the
diverging renormalization constants have been determined
perturbatively [perturbative calculations are usually performed using
  the Lorenz gauge fixing as in Eq.~(\ref{partfunc})], in powers of
$\varepsilon\equiv 4-d$~\cite{WK-74,HLM-74}.  In the perturbative
calculations dimensional regularization and the minimal-subtraction
(MS) renormalization scheme are generally used, see, e.g.,
Refs.~\cite{ZJ-book,PV-02}.  The corresponding MS $\beta$ functions
associated with the renormalized couplings $\alpha\equiv g^2$ and $u$
have been computed to four loops~\cite{IZMHS-19}.  The FPs of the RG
flow are obtained from the common zeroes of the $\beta$ function.
Their stability is controlled by the eigenvalues of the matrix
$\Omega_{ij} = \partial_{g_i} \beta_j$ (where $g_{i}$ indicates the
Lagrangian couplings, and $\beta_i$ the corresponding $\beta$
function)~\cite{ZJ-book,PV-02}: A FP is stable if all eigenvalues
$\lambda_i$ of $\Omega$, computed at the FP, are positive.

Here we report the one-loop $\beta$ functions~\cite{HLM-74}
\begin{eqnarray}
 \beta_\alpha \equiv \mu {\partial \alpha \over \partial \mu}
 = - \varepsilon \alpha + N \alpha^2,\qquad
 \beta_u \equiv \mu {\partial u\over \partial \mu}
 = - \varepsilon u + 
 (N+4) u^2 - 18 u \alpha + 54 \alpha^2.
 \label{betafuncAH}
\end{eqnarray}
The normalizations of the renormalized couplings $\alpha$ and $u$ have
been chosen to simplify the formulas (they can be easily inferred from
the above expressions).  The analysis of the one-loop $\beta$
functions shows that  a stable FP exists for
\begin{equation}
  N > N_4 = 90 + 24\sqrt{15} \approx 183.
  \label{nstar4dU1}
 \end{equation} 
It is given by
\begin{eqnarray}
  \alpha^* = {\varepsilon \over N},\qquad
  u^*  =  {N+18 + \sqrt{N^2-180N-540}\over 2 N(N+4)}\,\varepsilon.
\label{fipo}
\end{eqnarray}
We refer to this FP as {\em charged} FP, because $\alpha^*$ is
nonzero, thus implying nontrivial critical correlations of the gauge
field.

The value $N^*$ above which the theory has a stable FP
depends on the dimension $d$, thus, in the $\varepsilon$ expansion approach 
one defines a function $N^*(d)$ such that a
stable FP exists only for $N > N^*(d)$, with $N^*(4) = N_4$.
3D AH models may undergo a charged continuous transition
only if $N> N^*(3)$.  The critical number of components $N^\star(d)$
has been determined to four loops~\cite{IZMHS-19}:
\begin{equation}
N^*(4-\varepsilon) = N^*(4)\left[1 - 1.752 \,\varepsilon +
  0.789\, \varepsilon^2 + 0.362
  \,\varepsilon^3+O(\varepsilon^4)\right] .
\label{nceps}
\end{equation}
The determination of $N^*(3)$ from the asymptotic expansion
(\ref{nceps}) is quite difficult because of the large coefficients.
Moreover, the perturbative expansions for gauge theories are not Borel
summable, because of the presence of renormalons, see, e.g.,
Refs.~\cite{Lautrup-77,Parisi-78,Parisi-79a,Parisi-79b,David-83}, so
it is not clear whether the resummation methods ~\cite{ZJ-book} that
are successfully used for $\Phi^4$ theories\footnote{The
$\varepsilon$-expansion for $\Phi^4$ theories is conjectured to be
Borel summable \cite{ZJ-book}. } may provide reasonably accurate
estimates. In spite of these difficulties, by means of a resummation
method that makes the reasonable assumption that $N^*(d)$ vanishes
linearly for $d\to 2$ (this is suggested by the $D=2+\varepsilon$
perturbative results of Refs.~\cite{Hikami-79,March-Russell-92}),
Ref.~\cite{IZMHS-19} obtained $N^*(3) = 12(4)$ in three dimensions,
which confirms the absence of a stable FP---and therefore, of
continuous transitions---for small values of $N$.  The
field-theoretical estimate is in reasonable agreement with the
estimate~\cite{BPV-21-ncAH,BPV-22} $N^*(3) = 7(2)$ obtained by
numerical analyses of the LAH models, see also
Sec.~\ref{multicomphadia}.\footnote{A similar behavior has been
observed ~\cite{SZJSM-23} in quantum square-lattice SU($N$)
antiferromagnets at the transition between the SU($N$) N\'eel phase
and the valence-bond solid phase.  The numerical FSS analyses of the
R\'enyi entanglement entropy reported in Ref.~\cite{SZJSM-23} are
apparently compatible with the CFT predictions appropriate for
continuous transitions only for $N\ge 8$, leading to the conclusions
that systems with $N\le 7$ undergo (weak) first-order transitions.
The $N$-component AHFT (\ref{partfunc}) is a candidate field theory
for the critical behavior at these transitions, see, e.g.,
Refs.~\cite{RS-90,KS-12,Kaul-12,BMK-13,SZJSM-23}.  If this is the
case, then the results of Ref.~\cite{SZJSM-23} confirm the absence of
a FP for small values of $N$.  }

It is interesting to observe that, to all orders of perturbation theory, 
the $\beta$ function of the gauge coupling 
$\alpha$ can be expressed in terms of the anomalous dimension $\eta_A$ of the 
scalar field in the Lorenz gauge~\cite{HT-96,ZJ-book,BPV-23-gaufix}:
\begin{equation}\label{AH_beta_eta}
  \beta_\alpha= \alpha\, [d-4+\eta_A(\alpha,u)],\qquad
    \eta_A(\alpha,u) = \frac{\mathrm{d}\log Z_A}{\mathrm{d}\log\mu},
\end{equation}
where $Z_A$ is the MS renormalization constant associated with the
gauge field.  Therefore, at the charged FP ($\alpha^* \not=0$) the FP
condition $\beta_\alpha = 0$ implies the exact
relation~\cite{HT-96,Herbut-book}
\begin{equation}
\eta_A \equiv \eta_A(\alpha^*,u^*)= 4 - d.
\label{etaA-FT}
\end{equation}
In particular, $\eta_A = 1$ in three dimensions. 

The RG flow also presents an {\em uncharged} O($2N$)-symmetric FP with
vanishing gauge coupling, with $\alpha^*=0$ and $u^* = u^*_{{\rm
    O}(2N)}= \varepsilon/(N+4)$.  This FP exists for any $N$,
including $N=1$, and it is always unstable against gauge fluctuations.
Indeed, its stability matrix $\Omega_{ij} = \partial \beta_i/\partial
g_j$ has a negative eigenvalue $\lambda_\alpha$ that satisfies the
exact relation
\begin{equation}
\lambda_\alpha = 
\left. {\partial \beta_\alpha/\partial \alpha} \right|_{\alpha=0}
= - \varepsilon,
\label{lambdaresAH}
\end{equation}
to all orders in perturbation theory~\cite{PV-19-AH3d}.  Therefore,
the addition of U(1) gauge interactions is a relevant RG perturbation
of the ungauged O($2N$)-vector theory, with RG dimension
$y_\alpha=-\lambda_\alpha=1$ in three dimensions. As a consequence,
for small gauge couplings we always expect crossover effects with an
apparent O($2N$) behavior~\cite{ZJ-book, AM-book, Cardy-book,
  BPV-24-cAH}, independently of the existence of the stable charged
FP, which is only relevant for the eventual asymptotic behavior.

\subsubsection{Role of the gauge fixing}
\label{gaugefixAHFT}

As already mentioned above, the Lorenz gauge-fixing term ${\cal
  L}_{\rm gf}$ in Eq.~(\ref{partfunc}) is necessary to obtain a
well-defined functional path integral, and, in particular, a
well-defined perturbative expansion, see, e.g., Ref.~\cite{ZJ-book}.
Appropriate Ward identities guarantee that the gauge-invariant
observables do not depend on the parameter $\zeta$ appearing in the
gauge-fixing term ${\cal L}_{\rm gf}$. In particular, the MS
perturbative $\beta$ functions $\beta_\alpha$ and $\beta_u$, the
corresponding FPs, and the critical exponents are independent of
$\zeta$.  However, as discussed in Sec.~\ref{ncLAHMsec}, and, in more
detail, in
Refs.~\cite{KK-85,KK-86,BN-86,BN-86b,BN-87,BPV-23-gaufix,BPV-23-chgf},
the gauge fixing plays an important role also in lattice AH models. In
particular, different behaviors are observed for positive values of
$\zeta$ (in this case we speak of a soft Lorenz gauge fixing) and in
the limiting case $\zeta = 0$ (hard Lorenz gauge fixing). For
instance, only in the second case the nonlocal gauge-invariant charge
operators that condense in the low-temperature Higgs phase admit a
local field representation~\cite{BPV-23-gaufix,BPV-23-chgf}.  To
better understand the role of the gauge fixing in the lattice theory,
it is useful to discuss the RG flow of the Lorenz gauge parameter in
the corresponding AHFT.

As a consequence of the Ward identities that follow from gauge
invariance, which imply that the gauge-fixing term ${\cal L}_{\rm gf}$
in Eq.~(\ref{partfunc}) does not renormalize, see,
e.g., Ref.~\cite{ZJ-book}, the $\beta$ function of the parameter
$\zeta$ takes the simple form
\begin{equation}
  \beta_\zeta \equiv \mu {\partial \zeta\over \partial \mu}=
  - \zeta \, \eta_A(\alpha,u).
  \label{betazeta}
\end{equation}
Eq.~(\ref{betazeta}) shows that the value $\zeta = 0$ represents a FP
of the RG flow in the theory with a Lorenz gauge fixing. Moreover,
since $\eta_A > 0$, this FP is unstable. If we start the RG flow from
$\zeta > 0$, then $\zeta$ increases towards infinity, and the
fluctuations of the nongauge-invariant modes are expected to become
unbounded in this limit: nongauge-invariant correlations are expected
to become disordered on large scales as in the absence of a gauge
fixing. Therefore, we expect to be able to observe a critical behavior
in the nongauge invariant sector only for $\zeta = 0$, i.e., in the
presence of a hard Lorenz gauge fixing. This is confirmed by the
numerical results discussed in Refs.~\cite{BPV-23-gaufix,BPV-23-chgf}.

\subsubsection{Three-dimensional
  Abelian Higgs field theory in the large-$N$ limit}
\label{largeexp}

The existence of a critical transition for finite values of
$\varepsilon$, and in particular in three dimensions, and for
sufficiently large values of $N$ is confirmed by nonperturbative (with
respect to the couplings) computations in the large-$N$
limit~\cite{HLM-74,IKK-96,MZ-03,KS-08}.  These calculations provide
$1/N$ expansions of some critical exponents.  The correlation-length
exponent $\nu$ and the exponent $\eta_q$ defined in
Eq.~(\ref{gqcribeh}) are given by~\cite{HLM-74}:
\begin{eqnarray}
\nu &=& 1 - \frac{48}{\pi^2 N} + O(N^{-2}),
\label{nulargen} \\
\eta_q &=& 1 - \frac{32}{\pi^2 N} + O(N^{-2}).
\label{etalargen}
\end{eqnarray}
In the presence of a gauge fixing one can also consider correlation
functions that are not gauge invariant. For example, we may consider
the correlation function $G_\Phi({\bm x},{\bm y})\equiv \langle
\bar{\bm \Phi}({\bm x})\cdot{\bm \Phi}({\bm y}) \rangle$ of the
Lagrangian field ${\bm \Phi}({\bm x})$.  The anomalous dimension of
${\bm \Phi}({\bm x})$ depends on the gauge-fixing parameter $\zeta$
and is given by~\cite{HLM-74,KS-08}
\begin{equation}
  \eta_\phi = - {20+8\zeta\over \pi^2} {1\over N} + O(N^{-2}).
\label{etazlargeNzeta}
\end{equation}
As already discussed in Sec.~\ref{gaugefixAHFT}, only in the presence
of a hard Lorenz gauge fixing ($\zeta = 0$) nongauge invariant
correlations can become critical. In particular, as we shall see in
Sec.~\ref{nc-orderparameters}, only in this case the Higgs phase can
be characterized by the condensation of the scalar field, reflecting
the condensation of a nonlocal gauge-invariant charged operator. For
$\zeta=0$ we obtain
\begin{equation}
  \eta_\phi = - {20\over \pi^2} {1\over N} + O(N^{-2})
  \quad {\rm for}\;\;\zeta=0,
\label{etazlargeN}
\end{equation}
which allows one to compute the RG dimension of $\Phi$ in the hard
Lorenz gauge: $y_\phi = (1+\eta_\phi)/2$.

\subsection{Abelian Higgs field theories with multiparameter scalar potentials}
\label{exAHFT}

We can also consider AHFTs with more general scalar
potentials~\cite{Hikami-80,MU-90,BPV-23-mpp}, that are invariant under
a subgroup of SU($N$), that is still large enough to ensure the
presence of a single quadratic invariant term. If this were not the
case, the model would describe multicritical behaviors, as already
discussed in Sec.~\ref{LGWappro}.  In Secs.~\ref{OAHFT} and
\ref{fpsec} we consider SO($N$) invariant models. Further extensions
are presented in Sec.~\ref{fiextAHFT}.

\subsubsection{SO($N$)-symmetric Abelian Higgs field theories: 
Definition of the model}
\label{OAHFT}

The simplest extension of the SU($N$)-invariant AHFT 
is obtained by adding the quartic term $|{\bm\Phi}\cdot {\bm\Phi}|^2$
to the Lagrangian (\ref{AHFTL}). We obtain
\begin{eqnarray}
  {\cal L}_{\rm O} = \frac{1}{4 g^2} \sum_{\mu\nu} F_{\mu\nu}^2
  + \overline{D_\mu{\bm\Phi}} \cdot D_\mu{\bm\Phi}
  + r\, \bar{\bm\Phi} \cdot {\bm\Phi} +
   u \,(\bar{\bm\Phi} \cdot
  {\bm\Phi})^2 + v \,|{\bm\Phi}\cdot {\bm\Phi}|^2,
    \label{OAHFTL}
\end{eqnarray}
with $u\ge 0$ and $u+v\ge 0$, to guarantee the stability of the
potential. For $N=1$ the two quartic scalar terms are equivalent, thus
one recovers the standard one-component AHFT. For $N>1$,
the added term preserves the gauge invariance, but it breaks the
global SU($N$) symmetry, making the theory invariant under SO($N$)
transformations only. This symmetry group is large
enough to guarantee that $\bar{\bm\Phi} \cdot {\bm\Phi}$ is the only
quadratic invariant term allowed by the global and local symmetries.

To determine the possible ordered phases of the model with $N\ge 2$,
one can use the mean-field approximation,
since field fluctuations are only relevant along the
transition lines. This amounts to 
analyzing the minima of the scalar
potential ~\cite{BPV-23-mpp}. One finds two ordered phases, with
different symmetry, depending on the sign of $v$.
For $v < 0$, the ordered phase is invariant under SO($N-1$) transformations,
so the 
symmetry-breaking pattern at the transition 
is the same as that of the O($N$) vector
model, 
\begin{equation}
  {\rm SO}(N)\to {\rm SO}(N-1)\qquad {\rm for}\;\; v<0.
  \label{vs0symbr}
\end{equation}
For $v > 0$ the ordered phase is invariant under SO($N-2$) transformations,
thus leading to the symmetry-breaking pattern 
\begin{equation}
  {\rm SO}(N)\to {\rm SO}(N-2)\qquad {\rm for}\;\; v>0.
  \label{vg0symbr}
\end{equation}
Of course, for $v=0$ the model is SU($N$) invariant and thus the ordered phase 
is U($N-1$) invariant. The symmetry-breaking pattern is given in 
Eq.~(\ref{symbrUN}).

The appropriate order parameters for the two
different types of transitions are~\cite{BPV-23-mpp}
\begin{eqnarray}
  R^{ab}({\bm x}) = {\rm Re}\,Q^{ab}({\bm x}),\qquad T^{ab}({\bm x}) =
  {\rm Im}\,Q^{ab}({\bm x}),
  \label{wqoperft}
\end{eqnarray}
which transform under different representations of the SO($N$) group. Here
$Q^{ab}$ is the bilinear operator defined in Eq.~(\ref{qahft}).

As a final remark, it is important to note that the analysis of the possible 
phases and symmetry-breaking patterns only relies on the structure of the 
scalar potential. Therefore, it applies to any lattice model 
with SO($N$) global invariance, independently of the presence or
absence of gauge fields. Also the nature of the transition plays no role.

\subsubsection{SO($N$)-symmetric Abelian Higgs field theories: 
Renormalization-group flow}
\label{fpsec}

To determine the possible charged transitions in SO($N$) invariant
models, one can study the RG flow in the MS renormalization scheme
close to four dimensions, as it was done for the SU($N$) symmetric
AHFT in Sec.~\ref{epsexpAHFT}.  The one-loop $\beta$ functions
associated with the renormalized couplings $\alpha$, $u$, and $v$, are
given by~\cite{Hikami-80,MU-90}
\begin{eqnarray}
&&\beta_\alpha = - \varepsilon \alpha
  + N \alpha^2,
  \label{betafuncOAH}\\
&&\beta_u = - \varepsilon u + (N+4) u^2 + 4 u v + 4 v^2 - 18 u \alpha
  + 54 \alpha^2, \nonumber \\ &&
\beta_v = - \varepsilon v + N v^2 + 6
  v u -18 v \alpha.\nonumber
\end{eqnarray}
The normalizations of the renormalized couplings $\alpha, u,v$ have
been again chosen to simplify the formulas (they can be easily
inferred from the above expressions).  For large values of $N$ there
is a stable FP, with $\alpha^* \approx {\varepsilon/N}$, $u^* \approx
{\varepsilon/N}$, and $v^* \approx {\varepsilon/N}$. More precisely, a
stable charged FP exists for $N>N_o^*(d)$, where~\cite{MU-90}
\begin{equation}
N_o^*(d) = N_o^*(4) +
O(\varepsilon)\quad {\rm with} \quad N_o^*(4) \approx 210.
\label{Nostar}
\end{equation}
Note that $v^*>0$, so this FP is only relevant for charged
transitions characterized by the symmetry-breaking pattern given in
Eq.~(\ref{vg0symbr}).  There are no higher-order computations, that
allow one to determine $N_o^*(d)$ in $d$ dimensions and, in
particular, for $d=3$.  However, by analogy with the SU($N$)-symmetric
case, one expects $N_{o}^*(d=3)$ to be of order ten
~\cite{BPV-23-mpp}.

These results allow us to predict the critical behavior of the
transitions in generic SO($N$)-invariant AH systems, when gauge fields
are critical at the transition.  If the symmetry-breaking pattern is
the one given in Eq.~(\ref{vs0symbr}), so that the system is
effectively described by the SO($N$)-symmetric AHFT with $v < 0$, no
continuous transitions with critical gauge correlations are expected
to occur. If, instead, the symmetry-breaking pattern is the one
obtained for $v > 0$, see Eq.~(\ref{vg0symbr}), then the behavior
depends on the number of components of the scalar field.  For
$N>N_o^*(d=3)$, continuous charged transitions are possible, provided
the lattice system is effectively inside the attraction domain of the
stable charged FP. For smaller values of $N$, instead, transitions are
of first order.

The RG analysis reported above shows that SO($N$)-symmetric AH systems
may undergo charged continuous phase transitions for sufficiently
large values of $N$. Critical exponents in the large-$N$ limit can be
computed using the field-theoretical $1/N$ expansion.
Ref.~\cite{Hikami-80} computed the correlation-length exponent $\nu$
in a more general non-Abelian gauge theory with local O($M$) and
global O($N$) invariance, which is equivalent to the SO($N$)-symmetric
AHFT for $M=2$.  Using the results of Ref.~\cite{Hikami-80} for $M=2$
one obtains
\begin{equation}
\nu = 1 - {176 \over 3 \pi^2 N} + O(N^{-2}).
  \label{nulnoAHFT}
\end{equation}

\subsubsection{Further extensions of the Abelian Higgs field theories}
\label{fiextAHFT}

More general quartic scalar potential, with smaller global symmetry group,
have also been considered. For instance, one may consider the model
with Lagrangian~\cite{MU-90}
\begin{eqnarray}
  {\cal L}_{P} = \frac{1}{4 g^2} \,F_{\mu\nu}^2 +
  \overline{D_\mu{\bm\Phi}} \cdot D_\mu{\bm\Phi}
  + r\, \bar{\bm \Phi} \cdot {\bm \Phi} + 
  u  \,(\bar{\bm\Phi} \cdot {\bm\Phi})^2 
+ v \,|{\bm\Phi}\cdot {\bm\Phi}|^2 
+ w \,\sum_{a=1}^N (\bar\Phi^a\Phi^a)^2,
\label{AHFTP}
\end{eqnarray}
which is invariant  under the permutation group of $N$ elements ${\mathbb
  P}_N$.  For $N=2$ the quartic potential appearing in  the
Lagrangian (\ref{AHFTP}) is the most general one that is gauge invariant
and guarantees the uniqueness of the quadratic $\bar{\bm
  \Phi}\cdot {\bm\Phi}$ term.  For $N>2$ there are other quartic terms
satisfying these conditions. For instance, one can add
$\sum_{a=1}^N \bar\Phi^a \Phi^a
\bar\Phi^{a+1} \Phi^{a+1}$
(with the identification $\Phi^{N+1} = \Phi^1$), which
is only invariant under global ${\mathbb Z}_N$ transformations.

A one-loop $O(\varepsilon)$ analysis of the RG flow of the 
${\mathbb P}_N$-symmetric AHFT (\ref{AHFTP}) was reported in
Ref.~\cite{MU-90}, showing that, close to four dimensions, a stable FP
is present only for $N\ge 5494$.

\subsection{Summary}
\label{conclurem}

The analysis of the RG flow in the SU($N$) and SO($N$) invariant AHFTs
shows that a stable charged FP exists for sufficiently large $N$
values.  The existence of these FPs implies that statistical lattice
systems with the same local and global symmetries may undergo
continuous charged transitions, where both scalar and gauge fields are
critical.  In particular, SU($N$)-symmetric and O($N$)-symmetric AHFTs
provide the effective description of charged continuous transitions
characterized by the symmetry-breaking patterns ${\rm SU}(N) \to {\rm
  U}(N-1)$ and SO($N$)$\to$SO($N-2$), respectively. As we shall see,
such transitions are indeed observed in LAH models with noncompact
gauge variables~\cite{BPV-21-ncAH,BPV-23-mpp}, along the transition
line separating the Coulomb and Higgs phases. They are characterized
by the condensation of the local gauge-invariant bilinear tensor
operators defined above and also of nonlocal charged vector
operators~\cite{BPV-23-chgf,BPV-24-ncAH}.

%% file: ncLAHM.tex
\section{Lattice Abelian Higgs models with noncompact gauge variables}
\label{LAHMnc}

In this section we discuss the phase diagram and critical behavior of
3D lattice Abelian Higgs (LAH) models in which a noncompact Abelian
gauge field is coupled with an $N$-component scalar field.  We review
results for a specific nearest-neighbor model, but the main features
of the phase diagram and the universal properties of the transitions
are expected to be the same in a large class of lattice gauge models
characterized by a local U(1) gauge invariance.  We first consider LAH
models with global SU($N$) symmetry, then we discuss the
SO($N$)-symmetric case.

We present their phase diagrams, with several
phases, including Higgs phases, and their critical behavior along
the transition lines separating the different phases.  Different
scenarios emerge in the case of one-component and multi-component
scalar fields.  The observed critical behaviors can be interpreted 
using the approaches outlined in Sec.~\ref{gaugesymtr}.  In
particular, the one-component models undergo  topological
transitions belonging to the IXY universality class, see
Sec.~\ref{sec.IXY}, separating their Coulomb and Higgs
phases. Multi-component models show a more complex phase diagram,
arising from the nontrivial interplay between the global SU($N$) or
SO($N$) symmetry and the local U(1) gauge symmetry.  In particular,
for a sufficiently large number of scalar components the 
transitions along the line that separates the Coulomb and Higgs phases
are continuous and associated with the stable FP of 
the RG flow of the corresponding AHFT. We
also emphasize the role of the Lorenz gauge fixing, which allows one 
to probe the 
condensation of extended charged excitations in the Higgs phase.

\subsection{$N$-component SU($N$)-invariant lattice Abelian Higgs model}
\label{ncLAHMsec}

\subsubsection{Definition of the model}

In LAH models defined on a 3D cubic lattice the fundamental fields are
$N$-component unit-length complex vectors ${\bm z}_{\bm x}$ ($\bar{\bm
  z}_{\bm x} \cdot {\bm z}_{\bm x} =1$) defined on the lattice sites
${\bm x}$, and noncompact gauge variables $A_{{\bm x},\mu}\in {\mathbb
  R}$ ($\mu=1,2,3$) defined on the lattice links.  The simplest
nearest-neighbor lattice model is defined by
~\cite{KK-85,BN-86,BN-87,MV-04,BPV-21-ncAH}\footnote{We set $T=1$ and
rescale the coupling $J$ by a factor of $N$, to obtain a finite
$N\to\infty$ limit at fixed $J$, see, e.g., Ref.~\cite{MZ-03}.}
\begin{eqnarray}
  Z = \int [dA_{{\bm x},\mu} d\bar{\bm z}_{\bm x} d{\bm z}_{\bm x}]
e^{-H({\bm A},{\bm z})},\qquad
H({\bm A},{\bm z}) =
\frac{\kappa}{2} \sum_{{\bm x},\mu>\nu} F_{{\bm x},\mu\nu}^2
  - 2 N J \sum_{{\bm x},\mu} {\rm Re}\,( \lambda_{{\bm x},\mu} \,
  \bar{\bm z}_{\bm x} \cdot {\bm z}_{{\bm x}+\hat\mu}),  
  \qquad
  \label{ncLAHM}
\end{eqnarray}
where $\hat{\mu}$ is the unit spatial vector in the lattice direction
$\mu$, and
\begin{equation}
  \lambda_{{\bm x},\mu} = e^{iA_{{\bm x},\mu}}, \qquad
  F_{{\bm
    x},\mu\nu}= \Delta_{\mu} A_{{\bm x},\nu} - \Delta_{\nu}
A_{{\bm x},\mu}, \qquad \Delta_\mu A_{{\bm x},\nu} = A_{{\bm
    x}+\hat{\mu},\nu}- A_{{\bm x},\nu}.
\label{ncgaudef}
\end{equation}
The coupling $\kappa>0$ plays the role of inverse (square) gauge
coupling.  The model is invariant under the global transformations
${\bm z}_{\bm x} \to V {\bm z}_{\bm x}$ with $V\in\mathrm{SU}(N)$ and
under the local transformations
\begin{eqnarray}
  {\bm z}_{\bm x} \to e^{i\Lambda_{\bm x}} {\bm z}_{\bm x},
  \qquad A_{{\bm x},\mu} \to A_{{\bm x},\mu} + \Lambda_{\bm
    x}-\Lambda_{{\bm x}+\hat{\mu}},\qquad \Lambda_{\bm x}\in \mathbb{R}.
  \label{gaugeinv}
\end{eqnarray}
Model (\ref{ncLAHM}) can be
seen as a particular, quite straightforward, lattice regularization of
the AHFT defined by the Lagrangian (\ref{AHFTL}).

We should note that the partition function of the lattice gauge model
(\ref{ncLAHM}) is not well defined. Indeed it diverges, even on a
finite lattice, because of the zero modes due to the gauge invariance
of the model.  If periodic boundary conditions are used, this problem
cannot be solved even when a maximal gauge
fixing~\cite{Creutz-76,Creutz-book} is added.  This is due to the
invariance of the Hamiltonian $H$ under the noncompact group of
transformations $A_{{\bm x},\mu}\to A_{{\bm x},\mu} + 2\pi n_{\mu}$
(where $n_{\mu}\in\mathbb{Z}$ depends on the direction $\mu$ but is
independent of the point ${\bm x}$), which is also (at least
partially) present in the gauge-fixed theory, making the partition
function $Z$ ill-defined. In particular, the averages of functions of
the noncompact gauge-invariant Polyakov operators, i.e., the sums of
the fields $A_{{\bm x},\mu}$ along nontrivial paths winding around the
lattice, are ill-defined ~\cite{BPV-21-ncAH}. To obtain a well-defined
(rigorous) definition of the partition function of the finite-size
gauge-fixed theory, without breaking translation invariance, one may
consider the so-called $C^*$ boundary conditions~\cite{KW-91,
  LPRT-16,BPV-21-ncAH}. On a cubic lattice of size $L$, $C^*$ boundary
conditions are defined by
\begin{equation}
A_{{\bm x} + L\hat{\nu},\mu}= -A_{{\bm x},\mu},\qquad
{\bm z}_{{\bm x}+L\hat{\nu}}= \bar{\bm z}_{{\bm x}}.
\label{Cstarbc}
\end{equation}
These boundary conditions preserve the local gauge invariance of the
Hamiltonian~(\ref{ncLAHM}) provided $\Lambda_{\bm x}$, defined in
Eq.~\eqref{gaugeinv}, is antiperiodic.  They break the SU($N$) global
symmetry---they are only invariant under O($N$) transformations---but,
since the breaking only occurs on the boundaries, it is irrelevant for
the bulk critical behavior.

As we shall see, some properties of the model can be best analyzed
by adding a Lorenz gauge fixing. It is defined by
requiring~\cite{BPV-23-gaufix}
\begin{equation}
{\rm Lorenz}\;{\rm gauge}:
\qquad
\sum_{\mu}\Delta_\mu^{-} A_{{\bm x},\mu}=0,\qquad
\Delta_{\mu}^{-} A_{{\bm x},\nu}= A_{{\bm
x},\nu}-A_{{\bm x}-\hat{\mu},\nu},
\label{Lgauge} 
\end{equation}
for all lattice sites $\bm x$.  In the presence of C$^*$ boundary
conditions, the above Lorenz gauge condition represents  a complete gauge
fixing as it completely breaks gauge invariance and thus it makes
gauge correlations well defined.  As demonstrated in
Ref.~\cite{BPV-23-gaufix}, the lattice Lorenz gauge fixing
(\ref{Lgauge}) has an important property. When computed in the Lorenz
gauge, Fourier-transformed gauge correlations are also well defined in
the infinite-volume limit for any nonvanishing momentum, at variance
with what happens in other gauges.  For instance, axial gauges suffer
from the existence of an infinite two-dimensional family of quasizero
modes for $L$ large. These zero modes give rise to spurious
divergences, unrelated to the presence of long-range physical
correlations~\cite{BPV-23-gaufix}.

It is important to note that in the AHFT with action (\ref{partfunc})
gauge invariance is broken by 
adding ${\cal L}_{\rm gf}\sim
\zeta^{-1} (\partial_\mu A_\mu)^2$ to the gauge-invariant Lagrangian.
In the lattice case, this 
corresponds to adding 
\begin{equation}
  {1\over 2\zeta} \sum_{{\bm x},\mu}(\Delta_\mu^{-} A_{{\bm x},\mu})^2
  \label{sgaufixlah}
 \end{equation} 
to the lattice Hamiltonian, Eq.~(\ref{ncLAHM}). We name this gauge-symmetry
breaking a soft Lorenz gauge fixing. The gauge fixing condition 
(\ref{Lgauge}) (hard Lorenz
gauge fixing) is straightforwardly recovered in the $\zeta\to 0$ limit.
As discussed in Ref.~\cite{BPV-23-gaufix}
soft gauge fixings are not appropriate for the nonperturbative
analysis of the phase behavior of the LAH model, as they suffer from
the presence of propagating unphysical longitudinal modes, which may
hide the physical signal.  Similar conclusions are reached from the
analysis of scalar correlation functions, see Ref.~\cite{BPV-23-chgf}, 
and from the RG flow of the Lorenz parameter $\zeta$ 
in field theory, see Sec.~\ref{gaugefixAHFT}.

We finally mention that in the one-component ($N=1$) model one may
integrate out the scalar field by choosing the so-called unitary
gauge, that fixes $z_{\bm x} = 1$ everywhere. This allows us to write
the Hamiltonian in terms of the gauge variables only, as
\begin{equation}
      H_{\rm ug}= 
      \frac{\kappa}{2} \sum_{{\bm x},\mu>\nu} F_{{\bm x},\mu\nu}^2
      - 2 J \sum_{{\bm x},\mu} \cos A_{{\bm x},\mu}.
\label{Hug}
\end{equation}
The unitary gauge fixing is not complete, indeed the Hamiltonian
$H_{\rm ug}$ is still invariant under gauge transformations in which
$\Lambda_{\bm x}$ is a multiple of $2\pi$. As discussed in
Sec.~\ref{sec.ZQhiggs}, this residual gauge symmetry can induce
topological transitions: The model (\ref{Hug}) represents a soft
version of the inverted XY (IXY) gauge model defined in
Sec.~\ref{sec.IXY}, which is obtained for $J\to \infty$. Indeed, in
this limit, we have $\cos A_{{\bm x},\mu} = 1$, so 
$A_{{\bm x},\mu}$ takes only values that are multiples of $2\pi$,
i.e., $A_{{\bm x},\mu} = 2 \pi n_{{\bm x},\mu}$
with $n_{{\bm x},\mu} \in \mathbb Z$.

\begin{figure}[tbp]
  \begin{center}
  \includegraphics[width=0.45\columnwidth]{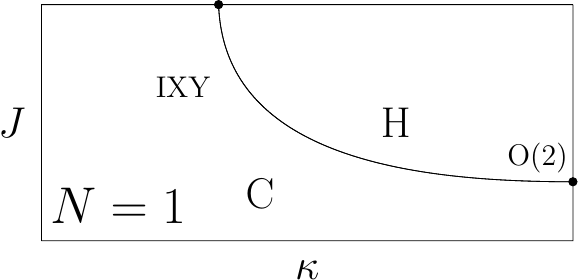}
  \hspace{1cm}
  \includegraphics[width=0.45\columnwidth]{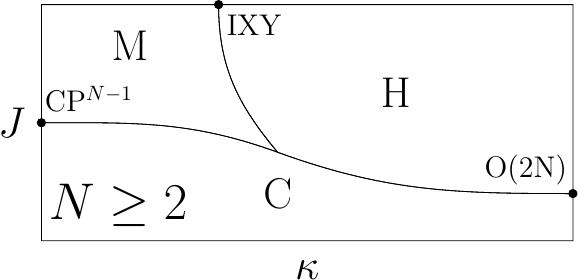}
  \caption{The $\kappa$-$J$ phase diagram of the $N$-component LAH
    model (\ref{ncLAHM}), for $N=1$ (left), and generic $N\ge 2$
    (right).  For $N=1$, there are two phases, the Coulomb (C) and
    the Higgs (H) phase, characterized by the confinement and
    deconfinement of charged gauge-invariant excitations,
    respectively.  For $N\ge 2$, the scalar field is disordered and
    gauge correlations are long ranged in the small-$J$ Coulomb (C)
    phase.  For large $J$ two phases occur, the molecular (M) and
    Higgs (H) ordered phases, in which the global SU($N$) symmetry is
    spontaneously broken.  The two phases are distinguished by the
    behavior of the gauge modes: the gauge field is long ranged in the
    M phase (small $\kappa$), while it is gapped in the H phase (large
    $\kappa$).  Moreover, while the C and M phases are confined
    phases, in the H phase charged gauge-invariant excitations are
    deconfined.}
  \label{phadiancLAH}
    \end{center}
\end{figure}

The qualitative features of the phase diagram of the model depend on
$N$, as sketched in Fig.~\ref{phadiancLAH}. For $N=1$ there are only
two phases, which differ in the topological properties of the gauge
fields and in the confinement properties of the charged nonlocal
excitations. For $N\ge 2$ the phase diagram is more complex, due to the
possibility of the spontaneous breaking of the global SU($N$)
symmetry. Note that a Higgs phase is always present for large $J$ and
$\kappa$.

For $N>1$ a line of topological transitions (it separates the M and H 
phases) is also present. The
mechanism underlying these transitions is the one discussed in
Sec.~\ref{sec.ZQhiggs}: in the broken-symmetry phase the minimum-energy
configurations
correspond (in the infinite-volume limit and up to gauge
transformations) to ${\bm z}_{\bm x}={\bm z}_{{\bm x}+\hat{\mu}}$ and
$\lambda_{{\bm x},\mu}=1$.  They are still invariant under
gauge transformations in which $\Lambda_{\bm x}$ is a multiple of
$2\pi$, exactly as for $N=1$, so we expect topological transitions analogous 
to the CH transitions for $N=1$.

\subsubsection{Order parameters}  \label{nc-orderparameters}

The critical behavior in LAH models can be characterized by using
three different types of order parameters. To characterize the Higgs
phase one can consider nonlocal charged operators--- charges are
deconfined in the Higgs phase and confined in the other phases---or
correlations of the gauge fields---they are gapped in the Higgs phase,
long-ranged in the other cases.  For $N\ge 2$, there are also
transitions where the global symmetry is spontaneously broken, which
can be signaled by appropriate gauge-invariant bilinear operators of
the scalar fields.

The condensation of the charged
excitations~\cite{KK-85,KK-86,BN-86,BN-86b,BN-87} can be monitored by
looking at the correlation functions of the nonlocal gauge-invariant
operator~\cite{Dirac-55,BPV-23-chgf}
\begin{equation}
{\bm \Gamma}_{\bm x} = {\bm z}_{\bm x} \exp\Bigl(i \sum_{{\bm y},\mu}
E_\mu({\bm y},{\bm x}) A_{{\bm y},\mu}\Bigr),
\label{phixdefmc}
\end{equation}
where 
$E_\mu({\bm y},{\bm x}) = V({\bm y} +\hat\mu,{\bm x}) - V({\bm y},{\bm x})$,
and $V({\bm y},{\bm x})$ is the lattice Coulomb potential associated
with a unit charge located at ${\bm x}$.  Note that this operator
cannot be defined for periodic boundary conditions, as $V(\bm{y},{\bm
  x})$ does not exist for these boundary conditions.  A rigorous
definition in a finite system is possible instead, if C$^*$ boundary
conditions are used.\footnote{The lattice Coulomb potential
$V({\bm x},{\bm y})$ is the solution of the lattice Poisson equation
$\sum_\mu \Delta^-_\mu \Delta_\mu V({\bm x},{\bm y}) = - \delta_{{\bm
    x},{\bm y}}$, where the lattice derivatives act on the ${\bm x}$
variable. A consistent definition of $\bm{\Gamma}_{\bm x}$ requires
the use of boundary conditions for which the lattice Poisson equation
has a unique solution. $C^*$ boundary conditions, but not periodic
boundary conditions, have this property. } The operator ${\bm
  \Gamma}_{\bm x}$ is invariant under local gauge transformations, but
transforms as ${\bm \Gamma}_{\bm x} \to e^{i\varphi} {\bm \Gamma}_{\bm
  x}$ under a global U(1) transformation ${\bm z}_{\bm x} \to
e^{i\varphi} {\bm z}_{\bm x}$, thus it behaves as a charged
gauge-invariant operator.\footnote{We are using here $C^*$ boundary
conditions, so ${\bm z}_{\bm x} \to e^{i\varphi} {\bm z}_{\bm x}$,
$A_{{\bm x},\mu} \to A_{{\bm x},\mu}$ is not a gauge transformation
(it would be for periodic boundary conditions). Indeed, if we set
$\Lambda_{\bm x} = \varphi$, $A_{{\bm x},\mu}$ transforms nontrivially
on the boundaries due the fact that $\Lambda_{\bm x}$ is
antiperiodic.}

The gauge-invariant nonlocal charged operator ${\bm \Gamma}_{\bm x}$
simplifies in the Lorenz gauge (\ref{Lgauge}).  Indeed, since
$\sum_{{\bm y},\mu} E_\mu({\bm y},{\bm x}) A_{{\bm y}, \mu} =
-\sum_{{\bm y},\mu} V({\bm y},{\bm x}) \Delta^{-}_{\mu} A_{{\bm
    y},\mu}=0$, we obtain ${\bm \Gamma}_{\bm x} = {\bm z}_{\bm x}$ and
thus
\begin{equation}
G_\Gamma({\bm x},{\bm y}) \equiv 
\langle \bar{\bm \Gamma}_{\bm x} \cdot {\bm \Gamma}_{\bm y}\rangle = 
G_z({\bm x},{\bm y}) \equiv \langle
\bar{\bm z}_{\bm x}\cdot {\bm z}_{\bm y}\rangle \qquad
({\rm Lorenz}\;\;{\rm gauge}).
\label{ggammagz}
\end{equation}
The Lorenz-gauge representation of the charged operator ${\bm
  \Gamma}_{\bm x}$ in terms of the local scalar field ${\bm z}_{\bm
  x}$ allows us to use the standard RG framework, which applies to
local
operators~\cite{Wilson-83,Fisher-74,Wegner-76,Privman-90,PV-02}. In
particular, we can predict the critical behavior of its correlation
functions. We define a critical exponent
$\eta_\Gamma$ from the critical-point relation
\begin{eqnarray}
  G_\Gamma({\bm x},{\bm y}) \sim
  {1\over |{\bm x}-{\bm y}|^{d-2+\eta_\Gamma}}.
    \label{etazdef}
\end{eqnarray}
In the Lorenz gauge $\eta_\Gamma$ can be computed from the critical
behavior of the correlation function of the local field ${\bm z}_{\bm
  x}$.

For $N=1$, in the unitary gauge 
the nonlocal scalar operator $\Gamma_{\bm x}$ becomes
\begin{equation}
 \widetilde{\Gamma}_{\bm x} = 
  \exp\Bigl(i \sum_{{\bm y},\mu} E_\mu({\bm y},{\bm x}) 
    A_{{\bm y},\mu}\Bigr),
\label{phixdef-IXY}
\end{equation}
which is invariant under gauge transformations in which $\Lambda_{\bm
  x}$ is a multiple of $2\pi$, i.e., under the discrete gauge
transformations that are appropriate for the IXY model or its soft
version (\ref{Hug}).  The operator $\widetilde{\Gamma}_{\bm x}$ is
well-defined also in the IXY model (it only depends on the gauge
fields), and therefore it represents a nonlocal order parameter for
the topological IXY transition.

To determine the critical behavior of the gauge correlations, one can
consider the gauge-invariant two-point correlation function of
$F_{{\bm x},\mu\nu}$. In the Lorenz gauge, it can be expressed in
terms of the correlation function
\begin{equation}
C_{\mu\nu}({\bm x},{\bm y}) = \langle A_{{\bm x},\mu} A_{{\bm
    y},\nu}\rangle
\label{cmmdef}
\end{equation}
of the gauge field~\cite{BPV-23-gaufix,BPV-23-chgf,BPV-24-ncAH}. At a
critical transition
\begin{eqnarray}
 C_{\mu\nu}({\bm x},{\bm y}) \sim
  {1\over |{\bm x}-{\bm y}|^{d-2+\eta_A}},
    \label{etaAdef}
\end{eqnarray}
which defines the critical exponent $\eta_A$.

Finally, in SU($N$) invariant models with $N\ge 2$,
to characterize the spontaneous breaking of the 
global symmetry, one can consider the 
gauge-invariant composite operator
\begin{equation}
  Q_{{\bm x}}^{ab} = \bar{z}_{\bm x}^a z_{\bm x}^b -\delta^{ab}/N,
  \label{QdefncAH}
\end{equation}
which transforms in the adjoint representation of the SU($N$) global
symmetry group. Its critical properties can be determined by studying 
the correlation function
\begin{equation}
  G_Q({\bm x},{\bm y})=\langle \,\mathrm{Tr}\,Q_{\bm x}Q_{\bm y}\,\rangle,
  \label{Qcorr}
\end{equation}
which, at the critical point, scales as 
\begin{eqnarray}
 G_Q({\bm x},{\bm y}) \sim {1\over |{\bm x}-{\bm y}|^{d-2+\eta_q}},
    \label{etaQdef}
\end{eqnarray}
with the critical exponent $\eta_q$.

\subsection{Phase diagram and critical behavior of
  the one-component lattice Abelian Higgs model}
\label{onecompphadia}

\subsubsection{The Coulomb and Higgs phases}

As shown in the left panel of Fig.~\ref{phadiancLAH}, the
one-component LAH model presents only two phases: a Coulomb (C) phase,
in which gauge correlators are gapless, and a Higgs (H) phase in which
gauge correlators are gapped; see, e.g., Ref.~\cite{BN-87}.  They can
also be characterized by the confinement/deconfinement of charged
gauge-invariant excitations~\cite{KK-85,KK-86,BN-87}, as we discuss
below.  The C and H phases are separated by a transition line
connecting the transition points occurring in the $J\to\infty$ and
$\kappa\to\infty$ limits, where the noncompact LAH model becomes
equivalent to the 3D IXY model discussed in Sec.~\ref{sec.IXY} and to
the standard O(2)-vector spin model, respectively.  As discussed in
Ref.~\cite{BPV-24-ncAH} and reviewed below, the critical behavior
along the whole CH line is the same as that of the IXY model.

\subsubsection{IXY critical behavior along the Coulomb-Higgs transition line}
\label{cboncLAHM}

To characterize the behavior of the LAH model along the CH line, we
first discuss the nature of the transitions in the two limiting cases,
$J\to\infty$ and $\kappa\to\infty$.

For $J\to\infty$ scalar fields obey the relation ${\bm{z}}_{\bm x} =
\lambda_{{\bm x},\mu}\, {\bm z}_{{\bm x}+\hat\mu}$.  Iterating this
condition along the sides of a plaquette we obtain
\begin{equation}
\lambda_{{\bm x},{\mu}} \,\lambda_{{\bm
    x}+\hat{\mu},{\nu}} \,\bar{\lambda}_{{\bm x}+\hat{\nu},{\mu}}
\,\bar{\lambda}_{{\bm x},{\nu}} = 1.
\label{largeJcons}
\end{equation}
By an appropriate gauge transformation, one can then set $A_{{\bm x},
  \mu} = 2 \pi n_{{\bm x}, \mu}$, where $n_{{\bm x}, \mu} \in {\mathbb
  Z}$, thus obtaining the IXY model, see Sec.~\ref{sec.IXY}.
Therefore, for $J\to\infty$ 3D LAH models undergo an IXY transition
located at $\kappa_c(J\to\infty) =
0.076051(2)$~\cite{NRR-03,BPV-21-ncAH}.

In the limit $\kappa\to\infty$, all plaquettes $F_{{\bm x},\mu\nu}$
vanish. Thus, in infinite volume we can set $A_{{\bm x},\mu} = 0$
everywhere up to a gauge transformation, obtaining the XY
model. Therefore, the one-component LAH model is expected to undergo
an XY transition at~\cite{Hasenbusch-19,CHPV-06,DBN-05}
$J_c(\kappa\to\infty)=0.22708234(9)$ for $\kappa\to\infty$ [this XY
  transition is denoted by O(2) in Fig.~\ref{phadiancLAH}].  Note
that, as discussed in Sec.~\ref{sec.IXY}, the continuous transitions
at the two endpoints have a different nature and belong to two
different universality classes, even if related by duality
\cite{DH-81,NRR-03}.  Indeed, duality transformations only relate
thermal quantities.

The behavior along the CH line has been discussed in
Ref.~\cite{BPV-24-ncAH}, finding that continuous transitions along the
CH line belong to the same universality class as that of the IXY
model, obtained for $J\to\infty$.  Note that it is not possible to
have finite-$\kappa$ transitions with the same critical features as
those of the spin XY model obtained in the limit
$\kappa\to\infty$. Indeed, gauge interactions are a relevant
perturbation of the spin XY model.  As discussed in
Sec.~\ref{epsexpAHFT}, the uncharged O(2) FP associated with the spin
XY universality class is unstable with respect to a perturbation
proportional to the gauge coupling [the corresponding eigenvalue of
  the stability matrix is negative, see Eq.~(\ref{lambdaresAH})].

\begin{figure}[tbp]
  \begin{center}
  \includegraphics[width=0.65\columnwidth]{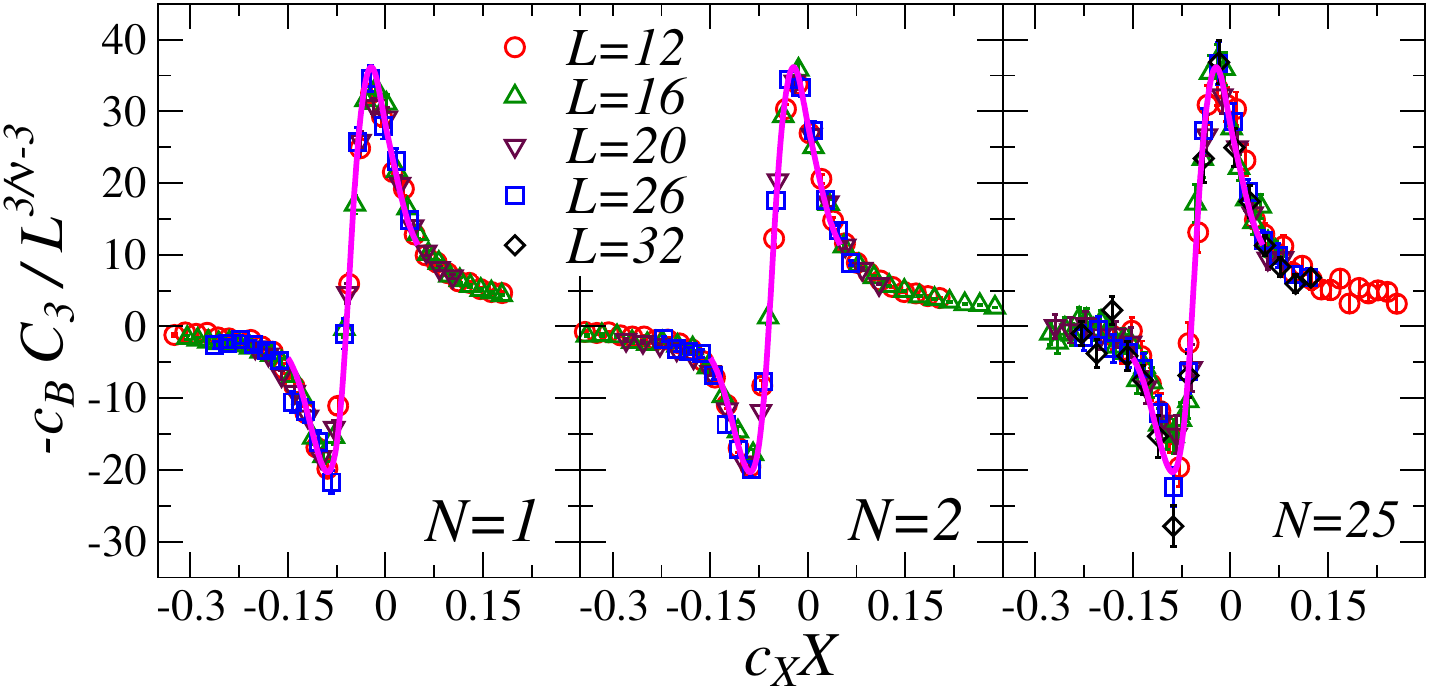}
  \caption{(Adapted from Ref.~\cite{BPV-24-ncAH}) Scaling plot of the
    third energy cumulant, defined by $C_3=-L^{-3}\langle (H-\langle
    H\rangle)^3\rangle$, for $N=1$ (data across the CH line at fixed
    $J=1$), $N=2$ (data across the MH line at fixed $J=1$), and $N=25$
    (data across the MH line at fixed $J=0.4$), from left to right.
    The FSS behavior of $C_3$ is discussed in \ref{fssencum}, see
    Eq.~\eqref{cksca}.  We plot $c_B C_3 L^{3-3/\nu}$ versus $c_X X$,
    where $X=(\kappa - \kappa_{c}) L^{1/\nu}$, setting $\nu=\nu_{\rm
      XY}\approx0.6717$, where $\nu_{\rm XY}$ is the XY value of the
    critical exponents (see \ref{univclass} for a list of estimates of
    $\nu_{\rm XY}$) and $\kappa_c$ is the critical-point value. The
    nonuniversal constants $c_X$ and $c_B$ are fixed by requiring the
    data to match the IXY curve (solid line). See
    Ref.~\cite{BPV-24-ncAH} for details.  The nice data collapse
    confirms that all transitions belong to the IXY universality
    class.}
  \label{B3figs}
  \end{center}
\end{figure}

The IXY behavior of the transitions is confirmed by the FSS behavior
of the energy cumulants~\cite{BPV-24-ncAH,NRR-03}, see also
\ref{fssencum}. Results for the third energy cumulant are shown in the
left panel of Fig.~\ref{B3figs}.  Data nicely scale if the XY critical
exponent $\nu_{\rm XY}=0.6717(1)$ is used.  Moreover, the scaling
curves match the curve obtained for the IXY model, reported as a
solid line in the left panel of Fig.~\ref{B3figs}.  These results
confirm that all CH transitions belong to the IXY, or gauge XY,
universality class.

\begin{figure}[t]
\begin{center}
\includegraphics[width=0.325\columnwidth]{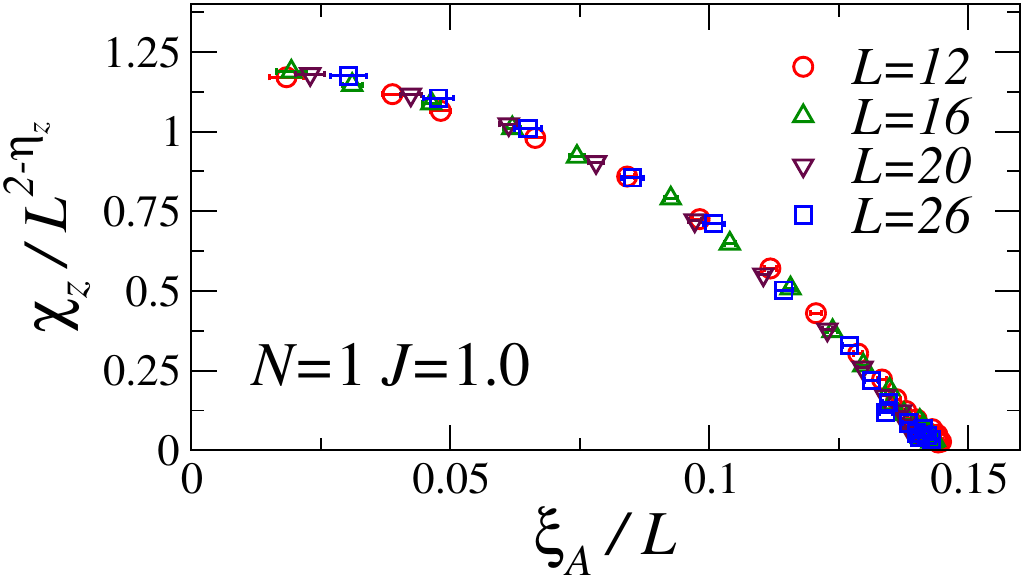}  
\includegraphics[width=0.325\columnwidth]{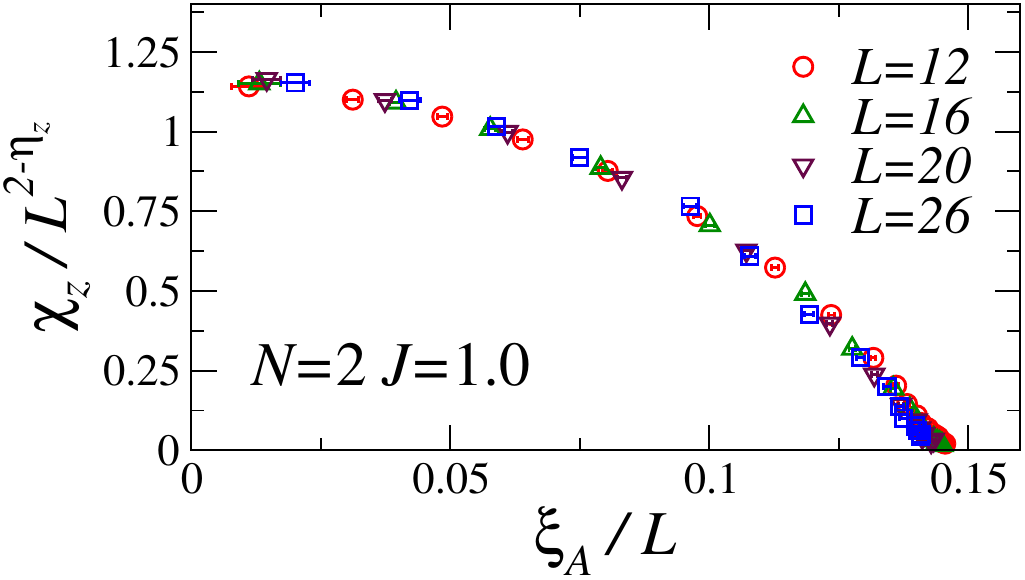} 
\includegraphics[width=0.325\columnwidth]{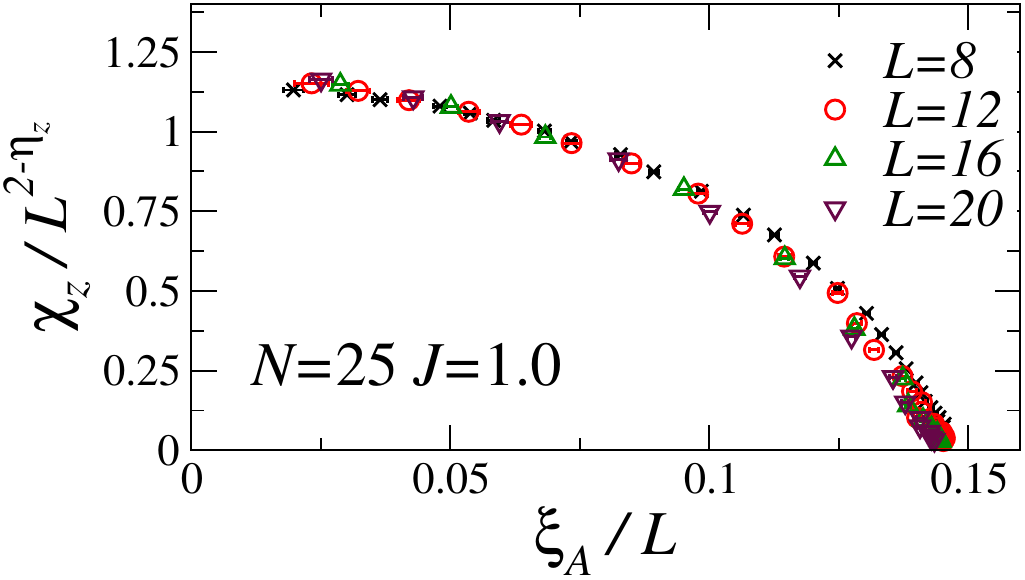} 
\caption{(Adapted from Ref.~\cite{BPV-24-ncAH}). Scaling plot of the
  susceptibility $\chi_z = V^{-1} \sum_{{\bm x},{\bm y}}G_z({\bm
    x},{\bm y})$, [$G_z$ is defined in \eqref{ggammagz}] as a function
  of the RG invariant ratio $\xi_A/L$, where $\xi_A$ is the
  second-moment correlation length $\xi_A$ defined in terms of the
  correlation function $C_{\mu\nu}({\bm x},{\bm y})$ defined in
  Eq.~(\ref{cmmdef}) with $\mu=\nu$. All correlation functions are
  computed in the Lorenz gauge.  Results at fixed $J=1$ for: a) $N=1$
  (CH line); b) $N=2$ (MH line); c) $N=25$ (MH line); see the phase
  diagram in Fig.~\ref{phadiancLAH}. We report $\chi_z/L^{2-\eta_z}$
  against $\xi_A/L$ [see the FSS relation, Eq.~\eqref{chisclaw2}]
  using $\eta_z = -0.74$. All data approach the same curve, apart from
  a multiplicative normalization and hence all systems belong to the
  same universality class.
\label{chizvsRxiA} 
}
\end{center}
\end{figure}

Ref.~\cite{BPV-24-ncAH} also studied the critical behavior of the
charged operator $\Gamma_{\bm x}$ and of the gauge field $A_{{\bm
    x},\mu}$. A numerical analysis of the susceptibility $\chi_z = V^{-1}
\sum_{{\bm x},{\bm y}} G_z({\bm x},{\bm y})$ in the Lorenz gauge
provides the estimate~\cite{BPV-24-ncAH}
\begin{equation}
\eta_\Gamma=\eta_z=-0.74(4), \qquad y_\Gamma = {d-2+\eta_\Gamma\over
  2}=0.13(2),
\label{etazest}
\end{equation}
where $y_\Gamma$ is the RG dimension that can be associated with the
nonlocal operator $\Gamma_{\bm x}$.  A scaling plot is reported in the
left panel of Fig.~\ref{chizvsRxiA}.

The RG anomalous dimension $\eta_A$ of the gauge field has been
determined in Ref.\cite{BPV-24-ncAH}, by performing a FSS analysis of
the gauge correlation function $C_{\mu\nu}({\bm x},{\bm y})$ defined
in Eq.~(\ref{cmmdef}) in the Lorenz gauge.  The numerical FSS analyses
\cite{BPV-24-ncAH} confirm the exact result $\eta_A = 1$ that can be
proved in the IXY model.\footnote{This exact result follows from
the correspondence between the zero-momentum $F_{{\bm x},\mu\nu}$
correlation function in the IXY model and the helicity modulus
$\Upsilon$ computed in the dual XY model~\cite{NRR-03}, and from the
fact that $\Upsilon$ scales as $L^{-1}$ in the XY model~\cite{FBJ-73},
see Ref.~\cite{BPV-24-ncAH} for details.}

Note that the critical behavior of the charged scalar correlations
along the CH line differs from that of the scalar correlations
in the XY model obtained for $\kappa\to\infty$, see
Fig.~\ref{phadiancLAH}.  Indeed, charged scalar correlations are
characterized by the critical exponent $\eta_\Gamma\approx -0.7$ along
the CH line at finite $\kappa$, definitely different from the XY
value~\cite{PV-02} $\eta_z\approx 0.038$ obtained for the limiting
model at $\kappa=\infty$.  Obviously, in a finite volume the
Lorenz-gauge correlations of the field ${\bm z}_{\bm x}$ along the CH
line converge to the scalar correlations of the XY model for
$\kappa\to\infty$.  However, this result does not imply that their
asymptotic infinite-size behavior is the same, since the
$\kappa\to\infty$ and the $L\to \infty$ limits do not commute: as
discussed in Sec.~\ref{epsexpAHFT} the uncharged XY FP, which controls
the critical behavior for $\kappa\to\infty$, is unstable with respect
to gauge fluctuations, giving rise to a nonanalytic crossover
behavior.

It is worth noting that the CH transitions are driven by the
condensation of the two local fields $z_{\bm x}$ and $A_{{\bm x},\mu}$
if one considers the LAH model in the Lorenz gauge. This allows us to
reinterpret the IXY transition as a conventional transition in a model
without gauge invariance---the equivalence holds in the Lorenz
gauge---with {\em local} order parameters. However, the effective
model requires an additional XY scalar field: the topological
properties of the IXY transition are now the result of the nontrivial
interplay between the field $A_{{\bm x},\mu}$ and the additional XY
scalar field $z_{\bm x}$ in the gauge-fixed $N=1$ LAH model.

We finally remark that, since the RG flow of the AHFT in the
$\varepsilon$-expansion does not show any evidence of the existence of
a stable charged FP for $N=1$, see Sec.~\ref{epsexpAHFT}, the IXY
critical behavior along the CH transition line is apparently unrelated
with the RG flow of the AHFT. However, as discussed above, the IXY
model has an effective description in terms of a model which appears
to be equivalent to the AHFT in the Lorenz gauge, with the same field
content and global symmetries.  To reconcile this interpretation of
the IXY transition with the results of Sec.~\ref{epsexpAHFT}, one may
conjecture that, for $N=1$, the 3D AHFT RG flow admits a stable FP
that is not smoothly connected with the large-$N$ and
$\varepsilon$-expansion regimes: as $N$ or the dimension $d$
increases, the FP apparently disappears, so it cannot be identified in
four dimensions or for large $N$.  This field-theoretical
interpretation is also consistent with the exact result $\eta_A=1$ for
the IXY universality class, since $\eta_A = 1$ holds at any charged FP
of the 3D AHFT, see Eq.~(\ref{etaA-FT}).  An analogous conjecture has
been put forward for other statistical systems, for example, for some
frustrated spin systems. In that case 3D field-theoretical high-order
perturbative analyses provided evidence of stable FPs not analytically
connected with those appearing close to four dimensions and in the
large-$N$ limit of the 3D theory~\cite{PV-02,CPPV-04,DPV-04}.
Field-theoretical approaches appropriate to describe the 3D critical behavior of
$N=1$ systems (superconductors) were also discussed in
Refs.~\cite{Kleinert-89,FH-96,KNS-02,KN-03,MHS-02}.

\subsection{Phase diagram and critical behavior of the
  multicomponent lattice Abelian Higgs model}
\label{multicomphadia}
 
\subsubsection{The Coulomb, Molecular, and Higgs phases}
\label{cmhphasesmc}

The $\kappa$-$J$ phase diagram of the LAH models for $N\ge 2$ (see
Refs.~\cite{MV-08,KMPST-08,BPV-21-ncAH}) is sketched in the right
panel of Fig.~\ref{phadiancLAH}. The model is invariant under SU($N$)
global transformations. Thus, transitions associated with the breaking
of the SU($N$) symmetry and phases characterized by standard, i.e.,
nontopological, order can be present.

The phase diagram is characterized by three different phases.  For
small $J$ there is a Coulomb (C) phase, which is SU($N$) symmetric and
in which the gauge field is gapless and charged scalar modes are
confined. For large $J$ values there are two phases in which the
SU$(N)$ symmetry is broken.  They are characterized by the different
behavior of the gauge and nonlocal charged modes. In the molecular (M)
phase the gauge field is long ranged and scalar charges are confined
(as in the C phase), while in the Higgs (H) phase gauge fields are
gapped and charges are deconfined, as it occurs in the one-component
model.

The existence of these three phases is consistent with the analysis of
the model behavior for $\kappa=0$, $J\to\infty$, and
$\kappa\to\infty$.  For $\kappa=0$ the LAH model reduces to the
lattice CP$^{N-1}$ model defined in Eq.~(\ref{altcpnmod}), which
undergoes a continuous O(3)-vector transition for $N=2$ and
discontinuous transitions for $N\ge 3$~\cite{PV-19-CP,PV-20-largeN}.
For $J\to\infty$ the model behaves as the one-component LAH model.
Indeed, for $J\to\infty$ the relation (\ref{largeJcons}) holds for any
$N$, therefore the gauge field $A_{{\bm x},\mu}$ takes only values
which are multiples of $2\pi$ in all cases.  As a consequence, for
$J\to\infty$ we have an IXY transition at $\kappa=\kappa_{c,IXY} =
0.076051(2)$~\cite{NRR-03,BPV-21-ncAH}, independently of $N$.  For
$\kappa=\infty$ all plaquettes vanish and the model reduces to the
standard O(2$N$) vector model, up to a gauge transformation (more
precisely, partition function and gauge-invariant correlations are the 
same).\footnote{This
equivalence allows us to estimate the value of the $J$ where the CH
line ends for $\kappa = \infty$: $J_c(\kappa\to\infty) =
0.23396363(6)$ for $N=2$~\cite{Hasenbusch-22}, and
$J_c(\kappa\to\infty) = J_{c,\infty} + a_{1} N^{-1} + O(N^{-2})$ in
the large-$N$ limit, with $J_{c,\infty}=0.252731...$ and $a_1 \approx
-0.234$~\cite{CPRV-96}.}

The three phases of the multicomponent LAH model are separated by
three different transition lines, the CM, CH, and MH transition lines,
which start from the transition points located at $\kappa=0$,
$J=\infty$, and $\kappa=\infty$, and are expected to meet at one point,
as shown in Fig.~\ref{phadiancLAH}. In particular, the phase diagram
in the large-$N$ limit is expected to have  a simple
shape~\cite{BPV-24-ncAH}: The transition lines separating the C,
M, and H phases become straight lines: the MH line is given by 
$\kappa = \kappa_c(J=\infty)$, while the CM and CH lines are given by
$J = J_c(\kappa=0) = J_c(\kappa\to\infty)$.

The three different transition lines can be characterized by using the
order parameters defined in Sec.~\ref{nc-orderparameters}.  The CM
transition line is uniquely determined by the spontaneous breaking of
the global SU($N$) symmetry. The appropriate order parameter is the
gauge-invariant composite operator defined in
Eq.~(\ref{QdefncAH}). This order parameter is also relevant along the
CH line, along which the global symmetry also breaks, but not along
the MH line, which is purely topological and separates two phases with
broken global symmetry.

The Higgs phase is a phase in which charges deconfine and gauge fields
are gapped.  Thus, the nonlocal gauge-invariant charged vector
operator~\cite{BPV-23-gaufix,BPV-23-chgf,BPV-24-ncAH} ${\bm
  \Gamma}_{\bm x}$ defined in Eq.~(\ref{phixdefmc}) is critical along
both the CH and the MH line. Analogously, along these two lines also
gauge correlation functions (for instance, the correlation function
defined in Eq.~(\ref{cmmdef}) computed in the Lorenz gauge) are
critical.

The transitions along the CM, CH, and MH lines may be of first order or
continuous and, in the latter case, belong to universality classes
that may depend on the number $N$ of scalar components.  The
continuous transitions are related to the stable (charged or
uncharged) FP of the RG flow of the corresponding effective theory,
each one with its own attraction domain in the model parameter space.

\begin{figure}[tbp]
\begin{center}
  \includegraphics[width=0.45\columnwidth]{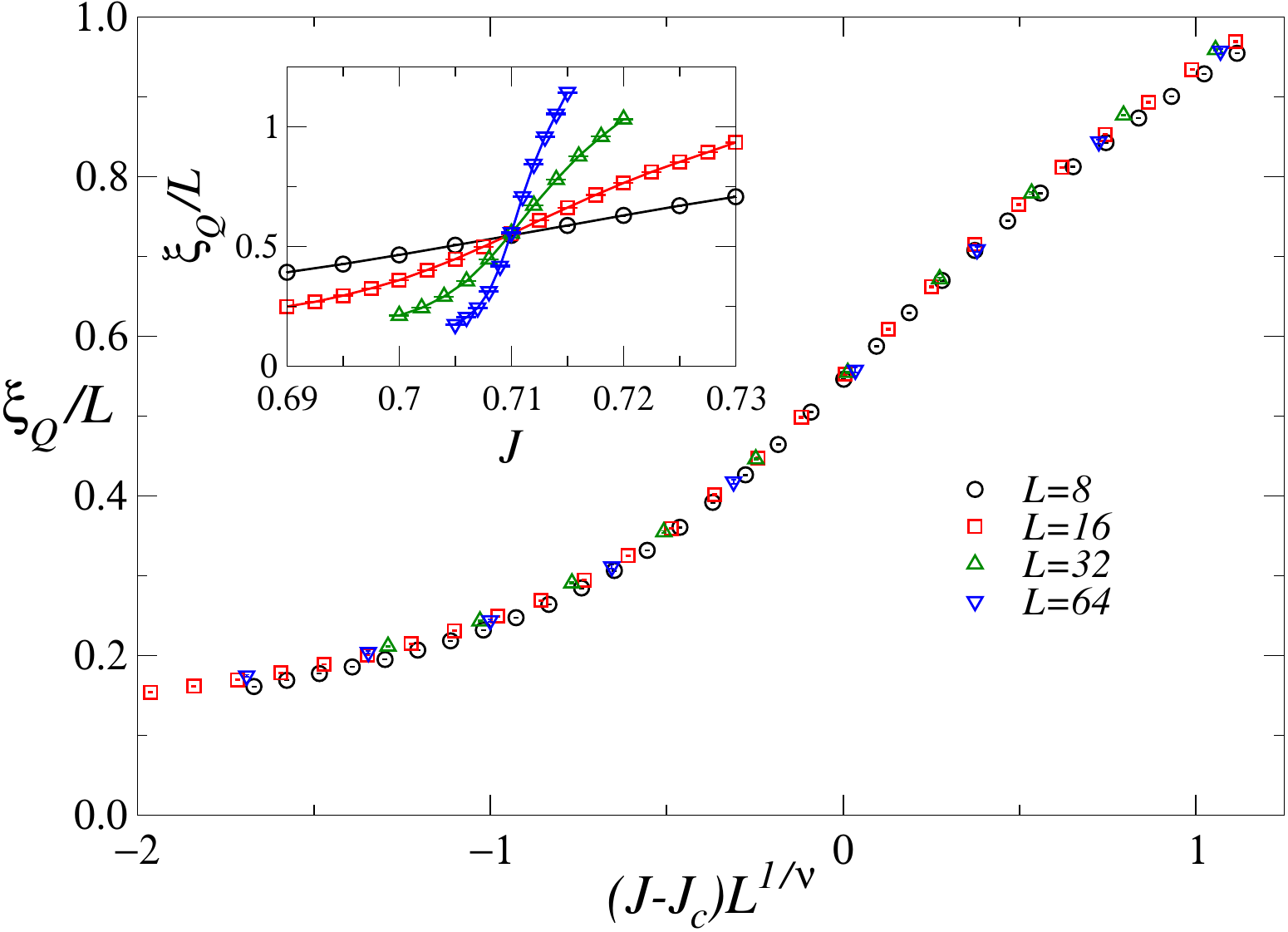}
  \quad
\includegraphics[width=0.45\columnwidth]{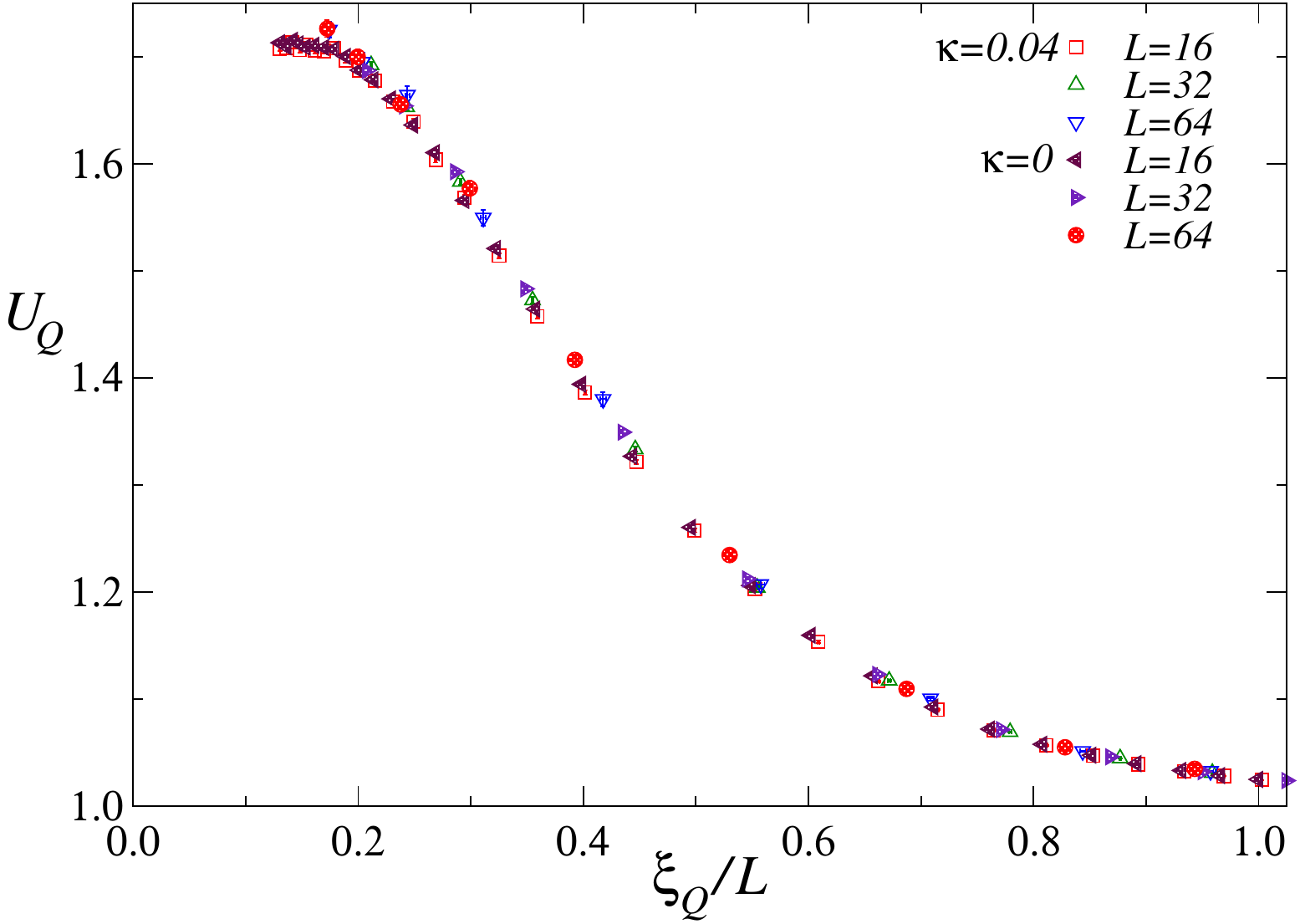} 
\caption{(Adapted from Ref.~\cite{BPV-21-ncAH}).  Numerical results
  for $N=2$ across the CM line.  (Left) Plot of $\xi_Q/L$, where
  $\xi_Q$ is the second-moment correlation length defined in terms of
  the correlation function (\ref{Qcorr}) (the FSS behavior of
  $\xi_Q/L$ is discussed in \ref{fsscorrfunc}), as a function of
  $(J-J_c)L^{1/\nu}$ (as a function of $J$ in the inset), with
  $J_c=0.7099$ and $\nu=\nu_{\mathrm{O(3)}}=0.7117$.  Data are
  obtained by varying $J$ for fixed $\kappa=0.04$.  The good collapse
  of the data provides evidence that the transition belongs to the
  O(3) universality class, as expected.  (Right) Plot of the Binder
  parameter $U_Q$ of the operator $Q_{\bm x}^{ab}$ [defined in
    Eq.~\eqref{QdefncAH}] as a function of the ratio $\xi_Q/L$ (the
  expected FSS behavior is discussed in \ref{fssrginv}); we report
  data obtained by varying $J$ for fixed $\kappa=0.04$ and $\kappa =
  0$.  The excellent collapse of the data for $\kappa=0$ and
  $\kappa=0.04$ onto the same curve provides a robust evidence that
  both transitions belong to the same O(3) universality
  class. \label{nccp1_kappa0p04}}
  \end{center}
\end{figure}

\subsubsection{The Coulomb-Molecular transition line}
\label{cmmmc}

Along the small-$\kappa$ CM line separating the C and M phases, see
the right panel of Fig.~\ref{phadiancLAH}, we expect the transitions
to have the same nature as the transition at $\kappa=0$.  For $\kappa
= 0$, the LAH model reduces to the lattice formulation of the 3D
CP$^{N-1}$ model that has been extensively discussed in
Sec.~\ref{cpnmodels}.  Accordingly, along the CM line only
gauge-invariant scalar modes are expected to be relevant, giving rise
to an {\em uncharged} critical behavior. The critical features of the
CM transitions can thus be discussed by using the same LGW model
appropriate for the 3D CP$^{N-1}$ model, see Sec.~\ref{cpnmodels}.

As discussed in Sec.~\ref{cpnmodels}, for $N\ge 3$ only first-order
transitions are possible.  Continuous transitions are possible for
$N=2$. In this case they belong to the Heisenberg universality class.
This is supported by the FSS analyses of numerical results shown in
Fig.~\ref{nccp1_kappa0p04}.

\subsubsection{The Coulomb-Higgs transition line}
\label{chmc}

We now discuss the nature of the transitions along the large-$\kappa$
line separating the Coulomb and Higgs phases. Numerical studies show
that they can be of first order or continuous, depending on the number
$N$ of components of the scalar
field~\cite{MV-08,KMPST-08,BPV-21-ncAH}.  In particular,
Ref.~\cite{BPV-21-ncAH} observed numerically continuous transitions
for $N=10,15,25$ and first-order transitions for $N=4$.  These results
indicate that the transitions along the CH line can be continuous for
$N > N_L^*$, and of first order in the opposite case, with
$N_L^*=7(2)$.

\begin{figure}[tbp]
\begin{center}
  \includegraphics[width=0.45\columnwidth]{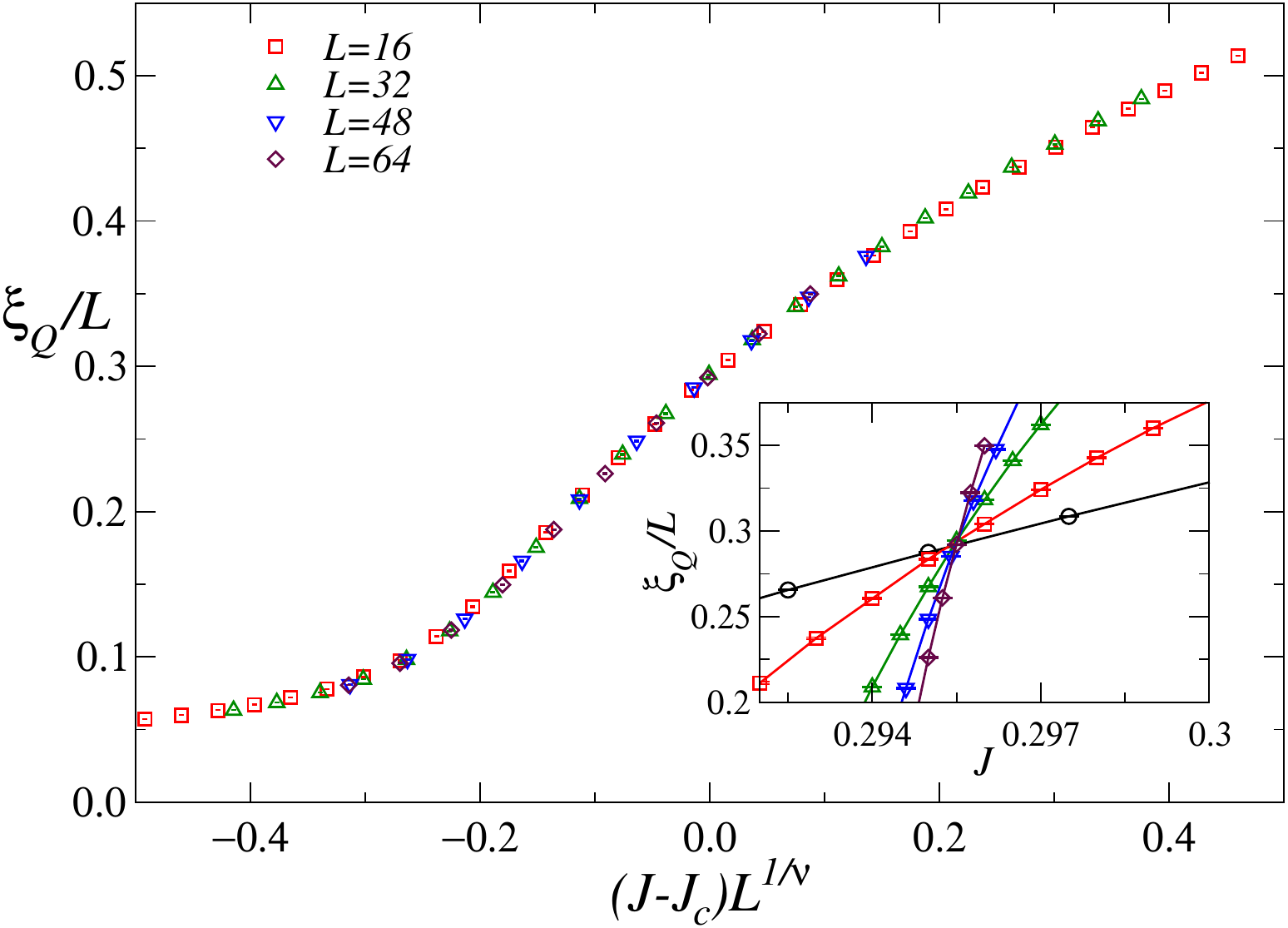}
  \quad
\includegraphics[width=0.45\columnwidth]{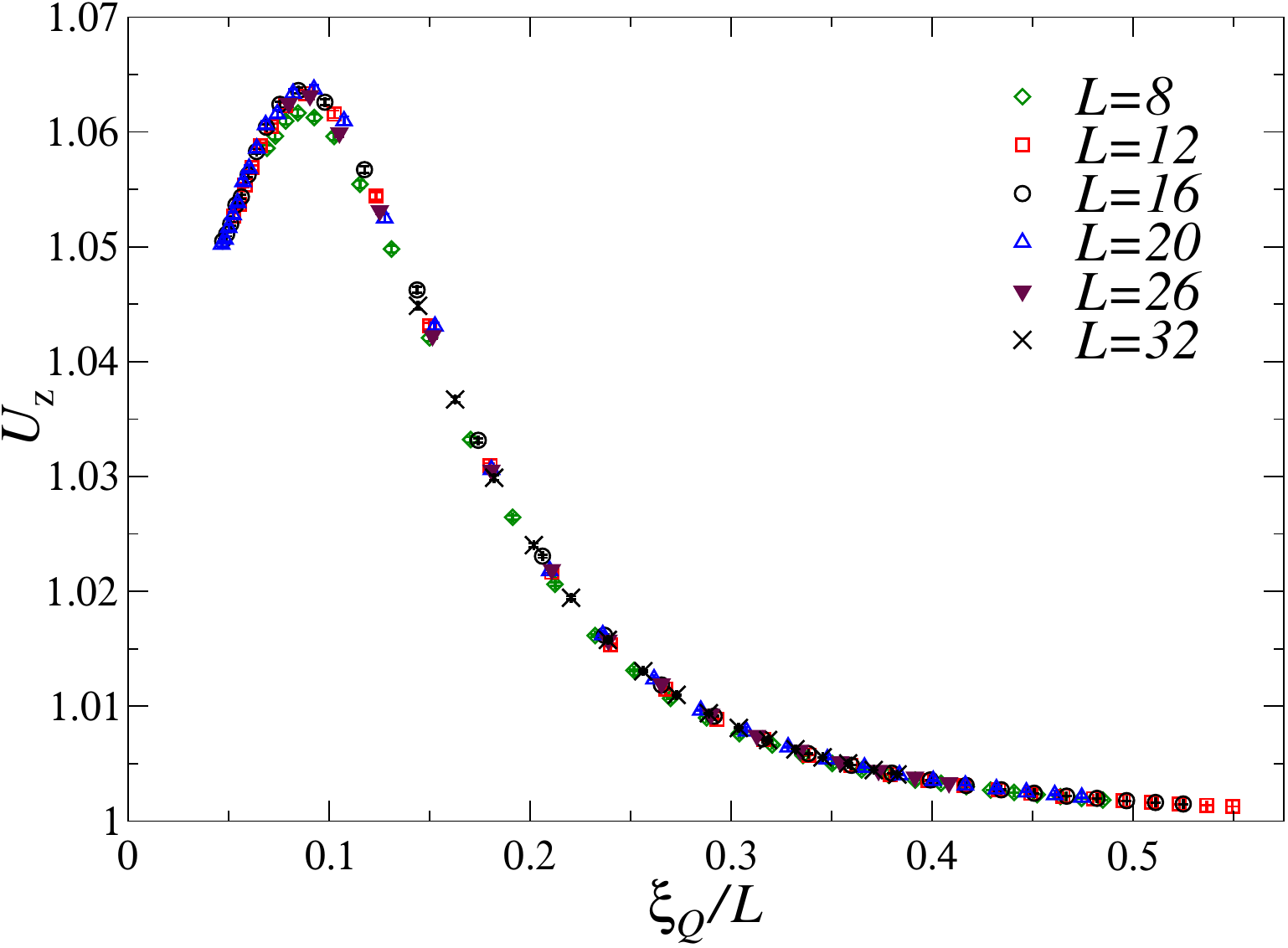} 
  \caption{Data for $N=25$ obtained by varying $J$ across the CH
    transition line, keeping $\kappa=0.4$ fixed.  (Left)(Adapted from
    Ref.~\cite{BPV-21-ncAH}) Plot of $\xi_Q/L$, where $\xi_Q$ is the
    second-moment correlation length determined from the two-point
    correlation function of $Q_{\bm x}^{ab}$, defined in
    Eq.~\eqref{QdefncAH}; in the main panel data are plotted versus of
    $(J-J_c)L^{1/\nu}$, with $J_c=0.2955$ and $\nu=0.802$ (the
    best-fit estimate), in the inset versus $J$.  (Right) (Adapted
    from Ref.~\cite{BPV-23-chgf}) Plot of the Binder parameter $U_z$
    of the scalar field ${\bm z}_{\bm x}$ computed in the Lorenz gauge
    as a function of the ratio $\xi_Q/L$, where $\xi_Q$ is defined
    above (see \ref{fssrginv} for a discussion of the expected FSS
    behavior).  The excellent data collapse signals that both the
    gauge-invariant correlation function $G_Q$ and the correlation
    function $G_z$ (computed in the Lorenz gauge) are critical along
    the CH transition line.
  \label{nccp24_kappa0p4}}
  \end{center}
\end{figure}

At the continuous CH transitions scalar, charged, and gauge modes are all
critical.  As an example, in Fig.~\ref{nccp24_kappa0p4} we report
the results~\cite{BPV-21-ncAH,BPV-23-chgf} 
of FSS analyses of numerical data (see \ref{fssapp} for a 
summary of the relevant FSS relations used in the numerical analysis),
that provide
evidence of a continuous transition for $N=25$ with critical scalar
modes.  For instance, the left panel of Fig.~\ref{nccp24_kappa0p4}
shows the scaling behavior of the correlation length $\xi_Q$ computed
in terms of the correlation function defined in Eq.~(\ref{Qcorr}),
while the right panel shows the scaling behavior of the Binder
parameter of the scalar field ${\bm z}_{\bm x}$ versus the RG
invariant ratio $\xi_Q/L$, computed in the Lorenz gauge (\ref{Lgauge})
(in this gauge the correlations of the local field ${\bm z}_{\bm x}$ are 
equal to
those of the nonlocal gauge-invariant charged operator ${\bm
  \Gamma}_{\bm x}$ defined in Eq.~(\ref{phixdefmc}), see
Sec.~\ref{nc-orderparameters}).  The excellent quality of the data collapse
confirms that the transition is continuous and that
correlations of both $Q_{\bm x}$ and ${\bm \Gamma}_{\bm x}$ are
critical. Similar results have been obtained for the gauge-field
correlation functions.

It is important to note that the choice of the hard Lorenz gauge
fixing (\ref{Lgauge}) turns out to be crucial to observe the
condensation of the scalar field ${\bm z}_{\bm x}$.  If, instead of
imposing the Lorenz constraint (\ref{Lgauge}), we add the gauge-fixing
term (\ref{sgaufixlah}) to the Hamiltonian (soft
Lorenz gauge fixing), then scalar-field correlations are always short
ranged and the gauge-field correlations show the typical Coulomb
behavior, for any positive value of the gauge parameter $\zeta$,
irrespective of the phase one is considering.  As already noted in
Sec.~\ref{gaugefixAHFT}, the hard Lorenz gauge fixing, which is obtained in 
the limit $\zeta \to 0$, is also singled out by the AHFT. Since $\zeta = 0$ 
is an unstable FP of the RG flow, see
Eq.~(\ref{betazeta}), if we start the RG flow from any finite
$\zeta>0$, then $\zeta$ increases towards infinity, so the
nongauge-invariant modes become unbounded. The instability of 
the $\zeta = 0$ FP, implies a
nontrivial crossover behavior for small values of $\zeta$,
characterized by a length scale $\xi_\zeta\sim \zeta^{-1}$.
This prediction has been verified numerically
in Ref.~\cite{BPV-23-chgf} for the noncompact LAH model.

The simultaneous criticality of gauge, scalar, and charged
correlations indicates that CH transitions are charged
transitions. Therefore, one expects them to be controlled by the
stable charged FP of the AHFT~\cite{BPV-21-ncAH}. Thus, one may
identify $N_L^*$ with the number $N^*$ of scalar components above
which the RG flow of the $N$-component AHFT admits a stable FP in
three dimensions,\footnote{Only the relation $N_L^*\ge N^*$ is
rigorously true, since for $N^*\le N< N^*_L$ the phase transition in
the lattice model could be outside the attraction domain of the stable
AHFT charged FP.} see Sec.~\ref{epsexpAHFT}. This conjecture has been
checked by comparing the numerical estimates of the critical exponents
along the CH transition line (obtained by numerical FSS analyses of MC
data~\cite{BPV-21-ncAH,BPV-23-chgf}), with the $1/N$ results reported
in Sec.~\ref{largeexp} that have been computed in the AHFT.

\begin{figure}[tbp]
\begin{center}
  \includegraphics[width=0.45\columnwidth]{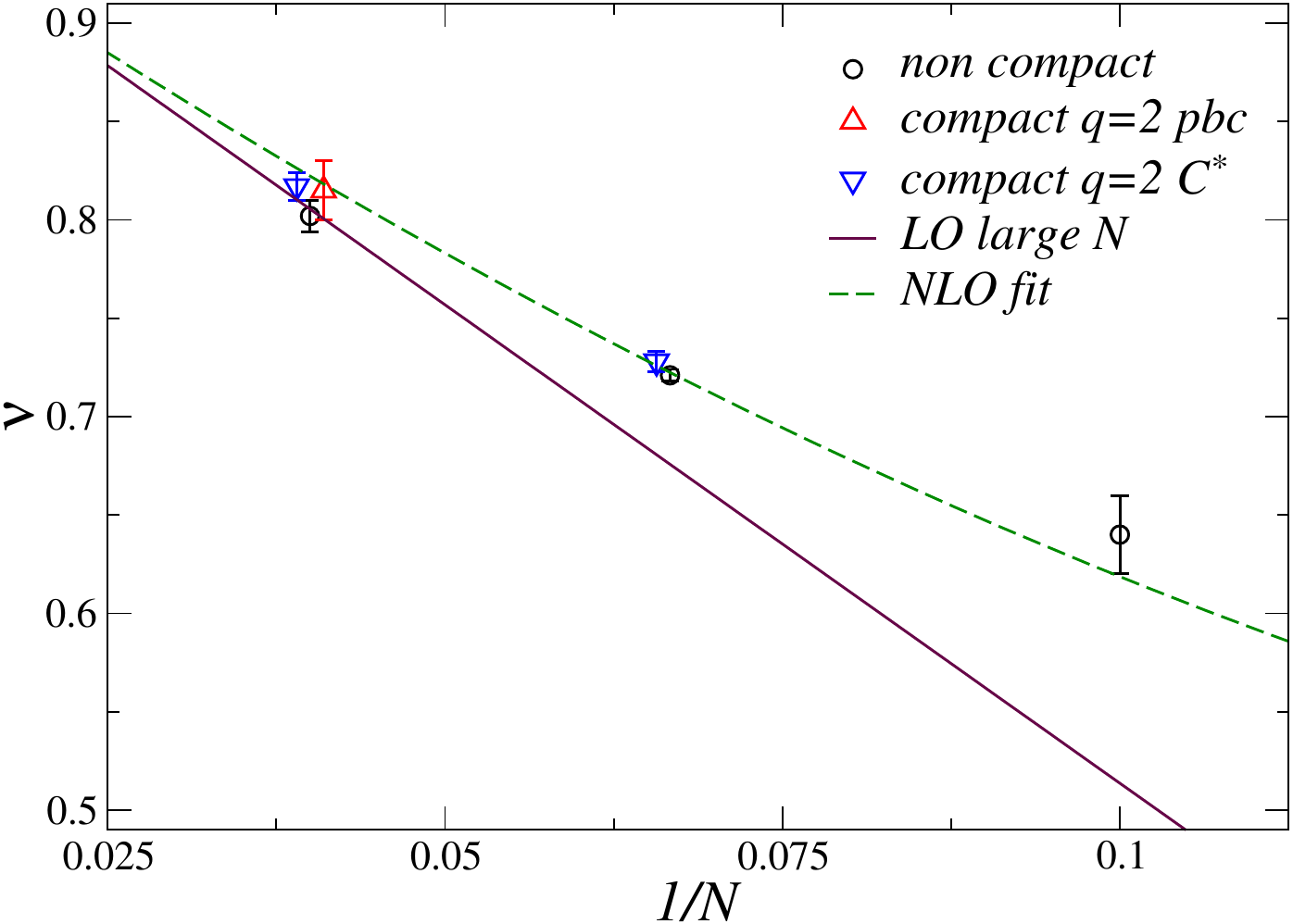}
  \hspace{1cm}
  \includegraphics[width=0.45\columnwidth]{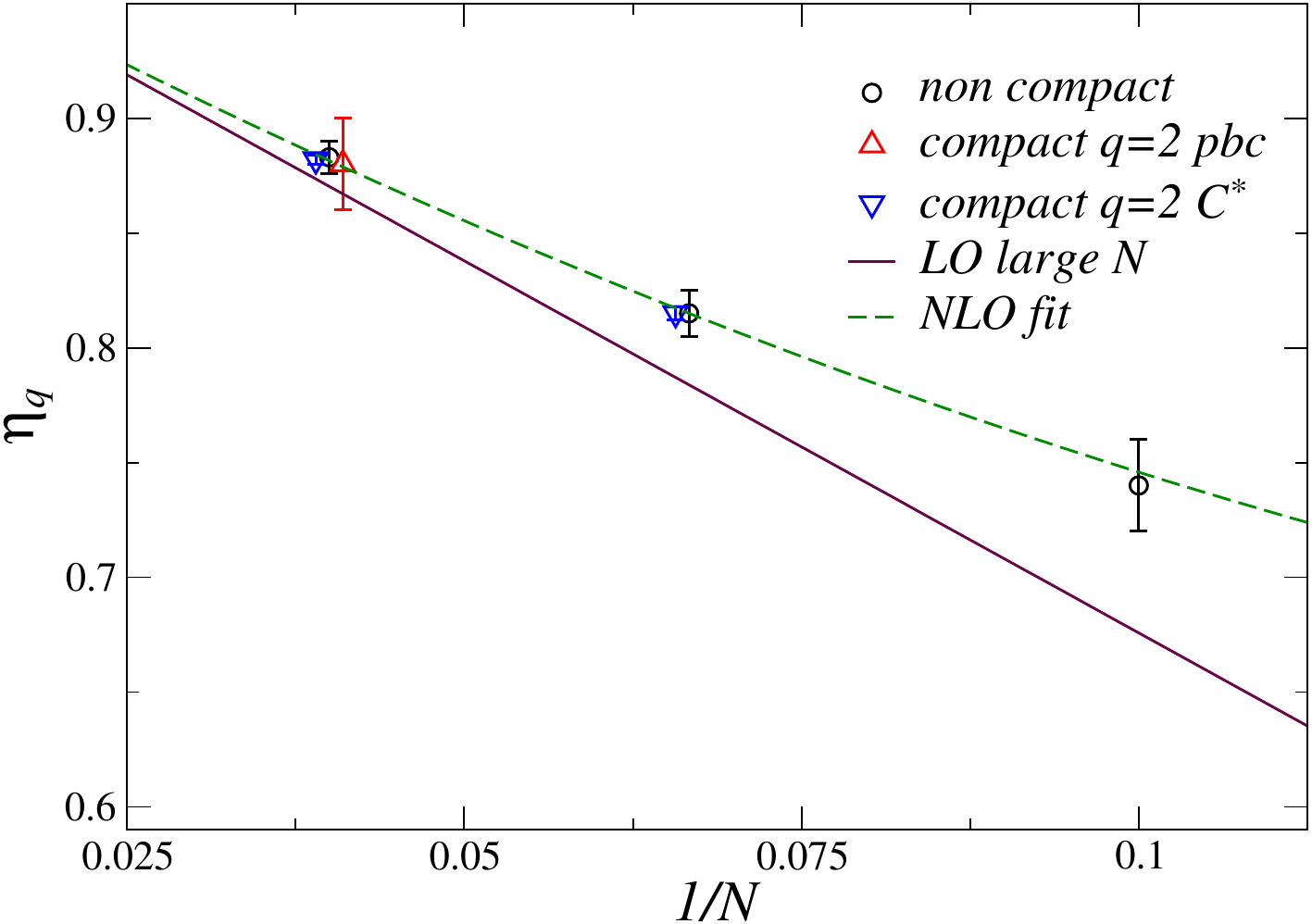}
  \caption{(Adapted from Ref.~\cite{BPV-22}) Summary of the results
    for the correlation-length critical exponent $\nu$ and for the
    anomalous dimension $\eta_q$ of the operator $Q_{\bm x}^{ab}$
    defined in Eq.~\eqref{QdefncAH}.  Results obtained in the
    noncompact (at CH transitions) and in the compact (at DC-OD
    transitions) model (these results will be discussed in
    Sec.~\ref{LAHMc}). Data from Refs.~\cite{BPV-21-ncAH, BPV-20-hcAH,
      BPV-22}.  (Left) We report the estimates of $\nu$, the
    leading-order large-$N$ estimate, Eq.~\eqref{nulargen} (LO large
    $N$), and an interpolation (NLO fit) obtained by adding a $1/N^2$
    correction term to Eq.~\eqref{nulargen}:
    $\nu=1-\frac{48}{\pi^2N}+\frac{10.5(5)}{N^2}$.  (Right) We report
    the estimates of $\eta_q$, the leading-order large-$N$ estimate,
    Eq.~\eqref{etalargen} (LO large $N$), and an interpolation (NLO
    fit) obtained by adding a $1/N^2$ correction term to
    Eq.~\eqref{etalargen}:
    $\eta_q=1-\frac{32}{\pi^2N}+\frac{7.0(5)}{N^2}$.
  \label{AH_crit_exp}}
  \end{center}
\end{figure}

In Fig.~\ref{AH_crit_exp} the numerical estimates of the exponent
$\nu$ and $\eta_q$ are compared with the $1/N$ expressions reported in
Sec.~\ref{largeexp}.  The agreement is very good, thus providing a
stringent check that the AHFT provides the correct effective
description of the continuous transitions along the CH line.  The same
check has been performed for the exponent $\eta_\Gamma$, which has
been determined by analyzing the susceptibility $\chi_z$ of the scalar
field ${\bm z}_{\bm x}$ computed in the Lorenz gauge (\ref{Lgauge}).
Ref.~\cite{BPV-23-chgf} estimated $\eta_\Gamma$ for $N=25$, which, in
turn provides an estimate of the RG dimension $y_\Gamma= (1 +
\eta_\Gamma)/2 $ of ${\bm \Gamma}_{\bm x}$. The result
$y_\Gamma=0.4655(5)$ compares satisfactorily with the large-$N$
field-theory formula~\cite{HLM-74,KS-08}
\begin{equation}
    y_\Gamma = {1\over 2} - {10\over \pi^2} {1\over N} + O(N^{-2}),
    \label{yzlargen}
\end{equation}
obtained from Eq.~(\ref{etazlargeN}), which reports $\eta_\phi$ in the
hard Lorenz gauge (in this case $\eta_\phi= \eta_\Gamma$). Indeed, 
for $N=25$ the large-$N$ formula (\ref{yzlargen}) predicts $y_\Gamma \approx
0.459$.  This result provides additional evidence of the charged
nature of the FP controlling the critical behavior along the CH
transition line for sufficiently large~$N$.

Finally, let us discuss the gauge-field correlations. As discussed in
Sec.~\ref{epsexpAHFT}, AHFT predicts $\eta_A=1$ at the charged FP.
Numerical results along the CH line are in good agreement with this
prediction. It is interesting to observe that the result $\eta_A = 1$
also holds at IXY transitions, see Sec.~\ref{cboncLAHM}.  However, the
IXY value cannot be explained by using the AHFT result, since the AHFT
is apparently only relevant for the charged CH transitions and not for
the topological IXY transition (see, however, the discussion reported at
the end of Sec.~\ref{cboncLAHM}).

\subsubsection{The Molecular-Higgs transition line}
\label{mhmc}

The MH line separates two ordered phases in which the SU($N$) symmetry
is broken to U($N-1$)~\cite{MZ-03}. MH transitions are characterized
by the condensation of the charged nonlocal excitations and of the
gauge modes. As shown in Ref.~\cite{BPV-24-ncAH} the critical
behavior along the MH line is the same as that along the CH line in the
one-component LAH model, see Fig.~\ref{phadiancLAH}.  Therefore, along
the MH line one expects IXY transitions.  This is confirmed by the
analyses of the energy cumulants and of the charged correlation
functions (computed in the Lorenz gauge) reported in
Ref.~\cite{BPV-24-ncAH}.  It is important to stress that the
decoupling of the scalar modes is not obvious, because of the presence
of $2N-2$ massless Goldstone bosons related with the spontaneous
breaking of the SU($N$) symmetry~\cite{MZ-03}. These massless
excitations turn out not to play any role in the critical behavior
along the MH transitions.

As an example, in Fig.~\ref{B3figs} we report a scaling plot of the
third energy cumulant for $N=2$ and $N=25$. Their FSS behavior is the
same as that at the CH transitions in the $N=1$ model,
demonstrating that the transitions belong to the IXY
universality class for any $N$.  Charged excitations deconfine at the
MH transitions.  The RG dimension $y_\Gamma$ of the charged operator
${\bm \Gamma}_{\bm x}$, estimated from the critical behavior of the
vector variable ${\bm z}_{\bm x}$ in the Lorenz gauge (see
Sec.~\ref{nc-orderparameters}), turns out \cite{BPV-24-ncAH} to be
independent of $N$ and perfectly consistent with the estimate
(\ref{etazest}) obtained along the CH line of the one-component LAH
model, see Fig.~\ref{chizvsRxiA}. Finally, also the critical behavior
of the gauge correlations (again computed in the Lorenz gauge) is
consistent with an IXY critical behavior.  For instance, one obtains
$\eta_A=1$ also along the MH line \cite{BPV-24-ncAH}.

\subsection{$N$-component SO($N$)-invariant lattice Abelian Higgs model}
\label{extLAH}

The above results can be extended to SO($N$)-symmetric LAH models,
which are the lattice analogue of the SO($N$)-symmetric AHFT discussed
in Sec.~\ref{OAHFT}. A lattice SO($N$)-invariant LAH model is
specified by the Hamiltonian~\cite{BPV-23-mpp,BPV-24-scson}
\begin{eqnarray}
  H_O({\bm A},{\bm z}) = \frac{\kappa}{2}
  \sum_{{\bm x},\mu>\nu} F_{{\bm x},\mu\nu}^2 - 2 N J
  \sum_{{\bm x},\mu} {\rm Re}\,( \lambda_{{\bm x},\mu} \bar{\bm
    z}_{\bm x} \cdot {\bm z}_{{\bm x}+\hat\mu}) +  v \,\sum_{\bm x}|{\bm
    z}_{\bm x}\cdot {\bm z}_{\bm x}|^2,
  \label{ncLAHMO}
\end{eqnarray}
where again we consider the unit-length constraint $\bar{\bm z}_{\bm
  x}\cdot {\bm z}_{\bm x}=1$.  One can easily check that, after
relaxing the unit-length constraint for the scalar field, this
Hamiltonian corresponds to the formal continuum limit of the
SO($N$)-symmetric AHFT with $\kappa=g^{-2}$.

The phase diagram for fixed values of $v$ is analogous to that of the
SU($N$)-symmetric LAH model, see the right panel of
Fig.~\ref{phadiancLAH}.  The qualitative features of the three phases
are the same, the only differences being the symmetry of the two
ordered phases and the nature of the transitions that depend on the
specific symmetry-breaking pattern. These two properties are
determined by the specific form of the scalar potential and, in
particular, by the sign of the Hamiltonian parameter $v$. The critical
behaviors along the various transition lines have been investigated in
Refs.~\cite{BPV-23-mpp,BPV-24-scson}. Here we summarize the main
results.

The MH transitions are expected to always belong to the IXY
universality class, as in SU($N$)-invariant LAH models.  This is due
to the fact that MH transitions are topological transitions driven by
the gauge modes, and scalar fields play no role at the transition, as
discussed in Sec.~\ref{mhmc}.  On the other hand, since the symmetry
of the ordered phases depends on the parameter $v$ (the ordered phases
are symmetric under SO($N-2$) and ${\rm SO}(N-1)$ transformations for
$v>0$ and $v<0$, respectively), the nature of the CH and CM
transitions depends on the sign of $v$.  Correspondingly, also the
relevant order parameter at the transition depends on the sign of the
parameter.  In particular, the transitions can be characterized  by the
condensation of the lattice bilinear operators $R_{L,{\bm x}}^{ab}$
and $T_{L,{\bm x}}^{ab}$ defined as their field-theoretical
counterparts, see Eq.~(\ref{wqoperft}).

The RG analysis of the SO($N$)-symmetric AHFT, reported in
Sec.~\ref{OAHFT}, indicates that the LAH model (\ref{ncLAHMO}) may
undergo charged continuous transitions along the CH line for a large
number $N$ of components. Indeed, the continuous or first-order nature
of the transitions is determined by the presence or absence of a
stable FP in corresponding AHFT.  As discussed in Sec.~\ref{OAHFT}, a
FP exists only for $N > N_o^*(d=3)$.  Moreover, the attraction domain
of the stable FP is limited to the parameter region $v > 0$, i.e., to
systems in which the ordered phase is SO($N-2$)-symmetric.  Therefore,
for $N < N_o^*(d=3)$ we expect only first-order transitions along the
CH line.  For $N > N_o^*(d=3)$, if the Higgs phase is symmetric under
SO$(N-1)$ transformations, then the CH line is again a line of
first-order transitions.  Instead, if the Higgs phase shows the
spontaneous symmetry breaking SO($N$)$\to$SO$(N-2)$, i.e., $v$ is
positive, then the CH transitions can be continuous, in the
universality class associated with the stable charged FP of the
SO($N$)-symmetric AHFT.

Along the CM line gauge modes do not develop critical correlations,
thus CM transitions admit a conventional LGW effective description in
terms of a gauge-invariant order parameter~\cite{BPV-23-mpp}.  For
negative $v$, transitions are driven by the condensation of $R_{\bm
  x}^{ab}$, defined in Eq.~(\ref{wqoperft}).  A straightforward LGW
analysis predicts first-order transitions for any $N\ge 3$. For $N=2$
continuous transitions in the XY universality class are possible. For
positive $v$, the critical behavior can be effectively described by
the O($N$)-symmetric LGW theory for an antisymmetric rank-two
tensor, which is the coarse-grained analogue of
$T_{\bm x}^{ab}$ defined in Eq.~(\ref{wqoperft}).  This allows us to
predict that, for $N=2$, continuous CM transitions belong to the Ising
universality class for all positive values of $v$. For $N=3$, instead,
one expects the existence of a tricritical value $v^*>0$, such that
the transition is of first order for $v < v^*$ and is continuous, in
the Heisenberg universality class, for $v > v^*$. The existence of a
tricritical value is due to the first-order nature of the CM
transitions in SU(3) invariant AH models, i.e., for $v = 0$.  The same
behavior, with critical Heisenberg transitions for large values of
$v$, occurs also for $N=4$.  For $N\ge 5$ transitions are always of
first order. A numerical check of these predictions is reported in
Ref.~\cite{BPV-24-scson}.

%% file: cLAHM.tex
\section{Lattice Abelian Higgs models with compact U(1) gauge variables}
\label{LAHMc}

In this section we consider an alternative lattice discretization of
the AHFT (\ref{AHFTL}) based on compact gauge variables. In compact
formulations there are topological features that are absent in
noncompact formulations, like the presence of monopole configurations
and a nontrivial dependence on the charge $Q$ of the scalar field.  As
we shall see, these features give rise to notable differences with
respect to the noncompact lattice formulation considered in
Sec.~\ref{LAHMnc}. In particular, the relation with the continuum AHFT
(\ref{AHFTL}) turns out to be much less straightforward.
The nontrivial
relation between the LAH models with noncompact gauge fields and the
compact LAH models with higher-charge scalar fields are discussed at
the end of the section.

Like noncompact LAH models, we must distinguish between one-component
and multi-component scalar models, which show qualitatively different
phase diagrams. As before, this is related to the 
presence of the global SU($N$) symmetry for $N\ge 2$.
Moreover, the behavior of unit-charge ($Q=1$) 
models differs substantially from that of models in which 
the scalar fields have charge $Q\ge 2$. Indeed, for $Q\ge 2$
the  models also undergo topological transitions analogous to
those in lattice ${\mathbb Z}_Q$ models (thus depending on the
charge $Q$), see Sec.~\ref{sec.ZQ}.

\subsection{The compact formulation of lattice Abelian Higgs models}
\label{coLAJHM}

In the compact formulation the gauge fields are
complex variables $\lambda_{{\bm x},\mu}$ satisfying $|\lambda_{{\bm
    x},\mu}|=1$, i.e., elements of the U(1) group, associated with the
lattice links. On a cubic lattice, 
the compact LAH model with $N$-component scalar fields
${\bm z}_{\bm x}$ of integer charge $Q$ is defined by the partition
function and lattice Hamiltonian
\begin{eqnarray}
Z = \sum_{\{{\bm z},\lambda\}} e^{-H_c},\qquad
H_c =  - 2 \,\kappa \sum_{{\bm x},\mu>\nu} {\rm Re}\, (\lambda_{{\bm
      x},{\mu}} \,\lambda_{{\bm x}+\hat{\mu},{\nu}}
  \,\bar{\lambda}_{{\bm x}+\hat{\nu},{\mu}} \,\bar{\lambda}_{{\bm
      x},{\nu}}) - 2 N J  \sum_{{\bm x}, \mu} {\rm Re}\,(
\lambda_{{\bm x},\mu}^Q \,\bar{\bm{z}}_{\bm x} \cdot
{\bm z}_{{\bm x}+\hat\mu}) ,
\label{clahmH}
\end{eqnarray}
where the two sums in $H_c$ run over all plaquettes and links of the
cubic lattice, respectively, and the site variables satisfy the
unit-length condition $\bar{\bm{z}}_{\bm x} \cdot {\bm z}_{{\bm
    x}}=1$.  The parameter $\kappa\ge 0$ plays the role of inverse
(square) gauge coupling, as can be seen by taking the naive continuum
limit in which the link variables are close to one.  The compact
formulation is well defined for all boundary conditions, since the
Hamiltonian is bounded and the configuration space is compact. While
in the noncompact case, the gauge group is the noncompact additive
group of real numbers, in the compact case the model is invariant
under U(1) gauge transformations. The nontrivial topology of the U(1)
group allows one to define an integer charge $Q$ and consider scalar
fields that transform under a charge-$Q$ representation of the U(1)
group. The charge-$Q$ model is invariant under the U(1) gauge
transformations
\begin{equation} 
 {\bm z}_{\bm x} \to U_{\bm x}^Q {\bm z}_{\bm x}, \qquad 
\lambda_{{\bm x},\mu} = U_{\bm x} \lambda_{{\bm x},\mu} \bar{U}_{{\bm
x}+\hat{\mu}}, 
\end{equation} 
where $U_{\bm x}$ is an ${\bm x}$-dependent phase, and under the
global transformation ${\bm z}_{\bm x} \to V {\bm z}_{\bm x}$, where
$V\in\mathrm{SU}(N)$.

Unlike what happens in noncompact formulations of LAH models, the
integer charge $Q$ cannot be absorbed in the definition of the gauge
field \cite{Yang-70}. Therefore, it should be considered as an
additional independent Hamiltonian parameter. As we shall see, the
phase diagram of compact LAH models crucially depends on the charge
$Q$.  In particular, the $\kappa$-$J$ phase diagram of models with
$Q=1$ differs from that of models with $Q\ge 2$.  Moreover, as in the
noncompact case, the behavior for $N=1$ is qualitatively different
from that obtained for $N\ge 2$.

For $N=1$ the Hamiltonian (\ref{clahmH}) can also be rewritten in
terms of the gauge fields only. Indeed, in the unitary gauge we can
set $z_{\bm x} = 1$ for all $\bm x$, obtaining the unitary-gauge
Hamiltonian
\begin{equation}
  H_{\rm ug} =
  - 2 \,\kappa \sum_{{\bm x},\mu>\nu} {\rm Re}\, (\lambda_{{\bm
      x},{\mu}} \,\lambda_{{\bm x}+\hat{\mu},{\nu}}
  \,\bar{\lambda}_{{\bm x}+\hat{\nu},{\mu}} \,\bar{\lambda}_{{\bm
      x},{\nu}})
  - 2 J \sum_{{\bm x}\mu} \hbox{Re}\ \lambda_{{\bm
    x},\mu}^Q.
\label{Hunitary}
\end{equation}
The unitary gauge is not complete and leaves a residual invariance
under local ${\mathbb Z}_Q$ gauge transformations, i.e., under the
transformations $\lambda_{{\bm x},\mu} = V_{\bm x} \lambda_{{\bm
    x},\mu} \bar{V}_{{\bm x}+\hat{\mu}}$, where $V_{\bm x}$ are
complex numbers satisfying $V_{\bm x}^Q=1$, i.e., elements of the
${\mathbb Z}_Q$ group. For $Q\ge 2$ topological transitions can be present,
as discussed in Sec.~\ref{sec.ZQhiggs}, since the model with
Hamiltonian (\ref{Hunitary}) is a soft version of the ${\mathbb Z}_Q$
gauge theory defined in Sec.~\ref{sec.ZQ}, which is obtained in the
limit $J\to \infty$. Indeed, for $J\to \infty$, we should have
$\lambda_{{\bm x},\mu}^Q = 1$ on all links, which implies that the
fields $\lambda_{{\bm x},\mu}$ are elements of the discrete ${\mathbb
  Z}_Q$ group.  For $Q=1$ the limit is obviously trivial.

\subsection{The compact lattice Abelian Higgs model with
  one-component scalar fields}
\label{onecoLAHM}

\begin{figure}[tbp]
    \begin{center}
  \includegraphics*[width=0.55\columnwidth]{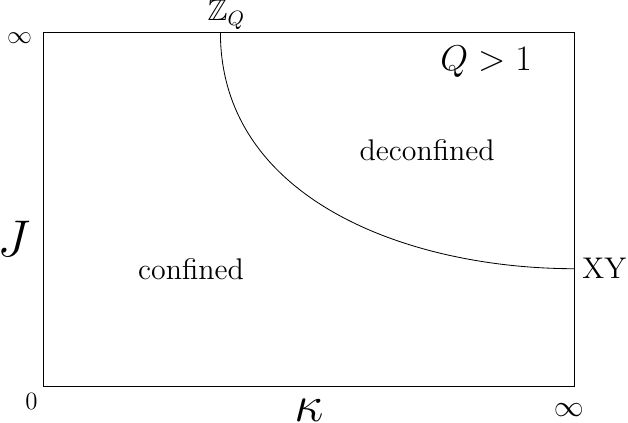}
  \caption{Sketch of the $\kappa$-$J$ phase diagram of the 3D
    one-component CLAH model, in which a compact U(1) gauge field is
    coupled with a (unit-length) complex scalar field of charge $Q\ge
    2$.  A confined phase (for small $J$ or small $\kappa$) and a
    deconfined phase (for large $J$ and $\kappa$) are present,
    separated by a deconfinement transition line.  For
    $\kappa\to\infty$ and $J\to\infty$ the model reduces to the XY
    vector model and to the lattice ${\mathbb Z}_Q$ gauge model with
    Wilson action, respectively. }
  \label{phadiacLAHN1}
    \end{center}
\end{figure}

The phase diagram of the one-component CLAH model with $Q=1$ is
trivial, as there is only one thermodynamic phase~\cite{FS-79}.  On
the other hand, models with any $Q\ge 2$ show two different
phases~\cite{FS-79,SSNHS-03,BPV-24-cAH}, as shown in
Fig.~\ref{phadiacLAHN1}. These two phases are distinguished by the
confinement properties of the charged excitations with $q<Q$. Indeed,
the transition line separating these two phases is related with the
deconfinement transition of the charged degrees of freedom with $q<Q$,
which can be probed by considering the behavior of the unit-charge
Wilson loops $W_{\cal C} = \prod_{\ell\in \cal C} \lambda_\ell$, where
$\cal C$ is a closed lattice loop. For $Q \ge 2$, the unit-charge
Wilson loops obey the area law in the confined phase and the perimeter
law in the deconfined phase. For $Q=1$ the area law never holds, due
to the screening of the charged scalar modes.

The deconfinement transition line for $Q\ge 2$ is expected to connect
the transition points of the models obtained for $J\to\infty$ and
$\kappa\to\infty$~\cite{FS-79,SSNHS-03}. As already mentioned, for
$J\to\infty$ we obtain the ${\mathbb Z}_Q$ gauge model with Wilson
action discussed in Sec.~\ref{sec.ZQ}.  For $\kappa\to\infty$, we can
set $\lambda_{{\bm x},\mu} = 1$ on all links, and thus we obtain the
standard XY model.  Given that the one-component model with charge $Q$
is equivalent to the gauge model with Hamiltonian ~(\ref{Hunitary}),
which is a ${\mathbb Z}_Q$ gauge-invariant model, the finite-$J$
deconfinement transitions are expected to belong to the ${\mathbb
  Z}_Q$ gauge universality class ~\cite{BPV-24-cAH} (see, however,
Ref.~\cite{SSNHS-03} for an alternative picture based on a line of
FPs, thus leading to continuously varying critical exponents).  As a
consequence (see Sec.~\ref{sec.ZQ} for a discussion of the critical
behavior of ${\mathbb Z}_Q$ gauge-invariant models), the continuous
transitions must belong to the Ising gauge universality class and to
the XY gauge universality class for $Q=2$ and $Q\ge 5$, respectively.
For $Q=3$ we expect first-order transitions.  For $Q=4$ the ${\mathbb
  Z}_4$ gauge theory that is obtained in the limit $J\to \infty$ has an
Ising gauge transition, but, as discussed in Sec.~\ref{sec.ZQ}, this
is not the generic behavior expected for ${\mathbb Z}_4$ gauge
invariant models. Generic ${\mathbb Z}_4$ gauge models should have
continuous transitions in the XY gauge universality class. For finite
$J$ we therefore expect XY gauge transitions also for $Q=4$.

\begin{figure}[tbp]
\begin{center}
  \includegraphics[width=0.45\columnwidth]{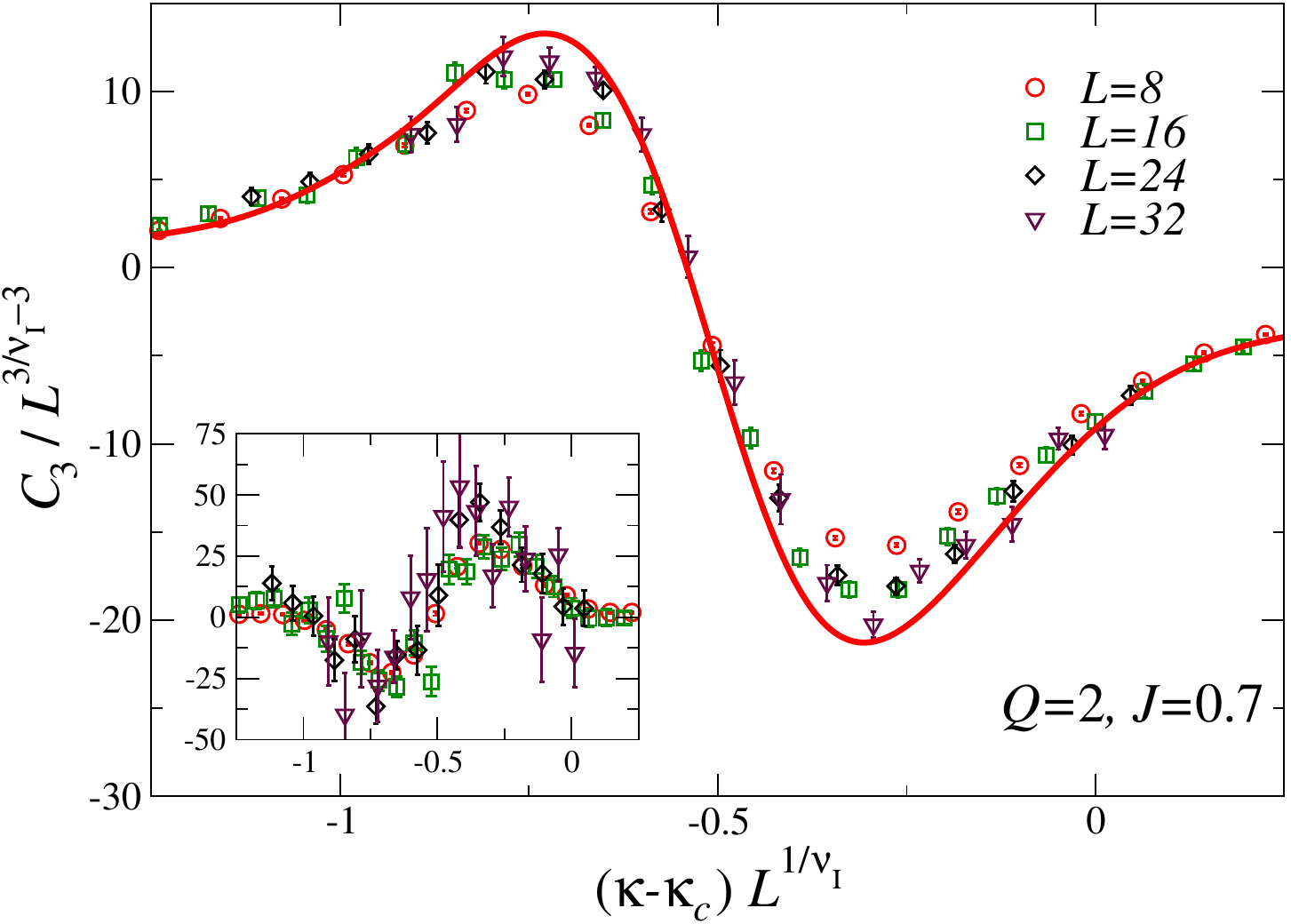}
  \hspace{1cm}
  \includegraphics[width=0.45\columnwidth]{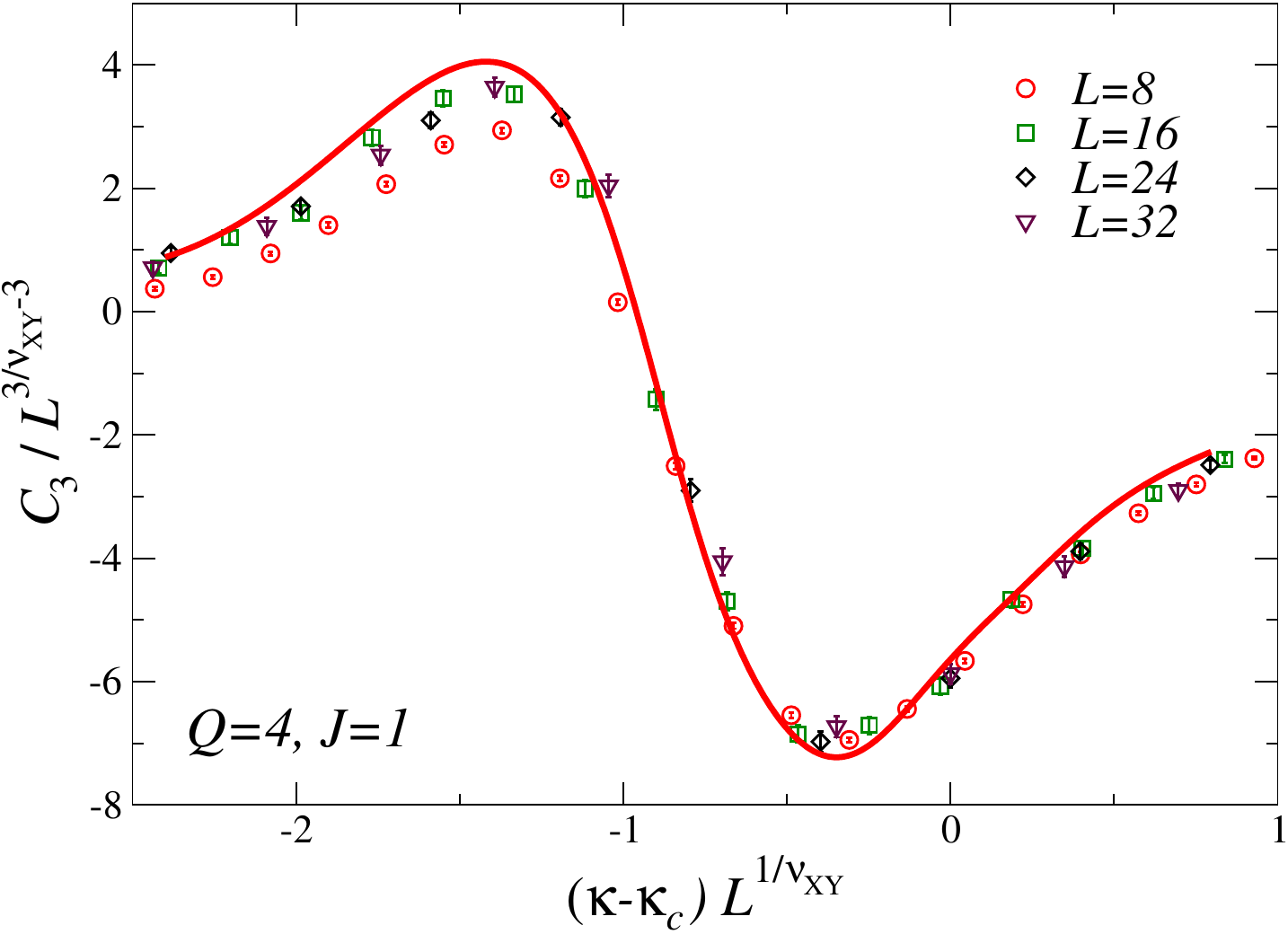}
  \caption{(From Ref.~\cite{BPV-24-cAH}) (Left) Plot of the third
    energy cumulant $C_3$ (its expected finite-size behavior is
    discussed in \ref{fssencum}) as a function of
    $(\kappa-\kappa_c)L^{1/\nu_I}$ for $N=1$ and $Q=2$, at fixed
    $J=0.7$. Here $\kappa_c=0.5880$ and $\nu_I$ is the Ising value
    $\nu_I=0.629971$ (see \ref{univclass} for a collection of
    estimates of $\nu_I$).  The continuous curve is (up to
    nonuniversal rescaling constants) the scaling curve for $C_3$ in
    the $\mathbb{Z}_2$ gauge model (see Ref.~\cite{BPV-24-cAH} for a
    parametrization).  Deviations can be explained by the presence of
    scaling corrections that vanish as $L^{-\omega_I}$ for large $L$,
    where $\omega_I = 0.83$ is the universal leading
    scaling-correction exponent for Ising systems.  This is confirmed
    by the plot shown in the inset, where the combination
    $L^{\omega_I}\left[ L^{3-3/\nu_I} C_3(J,K,L) - {\cal
        C}_3(X)\right]$ is reported as a function of
    $(\kappa-\kappa_c)L^{1/\nu_I}$, where ${\cal C}_3(X)$ (solid
    curve) is the properly rescaled 
    asymptotic curve computed in the $\mathbb{Z}_2$
    gauge model.  The collapse of the data onto a single curve
    indicates that the observed scaling corrections are consistent
    with those expected at an Ising transition.  (Right) Plot of the
    third energy cumulant $C_3$ as a function of
    $(\kappa-\kappa_c)L^{1/\nu_{\rm XY}}$, for the model with $N=1$
    and $Q=4$, at fixed $J=1$. Here $\kappa_c=1.0205$ and $\nu_{\rm
      XY}$ is the XY value $\nu_{\rm XY}=0.6717$ (see \ref{univclass}
    for estimates of $\nu_{\rm XY}$).  The continuous curve is (up to
    nonuniversal rescaling constants) the scaling curve for $C_3$ in
    the IXY gauge model (see Ref.~\cite{BPV-24-cAH} for a
    parametrization).
  \label{B3_highQ}}
  \end{center}
\end{figure}
 
Numerical FSS analyses of the energy cumulants~\cite{BPV-24-cAH} for
$Q=2,\,4,\,6$, see Fig.~\ref{B3_highQ}, fully support the above
predictions. It is worth mentioning that, for any value of the charge
$Q$, there are significant crossover effects for relatively large
values of $\kappa$, due to the presence of the XY spin FP that
controls the critical behavior in the $\kappa\to\infty$ limit, which
is unstable with respect to a nonzero gauge coupling $\alpha\equiv
g^2\sim 1/\kappa$, see Sec.~\ref{epsexpAHFT}.\footnote{The 3D RG
dimension $y_{\alpha}= 1$ of the gauge coupling $\alpha\sim
\kappa^{-1}$ provides the crossover exponent, which corresponds to the
energy dimension of the gauge coupling $\alpha$. Therefore, in the
large-$\kappa$ limit, the gauge field gives rise to an intrinsic
crossover scale $\xi_\alpha \sim \kappa$. If the correlation length
$\xi$ or the size of the system $L$ satisfies $\xi \lesssim
\xi_\alpha$ or $L \lesssim \xi_\alpha$, significant crossover effects
can be observed, with an apparent spin XY behavior.}

\subsection{The multicomponent compact lattice Abelian Higgs model}
\label{multicoLAHM}

\begin{figure}[tbp]
\begin{center}
  \includegraphics[width=0.45\columnwidth]{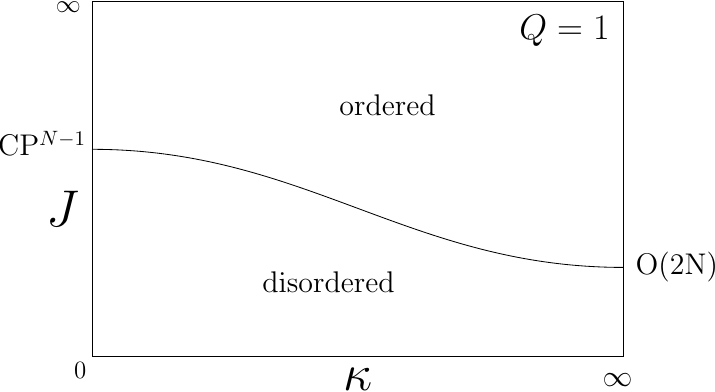}
  \hspace{1cm}
\includegraphics[width=0.45\columnwidth]{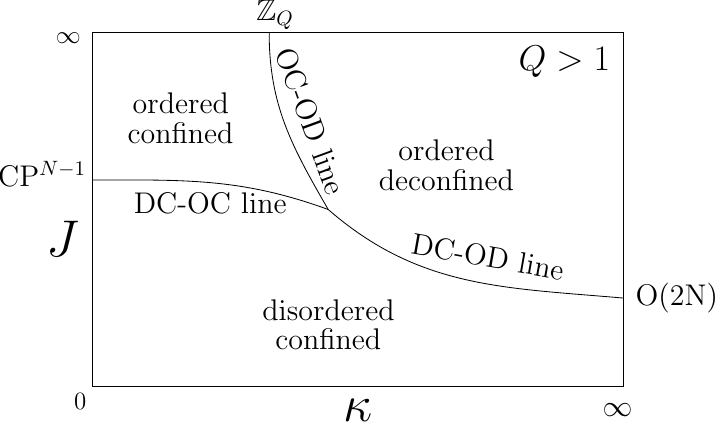}  
  \caption{Sketch of the phase diagram of the 3D LAH model
    (\ref{clahmH}) with charge $Q=1$ (left panel) and $Q\ge 2$ (right
    panel), for generic $N\ge 2$.  For $Q=1$ (left), the
    high-temperature (small $J$) and low-temperature (large $J$)
    phases are separated by a single transition line, characterized by
    the spontaneous breaking of the SU($N$) symmetry.  For $Q\ge 2$
    (right), the phase diagram presents three phases, the
    disordered-confined (DC), the ordered-deconfined (OD), and the
    ordered-confined (OC) phase.  The compact LAH model is equivalent
    to the CP$^{N-1}$ model for $\kappa=0$, to the O($2N$) vector
    model for $\kappa\to\infty$, and to the lattice ${\mathbb Z}_Q$
    gauge model for $J\to\infty$.  \label{phadiacLAHNQ}}
      \end{center}
\end{figure}

Multicomponent LAH models with compact gauge variables present a phase
diagram that is more complex than that of the one-component model,
because of the presence of transitions where the SU($N$) symmetry
spontaneously breaks.  The phase diagrams for any $N\ge 2$ are
sketched in the left and right panels of Fig.~\ref{phadiacLAHNQ}, for
$Q=1$ and $Q\ge 2$, respectively.

\subsubsection{Phase diagram for $Q=1$}
\label{q1phdia}

As sketched in the left panel of Fig.~\ref{phadiacLAHNQ}, only two
phases are present for $Q=1$: a disordered phase for small $J$ and an
ordered phase for large $J$. They are separated by a single transition
line, characterized by the breaking of the SU($N$) symmetry, signaled
by the condensation of the gauge-invariant bilinear operator $Q_{\bm
  x}^{ab}$, defined in Eq.~(\ref{QdefncAH}).  At the transition, gauge
fields only prevent gauge-dependent correlations, such as the vector
correlations $\langle \bar{\bm z}_{\bm x}\cdot {\bm z}_{\bm
  y}\rangle$, from becoming critical (actually, due to gauge
invariance $\langle \bar{\bm z}_{\bm x}\cdot {\bm z}_{\bm
  y}\rangle=\delta_{{\bm x},{\bm y}}$).

For $\kappa = 0$, the model is equivalent to the gauge CP$^{N-1}$
model, whose critical behavior has been discussed in
Sec.~\ref{cpnmodels}.  A natural hypothesis, that is confirmed by a
detailed numerical study \cite{PV-19-AH3d}, is that the critical
behavior does not depend on the value of $\kappa$. Thus, for any
finite $\kappa$ the $Q=1$ model undergoes a LGW transition with
gauge-invariant order parameter, as the CP$^{N-1}$ model.  It follows,
see Sec.~\ref{cpnmodels}, that continuous transitions (in the
Heisenberg universality class) are only possible for $N=2$.  For
larger values of $N$, we instead expect the transitions to be of first
order. Note that the same critical behavior occurs in noncompact LAH
models along the CM line (see the right panel of
Fig.~\ref{phadiancLAH}), as discussed in Sec.~\ref{cmmmc}.

In the limit $\kappa\to\infty$, the model reduces to an O($2N$) vector
model, which presents a continuous transition at a finite value of $J$
for any $N$. The corresponding FP is unstable with respect to a
nonzero gauge coupling $\alpha\sim \kappa^{-1}$, see
Sec.~\ref{epsexpAHFT}, but, as it occurs for the one-component compact
LAH model discussed in Sec.~\ref{onecoLAHM}, its presence gives rise
to significant crossover effects for large values of $\kappa$.

\subsubsection{Phase diagram for $Q\ge 2$}
\label{qge2phdia}

For $Q\ge 2$ the phase diagram is more complex, see the right panel of
Fig.~\ref{phadiacLAHNQ}, with three different phases~\cite{FS-79,
  SSSNH-02, SSNHS-03, NSSS-04, CFIS-05, CIS-06, WBJS-08, BPV-20-hcAH}.
They are characterized by the large-distance behavior of both scalar
and gauge observables. Beside the gauge-invariant bilinear order
parameter $Q_{\bm x}^{ab}$ [see. Eq.~(\ref{QdefncAH})], one may
consider Wilson loops, which signal the confinement or deconfinement
of external static sources with unit charge.

As shown in the right panel of Fig.~\ref{phadiacLAHNQ}, for small $J$
and any $\kappa \ge 0$, there is a phase in which scalar-field
correlations are disordered and single-charge modes are confined (the
Wilson loop obeys the area law).  For large values of $J$
(low-temperature region) scalar correlations are ordered and the
SU($N$) symmetry is broken. Two different phases occur here:
static single-charge test particles are confined for small $\kappa$ 
and deconfined for large $\kappa$.

The three different phases are separated by three transition lines:
the DC-OD transition line between the disordered-confined (DC) and the
ordered-deconfined (OD) phases, the DC-OC line between the
disordered-confined and ordered-confined (OC) phases, and the OC-OD
line between the ordered-confined and ordered-deconfined phases. They
are expected to meet at one point in the center  of the phase
diagram. The transition lines have different features that depend on
the number $N$ of components and on the charge $Q$ of the scalar
field.

The transitions along the DC-OC line are analogous to those occurring
along the unique transition line present for $Q=1$, see
Sec.~\ref{q1phdia}, and along the CM line in noncompact models. In
particular, they have the same nature as in the 3D lattice CP$^{N-1}$
model obtained for $\kappa = 0$. Therefore, continuous transitions are
only expected for $N=2$, with a critical behavior belonging to the O(3)
vector universality class.

In the limit $J\to\infty$, the model (\ref{clahmH}) becomes equivalent
to a ${\mathbb Z}_Q$ gauge model with Wilson
action~\cite{BPV-20-hcAH}, as it occurs for $N=1$, see
Sec.~\ref{onecoLAHM}. A natural hypothesis is that the transitions
along the OC-OD line belong to the universality class of the ${\mathbb
  Z}_Q$ gauge model, as in the $N=1$ compact LAH model, see
Sec.~\ref{onecoLAHM}.  This prediction has been verified numerically
for $Q=2$.  Ref.~\cite{BPV-20-hcAH} computed the energy cumulants,
finding that they have the same FSS behavior as in the ${\mathbb Z}_2$
gauge model.  These results confirm that the OC-OD transitions belong
to the Ising gauge universality class for this value of $Q$.

\begin{figure}[btp]
\begin{center}
  \includegraphics[width=0.45\columnwidth]{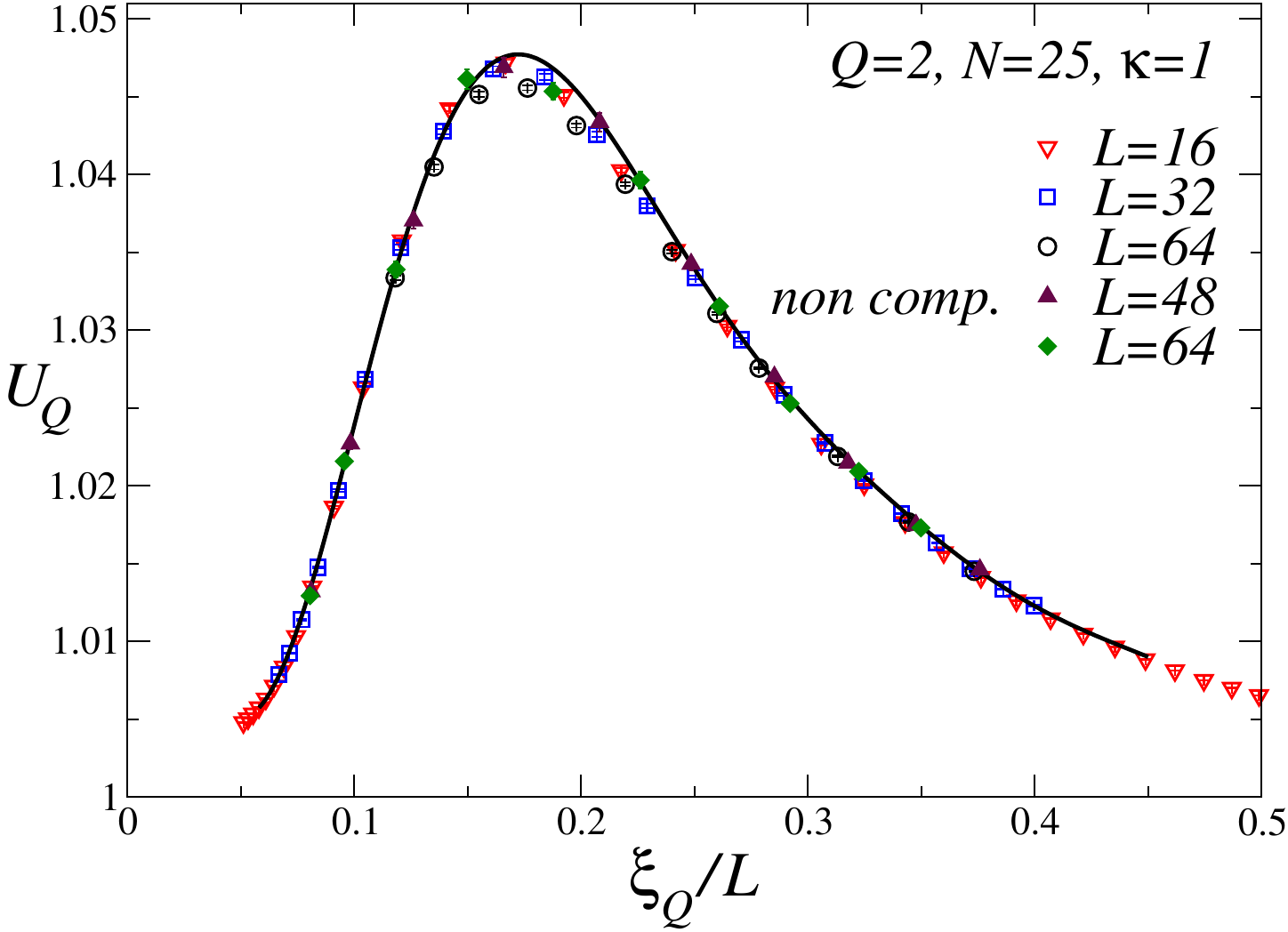}
  \hspace{1cm}
  \includegraphics[width=0.45\columnwidth]{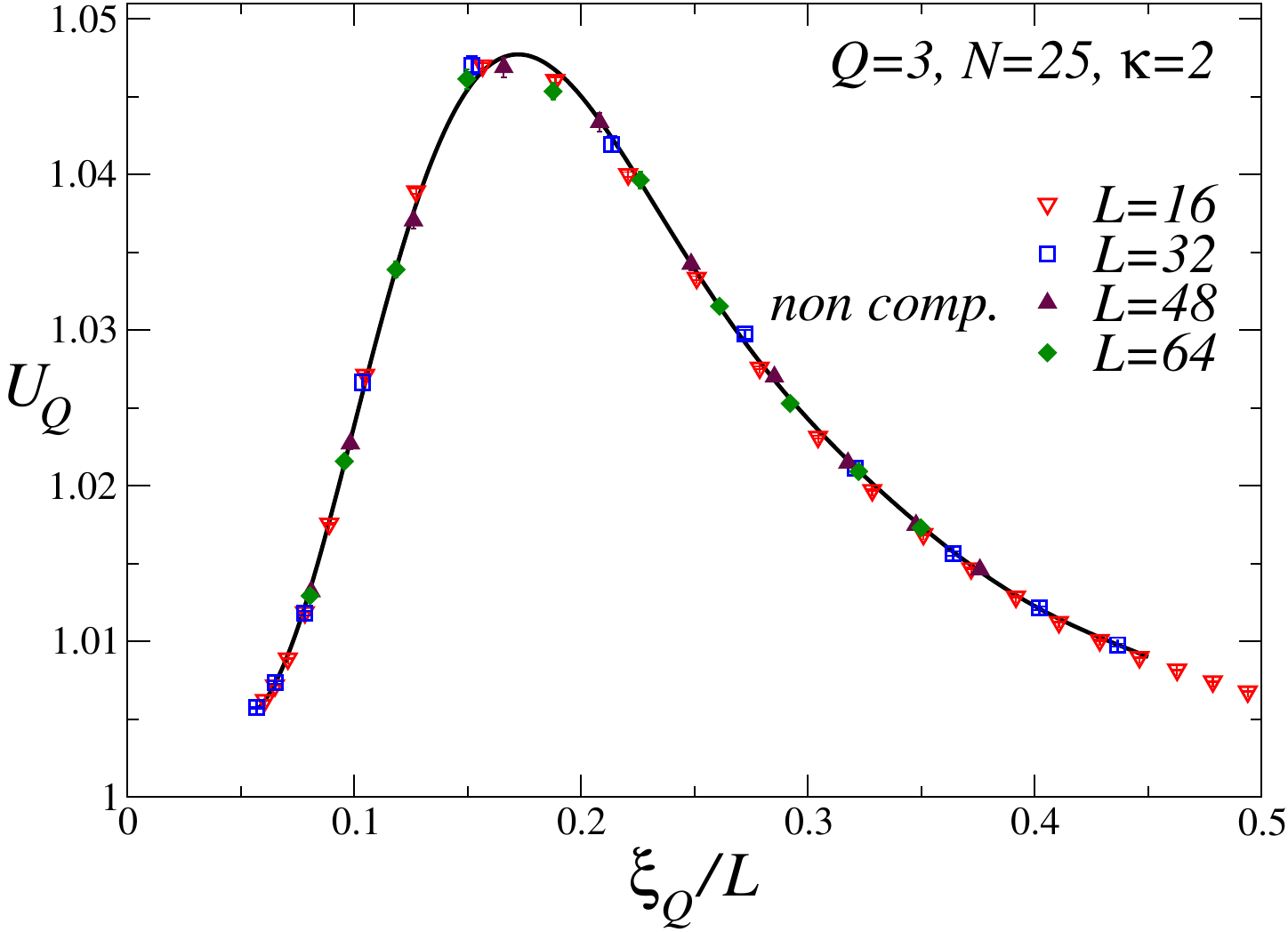}
  \caption{(Adapted from Ref.~\cite{BPV-22}) (Left) Plot of $U_Q$
    versus $\xi_Q/L$, where $U_Q$ and $\xi_Q$ are the Binder parameter
    and the second-moment correlation length defined in terms of the
    gauge-invariant operator $Q_{\bm x}^{ab}$ defined in
    Eq.~\eqref{QdefncAH}.  Results for the compact and noncompact
    model with $N=25$ and (compact model) $Q=2$.  Data are obtained in
    MC simulations varying $J$ across the DC-OD (compact) and CH
    (noncompact) line for $\kappa=1$.  The solid line is an
    extrapolation of the data for the noncompact LAH model.
    Since $U_Q$ and
    $\xi_Q/L$ are RG invariant quantities, the collapse of the data
    signals that the transitions in the two models belong to the same
    universality class (see \ref{fssrginv} for a discussion of the FSS
    behavior).  (Right) Same plot as in the left panel for the compact 
    model with $Q=3$, fixing $\kappa=2$.
   \label{CH_nc_highQ}}
  \end{center}
\end{figure}

The transitions along the large-$\kappa$ DC-OD line have been
investigated numerically in Ref.~\cite{BPV-20-hcAH}, which focused on
the compact model for $N=15$ and $25$ and two charge values, $Q=2$ and
$Q=3$, and analyzed the finite-size behavior of the correlation
function of the operator $Q_{\bm x}^{ab}$ (some data are reported in
Fig.~\ref{CH_nc_highQ}).  The analysis shows that the continuous DC-OD
transitions belong to the same universality as the transitions along
the CH line in the noncompact models discussed in Sec.~\ref{chmc}, for
all values of $N$ and $Q$ investigated. Some results taken from
Ref.~\cite{BPV-20-hcAH} are reported in Fig.~\ref{CH_nc_highQ}, which 
shows that the data for the compact and the noncompact model have the
same FSS behavior. Moreover, the estimated critical exponents computed
in the compact and in the  noncompact model for the same value of $N$ are
consistent within errors, see Fig.~\ref{AH_crit_exp}.  Therefore, the
stable charged FP point of the RG flow of the AHFT, see
Sec.~\ref{epsexpAHFT}, which exists for $N\gtrsim 7$, controls the
continuous transitions in multicomponent LAH models with both
noncompact and compact gauge variables for large values of $\kappa$;
more precisely, along the CH line in noncompact models and along the
DC-OD line in compact models with any charge $Q$ satisfying $Q\ge 2$.

\subsubsection{Relation between the compact and the noncompact
  lattice Abelian Higgs models}
\label{qinfty}

The previous results show that noncompact and compact ($Q\ge 2$) LAH
models have similar phase diagrams. For $N\ge 2$, the compact DC-OC
line and the noncompact CM line have the same nature, and so have the
compact DC-OD and noncompact CH line for any $Q\ge 2$. The MH line and
the OC-OD line both correspond to topological transitions. These
results may appear unexpected.  Therefore, it is worth discussing
the relation between the compact and noncompact models in more detail,
to explain the similarities of their phase diagram.

For this purpose, we note that the compact formulation of LAH models
is equivalent to the noncompact one in the limit $Q\to\infty$,
$\kappa\to\infty$ at fixed $\kappa/Q^2$.  Indeed, if we rewrite the
compact field $\lambda_{{\bm x},\mu}$ as $\lambda_{{\bm x},\mu} = e^{i
  A_{{\bm x},\mu}/Q}$ with $A_{{\bm x},\mu} \in [- \pi Q, \pi Q]$, the
compact-field Hamiltonian (\ref{clahmH}) becomes
\begin{eqnarray}
H_c = - 2 {\kappa}
\sum_{{\bm x},\mu>\nu} \hbox{Re } \exp\Bigl[-{i\over Q}
  (\Delta_{\hat\mu} A_{{\bm x},\nu} - \Delta_{\hat\nu} A_{{\bm
      x},\mu}) \Bigr]
- J N \sum_{{\bm x}, \mu} 2\, {\rm Re}\,(\bar{\bm{z}}_{\bm x}
\cdot e^{iA_{{\bm x},\mu}} \, {\bm z}_{{\bm x}+\hat\mu}).
\label{chham2}
\end{eqnarray}
For $Q\to \infty$, the gauge real variables $A_{{\bm x},\mu}$ become
unbounded, and the Hamiltonian becomes equivalent to that of the
noncompact formulation with $\kappa_{\rm nc}=2\kappa/Q^2$, where
$\kappa_{\rm nc}$ is the gauge coupling of the resulting noncompact
gauge-field Hamiltonian. Note that this equivalence trivially holds as
long as the fluctuations of $A_{{\bm x},\mu}$ on each plaquette are
bounded and uncorrelated for $Q\to \infty$, i.e., for any point of the
phase diagram. In particular, in Sec.~\ref{sec.ZQ} it has already been
discussed for the ${\mathbb Z}_Q$ gauge model (the limiting model of
the compact LAH model for $J\to \infty$), which is equivalent to the IXY
model (the limiting model of the noncompact LAH model for
$J\to\infty$) in the limit considered above.

The argument presented above shows that the compact model converges to
the noncompact one as $Q,\kappa\to \infty$, at fixed $\kappa_{\rm
  nc}=2\kappa/Q^2$.  Therefore, also their phase diagrams must become
identical for large $Q$.  As far as the nature of the critical lines,
the numerical evidence reported in Sec.~\ref{qge2phdia} shows that the
compactification of the model is an irrelevant perturbation of the
noncompact model obtained for $Q=\infty$, as already discussed for the
IXY model in Sec.~\ref{sec.ZQ}.  Thus, the critical behavior should become 
identical for $Q\to\infty$.  In practice, it is not
really required to take large values of $Q$ to observe the same IXY critical
behavior, as in the noncompact model.  
As we have already mentioned, the behavior along the DC-OC
and DC-OD is the same as that on the corresponding CM and CH lines for
any $Q\ge 2$, while the behavior along the OC-OD and MH lines is the
same for any $Q\ge 4$. Thus, differences only occur along the line of
topological transitions that end at $J=\infty$ for $Q=2$ and 3.  While
in the noncompact model these topological transitions belong to the
IXY universality class, in the compact one the IXY critical behavior
is only observed for $Q\ge 4$.  For $Q=3$ transitions are of first
order, while for $Q=2$ they belong to the Ising gauge universality
class.

\subsubsection{Relevance of monopole configurations}
\label{monosuppr}

In compact gauge models, the nontrivial topology of the gauge fields
allows one to define topological objects. Their role has been
investigated in several studies, see, e.g.,
Refs.~\cite{LD-89,MS-90,KM-93,MV-04,PV-20-mfcp,BPV-22-mpf}, which
indicate that their presence may significantly affect the nature of
the phase transitions and their critical
behavior. Refs.~\cite{PV-20-mfcp,BPV-22-mpf} performed a numerical
study to understand the role of monopole configurations, which
naturally emerge in LAH models with compact gauge fields, using the
monopole definition of Ref.~\cite{DT-80}.

In lattice systems with periodic boundary conditions one can define
monopoles and antimopoles using the prescription proposed in
Ref.~\cite{DT-80}.  This prescription starts from the noncompact
lattice curl $\Theta_{{\bm x},\mu\nu}= \theta_{{\bm x},\mu} +
\theta_{{\bm x} + \hat{\mu},\nu} - \theta_{{\bm x},\nu} - \theta_{{\bm
    x} + \hat{\nu},\mu}$ associated with each plaquette, where
$\theta_{{\bm x},\mu}\in [-\pi,\pi)$ is the phase associated with
the  compact lattice variable $\lambda_{{\bm x},\mu}$, $\lambda_{{\bm
      x},\mu} = e^{i\theta_{{\bm x},\mu}}$.  Note that $\Theta_{{\bm
      x},\mu\nu}$ is antisymmetric in $\mu$ and $\nu$, so this
  definition provides two different quantities that differ by a sign
  for each plaquette, depending on the orientation. One can easily
  verify that, for any closed lattice surface $S$ made of elementary
  plaquettes, the naively defined outgoing magnetic flux across $S$
  vanishes, i.e., $\sum_{P\in S} \Theta_{{\bm x},\mu\nu} = 0$, where
  the sum extends to all plaquettes in $S$, $\Theta_{{\bm x},\mu\nu}$
  is associated with plaquette $P = ({\bm x},\mu\nu)$, and the
  plaquette orientation is chosen in such a way that the unit vector
  $\hat{\mu}\times \hat{\nu}$ points outward with respect to the
  surface.\footnote{These rules are a discrete version of the
  integration rules of a differential two-form on a two-dimensional
  surface, which can be formalized using the formalism of lattice
  differential forms, see, e.g., Ref.~\cite{Guth-79}.} To define a
  nontrivial net number of monopoles (i.e., the difference between the
  number of monopoles and of antimonopoles) within the surface $S$,
  one must isolate the singular contribution from the smooth
  background corresponding to small values of $|\Theta_{{\bm
      x},\mu\nu}|$. This can be done by defining \begin{equation}
    N_{\rm mono}(S) = \sum_{P\in S} M\left({\Theta_{{\bm x},\mu\nu}
      \over 2\pi}\right), \qquad M(x) = x - \left\lfloor x + 1/2
    \right\rfloor,
\end{equation}
where $\lfloor \phantom{x}\rfloor$ denotes the floor function.  The
function $M(x)$ satisfies $-1/2\le M(x) < 1/2$, $M(x) = x$ for any $x$
in the interval $[-1/2,1/2)$; moreover, $M(x) - x$ is always an
  integer. The fact that $N_{\mathrm{mono}}$ is an integer number
  easily follows from these properties.  Note that $|\Theta_{{\bm
      x},\mu}|$ has to be larger than $\pi$ on some plaquettes for
  $N_{\rm mono}(S)$ to be nonvanishing.

A topic that has been investigated in the literature is whether the
suppression of the monopole configurations in the partition function
changes the nature of the transition, or whether it gives rise to new
universality classes somehow related to those found in noncompact
formulations, in which analogous monopole configurations are absent.
These issues have been addressed in
Refs.~\cite{PV-20-mfcp,BPV-22-mpf}.  In the lattice CP$^{N-1}$ model
defined by the Hamiltonian reported in Eq.~(\ref{altcpnmod}), a finite
density of monopoles is observed in the disordered low-$J$ phase up to
the critical point, while in the low-temperature regime $J>J_c$ only
isolated pairs of monopoles and anti-monopoles are typically present,
whose number decreases rapidly with increasing $J$, since
$\Theta_{{\bm x},\mu\nu}$ approaches zero in the large-$J$ limit.  Thus,
the monopole density is an appropriate order parameter of the
transition.  As a consequence, by considering only monopole-free
configurations, one is changing the nature of the small-$J$ phase, and
therefore one expects a different critical behavior or even the
absence of any transition.

To define a monopole-free version of the CP$^{N-1}$ model
(MFCP$^{N-1}$), one may change the configuration space, considering
only configurations for which $N_{\rm mono}(C)=0$ on any elementary
lattice cube $C$~\cite{PV-20-mfcp,BPV-22-mpf}.  The numerical analyses
of Refs.~\cite{PV-20-mfcp,BPV-22-mpf} show that the MFCP$^{N-1}$
models, as well as 
 some extensions obtained by relaxing the unit-length
condition on the scalar fields, undergo finite-$J$ transitions.
However, the features of these transitions are apparently unrelated
with those of the transitions in the standard CP$^{N-1}$ model. In
particular, there is no O(3) continuous transition for
$N=2$. Moreover, the numerical data definitely exclude that the
MFCP$^{N-1}$ model has a transition associated with the AHFT for
large values of $N$ ~\cite{PV-20-mfcp,BPV-22-mpf}.

Ref.~\cite{BPV-20-hcAH} also studied the critical behavior of
monopole-free ``higher-charge'' CP$^{N-1}$ models, obtained by setting
$\kappa=0$ in Eq.~\eqref{clahmH}.\footnote{If we set $\kappa=0$ in
Eq.~\eqref{clahmH} we obtain a charge-$Q$ CP$^{N-1}$ model. In the
absence of the monopole-free condition, this model is equivalent to
the standard $Q=1$ CP$^{N-1}$ model (it is enough to redefine the
gauge fields).  This equivalence does not hold in the monopole-free
version, as in this case it is not possible to replace
$\lambda_{{\bm x},\mu}^Q$ with $\lambda_{{\bm x},\mu}$, as the gauge
fields also appear in the monopole-free condition.  Thus, we obtain a
different monopole-free model for each $Q$.} The results show that,
for $Q\ge 2$ the monopole-free model has the same behavior as the
standard CP$^{N-1}$ model.  This means that in the compact LAH model
the topological properties of the gauge field are inessential for any
$Q\ge 2$, which is consistent with the fact that the compact and
noncompact model have analogous phase diagrams for $Q\ge 2$, and in
particular with the appearance (for $N\gtrsim 7$) of an AHFT critical
behavior both in the noncompact LAH model along the CH line and in the
compact LAH model with $Q\ge 2$ along the DC-OD line, see the
discussion in Sec.~\ref{qge2phdia}.

%% file: discretegauge.tex
\section{Lattice spin systems with discrete gauge symmetries}
\label{discgauge}

In this section we review results on lattice gauge spin systems with
discrete gauge group, focusing mainly on lattice ${\mathbb Z}_2$-gauge
$N$-vector models, obtained by minimally coupling $N$-component real
fields with ${\mathbb Z}_2$-gauge fields~\cite{BPV-24-z2Nv}.  As
discussed in Sec.~\ref{nonginvLGW} they are paradigmatic models
undergoing LGW$^{\times}$ phase transitions.

Lattice ${\mathbb Z}_2$-gauge $N$-vector models are relevant in
several different contexts, see, e.g., Refs.~\cite{JS-91, SV-99,
  SM-02, MS-02, SF-00, SR-91, LRT-93, CSS-94, LRT-95-a, LRT-95-b,
  SP-02, SSS-02, PS-02}. In particular, they are relevant for
transitions in nematic liquid crystal~\cite{LRT-93, LRT-95-a,
  LRT-95-b, LNNSWS-15, LNSWZ-16, BNWL-etal-19} and for systems with
fractionalized quantum numbers, see, e.g., Refs.~\cite{SSS-02,SM-02}.

We discuss their phase diagrams and critical behaviors, 
considering separately 
multicomponent ($N\ge 2$) and $N=1$ models.
Multicomponent models display three different phases,
characterized by the spontaneous breaking of the global O($N$)
symmetry and by the different topological properties of the ${\mathbb
  Z}_2$-gauge correlations, see, e.g., Refs.~\cite{FS-79,Sachdev-19}.
We also discuss the possibility of uncovering critical gauge-dependent
correlations at thier LGW$^\times$ transitions by implementing an
appropriate stochastic gauge fixing.  The particular case $N=1$
corresponds to the so-called ${\mathbb Z}_2$ Higgs model, which
satisfies an exact duality relation~\cite{BDI-75}.  Unlike
multicomponent models, it presents only two phases separated by two
Ising-like transition lines, which meet at an XY multicritical
point.

The final part of this section presents a survey of results for other
models with discrete gauge symmetries. In particular, we discuss 
whether there is an effective enlargement of the gauge symmetry at
the critical point, as it occurs in the lattice ${\mathbb
  Z}_Q$ gauge models for a sufficiently large $Q$, see
Sec.~\ref{sec.ZQ}.

\subsection{The ${\mathbb Z}_2$-gauge $N$-vector models}
\label{z2gaugenvectorm}

The ${\mathbb Z}_2$-gauge $N$-vector model is a lattice 
model with local ${\mathbb Z}_2$ gauge invariance.  From the point of
view of the symmetries, it can be interpreted as an $N$-vector model,
which is symmetric under O($N$) = ${\mathbb Z}_2\otimes
\mathrm{SO}(N)$ transformations, in which the ${\mathbb Z}_2$ symmetry
is gauged, giving rise to a ${\mathbb Z}_2$-gauge theory coupled with
scalar vector fields.  Its lattice Hamiltonian is reported in
Eq.~(\ref{hamz2N}).  It is invariant under global SO($N$)
transformations ${\bm s}_{\bm x} \to V {\bm s}_{\bm x}$ with $V\in
\mathrm{SO}(N)$, and local ${\mathbb Z}_2$ gauge transformations,
\begin{eqnarray}
{\bm s}_{\bm x}\to w_{\bm x} {\bm s}_{\bm x},\qquad
\sigma_{{\bm x},\nu}\to 
w_{\bm x} \sigma_{{\bm x},\nu} w_{{\bm x}+\hat{\nu}},\qquad w_{\bm x}=\pm 1.
\label{gautra}
\end{eqnarray}
Due to the ${\mathbb Z}_2$ gauge invariance, the vector correlation function
$G_s({\bm x},{\bm y}) = \langle
{\bm s}_{\bm x} \cdot {\bm s}_{\bm y} \rangle$, trivially vanishes for
${\bm x}\neq {\bm y}$. For $N\ge 2$, the spontaneous breaking of the
global SO($N$) symmetry is signaled by the condensation of the
gauge-invariant bilinear operator $R_{\bm x}^{ab}$,
\begin{eqnarray}
  R^{ab}_{\bm x} = s_{\bm x}^a s_{\bm x}^b - {1\over N}\delta^{ab},
  \label{Rabz2gN}
\end{eqnarray}
which is the analogue of the operator $Q^{ab}_{\bm x}$ defined in the 
LAH model, see Eq.~(\ref{QdefncAH}). Its 
correlation function
\begin{eqnarray}
  G_R({\bm x},{\bm y}) \equiv \langle \,{\rm Tr} \,R_{\bm x} R_{\bm
    y} \,\rangle
  \label{GRabz2gN}
\end{eqnarray}
orders at the transition. If the transition is continuous, 
at the critical point it behaves as 
\begin{eqnarray}
  G_R\left. ({\bm x},{\bm y})\right |_{T=T_c} \sim |{\bm x}-{\bm y}|^{-2Y_R},
    \label{gryr}
\end{eqnarray}
where $Y_R$ is the RG dimension of $R^{ab}_{\bm x}$, which may depend
on the number $N$ of components of the scalar field and also on the 
specific transition line (as we shall see, this occurs for $N=2$).

\subsection{Phase diagram and critical behavior of
   ${\mathbb Z}_2$-gauge $N$-vector models with $N\ge 2$}
\label{z2gNg2v}

\begin{figure}[tbp]
\centering
\includegraphics[width=0.65\columnwidth, clip]{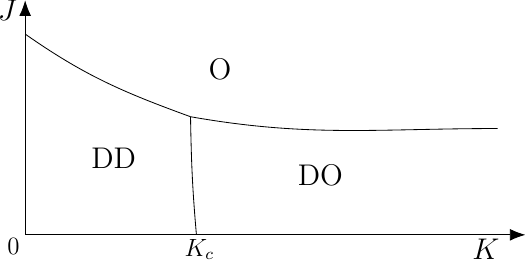}
\caption{Sketch of the $K$-$J$ phase diagram of the 3D ${\mathbb
    Z}_2$-gauge $N$-vector model with $N\ge 2$.  There are two
  spin-disordered phases for small $J$: a small-$K$ phase, in which
  both spin and ${\mathbb Z}_2$-gauge variables are disordered (we
  indicate it by DD), and a large-$K$ phase, in which the ${\mathbb
    Z}_2$-gauge variables order (we indicate it by DO).  For large $J$
  there is a single phase, in which both spins and gauge variables are
  ordered (we indicate it by O).  }
\label{phadiaz2gN}
\end{figure}

The phase diagram for $N\ge 2$ is sketched in Fig.~\ref{phadiaz2gN}, see, e.g.,
Refs.~\cite{FS-79,Sachdev-19,BPV-24-z2Nv, SC-24, CKT-25}.
Two spin-disordered phases are
present for small $J$: a small-$K$ phase, in which both spin and ${\mathbb
Z}_2$-gauge variables are disordered (DD), and a large-$K$ phase in which the
${\mathbb Z}_2$-gauge variables order (DO).  For large $J$ there is a single
ordered phase in which both spins and gauge variables order (O). These phases
are separated by three transition lines, which meet at one point, see
Fig.~\ref{phadiaz2gN}.\footnote{An estimate of the intersection point
$(K_\star,J_\star)$ of the three lines for small values of $N$ ($N\lesssim 5$
say) is reported in Ref.~\cite{BPV-24-z2Nv}: $K_\star\approx 0.761 - 0.003 N
\approx 0.75$, $J_\star\approx 0.23$.} Note that in the ordered
phase, once a representative of the minimum energy configuration is chosen,
i.e., ${\bm s}_{\bm x}={\bm s}_{{\bm x}+\hat{\mu}}$ and $w_{{\bm x},\mu}=1$ (up
to gauge transformations and in the thermodynamic limit), there is no residual
gauge symmetry. Therefore, 
we do not expect the presence of topological phases or
transitions, see the discussion reported in Sec.~\ref{sec.ZQhiggs}.

In the following we review the existing theoretical and numerical
results for the nature of the transitions along the three transition
lines. More details can be found in
Refs.~\cite{BPV-24-z2Nv,BPV-24-unco}.

\subsubsection{Topological ${\mathbb Z}_2$-gauge transitions
  along the DD-DO transition line}
\label{dddo}

For $J=0$ the model reduces to the ${\mathbb Z}_2$ gauge
model~\cite{Wegner-71}, reviewed in Sec.~\ref{sec.ZQ}, for any
$N$. Therefore, there is a continuous topological phase
transition~\cite{BPV-20-hcAH,FXL-18} at $J=0$, $K_{{\mathbb
    Z}_2}=0.761413292(11)$, separating a small-$K$ confined phase from
a large-$K$ deconfined phase. The $J=0$ critical point is the starting
point of a transition line, which separates two phases in which the
spin variables are disordered.  For $J>0$ these phases can no longer
be characterized by the size behavior of the Wilson loops, which
always obey the perimeter law because of the screening of the charged
fields. Nonetheless, they can still be distinguished by the different
topological properties of the gauge modes~\cite{FS-79,Sachdev-19}. In
particular, the gauge field is disordered for small $K$ and ordered
for large $K$.  This result follows from the observation that, since
the spin variables are not critical for sufficiently small values of
$J$, they can be integrated out. For the values of $J$ for which the
strong-coupling expansion converges, one obtains an effective
${\mathbb Z}_2$ gauge theory with local couplings, which is expected
to have the same critical behavior as the model for $J=0$.\footnote{At
leading order in $J$, one again obtains the ${\mathbb Z}_2$ gauge
model~\cite{Wegner-71,FS-79,Kogut-79,LRT-93}, with renormalized gauge
coupling $K + N J^4$.  This implies $K_c(J) = K_c(J=0) - N J^4 +
O(J^6)$.}

\subsubsection{RP$^{N-1}$-like transitions along the 
  DD-O transition line}
\label{ddo}

The transitions along the DD-O line have the same nature as the
transition in the RP$^{N-1}$ model obtained for
$K=0$~\cite{BPV-24-z2Nv}.  As discussed in Sec.~\ref{RPN}, the
RP$^{N-1}$ model undergoes a LGW transition, which may be continuous
only for $N=2$---in this case it belongs to the XY universality
class. For $N\ge 3$ the transitions are of first order.

The order parameter of the DD-O transitions is the bilinear operator
$R^{ab}_{\bm x}$ defined in Eq.~\eqref{Rabz2gN}.  For $N=2$, if the
transition is continuous, it behaves as an XY vector field, so its RG
dimension $Y_R$ coincides with the RG dimension $Y_{V,{\rm XY}} =
(1+\eta_{\rm XY})/2$ of the vector field in the standard XY model.
Thus, at continuous transitions along the DD-O line for $N=2$, we have
$Y_R = Y_{V,{\rm XY}}=0.519088(22)$ (using the estimates of $\eta_{\rm
  XY}$ reported in \ref{univclass}). The transitions along the DD-O
line for $N=2$ have also been investigated
numerically~\cite{BPV-24-z2Nv}, confirming that continuous transitions
behave as predicted by the LGW approach, but also observing that the
continuous transitions turn into first-order ones for $K = K_{\rm
  tri}$, with $K_{\rm tri} < K^*$, i.e., before the intersection point
(at $K=K^\star$) of the transition lines.  Some numerical results are
reported in the left panel of Fig.~\ref{chi_R_N2}.

\subsubsection{O($N$)$^{\times}$ transitions along the DO-O transition line}
\label{onstar}

The transitions along the DO-O line, at least for large enough values of $K$,
belong to the O($N$)$^{\times}$ universality
class~\cite{BPV-24-unco,Senthil-23,SWHSL-16,IMH-12}.\footnote{Note that in
these references the O($N$)$^{\times}$ universality class was denoted by the
symbol O($N$)$^*$.}  As the standard O($N$) vector transitions, they are
characterized by the symmetry-breaking pattern SO($N$)$\to$O($N-1$).  However,
the vector field that is supposed to be the order parameter of the transition
is not gauge invariant. They are therefore LGW$^{\times}$ transitions in the
classification of Sec.~\ref{difftype}.

For $K\to\infty$ the plaquette term in the Hamiltonian (\ref{hamz2N})
converges to one.  Therefore, in infinite volume, we can set
$\sigma_{{\bm x},\mu} = 1$ modulo gauge transformations, obtaining the
same partition function as that of the standard $N$-vector model.  It
follows that the model undergoes a continuous transition at a finite
$J_{c}(K=\infty)$ belonging to the O($N$) vector universality
class.\footnote{Estimates of the critical point of the standard
$N$-vector models defined in Eq.~(\ref{Nvectormod})---therefore, of
$J_{c}(K=\infty)$---can be found in
Refs.~\cite{Hasenbusch-19,DBN-05,BFMM-96,Hasenbusch-22,DPV-15,BC-97,CPRV-96}.}

As discussed in Sec.~\ref{z2gaugenvectorm}, the breaking of the
SO($N$) symmetry along the DO-O line is signaled by the condensation
of the operator $R^{ab}_{\bm x}$ defined in Eq.~(\ref{Rabz2gN}). Along
the DO-O line this operator, as well as all gauge-invariant operators,
have the same critical behavior as in the conventional $N$-vector
model without gauge invariance \cite{BPV-24-z2Nv}. The equivalence of
the gauge-invariant correlations in O($N$) models and along the DO-O
line implies that the gauge modes do not drive the critical behavior
for finite large values of $K$. Moreover, these transitions have the
same effective description as the conventional O($N$) ones, although
there is no appropriate gauge-invariant order parameter.  These
transitions are therefore O($N$)$^{\times}$ transitions, according to
the classification reported in Sec.~\ref{difftype}. Note that these
transitions differ from the RP$^{N-1}$  transitions that occur
along the DD-O line, which are also LGW transitions but with a
different, gauge-invariant, order parameter, see Sec.~\ref{ddo}.

\begin{figure}[tbp]
\centering
\includegraphics[width=0.45\columnwidth, clip]{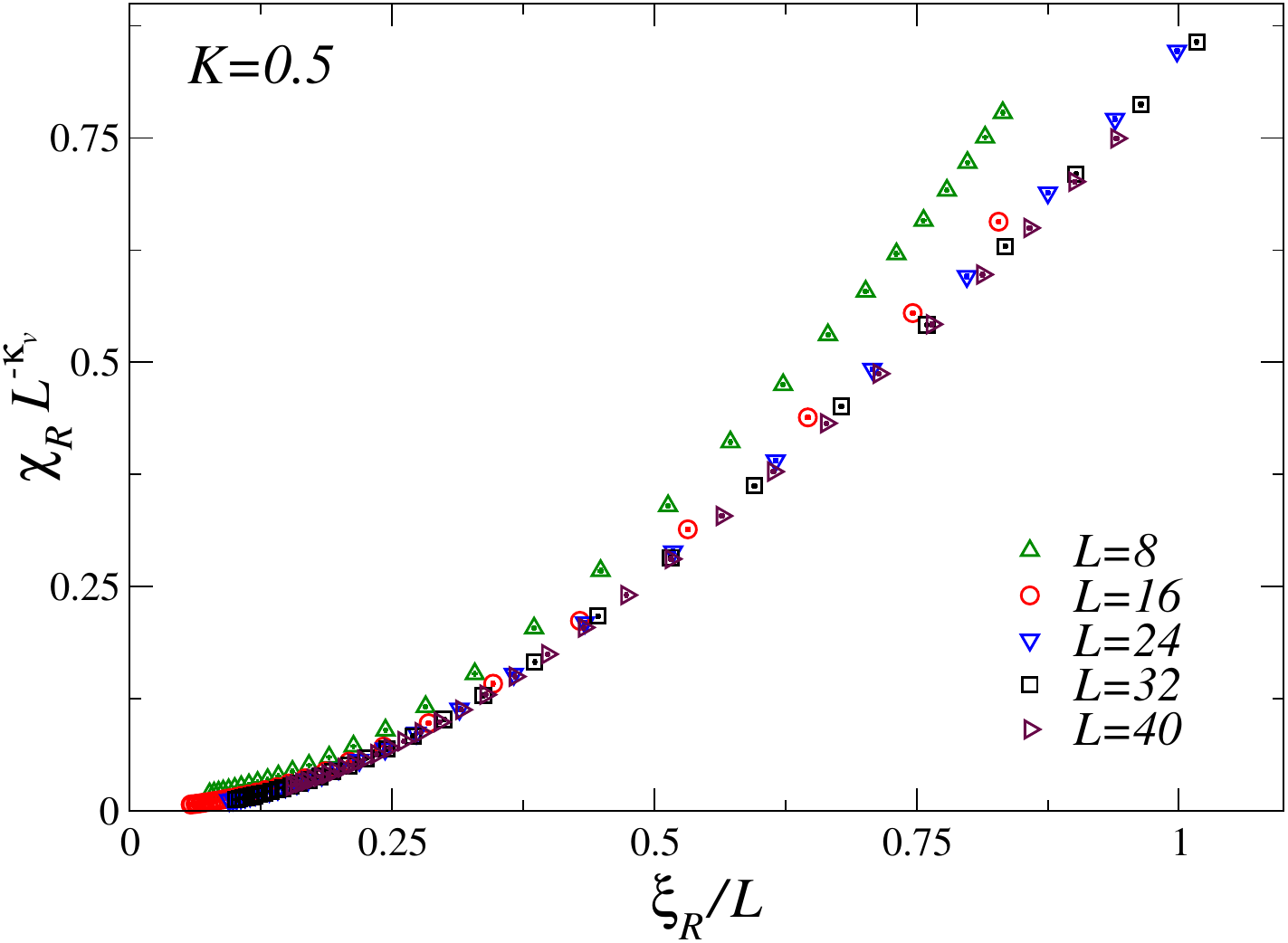}
\hspace{1cm}
\includegraphics[width=0.45\columnwidth, clip]{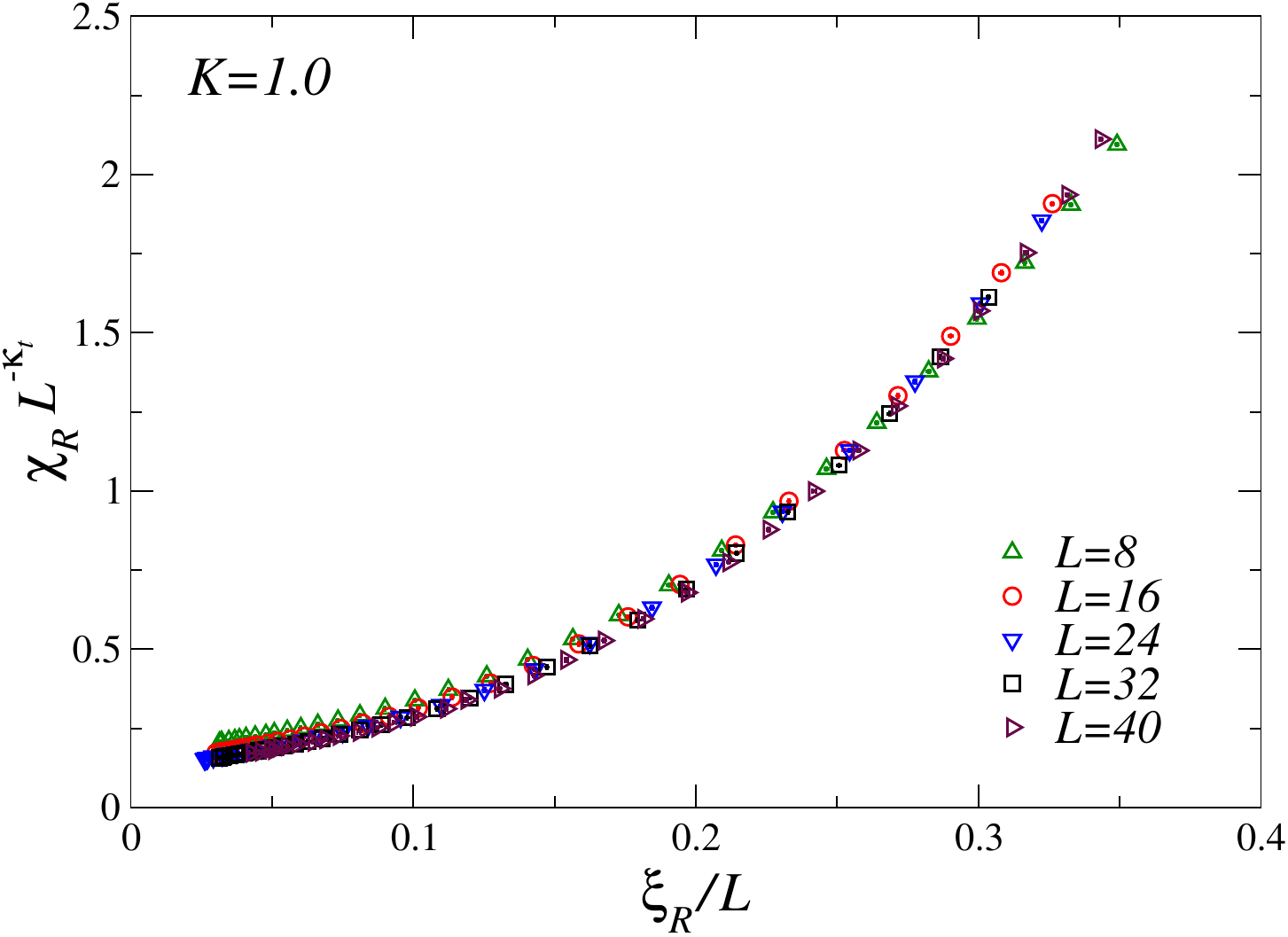}
\caption{(Adapted from Ref.~\cite{BPV-24-z2Nv}) Plot of the rescaled
  susceptibility $\chi_R$ versus $\xi_R/L$, where $\chi_R$ and the
  second-moment correlation length $\xi_R$ are defined in terms of
  two-point function $G_{R}({\bm x},{\bm y})$ defined in
  Eq.~(\ref{gryr}).  Results for $N=2$. Left: results for $K=0.5$
  (DD-O line); we set $\kappa_v=3-2Y_{V,XY}=1.96182$. Right: results
  for $K=1.0$ (DO-O line); we set $\kappa_t=3-2Y_{T,XY}=0.5274$.  The
  nice data collapse shows that $Y_{V,XY}$ and $Y_{T,XY}$ are the
  correct scaling dimensions of the operator $R_{\bm x}^{ab}$ along
  the DD-O and DO-O transition lines, respectively.  \label{chi_R_N2}}
\end{figure}

The existence of the DO-O O($N$)$^{\times}$ transition line implies
the stability, against gauge fluctuations, of the O($N$) vector
transition occurring for $K\to\infty$. Note that this is only possible
if the gauge group is discrete and is related to the fact that the
DO-O line separates two phases, in which the gauge variables are
topologically ordered.  Indeed, gauge fluctuations are expected to be
generally relevant in models with continuous Abelian and non-Abelian
gauge symmetries.  In that case, gauge interactions destabilize the
vector critical behavior, leading to transitions of different nature,
as discussed in Section~\ref{epsexpAHFT}.

In O($N$)$^{\times}$ transitions, the main issue is the identification
of the non-gauge-invariant order parameter.  In noncompact models,
local order parameters are identified by introducing an appropriate
gauge fixing, the Lorenz gauge, see Sec.~\ref{LAHMnc} and, particular,
Sec.~\ref{nc-orderparameters}. In the context of compact models, one
can exploit a stochastic gauge-fixing
procedure~\cite{BPV-24-unco,BPV-24-z2Nv}, as discussed below in
Sec.~\ref{unconstar}.  The introduction of this peculiar gauge fixing
allows one to identify the spin variable ${\bm s}_{\bm x}$ as the
order parameter of the transition, as in standard $N$-vector systems.
Numerical results for $N=1,2,3$ reported in
Refs.~\cite{BPV-24-unco,BPV-24-z2Nv} fully confirm this scenario.
In the gauge-fixed theory, along the DO-O line the vector field ${\bm
  s}_{\bm x}$ behaves as in the standard $N$-vector model.

Let us finally focus on the ${\mathbb Z}_2$-gauge XY ($N=2$) model,
which is the only case in which continuous transitions may occur along
all three transition lines.  It is interesting to note that the
continuous transitions along the DD-O and DO-O lines are all expected
to belong to the 3D XY universality class. However, this does not
imply that the critical behavior is the same, as the two transition
lines have different effective LGW descriptions.  This is clearly
demonstrated by the different critical behavior of the operator
$R^{ab}_{\bm x}$ defined in Eq.~(\ref{Rabz2gN}).  As discussed in
Sec.~\ref{ddo}, the bilinear operator $R_{\bm x}^{ab}$ is the
effective XY order parameter at continuous DD-O transitions, so its RG
dimensions is the same as that of the vector field in the XY model:
$Y_R = Y_{V,{\rm XY}}= (1 + \eta_{\rm XY})/2 \approx 0.519$.  On the
other hand, along the DO-O line, gauge-invariant quantities behave as
in the O($N$) vector model. Therefore, the operator $R_{\bm x}^{ab}$
behaves as a spin-two tensor operator at a standard XY transition. It
follows that its RG dimension equals the RG dimension $Y_{T,{\rm XY}}$
of the spin-2 operator in the XY model (numerical estimates of this
quantity are reported in Refs.~\cite{CLLPSSV-20,HV-11,CPV-03,CPV-02}):
$Y_R = Y_{T,{\rm XY}} = 1.23629(11)$. These predictions have been
verified numerically in Ref.~\cite{BPV-24-z2Nv}.  As an example, some
results are reported Fig.~\ref{chi_R_N2}, which confirm the
identification of the DO-O transitions as O($N$)$^{\times}$ continuous
transitions with a gauge-dependent vector order parameter, and of the
DD-O transitions as LGW transitions with a gauge-invariant spin-two order
parameter.

\subsection{The stochastic gauge fixing}
\label{unconstar}

In noncompact LAH models the critical behavior of gauge and charged
excitations can be determined by studying local correlation functions
in an appropriately gauge-fixed theory, see
Sec.~\ref{nc-orderparameters}.  Moreover, as discussed in
Refs.~\cite{BPV-23-gaufix,BPV-23-chgf} not all gauge fixings can be
used: local gauge-dependent operators may be used to uncover the
critical behavior of charged excitations in the Lorenz gauge, but
not, e.g., in the axial gauge.

By analogy, one expects that the identification of the vector order
parameter of O($N$)$^{\times}$ transitions with a local operator requires the
introduction of a gauge fixing in the theory.  A standard way of
fixing the gauge in compact models consists in setting the bond
variables on a maximal lattice tree equal to the
identity~\cite{Creutz-76, OS-78}. A particular case of this procedure
is the axial gauge, in which all bonds in a given lattice direction
are set equal to the identity (paying attention to the boundary
conditions and adding some additional constraints on the boundaries).
However, as discussed in Ref.~\cite{BPV-24-unco}, these gauge fixings
do not solve the problem, i.e., they do not allow one to identify the
critical vector modes that characterize the O($N$)$^{\times}$ universality
class. As we already mentioned, this is not unexpected, as the axial
gauge is also not appropriate in noncompact gauge
models~\cite{BPV-23-gaufix,BPV-23-chgf}. We also mention that
Ref.~\cite{BPV-24-unco} also considered a Lorenz-like complete gauge
fixing, again with no success.

To define a local O($N$)$^{\times}$ order parameter, a 
novel procedure, which is conceptually different
 from the usual gauge-fixing strategies, was
proposed in Refs.~\cite{BPV-24-unco,BPV-24-z2Nv}.
It generalizes the
variational approach that is used in the context of lattice gauge
theories and is usually called Landau gauge fixing.\footnote{In a
lattice gauge theory with gauge fields $U_{{\bm x},\mu}$, the Landau
gauge is defined as the set of gauge fields that maximize $\sum_{{\bm
    x}\mu} \hbox{Re Tr }U_{{\bm x},\mu}$, see, e.g.,
Ref.~\cite{Davies-etal-88}.  This approach is numerically difficult to
apply and, in general, it suffers from the so-called Gribov
problem~\cite{Gribov-78,Singer-78}: there may be several degenerate or
quasidegenerate maxima. In the continuum limit, it goes over to the
gauge-fixing condition $\partial_\mu A_\mu = 0$, that we have named
Lorenz gauge, but which is also referred to as Landau gauge.  In
continuum theories Landau gauge and Lorenz gauge are often used
interchangeably, although the name ``Landau gauge" is often preferred
for Euclidean field theories, ``Lorenz gauge" for theories defined in
Minkowski space. To be consistent with the definitions of
Sec.~\ref{LAHMnc}, we would call it Lorenz
gauge. \label{footnote:Landau}} We review it below, in the context of
the ${\mathbb Z}_2$-gauge $N$-vector model.

The basic idea is to average non-gauge invariant quantities over all
possible gauge transformations with a properly chosen, not
gauge-invariant, weight. This is achieved by introducing ${\mathbb
  Z}_2$ fields $w_{\bm x}=\pm 1$ defined on the lattice sites, and an
ancillary Hamiltonian $H_w$ that generally depends on $w_{\bm x}$,
${\bm s}_{\bm x}$, and $\sigma_{{\bm x},\mu}$. If $A({\bm s}_{\bm
  x},\sigma_{{\bm x},\mu})$ is a function of the fields, one defines
its weighted average over the gauge transformations as
\begin{eqnarray} 
[A({\bm s}_{\bm x},\sigma_{{\bm x},\mu})] = 
{\sum_{\{w\}} A(\hat{\bm s}_{\bm x},
  \hat\sigma_{{\bm x},\mu}) e^{-H_w} \over
  \sum_{\{w\}}  e^{-H_w} },\qquad
  \hat{\bm s}_{\bm x} = w_{\bm x}{\bm s}_{\bm x}, \qquad
  \hat\sigma_{{\bm x},\mu} = w_{\bm x}\sigma_{{\bm x},\mu} w_{{\bm
      x}+\hat{\mu}},
\label{gfav}
\end{eqnarray}
where $\hat{\bm s}_{\bm x}$ and $\hat\sigma_{{\bm x},\mu}$ correspond
to the fields obtained by performing a gauge transformation with gauge
function $w_{\bm x}$ as in Eq.~\eqref{gautra}. The average 
$[A({\bm s}_{\bm x},\sigma_{{\bm x},\mu})]$ is then averaged over the fields
${\bm s}_{\bm x}$ and $\sigma_{{\bm x},\mu}$ using the original
Hamiltonian (\ref{hamz2N}), i.e.,
\begin{eqnarray}
  \langle [A({\bm s}_{\bm x},\sigma_{{\bm x},\mu})] \rangle =
{\sum_{\{{\bm s},\sigma\}} [A({\bm s}_{\bm x},
  \sigma_{{\bm x},\mu})] e^{-H} \over
  \sum_{\{{\bm s},\sigma\}}  e^{-H} }.
\label{finav}
\end{eqnarray}
One can easily see that gauge-invariant observables are invariant
under this gauge-fixing procedure. In this approach, we can define
the vector correlation function as
\begin{equation}
G_V({\bm x},{\bm y}) = \langle [{\bm s}_{\bm x} \cdot {\bm
    s}_{\bm y}] \rangle.
\label{gvgf}
\end{equation}
The choice of the ancillary Hamiltonian $H_w$ is a crucial point. To
obtain critical vector correlations, one would like to work in a gauge
which maximizes the number of bonds with $\sigma_{{\bm x},\mu} =
1$. Indeed, this implies that the Hamiltonian for the
gauge-transformed fields $\hat{\bm s}_{\bm x}$ is almost
ferromagnetic. Therefore, these fields display the same critical
behavior as vector fields in the O($N$) model.  With this idea in
mind, an optimal choice is provided by the simple Hamiltonian
\begin{equation}
  H_w(\gamma)=- \gamma \sum_{{\bm x},\mu} \hat\sigma_{{\bm x},\mu} = -
  \gamma \sum_{{\bm x},\mu} w_{\bm x} \sigma_{{\bm x},\mu} w_{{\bm
      x}+\hat\mu},
\label{Hw}
\end{equation}
where $\gamma$ is a positive number that should be large enough to
ensure that the minima of $H_w(\gamma)$ dominate in the average over
the gauge transformations.  It is worth noting that the global theory
including the quenched stochastic gauge fixing is invariant under an
extended set of local transformations with ${\mathbb Z}_2$-gauge
parameter $v_{\bm x}=\pm 1$ given by\footnote{For $N=1$ i.e., in the
${\mathbb Z}_2$-gauge Higgs model, one can fix this gauge invariance
by working in the unitary gauge, $s_{\bm x}=1$ for all points ${\bm
  x}$. \label{foot:unitary} }
\begin{eqnarray}\label{gaugetr-2}
{\bm s}_{\bm x} \to v_{\bm x} {\bm s}_{\bm x}, \qquad
\sigma_{{\bm x},\mu}  \to
v_{\bm x} \sigma_{{\bm x},\mu} v_{{\bm x}+\hat{\mu}}, \qquad
w_{\bm x} \to v_{\bm x} w_{\bm x}.
\end{eqnarray}
Thus, only observables that are invariant under this set of
transformations, such as $\hat{s}_{\bm x}$ and $\hat{\sigma}_{{\bm
    x},\mu}$, have nonvanishing correlation functions and a nontrivial
critical behavior.

The stochastic gauge-fixing procedure mimics what is done in random
systems with quenched disorder, for instance in spin glasses.  The
variables ${\bm s}_{\bm x}$ and $\sigma_{{\bm x},\mu}$ are the
disorder variables and $e^{-H}/Z$ represents the disorder
distribution, while $w_{\bm x}$ are the system variables that are
distributed with Gibbs weight $e^{-H_w}/Z_w$ at fixed disorder.  In
the language of disordered systems, the average $[\cdot ]$ therefore
represents the thermal average at fixed disorder, while $\langle
\cdot\rangle$ is the average over the different disorder
realizations.\footnote{To avoid confusion, note that symbols $[\cdot
]$ and $\langle \cdot\rangle$ have typically the opposite meaning in
the random-system literature: the former represents the disorder
average and the latter the thermal average.}  This analogy allows one
to exploit known results for the properties of quenched random
systems. In particular, the present procedure is thermodynamically
consistent and, when the low-temperature (large $\gamma$) phase is not
a spin-glass phase, it admits a local field-theory representation,
see, Refs.~\cite{GL-76,PV-00}.  Thus, the standard RG machinery
can be applied to correlations computed in the gauge-fixed theory.

The resulting model with the added variables $w_{\bm x}$ is a quenched
random-bond Ising model~\cite{Harris-74,EA-75} with a particular
choice of bond distribution.  Quenched random-bond Ising models have
various phases---disordered, ferromagnetic, and glassy
phases---depending on the temperature (whose role is played here by
$1/\gamma$), the amount of randomness of the bond distribution, and
its spatial correlations, see, e.g.,
Refs.~\cite{Harris-74,EA-75,Nishimori-81,LH-88,PC-99,Betal-00,Nishimori-book,
KKY-06,HPPV-07-a,HPPV-07-b,HPV-08-a,HPV-08-b,CPV-11,Janus-13,LPP-16,CP-19,Nishimori-24,
Nishimori-25}.
In particular, the present model is expected to undergo a quenched
transition at $\gamma = \gamma_c(J,K)$ for any $J$ and $K$. 
The transition separates a
disordered phase for $\gamma < \gamma_c(J,K)$ from a large-$\gamma$
phase, which {\em a priori} can be ferromagnetic or glassy. If $J$ and
$K$ belong to the DO-O transition line, the large-$\gamma$ phase turns
out to be ferromagnetic.\footnote{
Note that the two-point function of
$w_{\bm x}$ vanishes for non-coincident points, due to the generalized
gauge invariance, Eq.~\eqref{gaugetr-2}. The nature of the
large-$\gamma$ phase can be determined by considering correlation
functions of the so-called overlap $O_{\bm x}=w_{\bm x}^{(1)}w_{\bm
  x}^{(2)}$, where $w_{\bm x}^{(1)}$ and $w_{\bm x}^{(2)}$ represent
two different configurations sampled with the same probability
$e^{-H_w}/Z_w$, i.e. corresponding to the same values of
$\sigma_{{\bm x},\mu}$ and $s_{\bm x}$.  Along the DO-O line, for
$\gamma > \gamma_c(J,K)$ the Binder cumulant of the overlap variable
approaches $1$, and the overlap susceptibility diverges linearly with
the volume, indicating that the large-$\gamma$ phase is a standard
ferromagnetic phase.}  
Thus, the long-distance behavior of the
variables $w_{\bm x}$ is the same for all $\gamma > \gamma_c(J,K)$:
The variables $w_{\bm x}$ simply make uncorrelated short-range
fluctuations around the minimum configurations obtained for $\gamma
\to \infty$. It is thus natural to conjecture that $\gamma$ is an
irrelevant parameter, i.e., that the critical behavior of the
gauge-fixed quantities is the same for any $\gamma>\gamma_c(J,K)$
along the DO-O transition line.  In the RG language, $1/\gamma$ represents
an irrelevant perturbation of the $\gamma=\infty$ fixed point.  The
irrelevance of $\gamma$ is conjectured to be a general feature of the
stochastic gauge fixing, which holds for any $N$ (the numerical
analyses of Refs.~\cite{BPV-24-unco,BPV-24-z2Nv} confirm this
conjecture).  It is important to stress the significant advantage of
the stochastic gauge fixing with respect to traditional gauge-fixing
approaches.  As we can work at fixed $\gamma$, there is no need to
determine the minima or maxima of some gauge-fixing function,
bypassing the problem of the Gribov copies~\cite{Gribov-78,Singer-78}.

\begin{figure}[tbp]
\centering
\includegraphics[width=0.45\columnwidth, clip]{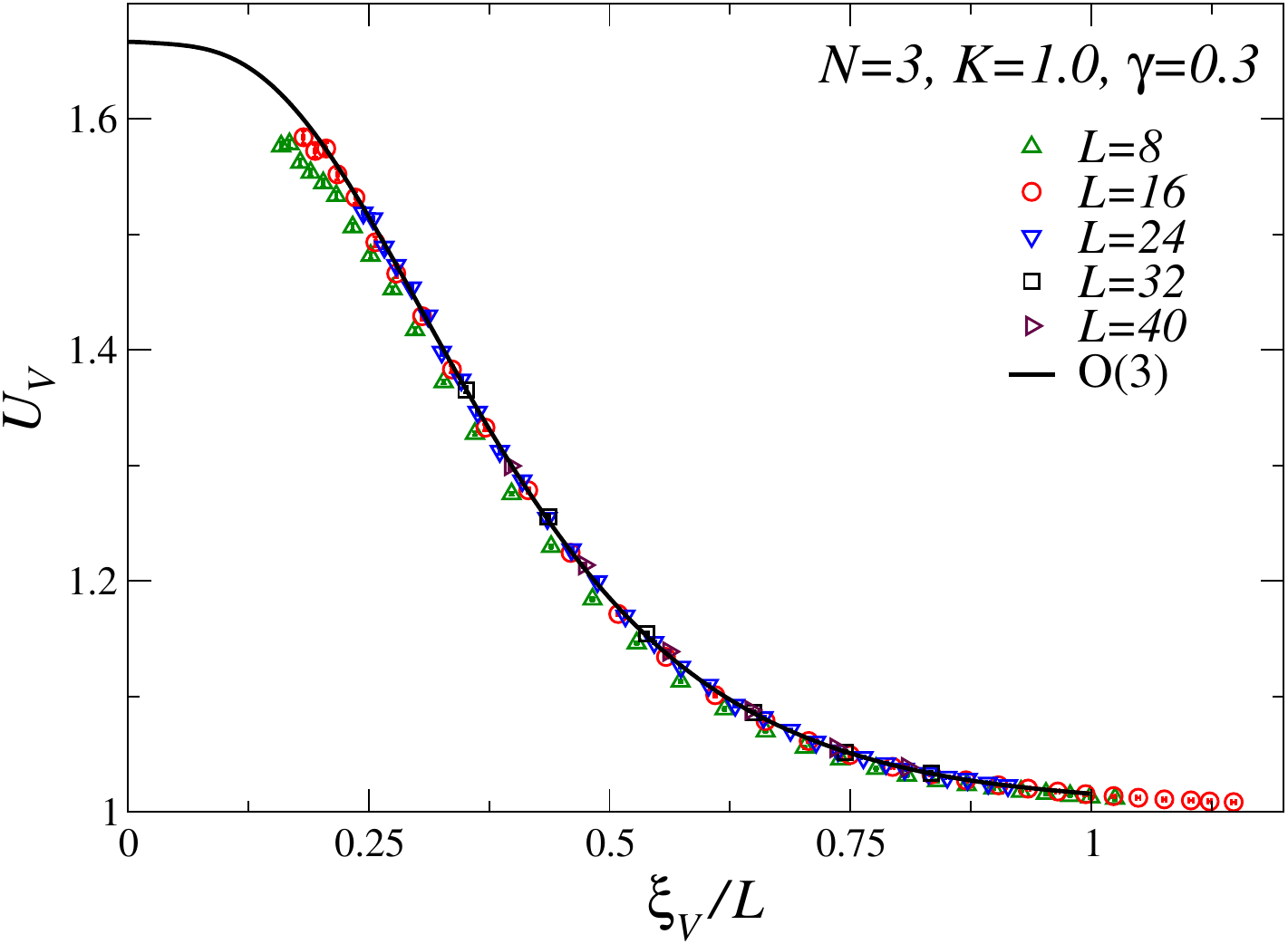}
\hspace{1cm}
\includegraphics[width=0.45\columnwidth, clip]{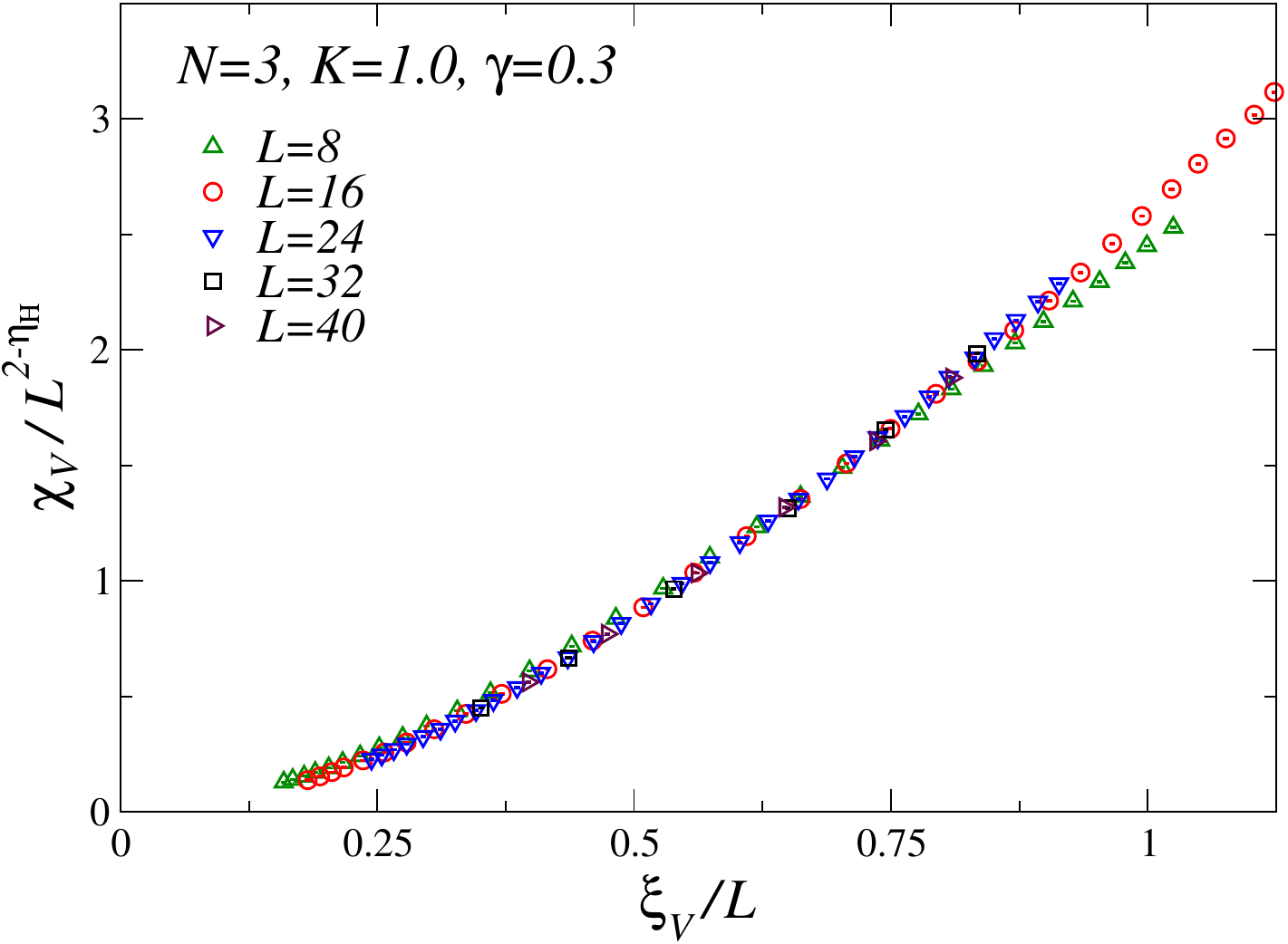}
\caption{(Adapted from Ref.~\cite{BPV-24-unco}) FSS analyses (see
  \ref{fssapp} for a discussion of the expected scaling behavior) 
  of MC data for the $\mathbb{Z}_2$-gauge $N$-vector
  model with $N=3$, using the stochastic gauge fixing with parameter
  $\gamma=0.3$. Simulations at fixed $K=1$ across the DO-O transition
  line.  (Left) Plot of the vector Binder parameter $U_V$ defined in
  terms of ${\bm s}_{\bm x}$, versus $\xi_V/L$, where $\xi_V$ is the
  second-moment correlation length computed by using the vector
  correlation function $G_V$. Data collapse on the $N$-vector ($N=3$)
  universal curve (see Ref.~\cite{BPV-21-bgi2} for a parametrization),
  showing that the vector degrees of freedom behave as at a standard
  O(3) transition.  (Right) Plot of the vector susceptibility $\chi_V$
  defined in terms of $G_V$ versus the ratio $\xi_V/L$. The
  susceptibility is rescaled according to Eq.~(\ref{chisclaw2}), using
  the exponent $\eta_H=0.03784$ of the Heisenberg O(3) universality
  class (see \ref{univclass} for a list of estimates of $\eta_H$). The
  nice collapse of the data supports the O(3) nature of the vector
  modes in the presence of a stochastic gauge fixing.  \label{V_N3}}
\end{figure}

The ancillary Hamiltonian $H_w$ given in Eq.~(\ref{Hw}) is simple and
allows us to reinterpret the gauge-fixed model as a quenched
random-bond model.  However, other choices may work as well.  It is
important to note that the emerging critical behavior of the vector
field is expected to be universal, i.e., independent of the ancillary
Hamiltonian $H_w$, provided that $H_w$ has been properly chosen to
make the spin-spin interactions ferromagnetic. This is essentially due
to the fact that the critical behavior of all gauge invariant
quantities---for instance, the spin-two operator $R_{\bm x}^{ab}$ or
the gauge-invariant energy---is independent of $H_w$: they all
behave as in the $N$-vector model. Therefore, along the DO-O line,
$G_V({\bm x} - {\bm y})$ should also behave as in the $N$-vector
model, if it is critical.

The theoretical predictions have been confirmed numerically in
Refs.~\cite{BPV-24-z2Nv,BPV-24-unco}. For both $N=2$ and 3, the
stochastically gauge-fixed correlation function $G_V$ defined in
Eq.~(\ref{gvgf}), as well as other vector observables like the vector
Binder parameter, show the same critical behavior as in the $N$-vector
model. Some data for $N=3$ are reported in Fig.~\ref{V_N3}.

We finally mention that the above ideas can be straightforwardly
extended to lattice gauge models with other discrete groups or with
continuous compact gauge variables.

\subsection{Phase diagram and critical behavior
   of the ${\mathbb Z}_2$-gauge Higgs model}
\label{z2gHim}

We now focus on the phase diagram of the ${\mathbb Z}_2$-gauge Higgs
model, i.e., of the lattice model (\ref{hamz2N}) for $N=1$. In this case
the spin variables take the integer values $s_{\bm x}=\pm 1$, as the
link variables.

The model satisfies an exact duality relation~\cite{BDI-75}. If we 
redefine  the Hamiltonian parameters as 
\begin{equation} 
 \left(J^\prime, K^\prime\right)=
  \left( -{1\over 2} {\rm ln}\,{\rm
      tanh}\,K\,, -{1\over 2} {\rm ln}\,{\rm tanh}\, J \right),
  \label{dualitymap}
\end{equation}
the free-energy density $F(J,K) = - {T\over L^d} \ln
Z$ satisfies the relation
\begin{eqnarray}
  F(J^\prime, K^\prime) = F(J, K) - {3\over 2}
  \ln[\sinh(2J)\sinh(2K)].
  \label{dualitymapF}
\end{eqnarray}
One can also define a self-dual line, 
requiring $J^\prime = J$ and $K^\prime = K$, which can be parametrized 
by one of the two equivalent equations
\begin{equation}
D(J,K) = J + {1\over 2} {\rm ln}\,{\rm tanh}\,K = 0, \qquad 
D'(J,K) = K + {1\over 2} {\rm ln}\,{\rm tanh}\,J = 0. \qquad 
\label{selfdual}
\end{equation}

\subsubsection{The phase diagram of the three-dimensional
  ${\mathbb Z}_2$-gauge Higgs model}
\label{phadiaz2higgs}

\begin{figure}[tbp]
\centering
\includegraphics[width=0.65\columnwidth, clip]{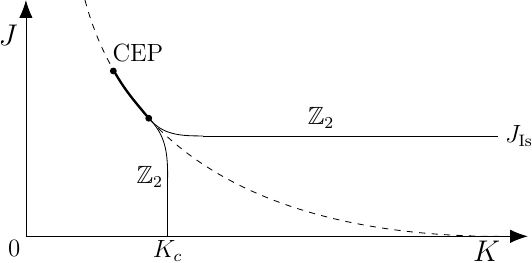}
\caption{Sketch of the phase diagram of the 3D ${\mathbb Z}_2$ gauge
  Higgs model, see, e.g., Refs.~\cite{BPV-22-z2g,SSN-21,TKPS-10}. The
  dashed line is the self-dual line, the thick line is a finite
  stretch of the self-dual line corresponding to first-order
  transitions.  The two lines labeled ``${\mathbb Z}_2$" are related
  by duality, and correspond to Ising continuous transitions.  They
  end at $[J = J_{\rm Is} \approx 0.221655, K=\infty]$ and at $[J =0,K
    = K_{{\mathbb Z}_2} \approx 0.761413]$.  The three lines meet at a
  MCP~\cite{JSJ-80,TKPS-10,SSN-21,BPV-22-z2g,OKGR-23,XPK-24} at
  $[K_\star=0.7525(1),J_\star= 0.22578(5)]$. The corresponding
  multicritical behavior turns out to be controlled by the
  multicritical XY FP~\cite{BPV-22-z2g,BPV-24-com,XPK-24}. The second
  endpoint (CEP) of the first-order transition line, at $[K\approx
    0.688,J\approx 0.258]$, is expected to be an Ising critical
  endpoint.}
\label{phadiaz2gN1}
\end{figure}

A sketch of the phase diagram is shown in Fig.~\ref{phadiaz2gN1}.  For
$K\to\infty$ an Ising transition occurs at~\cite{FXL-18} $J_{\rm Is} =
0.221654626(5)$.  By duality, in the pure ${\mathbb Z}_2$ gauge model
a transition occurs in the corresponding point, $J=0$ and $K_c =
-{1\over 2} {\rm ln}\,{\rm tanh}\,J_{\rm Is} = 0.761413292(11)$.  Two
Ising continuous transition lines, related by the duality
transformation (\ref{dualitymap}), start from these
points~\cite{FS-79, Nussinov-05} (the argument is analogous to that
discussed in Sec.~\ref{onstar}) and intersect on the self-dual
line~\cite{SSN-21,BPV-22-z2g} at $K_\star=0.7525(1)$, $J_\star=
0.22578(5)$.  Note that the transitions along both lines are
Ising-like transitions and that in both cases there is no global
symmetry. Nonetheless, as we shall discuss, the two lines have
different nature: the line that starts at $J=0$ is a topological
transition line, while the line that starts at $K = \infty$ is an
Ising$^{\times}$ line, i.e., it admits an Ising order parameter that
can only be identified once a proper gauge fixing is introduced.
Note that, in the absence of a gauge fixing, only the cumulants of the energy
density (see Sec.~\ref{fssencum} for definitions) 
can be used to characterize these
transitions in a gauge-invariant way ~\cite{SSN-21, BPV-22-z2g} (note
that the tensor $R^{ab}$ defined in Eq.~\eqref{Rabz2gN} trivially
vanishes for $N=1$). 

 Finally, numerical studies~\cite{JSJ-80,TKPS-10,
  GGRT-03} have provided evidence of first-order transitions along the
self-dual line, in the relatively small interval starting from the MCP
and ending at $J_{\star}\approx 0.258$ and $\kappa_{\star}\approx
0.688$. This endpoint is expected to correspond to a continuous
transition belonging to the Ising universality class (see
Ref.~\cite{SSN-21} for a numerical study).

Since the first-order transition line is limited to a finite interval
along the self-dual line, only two thermodynamic phases exist, see
Fig.~\ref{phadiaz2gN1}. For small $J$ and large $K$ there is a
topological deconfined phase. The remaining part of the phase diagram
corresponds to a single phase that extends from the disordered
small-$J,K$ region to the whole large-$J$ region. In particular, no
phase transition occurs along the line $K=0$, where the model becomes
trivial, as can be seen in the unitary gauge $s_{\bm x}=1$.  We may
only distinguish two different {\em regimes}: a Higgs-like regime in
the large $J$ and $K$ region, and a confined regime in the small $J$
and $K$ region (see, e.g., Ref.~\cite{CG-08}). However, these two regions
are connected by continuous paths in the phase diagram, along which
all local gauge-invariant operators (more precisely the gauge
invariant operators depending on a finite number of $s_{\bm x}$
and $\sigma_{{\bm x},\mu}$ variables) are analytic functions of the
Hamiltonian parameters~\cite{FS-79, Nussinov-05}.

\subsubsection{Order parameters driving the Ising$^{\times}$ transitions}
\label{ordpara}

Along the two transition lines that start at $J=0$ and $K=\infty$ and
are related by duality, thermal observables behave as in the Ising
universality class.  However, as already discussed in
Sec.~\ref{sec.ZQ}, the corresponding universality classes are
different. Even if there is no ${\mathbb Z}_2$ global
symmetry, the transitions along the line that starts at $K=\infty$ are
Ising$^{\times}$ transitions and admit, as we shall describe below, an
effective description in terms of a standard LGW theory with ${\mathbb
  Z}_2$ global invariance, once an appropriate gauge fixing has been
introduced. On the other hand, along the line that starts at $J=0$,
transitions are topological as in the pure ${\mathbb Z}_2$ gauge
model.\footnote{Note that the topological transitions along the
transition line starting from the $J=0$ critical point cannot be
characterized by the size behavior of the Wilson loops, at variance with 
what occurs 
in the 3D ${\mathbb Z}_2$ gauge model, see Sec.~\ref{sec.ZQ},
because Wilson loops satisfy the perimeter law for any $J>0$, due
to the screening of the site spin variables.  A more general
interpretation of these topological transitions is discussed in
Ref.~\cite{Sachdev-19} in a quantum setting, see also
Refs.~\cite{RS-91,Wen-91,BVD-92,MMS-01,Kitaev-03,FNSWW-03,HOS-04}.
They correspond to topological transitions
between a small-$K$ trivial phase and a
large-$K$ phase with topological order. These two phases
correspond to the confined and deconfined phases at $J=0$, respectively. 
We also
mention that these transitions are sometimes referred to as Ising$^*$
transitions; see, e.g.,
Refs.~\cite{TKPS-10,SWHSL-16,SSN-21,BPV-22-z2g}. As discussed in
Sec.~\ref{sec.ZQ}, we name them Ising gauge transitions.}

To verify the Ising$^{\times}$ critical behavior, Ref.~\cite{BPV-24-unco}
studied the ${\mathbb Z}_2$-gauge Higgs model with the
stochastic gauge fixing introduced in Sec.~\ref{unconstar}, using the
ancillary Hamiltonian $H_w$ defined in Eq.~(\ref{Hw}). Numerical FSS
analyses~\cite{BPV-24-unco} along the large-$K$ transition line show
that the critical behavior of the correlation function $G_V({\bm
  x},{\bm y})$ defined in Eq.~(\ref{gvgf}) is the same as in the
standard Ising model, with the large-$J$, large-$K$ region
corresponding to the magnetized phase. Therefore, as in multicomponent
models the stochastic gauge fixing allows us to identify the universal
gauge-dependent spin correlations which characterize standard Ising
transitions. The same conclusions were reached in
Ref.~\cite{Grady-21}, in what was called Landau gauge (see footnote
\ref{footnote:Landau}), which essentially corresponds to taking the
limit $\gamma\to\infty$ in the stochastic gauge fixing approach (but
note that the multiplicity of the minima, the so-called Gribov copies,
could not be taken into account).  The same analysis, using the
stochastic gauge fixing, was also performed along the small-$J$
transition line.  In this case the correlation function $G_V({\bm
  x},{\bm y})$ is not critical at the transition. As expected, these
transitions are not LGW$^{\times}$ transitions.

In the stochastic gauge-fixed theory, the spin $s_{\bm x}$ magnetizes
along the Ising$^{\times}$ transition line, i.e., there is a
ferromagnetic phase in the large-$J$, large-$K$ Higgs phase. On the
other hand, for small values of $J$, $s_{\bm x}$ is always
disordered. This implies the presence of an additional transition line in the
gauge-fixed theory,which separates the single phase~\cite{FS-79} in
which the confining and the Higgs regime both occur.  The line most
probably starts at the CEP, see Fig.~\ref{phadiaz2gN1}, and
necessarily ends on the $K=0$ axis.\footnote{For $K=0$ the
gauge-fixed model is equivalent to the so-called $\pm J$ Ising
model~\cite{Harris-74,EA-75,Nishimori-81,LH-88,PC-99,Betal-00,
  Nishimori-book,KKY-06,HPPV-07-a,HPPV-07-b,HPV-08-a,HPV-08-b,CPV-11,
  Janus-13,LPP-16,CP-19}, with spatially-independent random links (it
is immediate to verify it in the so-called unitary gauge $s_{\bm
  x}=1$, see footnote \ref{foot:unitary}).  In the notation of
Refs.~\cite{HPPV-07-b,HPV-08-a,HPV-08-b}, the probability $p$ of the
random link variable $J_{\rm xy}$ ($\sigma_{{\bm x},\mu}$ in our model
in the unitary gauge) is $P(J_{\rm xy}) = p \delta(J_{\rm xy}-1) +
(1-p)\delta(J_{\rm xy}+1)$ with $p=e^{J}/(e^J+e^{-J})$. The
temperature $T$ in the $\pm J$ Ising model corresponds to
$T=\gamma^{-1}$, so the line $J=\gamma$ corresponds to the so-called
Nishimori line~\cite{Nishimori-book}. Neglecting the small reentrance
of the ferromagnetic-glassy transition line---see
Refs.~\cite{HPPV-07-b,HPV-08-a,HPV-08-b} for a discussion of the phase
diagram of the $\pm J$ Ising model---,the ancillary model has a
large-$\gamma$ ferromagnetic phase only for $J > 1/T_{\rm N} = J_{N} =
0.5991(1)$ ($T_N$ is the Nishimori temperature).  Moreover, in the
ferromagnetic phase the correlation function $G_V$ orders (note that
the ancillary model has a ferromagnetic phase also close to the
small-$J$ topological line, but here $G_V$ does not order).  This
implies that the transition line ends in $K=0$,
$J=J_N$.} This is confirmed by numerical
analyses, which observe continuous transitions
in the small-$K$ region, on the left of the first-order transition
line, see Fig.~\ref{phadiaz2gN1}.  This line, which separates the
Higgs regime, in which the gauge-fixed spin correlations (in the generalized 
unitary gauge they are equivalent to 
correlations of the ancillary fields) magnetize,
from the confined regime, is
related to the specific form of the ancillary model and therefore does
not have a direct physical significance. However, it signals that the
thermodynamic description does not provide the full picture.

To distinguish the confinement/deconfinement properties of the model,
several nonlocal order parameters have been proposed, see, e.g.,
Refs.\cite{RS-91,Wen-91,BVD-92,BFZ-98,MMS-01,Kitaev-03,FNSWW-03,HOS-04,
  Sachdev-19,XPK-24}.  Note that the size dependence of Wilson loops
does not provide a good order parameter.  In the presence of dynamical
matter fields transforming in the fundamental representation, charge
screening is present and Wilson loops always scale with the perimeter
law. Only for $J=0$ does the transition at $K=K_{\mathbb{Z}_2}$
separate regions characterized by the area law and the perimeter law.
A possible order parameter is the so-called
Fredenhagen-Marcu~\cite{FM-83, FM-86-conf} order parameter, which is
defined by using a gauge-invariant two-point function built by joining
two spins $s_{\bm x}$ with a staple-shaped string of gauge fields,
normalized by the square root of a Wilson loop. The behavior of this
order parameter in the $\mathbb{Z}_2$-gauge Higgs model has been
studied in Refs.~\cite{XPK-24, ABP-24}. It correctly identifies the
confinement/deconfinement transition in the model and its
Ising$^{\times}$ universality class.  A different order parameter has
been advocated in Ref.~\cite{PK-90} (see also Ref.~\cite{BFZ-98}),
defined by using a properly normalized correlator between a Wilson and
't~Hooft loop, and studied numerically in Ref.~\cite{ABP-24}.  Another
order parameter, whose definition is strictly related with the
strong-coupling surface formulation of the $\mathbb{Z}_2$-gauge Higgs
model (see, e.g., Refs.~\cite{HL-91,SSN-21}), has been proposed and
numerically tested in Ref.~\cite{SSN-24}. The possibility that the
crossover between the confinement and the Higgs regime could induce
superficial phase transitions in quantum many-body systems has instead
been investigated in Refs.~\cite{VBVMT-22, TRVV-23}.

\subsubsection{Multicritical behavior at the intersection point of the
  transition lines}
\label{mcxybeh}

The first-order and the two continuous Ising transition lines
intersect in a multicritical point (MCP) located along the self-dual 
line~\cite{BPV-22-z2g,SSN-21,TKPS-10}.\footnote{See
\ref{fssmcp} for a brief review of the main universal features of the
scaling behavior close to a MCP.} Critical exponents have been
numerically computed at the
MCP~\cite{BPV-22-z2g,SSN-21,OKGR-23,XPK-24}, obtaining results
consistent with a multicritical XY behavior.

The apparent XY nature of the multicritical behavior can be naturally
explained \cite{BPV-22-z2g} by assuming that the effective description
of the transition is provided by the multicritical ${\mathbb
  Z}_2\oplus{\mathbb Z_2}$ LGW theory presented in
Sec.~\ref{sec:multi}, describing the competition of two one-component
($N_1=N_2=1$) scalar fields. This conjecture is made plausible by the
following arguments. The multicritical behavior arises from the
competition of the order parameters that characterize the critical
behavior along the two Ising-like lines. As we have discussed in the
previous section, along the large-$K$ transition line, we have
Ising$^{\times}$ transitions, with a scalar order parameter. Thus,
they admit a LGW description in terms of a local scalar nongauge-invariant
field $\phi_1$.  The transitions along the small-$J$ line are instead
topological transitions, characterized by an order parameter that is a
nonlocal function of the gauge degrees of freedom. Assuming that
duality can be extended from thermodynamic quantities to operators,
Ref.~\cite{BPV-22-z2g} argued that the nonlocal order parameter is
dual to a {\em local} order parameter, since the topological
transition line is mapped onto an Ising$^{\times}$ transition line
under duality. Thus, as a working hypothesis, one may associate a
scalar field $\phi_2$ with the topological Ising gauge transitions,
which locally interacts with the Ising$^{\times}$ order-parameter
field $\phi_1$.  If this conjecture holds, the multicritical behavior 
has an effective description in terms of a local Lagrangian for 
the two fields $\phi_1$ and $\phi_2$. We thus obtain 
the LGW Lagrangian  (\ref{on1on2}) discussed in
Sec.~\ref{sec:multi} with $N_1= N_2 = 1$.\footnote{We do not have
sound arguments to argue that these considerations can also be applied to the
${\mathbb Z}_2$-gauge $N$-vector models with $N>1$, to determine the
behavior at the intersection point of the three
transition lines, see Fig.~\ref{phadiaz2gN}.  Nevertheless, if, in analogy
with the ${\mathbb Z}_2$-gauge Higgs model, we assume that the
competition of the Ising critical modes (DD-DO line) and
the $N$-vector critical modes (DO-O line) can somehow be
represented by the competition of corresponding local order
parameters, we can use the multicritical LGW theory
(\ref{on1on2}) with $N_1=1$ and $N_2=N>1$ to understand the behavior 
close to the intersection point of the transition lines. 
Under this assumption, the
results for the 3D RG flow of the multicritical
O($N_1$)$\oplus$O($N_2$) LGW theory, outlined in Sec.~\ref{sec:multi},
predict that the intersection point would generally correspond to a
first-order transition for any $N>1$, with a  phase diagram analogous to the 
one shown in
the right panel of Fig.~\ref{multicri}.}

\begin{figure}[tbp]
\centering
\includegraphics[width=0.45\columnwidth, clip]{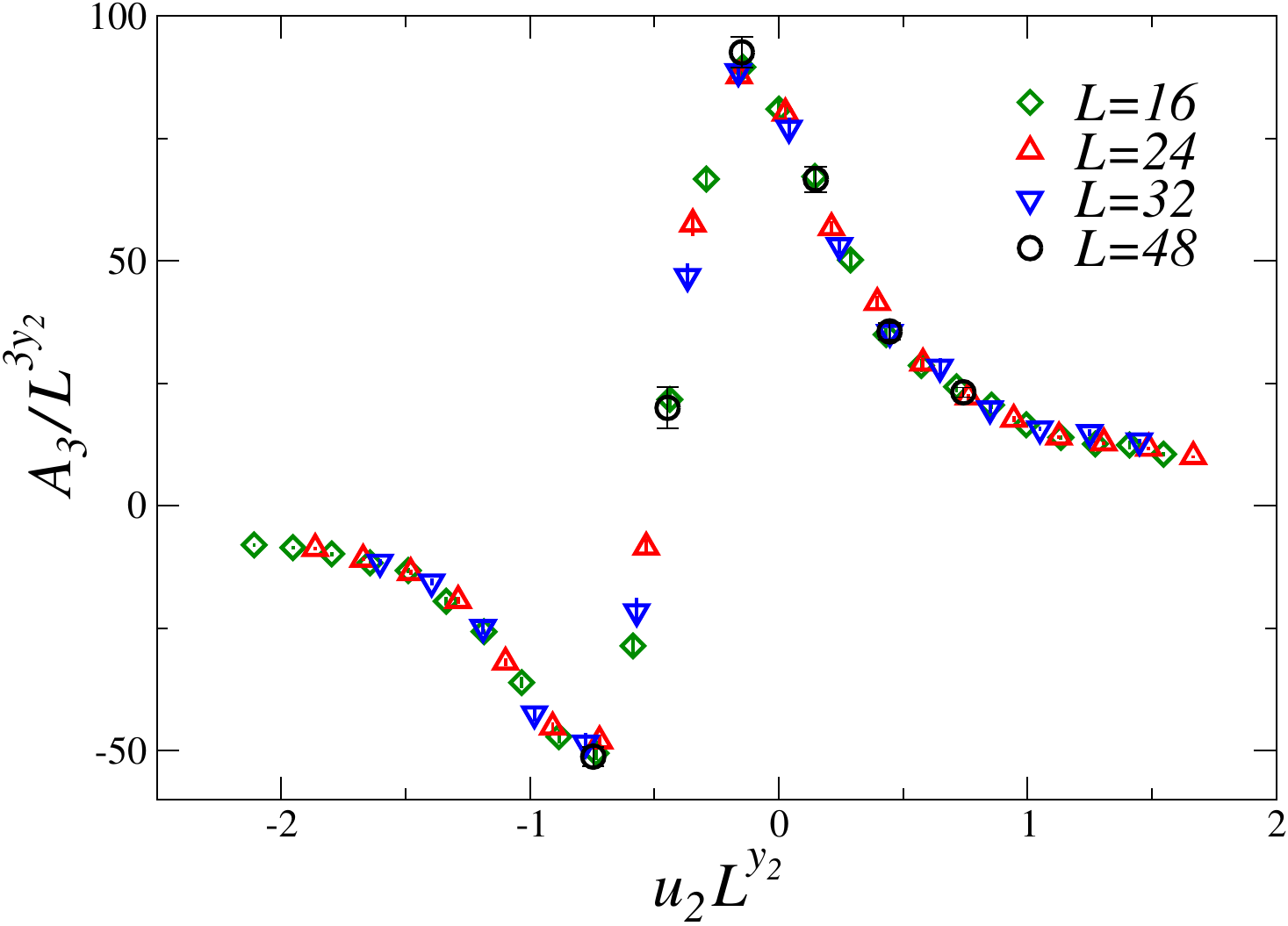}
\hspace{1cm}
\includegraphics[width=0.45\columnwidth, clip]{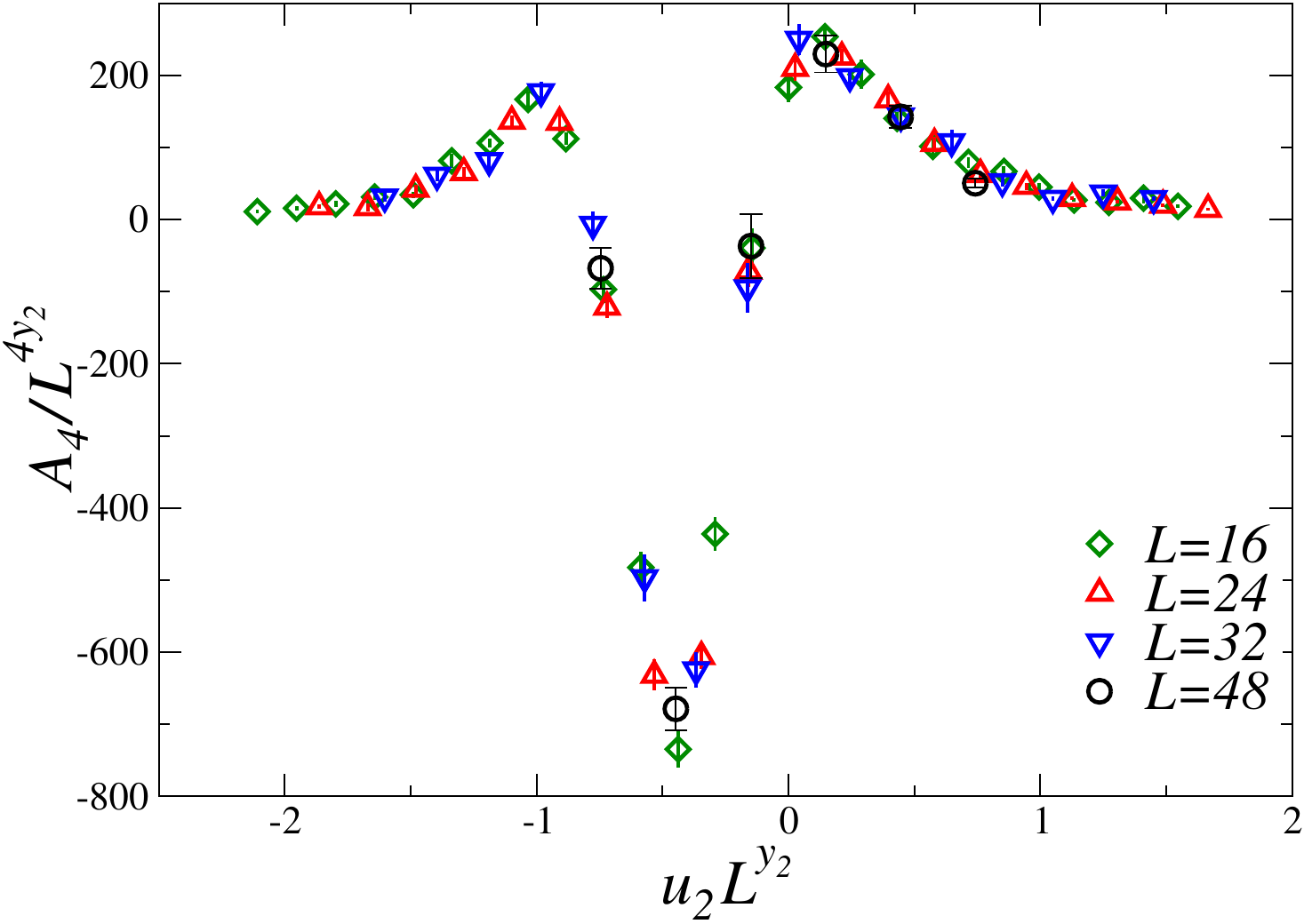}
\caption{(From Ref.~\cite{BPV-22-z2g}) FSS behavior of the third and
  fourth cumulant of an energy-like variable that scales with critical
  exponent $y_2$ when measured along the self-dual line (see
  \ref{fssmcp} for a discussion of FSS at MCPs).  See
  Ref.~\cite{BPV-22-z2g} for details.  The nice collapse of the data
  when using the XY critical exponent $y_2=1/\nu_{\rm XY}$ provides an
  accurate check of the conjectured multicritical XY behavior at the
  MCP point.
\label{fig_z2g}}
\end{figure}

The multicritical model defined in Sec.~\ref{sec:multi} has been
extensively
studied~\cite{FN-74,NKF-74,PV-02,CPV-03,HV-11,BPV-22-z2g}. In
particular, for $N_1=N_2=1$ it admits a bicritical point (see the left
panel of Fig.~\ref{multicri}) with a critical behavior controlled by a
stable fixed point with enlarged O(2) global symmetry.  The main
conjecture put forward in Ref.~\cite{BPV-22-z2g} is that this stable
multicritical XY FP is the one that controls the multicritical XY
behavior in the $\mathbb{Z}_2$-gauge Higgs theory. In this scenario,
the two relevant operators at the MCP are the quadratic spin-2 and
spin-0 field combinations, whose RG dimensions are~\cite{CPV-03,HV-11}
$y_1 = 1.7639(11)$ and~\cite{CHPV-06} $y_2 = 1/\nu_{\rm XY}=
1.4888(2)$ respectively. The numerical
estimates~\cite{SSN-21,BPV-22-z2g,OKGR-23,XPK-24} of $y_1$ and $y_2$
are in good agreement (errors are of the order of 1\%) with the more
accurate XY estimates reported above.  To give an idea of the accuracy
of the multicritical XY scenario, in Fig.~\ref{fig_z2g} we show the
FSS behavior of some particular third- and fourth-order energy
cumulants using the XY critical exponent $y_2=1/\nu_{\rm XY}$.  The
agreement with the multicritical XY scenario is globally very good and
strongly supports it.  However, we also note that the multicritical XY
theory is based on some unproved assumptions, the main one being the
existence of a (dual) local order parameter for the topological
transitions.  We should also mention that this assumption is still
controversial: Refs.~\cite{SSN-21, SSN-24, OKGR-23} consider it
implausible, and interpret the agreement between numerical results and
XY estimates as a mere coincidence.

\subsection{Other models with discrete symmetries}
\label{otherdismod}

In Secs.~\ref{LAHMnc} and \ref{LAHMc} we discussed the
phase diagram and critical properties of U(1) gauge theories with
global U($N$) symmetry, while in this section we considered gauge
models with discrete $\mathbb{Z}_2$ gauge symmetry and O($N$) global
symmetry. We now present some additional models with discrete gauge
and/or global symmetry, in which the gauge and/or the scalar fields take
only a discrete set of values.  These models can be considered as
approximations of models with continuous groups. However, classical
theorems by Jordan and Turing \cite{Turing-38} show that the only
continuous groups of transformations that can be approximated with
arbitrary precision by using their discrete subgroups are the Abelian
ones.  Thus, in the compact case, the only possibility is replacing
the U(1) group with the cyclic finite group ${\mathbb Z}_N$. In the
noncompact case, the additive group ${\mathbb R}$ can be replaced by
the discrete, but still infinite, additive group ${\mathbb Z}$.  As we
shall see, in some cases these discretized models undergo transitions
that are analogous to those occurring in models with continuous fields
and symmetry groups, with an enlargement of the global and/or local
gauge symmetry.

\subsubsection{Compact models with ${\mathbb Z}_{\cal N}$ local symmetry}
\label{otherdis1}

\begin{figure}[tbp]
\centering
\includegraphics[width=0.65\columnwidth, clip]{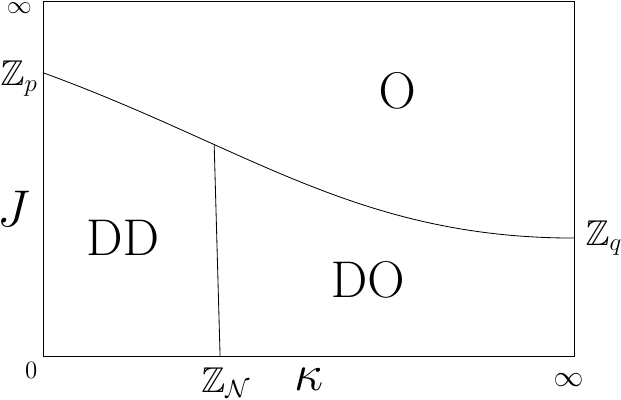}
\caption{Sketch of the phase diagram of the 3D model with Hamiltonian
  Eq.~\eqref{disH1}, for $p=q/\mathcal{N}>1$.  For small $J$ there are
  two phases: a spin- and gauge-disordered phase (DD) for small
  $\kappa$ (corresponding to the confined phase for $J=0$) and a spin-
  and gauge-ordered phase (DO) for large $\kappa$ (corresponding to
  the deconfined phase for $J=0$). For large $J$ a single spin- and
  gauge-ordered phase is present (O).  }
\label{ph_diag_disc1}
\end{figure}

We consider a model with global $\mathbb{Z}_q$ symmetry and local
$\mathbb{Z}_{\mathcal{N}}$ gauge symmetry. We assume that $q$ is an
integer multiple of $\mathcal{N}$ and we define $p=q/\mathcal{N}$. The
scalar fields are complex phases $z_{\bm x}$, which take $q$ different
values, $z_{\bm x}=\exp(2\pi i m/q)$ with $m=1,\ldots q-1$, while gauge
fields are complex phases $\lambda_{{\bm x},\nu}=\exp(2\pi n
i/\mathcal{N})$, $n=0,\ldots,{\mathcal{N}} -1$. The Hamiltonian and
the partition function are given by
\begin{equation}\label{disH1}
H= - 2J\ \sum_{{\bm x}, \mu} \mathrm{Re}\,
\bar{z}_{\bm x}\,\lambda_{{\bm x},\mu}\,
       {z}_{{\bm x}+\hat\mu}
- 2 \kappa \sum_{{\bm x},\mu>\nu} \hbox{Re}\,
\lambda_{{\bm x},\mu} \lambda_{{\bm x} + \hat{\mu},\nu}
   \bar{\lambda}_{{\bm x} + \hat{\nu},\mu} \bar{\lambda}_{{\bm x},\nu},\qquad
Z=\sum_{\{z,\lambda\}}e^{-H/T}.
\end{equation}
In the following we conventionally set $T=1$.  For $\mathcal{N}=q=2$
this model is the $\mathbb{Z}_2$-gauge Higgs model discussed in
Sec.~\ref{z2gHim}. Here we review the results of
Ref.~\cite{BPV-22-dis}, where these models for $p=q/\mathcal{N}>1$
were studied, obtaining the phase diagram reported in
Fig.~\ref{ph_diag_disc1}.

The nature of the different transition lines can be understood by
looking at the critical behavior when $\kappa$ or $J$ are zero or
infinity.  In the limit $\kappa\to\infty$ all plaquettes approach one,
and (in the thermodynamic limit) one can set $\sigma_{{\bm x},\mu} =
1$ on all links using a gauge transformation, thus obtaining the
ferromagnetic ${\mathbb Z}_q$ clock model. The critical properties of
this model have been discussed in Sec.~\ref{sec.ZQ}. We summarize here
the conclusions: for $q=2$ and 4 the $\mathbb{Z}_q$ clock model
undergoes a transition in the Ising universality class, for $q=3$ it
undergoes a first-order transition, while for $q\ge 5$ its critical
behavior is the same as in the XY model.

For $J = 0$, there are no scalar fields and one obtains the pure gauge
${\mathbb Z}_{\mathcal{N}}$ model with Wilson action, see
Eq.~\eqref{zqgaumodels}, whose critical properties have already been
discussed in Sec.~\ref{sec.ZQ}. We thus expect an Ising gauge
transition for $\mathcal{N}=2$ and 4, a discontinuous transition for
$\mathcal{N}=3$, and XY gauge transitions for $\mathcal{N}\ge 5$.

For $J\to\infty$ no transition is expected (fixing the global and
local invariance we can set $z_{\bm x} = \lambda_{{\bm x},\mu} = 1$ on
all links), while for $\kappa=0$ one can show, using universality
arguments, the transition to be a LGW transition in the $\mathbb{Z}_p$
universality class, see Ref.~\cite{BPV-22-dis} for more details.
Thus, no transition occurs for $p=1$. The model undergoes an Ising and
XY transitions for $p=2$ and $p\ge 5$, respectively, and a first-order
transition for $p = 3$.  An XY transition is also expected for
$p=4$. Indeed, in the latter case the model obtained for $\kappa=0$ is
a generic model with ${\mathbb Z}_4$ global symmetry, and thus it
should have an XY transition, as discussed in Sec.~\ref{sec.ZQ} and,
in particular, in footnote~\ref{clockfoot}. The order parameter at
these transitions is the gauge-invariant operator $s_{\bm
  x}^{\mathcal{N}}$.  This quantity transforms nontrivially only under
the ${\mathbb Z}_p$ subgroup of the global invariance group, so the
effective description is given by the LGW Lagrangian (\ref{lgwzq})
with $Q = p$.

From the discussion of the limiting cases it follows that the phase
diagram for $p>1$ has the form sketched in Fig.~\ref{ph_diag_disc1},
while the phase diagram for $p=1$ is qualitatively similar to the one
reported in Fig.~\ref{phadiaz2gN1}, with no transition for $\kappa =
0$.  It is reasonable to assume the transitions along the DD-O line to
be analogous to those occurring for $\kappa = 0$, while the
transitions along the DD-DO are expected to be topological transitions
as for $J=0$. Again, the case ${\cal N} = 4$ should be discussed
separately. Indeed, while the transition for $J = 0$ is an Ising gauge
transition, for $J\not=0$ we expect an XY gauge phase transition (see
the discussion in Sec.~\ref{sec.ZQ} and, in particular,
footnote~\ref{clockfoot}).

The behavior along the DO-O line is similar to what is observed in the
$\mathbb{Z}_2$-gauge $N$-vector model. Gauge fluctuations, due to the
discrete nature of the gauge group, are not able to destabilize the
${\mathbb Z}_q$ critical behavior observed for $\kappa=\infty$. The
scalar field---the order parameter in ${\mathbb Z}_q$ transitions---is
not gauge invariant, so the transition is of the LGW$^{\times}$ type.  We
thus expect Ising$^{\times}$ transitions for $q=2$, discontinuous transitions
for $q=3$, and XY$^{\times}$ transitions for $q\ge 5$.  For $q=4$ we expect
an XY$^{\times}$ transition if the gauge fluctuations are strong enough to
destabilize the decoupled Ising transition at $\kappa=\infty$, and an
Ising$^{\times}$ transition if this is not the case.

These predictions have been confirmed in Ref.~\cite{BPV-22-dis} by
numerical simulations, studying the behavior of the gauge-invariant
correlation function
\begin{equation}\label{Gdisc1}
G_{\mathcal{N}}({\bm x},{\bm y}) = 
   \mathrm{Re} \, \langle (\bar{z}_{\bm x} {z}_{\bm y})^{\mathcal{N}} \rangle\ .
\end{equation}
In particular, for $q=4$ 
the transitions along the DO-O line are numerically
compatible with an Ising$^{\times}$ behavior. Thus, gauge interactions
do not destabilize the 
accidental symmetry that is present
in the $\mathbb{Z}_4$ clock model at $\kappa\to\infty$.

It is interesting to note that the model with $\mathcal{N}=2$ and any
$p\ge 5$ has the same phase diagram and critical behaviors as the
$\mathbb{Z}_2$-gauge $N$-vector model with $N=2$, which can be
obtained by performing a naive limit $q\to\infty$ of the
Hamiltonian~\eqref{disH1}.\footnote{ For $\mathcal{N}=2$ the
correlation function $G_{\mathcal{N}}({\bm x},{\bm y})$ defined in
Eq.~\eqref{Gdisc1} is equal to $2 G_R({\bm x}, {\bm y})$ [see
  Eq.~\eqref{GRabz2gN}] for $N=2$, if we identify ${\bm s}_{\bm
  x}=(\mathrm{Re}\,z_{\bm x},\mathrm{Im}\,z_{\bm x})$ in
Eq.~\eqref{Rabz2gN}. Therefore, for $q\to\infty$ the function
$G_{\mathcal{N}}({\bm x},{\bm y})$ is equivalent to $G_R({\bm x}, {\bm
  y})$ in the $\mathbb{Z}_2$-gauge XY model.}  Thus, for $p \ge 5$ we
observe a symmetry enlargement of the global symmetry along the DD-O
and DO-O lines. Moreover, for ${\cal N} \ge 4$, correlation functions
display an enlarged gauge symmetry along the topological DD-DO
transition line.

\subsubsection{Gauge symmetry enlargement}
\label{otherdis2}

As we have already discussed several times, see, e.g.,
Sec.~\ref{otherdis1}, at transitions of LGW and LGW$^{\times}$ type,
gauge models may display an enlargement of the global
symmetry. Analogously, at topological transitions, one may observe an
enlargement of the gauge symmetry---this occurs in ${\mathbb Z}_Q$
gauge models, see Sec.~\ref{sec.ZQ}.  The models we are now going to
discuss have instead been investigated to understand whether there are
GFT transitions, effectively described by a continuum GFT, that 
show a gauge-symmetry enlargement.
In the present context, one
is looking for models with discrete gauge symmetry that undergo
charged transitions that are associated with the AHFT, with an
enlarged U(1) gauge symmetry.

Charged transitions controlled by the AHFT occur in the $N$-component
LAH model with noncompact Abelian gauge field (along the CH line) and in
the higher-charge ($Q>1$) $N$-component LAH model with compact
Abelian gauge field (along the DC-OD line).  In both cases the gauge
group can be approximated with arbitrary accuracy by considering
discrete compact or noncompact subgroups and one can define models
that are only invariant under these discrete subgroups.  The question
is whether these discrete models also have charged transitions as
their continuous counterparts.

We begin by discussing the discretized version~\cite{BF-23} of the
noncompact $N$-component LAH model with Hamiltonian~\eqref{ncLAHM}
defined in Sec.~\ref{ncLAHMsec}. The discretized model is defined in
the same way, the only difference being the set of values that the
gauge field $A_{{\bm x},\mu}$ takes. In the discretized model, the
gauge field takes only the values $2\pi m/q$, with
$m\in\mathbb{Z}$. The model is invariant under a restricted set of
gauge transformations (we indicate with ${\mathbb Z}_q^{(nc)}$ the
corresponding group)
\begin{eqnarray}
  {\bm z}_{\bm x} \to e^{i\Lambda_{\bm x}} {\bm z}_{\bm x},
  \qquad A_{{\bm x},\mu} \to A_{{\bm x},\mu} + \Lambda_{\bm
    x}-\Lambda_{{\bm x}+\hat{\mu}},
  \qquad \Lambda_{\bm x} = 2\pi m/q, \quad m\in\mathbb{Z}.
\end{eqnarray}
The original noncompact U(1) gauge theory is formally recovered in the
limit $q\to\infty$. Note that the gauge group is discrete, but
infinite. As already discussed in Sec.~\ref{ncLAHMsec}, a proper
definition of the model requires the introduction of $C^*$ boundary
conditions.

\begin{figure}[tbp]
\centering
\includegraphics[width=0.65\columnwidth, clip]{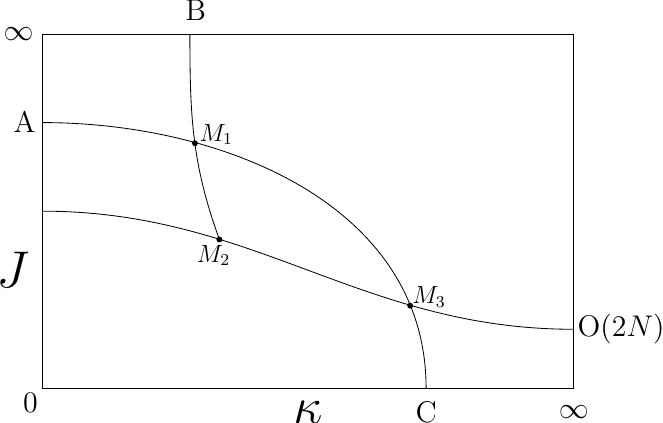}
\caption{Sketch of the phase diagram of the 3D $N$-component LAH model
  with discrete gauge symmetry. In the noncompact case the transition
  in $A$ belongs to the XY universality class, while the transitions at 
  points $B$ and $C$ belong to the IXY universality class \cite{BF-23}.
  In the compact case the transition in $A$ is in the XY universality
  class, while the transitions at points $B$ and $C$ are in the
  $\mathbb{Z}_Q$ and $\mathbb{Z}_q$ universality class, respectively
  \cite{BP-23}.  Both in the compact and in the noncompact case,  we
  have $J_c(A), \kappa_c(C)\to\infty$ in the limit $q\to\infty$.  }
\label{ph_diag_disc2}
\end{figure}

The model is invariant under global SU($N$) transformations as the
usual LAH model, the order parameter being the operator $Q_{\bm
  x}^{ab}$ defined in Eq.~\eqref{QdefncAH}.  Moreover, because of the
smaller local gauge symmetry, it is also invariant under an additional
set of global transformations with
$\mathbb{R}/\mathbb{Z}_q^{(nc)}=\mathrm{U(1)}/\mathbb{Z}_q$ symmetry
group.  A gauge-invariant order parameter associated with this
symmetry is the operator $\Sigma^{i_1\cdots i_q}_{\bm x}=z_{\bm
  x}^{i_1}\cdots z_{\bm x}^{i_q}$, where $z_{\bm x}^{i}$ denotes the
$i$-th component of the complex $N$-vector ${\bm z}_{\bm x}$.  Note
that this order parameter transforms nontrivially under the global
SU($N$) symmetry---thus, it acquires a nonvanishing expectation value
only if both symmetries are broken---while the order parameter $Q_{\bm
  x}^{ab}$ is invariant under the $\mathrm{U(1)}/\mathbb{Z}_q$
symmetry, and thus it is only sensitive to the breaking of the SU($N$)
symmetry.  As a consequence, we expect the
$\mathrm{U(1)}/\mathbb{Z}_q$ symmetry to be spontaneously broken only
in a phase in which the SU($N$) symmetry is also broken.

Due to its larger global symmetry, the phase diagram of the discrete
$N$-component LAH model is significantly more complex than that of the
$N$-component LAH with U(1) gauge symmetry. A sketch is shown in
Fig.~\ref{ph_diag_disc2}. Using the same arguments presented in
Sec.~\ref{onecoLAHM}, it is possible to show (see Ref.~\cite{BF-23}
for more details) that an IXY transition occurs both for $J=\infty$
(as in the noncompact U(1) gauge AH model) and for $J=0$. Moreover,
the transition on the $J=0$ axis occurs at $\kappa_c(J=0)=q^2
\kappa_c(J=\infty)$, where $\kappa_c(J=\infty)= 0.076051(2)$, so
$\kappa_c(J=0)\to\infty$ for $q\to\infty$.  This should not come as a
surprise, since we should recover the phase diagram of the model with
U(1) symmetry with no transition on the $J=0$ line, see
Fig.~\ref{phadiancLAH}, for $q\to\infty$.  Moreover, two transitions
are present on the $\kappa=0$ axis: a transition for
$J=J_{\mathrm{SU}(N)}$, where the global SU($N$) symmetry is
spontaneously broken, and a transition at $J=J_q$, associated with the
spontaneous breaking of the U(1)$/\mathbb{Z}_q$ symmetry. As discussed
above, we expect the latter symmetry to break in a phase in which the
SU($N$) symmetry is already broken, implying $J_q >
J_{\mathrm{SU}(N)}$ (therefore the point $A$ in
Fig.~\ref{ph_diag_disc2} corresponds to $J=J_q$).  Note that $J_q\to
\infty$ as $q\to \infty$, since this transition is not present in the
phase diagram of the noncompact U(1) gauge model, see
Fig.~\ref{phadiancLAH}.

In Ref.~\cite{BF-23} the phase diagram of the $N$-component
$\mathbb{Z}_q^{(nc)}$ gauge model has been numerically investigated,
studying in detail the critical behavior along the transition line
$M_2M_3$ in Fig.~\ref{ph_diag_disc2}. Indeed, for $q\to\infty$, this
line corresponds to the CH line (see Fig.~\ref{phadiancLAH}) of the
noncompact U(1) gauge model, which is the line where the AHFT critical
behavior emerges if the number $N$ of scalar components is large
enough, as discussed in Sec.~\ref{chmc}. The numerical results show
that, for $N=25$ and $q\ge 5$, the transitions along the line $M_2M_3$
are continuous, and that their critical behavior is the same as along
the CH line in the noncompact U(1) gauge model.  Thus, for $q\ge 5$
and large values of $N$, the transitions along the line $M_2M_3$ are
effectively associated with the stable charged FP of the AHFT, with a
corresponding enlargement of the gauge symmetry.  On the right of the
intersection point $M_3$ we instead have O($2N$)$^{\times}$ critical
transitions.

A similar analysis \cite{BP-23} can be performed for the compact model
defined in Sec.~\ref{qge2phdia}. One considers the multicomponent LAH
model with charge-$Q$ scalar fields and Hamiltonian~\eqref{clahmH} and
defines a discretized version by requiring the gauge fields to take
the values $\lambda_{{\bm x},\mu} = \exp(i 2\pi n/q)$, $n=0,\ldots
q-1$. Since $\lambda_{{\bm x},\mu}^q=1$ and the theory is invariant
under charge conjugation, it is sufficient to consider values of $Q$
satisfying $1\le Q\le q/2$. The model is invariant under the local
${\mathbb Z}_q$ transformations
\begin{equation}
{\bm z}_{\bm x}\to \alpha_{\bm x}^Q{\bm z}_{\bm x},\quad 
\lambda_{{\bm x},\mu}\to \alpha_{\bm x} \lambda_{{\bm x},\mu}
\bar{\alpha}_{{\bm x}+\hat{\mu}},
\end{equation}
where $\alpha_{\bm x}\in\mathbb{Z}_q$, and under the global
transformation ${\bm z}_{\bm x}\to V{\bm z}_{\bm x}$,
where\footnote{If $q/Q=2$ the global symmetry is larger: the
invariance group is O($2N_f$), see Ref.~\cite{BP-23}.}
$V\in\mathrm{SU}(N)$.  As in the noncompact $\mathbb{Z}_q^{(nc)}$
model, also in the $\mathbb{Z}_q$ gauge model there is an additional
U(1)$/{\mathbb Z}_q$ global symmetry.

The phase diagram of the $N$-component $\mathbb{Z}_q$-gauge model has
been studied in Ref.~\cite{BP-23}. For $Q>1$ it is the same as the
phase diagram for the noncompact model reported in
Fig.~\ref{ph_diag_disc2}. Only the nature of some transition lines
differs.  For $J\to \infty$ and $J=0$ the model is equivalent to a
$\mathbb{Z}_Q$ and $\mathbb{Z}_q$ model, respectively, so the
transitions along the $BM_2$ (the numerical results of
Ref.~\cite{BP-23} show that the behavior along $M_1M_2$ is the same as
along the $BM_1$ line) and $CM_3$ lines are expected to belong to the
gauge $\mathbb{Z}_Q$ and $\mathbb{Z}_q$ universality classes,
respectively. The transitions along the $AM_1$ line are instead
expected to be in the XY universality class, as in the
$\mathbb{Z}_q^{(nc)}$ model.

In Ref.~\cite{BP-23} the universality class of the transitions along
the line $M_2M_3$ has been investigated for $Q=2$ and $N=15$. In the
U(1) invariant model, recovered for $q\to \infty$, these transitions
are charged transitions associated with the stable FP of the AHFT, see
Sec.~\ref{qge2phdia}. On the other hand, only discontinuous
transitions were identified along the $M_2M_3$ line for values of
$q$ up to $q=10$.  Apparently, in the discretized model there are no
charged transitions.  It is still an open issue to understand why the
two apparently similar discretized models, the compact and the
noncompact model, behave differently.  The origin may lie in the fact
that in the noncompact model the gauge group is discrete but still
infinite, while in the compact case the gauge group is discrete and
finite.

%% file: NAHSFT.tex
\section{Non-Abelian Higgs field theories}
\label{NAHSFT}

Non-Abelian gauge symmetries play a fundamental role in the
construction of quantum and statistical field theories that describe
phenomena in various physical contexts.  In high-energy physics they
provide the framework to formulate the fundamental theories of the
strong and electroweak interactions, see, e.g.,
Refs.~\cite{Weinberg-book,ZJ-book,Georgi-book}, while in
condensed-matter physics they provide an effective description of
classical and quantum critical phenomena characterized by the
emergence of non-Abelian gauge fields, see, e.g.,
Refs~\cite{GASVW-18,SSST-19,Sachdev-19,STS-21,Sachdev-book2}.  In the
presence of low-energy modes associated with scalar fields, they give
rise to the non-Abelian Higgs
mechanism~\cite{SSBgauge1,SSBgauge2,SSBgauge3}, which, for instance,
is one of the main features of the theory of electroweak
interactions~\cite{Weinberg-book,ZJ-book,Georgi-book}.  The emerging
Higgs phases in non-Abelian Higgs (NAH) theories crucially depend on
the interplay between local and global symmetries, and on the pattern of 
the spontaneous symmetry breaking, see, e.g.,
Refs.~\cite{GG-72,OS-78,FS-79,DRS-80,Hikami-80,SSSNH-02,MZ-03,
  NRR-03,SBSVF-04,BP-14,SSST-19,BPV-19,SPSS-20,BPV-20-on,
  BFPV-21-adj,BFPV-21-mpsun}.

Three-dimensional non-Abelian gauge models can be used to effectively
describe the critical behavior of 3D statistical non-Abelian gauge
models at finite-temperature phase transitions~\cite{Nadkarni-90,
  KRS-93, AY-94, BP-95, Laine-95, KLRS-96, HPST-97, KLRS-97, HP-00, GGR-22},
and to investigate some aspects of fundamental mechanisms, such as
confinement~\cite{GG-72, tHooft-74, Polyakov-74,Polyakov-77}.  The
peculiar critical behaviors occurring in  3D NAH models also provide
additional possible scenarios for finite-temperature continuous
transitions in $D=3+1$ non-Abelian gauge QFTs. In this case, temperature
is related to the (inverse) finite size along the fourth Euclidean
direction in the corresponding  path-integral formulation, see, e.g.,
Refs.~\cite{PW-84, Nadkarni-90, KRS-93, AY-94, BP-95, Laine-95,
  KLRS-96, Meyer-Ortmanns-96, HPST-97, Wilczek-00, Karsch-02, BPV-03,
  STAR-05, BVS-06, PV-13}.  
NAH theories have also been considered to explain some aspects of the rich
phenomenology of high-temperature cuprate superconductors near optimal doping,
with display the simultaneous disappearance of nematic and charge-density
orders, the reconstruction of the Fermi surface, and the
restoration of time-reversal symmetry; see, e.g.,
Refs.~\cite{Xia-etal-08, Fujita-etal-14, He-etal-14, Michon-etal-19}, and
Ref.~\cite{PT-19} 
for a review. In particular, the possible emergence at criticality 
of a non-Abelian gauge symmetry in the presence of scalar fields 
has been proposed in
Refs.~\cite{GASVW-18,SSST-19}; see also 
Refs.~\cite{ZOKNT-01, ZDS-02, SaM-02, NZ-02,
ZN-03, MS-12, MS-12-b} for earlier studies.

In this section we introduce the NAH field theories (NAHFTs), i.e.,
field theories in which scalar fields are coupled with
non-Abelian gauge fields. In particular, we discuss statistical field
theories with local SU($N_c$) and SO($N_c$) gauge symmetry, and scalar
fields transforming according to different representations of the gauge group
(we consider the fundamental and the adjoint representation). We outline the
known features of the RG flow, whose stable infrared FPs are
expected to describe the charged critical behaviors occurring in  the
corresponding lattice systems.

\subsection{SU($N_c$) gauge Higgs theory with
scalar fields in the fundamental representation}
\label{suNAHFT}

\subsubsection{The statistical field theory}
\label{NAHSFTl}

We consider the NAHFT in which SU($N_c$) gauge fields interact with
$N_f$ degenerate scalar fields in the fundamental representation of
the gauge group.  The fundamental fields are a complex scalar field
$\Phi^{af}(\bm{x})$, where $a=1,...,N_c$ and $f=1,...,N_f$, and the
gauge field $A_{\mu}^c({\bm x})$, where $c=1,...,N_c^2-1$.
 The most
general renormalizable Lagrangian that is invariant under local
SU($N_c$) {\em color} transformations and under global U($N_f$) {\em
  flavor} transformations reads~\cite{Hikami-80,BFPV-21-mpsun}
\begin{eqnarray}
    \label{cogausun} 
    &&{\cal L}_{\rm SU} = {1\over 2 g^2} {\rm Tr}\,{\cal F}_{\mu\nu}^2
    + {\rm Tr}
  \, [(D_\mu \Phi)^\dagger (D_\mu \Phi)] + V(\Phi),\\
&&V(\Phi) =
  r\,{\rm Tr}\,\Phi^\dagger\Phi
+ \,u \,({\rm Tr}\,\Phi^\dagger\Phi)^2 \,+ \,v\, {\rm
  Tr}\,(\Phi^\dagger\Phi)^2,
\nonumber
\end{eqnarray}
where ${\cal F}_{\mu\nu} = \partial_\mu {\cal A}_\nu -\partial_\nu
{\cal A}_\mu -i[{\cal A}_\mu, {\cal A}_\nu]$, ${\cal A}_\mu =
\sum_{c} A_\mu^c T^c$, $D_{\mu}$ is the covariant derivative
in the fundamental representation, i.e., $D_{\mu, ab} =
\partial_\mu\delta_{ab} -i \sum_c T_{F,ab}^c A_\mu^c$, where $T_F^c$
are the $N_c^2-1$ Hermitean generators in the fundamental representation of
the SU($N_c$) algebra,
normalized as $\mathrm{Tr}(T^a T^b)=\frac{1}{2}\delta^{ab}$. 
The global U($N_f$) group acts as $\Phi({\bm x})\to
\Phi(\bm x)W$, with $W\in\mathrm{U}(N_f)$, while the local SU($N_c$)
gauge group acts as
\begin{equation}
\Phi({\bm x})\to V({\bm x})\Phi(\bm x), \qquad 
{\cal A}_{\mu}({\bm x})\to 
     V({\bm x}){\cal A}_{\mu}({\bm x})V^{\dag}({\bm x})-
     i\big(\partial_{\mu}V({\bm x})\big)V^{\dag}({\bm x}), \qquad
     V(\bm x)\in\mathrm{SU}(N_c).
\label{gaugetransfsun}
\end{equation}
Since the group U($N_f$) is not simple, we may separately consider
SU($N_f)$ and U(1) transformations, that correspond to $\Phi^{af} \to
\sum_g V^{fg} \Phi^{ag}$, $V\in$ SU($N_f$), and $\Phi^{af} \to
e^{i\theta} \Phi^{af}$, $\theta \in [0,2\pi)$, respectively.  For
  $N_f<N_c$ the U(1) symmetry is only apparent, as a generic U(1)
  global transformation of the field $\Phi_{\bm x}$ can be reabsorbed
  by composing an SU($N_f$) global transformation and an SU($N_c$) local
  transformation, see Ref.~\cite{BPV-20-sun} for details. For $N_f\ge
  N_c$, since the diagonal matrices $e^{2\pi i/N_c}$ are elements of
  the SU($N_c$) group (they correspond to its center), $\theta$ can be
  restricted to $[0,2\pi/N_c)$, and the global symmetry group is
    actually U$(N_f)/\mathbb{Z}_{N_c}$.

For $N_c=2$, because of the pseudoreality of
the SU(2) gauge group, the NAHFT Hamiltonian 
(\ref{cogausun}) with $v=0$ turns out
to be invariant under Sp($N_f$)/${\mathbb Z}_2$ transformations, i.e., 
under a group that  is larger than 
U($N_f$),~\cite{Georgi-book, AY-94, BP-14, WNMXS-17, BPV-19, BPV-20-sun,
BFPV-21-mpsun}. In particular, it is invariant under Sp(2)/${\mathbb
  Z}_2$ = SO(5) transformations for $N_f=2$~\cite{BPV-19, BPV-20-sun}.
Therefore, the $N_c=2$ NAHFT (\ref{cogausun}) with $v=0$ is stable
under RG transformations, defining a distinct theory characterized by
the global symmetry Sp($N_f$)/${\mathbb Z}_2$.

We finally recall that the standard Faddeev-Popov
quantization~\cite{FP-67} of a non-Abelian gauge field theory requires
the addition of a gauge fixing.  If we consider the Lorenz gauge
fixing used in the AHFT defined in Eq.~(\ref{partfunc}), one should
also introduce appropriate scalar fields with fermionic statistics
(ghosts), see, e.g., Refs.~\cite{Weinberg-book,ZJ-book}.

\subsubsection{Higgs phases and breaking patterns of 
the global symmetry} \label{sec:NAHSUfasi}

The global symmetry of the Higgs phases, which is nontrivial only for
$N_f \ge 2$, can be studied by performing an analysis of the minima of
the scalar potential.  We report here the main results. See
Ref.~\cite{BFPV-21-mpsun} for more details, and \ref{meanfieldHiggs}
for a sketch of the derivations in the corresponding lattice models.
The results of these analyses depend on the sign of $v$ and, for $v >
0$, also on the number of colors $N_c$ and flavors $N_f$.

For $v < 0$ the Higgs phase is invariant under ${\mathrm U}(1) \oplus
{\mathrm U}(N_f-1)$ transformations, leading to the spontaneous
symmetry breaking pattern
\begin{eqnarray}
\mathrm{U}(N_f) \to {\mathrm U}(1) \oplus {\mathrm U}(N_f-1).
 \label{gsbp-v-less-0}
\end{eqnarray}
Note that the U(1) symmetry is not broken, so only the SU($N_f$)
symmetry plays a role. The relevant symmetry breaking pattern is
therefore
\begin{eqnarray}
  \mathrm{SU}(N_f) \to {\mathrm U}(N_f-1).
 \label{eq:gsbp-v-less-0}
\end{eqnarray}
For $v > 0$ one should distinguish three different cases.  For $N_f <
N_c$ there is no symmetry breaking and therefore no Higgs ordered
phase.  For $N_f\ge N_c$ we have instead
\begin{eqnarray}
& \mathrm{U}(N_f)\rightarrow \mathrm{SU}(N_f) \quad &{\rm for}\;\; N_f
  = N_c,
 \label{glsymbrpatNfNc}
\\ & \mathrm{U}(N_f)\rightarrow
\mathrm{SU}(N_c)\otimes\mathrm{U}(N_f-N_c) \quad &{\rm for}\;\; N_f >
N_c.
 \label{glsymbrpat}
\end{eqnarray}
It is important to note
that for $N_f = N_c$ the Higgs phase is symmetric under SU($N_f$)
transformations. In this case transitions in the corresponding lattice
models are only associated with the breaking of the U(1) symmetry. On
the other hand, for $N_f > N_c$, in the Higgs phase both the U(1) and
the SU($N_f$) symmetry are broken.

\subsubsection{Renormalization-group flow}
\label{epsexpsuNAHFT}

A nonperturbative formulation of the NAHFT is provided by the
continuum (critical) limit of corresponding lattice models with
the same local and global symmetries, as it is done for the
strong-interaction theory, quantum chromodynamics
(QCD)~\cite{Wilson-74,MM-book}. For this purpose it is crucial that
the corresponding lattice discretizations have a continuous charged
transition where the gauge fields play a relevant role. Therefore, it
is important to identify the universality classes of the charged
transitions in this class of systems. They are related to the 
stable FPs of the RG flow of the NAHFT.

The RG flow of the NAHFT (\ref{cogausun}) can be investigated
perturbatively, in powers of $\varepsilon\equiv 4-d$~\cite{WK-74},
using dimensional regularization and the minimal-subtraction (MS)
renormalization scheme, see, e.g., Ref.~\cite{ZJ-book,PV-02}.  The RG
flow close to four dimensions is determined by the MS $\beta$
functions associated with the Lagrangian couplings $\alpha = g^2$,
$u$, and $v$. At one-loop order, we have~\cite{BFPV-21-mpsun,Hikami-80}
\begin{eqnarray}
&&\beta_\alpha \equiv \mu{\partial\alpha\over \partial\mu} = -
  \varepsilon \alpha + (N_f-22N_c)\,\alpha^2,
\label{betasunc}\\
&&\beta_u \equiv \mu {\partial u\over \partial\mu} 
= -\varepsilon u + (N_f N_c + 4) u^2   
+ 2 (N_f+N_c) u v + 3 v^2  
 - {18 \,(N_c^2 -1)\over N_c} \, u \alpha 
+ {27 (N_c^2 + 2)\over N_c^2}\,\alpha^2,
\nonumber\\
&&\beta_v \equiv \mu {\partial v\over \partial\mu}
= - \varepsilon v + (N_f+N_c)v^2 + 6uv
- {18 \,(N_c^2 -1)\over N_c} \, v \alpha
+ {27 (N_c^2 - 4)\over N_c}\,\alpha^2.
\nonumber
\end{eqnarray}
Some numerical factors have been reabsorbed in the normalizations of
the renormalized couplings, in order to simplify the previous expressions.

The analysis of the common zeroes of the $\beta$ functions shows that
the RG flow close to four dimensions has a stable charged FP with a
nonzero FP value of the gauge coupling $\alpha$, for a sufficiently
large number of flavors; more precisely, for $N_f > N_f^*$, with,
e.g., $N_f^*= 375.4+O(\varepsilon)$ for $N_c=2$, and $N_f^* =
638.9+O(\varepsilon)$ for $N_c=3$.  The values of $N_f^*$ are large in
four dimensions, but they are expected to decrease significantly in
three dimensions for any $N_c$, as it also happens in the AHFT, where
$N_f^*$ varies from $N_f^*\approx 183$ in four dimensions to
$N_f^*\approx 7$ in three dimensions.  In particular, for $N_c=2$
numerical MC studies of lattice SU(2)-gauge Higgs
models~\cite{BPSV-24} indicate $N_f^*=25(4)$ in three dimensions, see
Sec.~\ref{rNASFT}. We also mention that in the case of the SU(2)-gauge
NAHFT (\ref{cogausun}) for $v=0$, for which the global symmetry
enlarges to Sp($N_f$)/${\mathbb Z}_2$, a charged FP exists for $N_f >
N_f^* \approx 359 + O(\varepsilon)$~\cite{BPV-19}. This FP turns out
to be unstable in the full theory, i.e, with respect to nonzero values
of the Lagrangian coupling $v$.

The stable charged FP of this class of NAHFTs is located in the
positive $v$ region for any $N_c$.  This result indicates that only
transitions characterized by the $\mathrm{U}(N_f)\rightarrow
\mathrm{SU}(N_c)\otimes\mathrm{U}(N_f-N_c)$ symmetry-breaking pattern
may correspond to charged continuous transitions. Instead, 
those characterized
by the symmetry-breaking pattern $\mathrm{U}(N_f) \to {\mathrm U}(1)
\oplus {\mathrm U}(N_f-1)$, corresponding to $v < 0$, cannot
be charged continuous transitions.  As we have already discussed for
the AH models, the FPs of the NAHFTs are
relevant for the existence of charged continuous transitions, i.e., of
continuous transitions where both scalar and
gauge fields are critical. As discussed in Sec.~\ref{difftype}, gauge
models can also undergo other types of transitions that admit
different effective descriptions. Therefore, it is {\em a priori}
possible to have continuous transitions, which may be LGW or
topological transitions, also for $N_f < N_f^*$ and for $v < 0$. Some
examples will be discussed in Sec.~\ref{NAHLM}.

As it occurs in models with Abelian gauge symmetry, the uncharged
FP with vanishing gauge coupling $\alpha=0$ is unstable with respect
to the gauge coupling, since its stability matrix $\Omega_{ij} =
\partial \beta_i/\partial g_j$ has a negative
eigenvalue~\cite{BFPV-21-mpsun}
\begin{equation}
\lambda_\alpha = \left. {\partial \beta_\alpha/\partial \alpha}
\right|_{\alpha=0} = - \varepsilon + O(\varepsilon^2).
\label{lambdares}
\end{equation}
This result suggests that also in three dimensions gauge
fluctuations make the uncharged FP unstable, as also confirmed by the
numerical studies of 3D SU(2)-gauge LNAH models~\cite{BFPV-21-mpsun},
see also Sec.~\ref{NAHLM}. Note that continuous transitions described
by the uncharged FP of the NAHFT for $\alpha=0$ would be characterized
by an asymptotic critical behavior in which gauge-field correlations
are not critical, thus they would correspond to LGW$^{\times}$
transitions, effectively described by a LGW theory with a
gauge-dependent complex order-parameter field $\Phi^{af}(\bm{x})$ with
$a=1,...,N_c$ and $f=1,...,N_f$.

An alternative approach to the study of the SU($N_c$)-gauge NAHFT is
provided by the $1/N_f$ expansion at fixed
$N_c$~\cite{Hikami-80}. These computations confirm the existence of a
charged critical behavior in three dimensions. For example, a
$O(1/N_f)$ computation of the length-scale critical exponent $\nu$
predicts~\cite{Hikami-80}
\begin{eqnarray}
  \nu =  1 +  {a_1\over N_f} + O(N_f^{-2}),\qquad 
      a_1 = -{48 N_c\over \pi^2 N_f},
  \label{nulargensun}
  \end{eqnarray}
in the 3D SU($N_c$)-gauge NAHFT with Lagrangian (\ref{cogausun}).

\subsection{SO($N_c$) gauge Higgs theory with
  scalar fields in the fundamental representation}
\label{soNAHFT}

The NAHFT with gauge group SO($N_c$) and multiflavor scalar fields in
the fundamental representation is obtained by considering a real
scalar field $\Phi^{af}({\bm x})$ (with $a=1,...,N_c$ and
$f=1,\ldots,N_f$) and the gauge field $A_{\mu}^c({\bm x})$,
$c=1,\ldots,N_c(N_c-1)/2$. 
The most general
renormalizable Lagrangian is~\cite{BPV-20-on,Hikami-80,PRV-01}
\begin{eqnarray}
\label{cogausonc}
      && {\cal L}_{\rm SO} =
      {1\over 2 g^2} {\rm Tr}\,{\cal F}_{\mu\nu}^2 +
      + {\rm Tr}
  \, [(D_\mu \Phi)^\dagger (D_\mu \Phi)] + V(\Phi),
\\ && V(\Phi) = \frac{r}{2}\mathrm{Tr}\,\Phi^t\Phi+ u
(\mathrm{Tr}\,\Phi^t\Phi)^2 + v
\big[\mathrm{Tr}\,(\Phi^t\Phi)^2-(\mathrm{Tr}\,\Phi^t\Phi)^2\big],
\nonumber
\end{eqnarray}
where ${\cal F}_{\mu\nu}$ is the non-Abelian field strength of the
gauge field ${\cal A}_{\mu}=\sum_c A_\mu^c T^c$, $D_{\mu}$ is the
covariant derivative in the fundamental representation of
SO($N_c$), and $T_c$ are the generators of the SO($N_c$) algebra. 
For any value of $N_c$ and $N_f$, the Lagrangian
(\ref{cogausonc}) is invariant under the local SO($N_c$) gauge
transformations,
\begin{equation}
\Phi({\bm x})\to V({\bm x})\Phi({\bm x}), \qquad 
A_{\mu}({\bm x})\to V({\bm x}){\cal A}_{\mu}({\bm x})V^{t}({\bm
  x})-\big(\partial_{\mu}V({\bm x})\big)V^{t}({\bm x}),
\qquad
V({\bm x})\in\mathrm{SO}(N_c),
\label{gaugetransfon}
\end{equation} 
and under the global O($N_f$) transformations, $\Phi({\bm x})\to
\Phi({\bm x})W$ with $W\in$ O($N_f$).  For $N_c=2$, we have an Abelian
O(2) gauge model with SO($N_f$) global symmetry, which is equivalent
to the one discussed in Sec.~\ref{OAHFT}.\footnote {If we define
$\Phi_{A}^f = (\Phi^{1f} + i \Phi^{2f})/\sqrt{2}$, $u_{A} = 4 u - 2v$,
$v_{A} = 2v, $ $r_{A} = r$, the Lagrangian (\ref{cogausonc}) takes the
form given in Eq.~(\ref{OAHFTL}), provided we indicate with $\Phi_{A}$
the complex field and with $r_{A}$, $u_{A}$, and $v_{A}$ the
Hamiltonian parameters of the AHFT.  Note that for $v=0$ its global
symmetry enlarges to U($N_f$), see Sec.~\ref{OAHFT}.}

The RG flow determined by the $\beta$ functions associated with the
Lagrangian quartic couplings provides information on the nature of the
transitions described by the continuum SO($N_c$) gauge theory
(\ref{cogausonc}). In the $\varepsilon$-expansion framework, the one-loop
MS $\beta$ functions are~\cite{BFPV-21-adj,Hikami-80,BPV-20-on}
\begin{eqnarray}
&&\beta_\alpha \equiv \mu{\partial\alpha\over \partial\mu} = 
-\varepsilon \alpha  + {N_f  - 22(N_c-2) \over 12}\alpha^2,
\label{betasonc}\\
&&\beta_u \equiv \mu{\partial u\over \partial\mu} = 
-\varepsilon u + {N_c N_f + 8\over 6} u^2   
+ {(N_f-1)(N_c-1)\over 6} (v^2 - 2 u v) 
- {3\over 2} (N_c-1)u \alpha  + {9\over 8}(N_c-1)\alpha^2 ,
\nonumber\\
&& \beta_v \equiv \mu{\partial v\over \partial\mu} = 
- \varepsilon v +
     {N_c+N_f-8\over 6}v^2 + 2uv
- {3\over 2}(N_c-1) v\alpha 
    + {9\over 8} (N_c-2) \alpha^2,
\nonumber
\end{eqnarray}
where the renormalized couplings $\alpha = g^2$, $u$, and $v$ have
been normalized to simplify the expressions.\footnote{Because of the
equivalence of the present model with $N_c=2$ with the
SO($N$)-symmetric AHFT, their $\beta$ functions are related. If we
redefine the couplings in Eq.~(\ref{betasonc}) as $\alpha \to 12
\alpha$, $u \to 3( u + v)$ and $v\to 6 v$, we obtain the $\beta$
functions reported in Eq.~(\ref{betafuncOAH}).}

The RG flow of the theory does not have stable charged FPs with
$\alpha>0$, unless $N_f>N_f^*(d)$, where $N_f^*(d)$ is very large in
four dimensions.  Indeed, we find $N^*_f(4) = 210.5$ for $N_c =2$,
and $N^*_f(4) \approx 443.8$, for $N_c =3$. The stable FPs always lie
in the region $u,v> 0$. More precisely, one finds $v^* \approx
6\varepsilon/N_f$, $u^* \approx 6\varepsilon/N_f$, and $\alpha^* =
12\varepsilon/N_f$ for large values of $N_f$.  Note that the existence
of a critical theory for large values of $N_f$ is also confirmed by
the analysis of the model using the large-$N_f$
expansion~\cite{Hikami-80}.

The results reviewed above show that continuous transitions with the
effective field-theory description (\ref{cogausonc}) are only possible
for a large number of components. Again, we remind the reader that
this result does not exclude the presence of other types of continuous
transitions (see Sec.~\ref{difftype}) for small values of $N_f$.

\subsection{SU($N_c$) gauge Higgs theory with 
  scalar fields in the adjoint representation}
\label{suNAHFTadj}

We now consider the SU($N_c$) gauge NAHFT with scalar fields in the
adjoint representation. As argued in Ref.~\cite{SSST-19}, 3D
SU(2)-gauge theories with adjoint scalar fields with $N_f=1$ or $N_f=4
$ are expected to describe some aspects of the optimal doping
criticality in cuprate superconductors, as emerging from experiments,
see, e.g., Refs.~\cite{PT-19, Fujita-etal-14, He-etal-14}.

To define the model we consider real scalar fields $\Phi^{af}({\bm
  x})$ with $a=1,...,N_c^2-1$ ({\em color} index) and $f=1,...,N_f$
({\em flavor} index), and $N_c^2-1$ real gauge fields $A_{\mu}^c({\bm
  x})$.  The most general renormalizable Lagrangian is
\begin{eqnarray}
\label{cogausuadj}
&&      {\cal L}_{\rm SUadj}=
{1\over 2 g^2} {\rm Tr}\,{\cal F}_{\mu\nu}^2 +
{\rm Tr}
  \, [(D_\mu \Phi)^\dagger (D_\mu \Phi)] + V(\Phi),\\
&& V(\Phi) =
   {r\over 2} \,{\rm
  Tr}\,\Phi^t\Phi + u ({\rm Tr}\,\Phi^t\Phi)^2 + v \left[ {\rm
    Tr}\,(\Phi^t\Phi)^2 - ({\rm Tr}\,\Phi^t\Phi)^2 \right], \nonumber
\end{eqnarray}
where ${\cal F}_{\mu\nu}$ is the field strength of the gauge field
${\cal A}_{\mu} \equiv \sum_c A_{\mu}^c T^c$, $D_{\mu}$ is the
covariant derivative in the adjoint representation, i.e.,
$\partial_\mu + i \sum_c A_{\mu}^c T_{A}^{c}$, and $T_{A}^{abc}=-i
f^{abc}$ are the Hermitian SU($N_c$) generators in the adjoint
representation ($f^{abc}$ are the structure constants of the SU($N_c$)
group).  The model is invariant under left SU($N_c$) local
transformations and under right O($N_f$) global transformations.

For $N_c=2$ the model is equivalent to the SO(3) gauge model
(\ref{cogausonc}) with $N_f$ flavor, which allows one to use the
results reported in Sec.~\ref{soNAHFT}.  For instance, the $\beta$
functions of the SU(2) gauge theory with adjoint scalar matter
are~\cite{SSST-19}
\begin{eqnarray}
&&\beta_\alpha =
-\varepsilon \alpha  + {N_f  - 22 \over 12}\alpha^2,
\label{betassuncadj}\\
&&\beta_u = -\varepsilon u + {3 N_f + 8\over 6} u^2   
+ {N_f-1\over 3} (v^2 - 2 u v)
- 3 u \alpha  + {9\over 4}\alpha^2,
\nonumber\\
&&\beta_v = - \varepsilon v +
{N_f-5\over 6}v^2 + 2uv - 3 v\alpha + {9\over 8} \alpha^2.
\nonumber
\end{eqnarray}
They have a stable FP for sufficiently large $N_f$, more precisely for
$N_f > N^* + O(\varepsilon)$ with $N^*\approx 443.8$.  In particular,
in the large-$N_f$ limit the stable FP of the $\beta$ functions is
located at $\alpha^* \approx 12\varepsilon/N_f$, $u^* \approx
6\varepsilon/N_f$, $v^* \approx 6\varepsilon/N_f$.  Note that, also in this
case, the stable large-$N_f$ charged FP is located in the region
$v>0$.

%% file: NAHL.tex
\section{Lattice non-Abelian Higgs models}
\label{NAHLM}

In this section we consider 3D lattice non-Abelian Higgs (LNAH)
models, i.e., 3D lattice non-Abelian gauge models with multicomponent
degenerate scalar fields, and review the known results regarding their
phase diagram, the main properties of their low-temperature Higgs
phases, and the nature of their phase transitions and critical
behaviors. We mainly consider SU($N_c$) gauge models with scalar
fields transforming in the fundamental and adjoint representations of
the gauge group.  We also briefly discuss SO($N_c$) gauge models with
multicomponent scalar fields in the fundamental representation.

As we shall see, LNAH models present various phases and phase
transitions, whose nature depends of 
the global and local symmetries. As already discussed for 
LAH theories, see Secs.~\ref{LAHMnc} and \ref{LAHMc}, one of the
main issues is the identification of the effective field-theoretical
description of the phase transitions present in LNAH models.  As
already remarked before, to define nonperturbatively the NAHFT as the
continuum limit of a lattice model, it is crucial that the
corresponding LNAH models, with the same local and global symmetries,
undergo charged continuous transitions where both scalar and gauge
modes are critical.

The phase transitions of lattice SU($N_c$) gauge Higgs models exhibit
critical behaviors of various types, as discussed in
Sec.~\ref{difftype}. There are transitions that admit a LGW
description in terms of a gauge-invariant order-parameter field---this
is appropriate when gauge modes are not critical---and GFT transitions
with critical gauge modes, associated with the stable charged FPs of
the NAHFTs discussed in Sec.~\ref{NAHSFT}.  We mostly focus on lattice
models with scalar matter in the fundamental representation of the
gauge group. But we also discuss the case of scalar fields that
transform in the adjoint representation of the SU($N_c$) gauge group,
where also topological transitions appear, which are not associated
with the breaking of the global symmetry. Finally, we briefly mention
the main features of the phase diagrams and phase transitions of LNAH
models with SO($N_c$) gauge symmetries.

\subsection{Lattice SU($N_c$) gauge Higgs models with multiflavor scalar fields}
\label{modelsubfu}

We consider LNAH models with multiflavor scalar fields, which are
invariant under SU($N_c$) gauge transformations and U($N_f$) global
transformations.  Scalar fields transform under the fundamental
representation of the local and global symmetry groups. The
Hamiltonian can be straightforwardly obtained by taking a lattice
discretization of the Lagrangian (\ref{cogausun}), using the compact
Wilson formulation in which gauge fields are replaced by link
variables taking values in the gauge group ~\cite{Wilson-74}.  In the
lattice model---for simplicity we consider cubic lattices---the
fundamental fields are complex matrices $\Phi^{af}_{\bm x}$, with
$a=1,...,N_c$ ({\em color} index) and $f=1,...,N_f$ ({\em flavor}
index), defined on the lattice sites, and ${\rm SU}(N_c)$ matrices
$U_{{\bm x},\mu}$, defined on the lattice links.  The partition
function is
\begin{eqnarray}
  Z = \sum_{\{\Phi,U\}} e^{-\beta H}, \qquad \beta=1/T,
  \qquad H = H_K(\Phi,U) + H_V(\Phi) + H_G(U). 
\label{hgaugesun}
\end{eqnarray}
The Hamiltonian $H$ is the sum of the scalar-field kinetic term $H_K$,
of the local scalar potential $H_V$, and of the pure-gauge Hamiltonian
$H_G$.  As usual, we set the lattice spacing equal to one, so all
lengths are measured in units of the lattice spacing.  The kinetic
term $H_K$ is given by
\begin{eqnarray}
  H_K(\Phi,U) =   - J
  N_f \sum_{{\bm x},\mu} {\rm Re} \, {\rm Tr} \,\Phi_{\bm x}^\dagger \,
 U_{{\bm x},\mu} \, \Phi_{{\bm x}+\hat{\mu}}^{\phantom\dagger}\,.
  \label{Kinterm}
\end{eqnarray}
In the following we set $J=1$, so that energies are measured in units
of $J$. The local scalar potential $H_V$ is given by
\begin{eqnarray}
  H_V(\Phi) = \sum_{\bm x} V(\Phi_{\bm
    x}),
  \qquad V(\Phi)= {r\over 2} \, {\rm
    Tr}\,\Phi^\dagger\Phi + {u\over 4} \left( {\rm
    Tr}\,\Phi^\dagger\Phi\right)^2 + {v\over 4} \, {\rm
    Tr}\,(\Phi^\dagger\Phi)^2, \label{potentialsun}
\end{eqnarray}
where $V(\Phi)$ is the most general quartic polynomial that is
symmetric under [U($N_f$)$\otimes$U($N_c$)]/U(1) transformations.
Note that for $N_f=1$ the two quartic terms in $V(\Phi)$ are
equivalent.  Finally, the pure-gauge Hamiltonian $H_G(U)$ is the sum
of plaquette terms~\cite{Wilson-74},
\begin{eqnarray}
H_G(U) = - {\gamma\over N_c} \sum_{{\bm x},\mu>\nu} {\rm Re} \, {\rm
    Tr}\, \Pi_{{\bm x},\mu\nu},\qquad
\Pi_{{\bm x},\mu\nu}=
      U_{{\bm x},\mu} \,U_{{\bm x}+\hat{\mu},\nu} \,U_{{\bm
    x}+\hat{\nu},\mu}^\dagger \,U_{{\bm x},\nu}^\dagger,
\label{plaquette}
\end{eqnarray}
where the parameter $\gamma$ plays the role of inverse gauge coupling.

The Hamiltonian $H$ is invariant under global U($N_f$) transformations
that only act on the scalar field, $\Phi \to \Phi W$,
$W\in\hbox{U($N_f$)}$, and under local SU($N_c$) gauge transformations
\begin{eqnarray} 
\Phi_{\bm x} \to V_{\bm x} \Phi_{\bm x}, \qquad 
U_{{\bm x},\mu} \to V_{\bm x} U_{{\bm x},\mu} V_{{\bm x}+\hat{\mu}}^\dagger,
\qquad V_{\bm x} \in {\rm SU}(N_c).
\end{eqnarray}
The lattice model (\ref{hgaugesun}) can be simplified by considering
the {\em fixed-length} limit of the scalar field, i.e., by 
requiring ${\rm Tr}\, \Phi_{\bm x}^\dagger \Phi_{\bm x}^{\phantom\dagger} = 1$, which can be
formally obtained by considering the limit $u,r\to\infty$ keeping the
ratio $r/u=-1$ fixed.  In the unit-length limit the lattice potential
(\ref{potentialsun}) reduces to
\begin{eqnarray}
  V(\Phi)= {v\over 4} \, {\rm
    Tr}\,(\Phi^\dagger\Phi)^2,\qquad
  {\rm Tr}\, \Phi^\dagger \Phi^{\phantom\dagger} = 1.
  \label{flpotentialsun}  
\end{eqnarray}
The model in the unit-length limit is generally expected to have the
same features as  models with generic $r$ and $u$.

As already discussed in Sec.~\ref{sec:NAHSUfasi}, phase transitions
can be characterized by the breaking of the global SU($N_f$) symmetry
or by the breaking of the U(1) symmetry.  The spontaneous breaking of
the global SU($N_f$) symmetry is signaled by the condensation of the
gauge-invariant bilinear operator
\begin{eqnarray}
  P_{\bm x}^{fg} = \sum_a {\bar\Phi}_{\bm x}^{af} \Phi_{\bm x}^{ag} -
  {1\over N_f} \delta^{fg},\qquad
  G_P({\bm x},{\bm y}) \equiv \langle {\rm Tr}\, P_{\bm x} P_{\bm y} \rangle.
\label{pdefsun}
\end{eqnarray}
The bilinear operator $P_{\bm x}^{fg}$ is invariant under the U(1)
global transformations and satisfies ${\rm Tr} \, P_{\bm x}=0$
(because of the unit-length constraint of $\Phi_{\bm x}$).  To monitor
the breaking of the U(1) global symmetry, one may consider
gauge-invariant operators that transform nontrivially under U(1)
transformations.  For example, for $N_c = 2$, one may consider the
bilinear operator
\begin{equation}
Y_{\bm x}^{fg} = \epsilon^{ab}\, \Phi_{\bm x}^{af}\, \Phi_{\bm x}^{bg},
\qquad G_Y({\bm x},{\bm y})\equiv
\langle \, {\rm Tr}
\,Y_{\bm x}^\dagger Y_{\bm y} \rangle,
\label{Yphase}
\end{equation}
where $\epsilon^{ab}$ is the completely antisymmetric tensor in the
color space, with $\epsilon^{12}=1$. Note that, if $N_f>N_c=2$, the
composite operator $Y_{\bm x}^{fg}$ transforms nontrivially under the
SU($N_f$) group, therefore the U(1) symmetry can be spontaneously
broken only in phases characterized by a broken SU($N_f$)
symmetry. Instead, for $N_f=N_c=2$, the operator $Y_{\bm x}^{fg}$ can
be rewritten as $Y_{\bm x}^{fg}=\epsilon^{fg}\det \Phi_{\bm x}$, and
is therefore invariant under SU(2) global transformations.  Thus,
the U(1) symmetry can be broken irrespectively of the
breaking of the SU($N_f$) symmetry. Analogous results hold for generic
models with $N_f=N_c$, irrespective of the 
value of $N_c$, see Refs.~\cite{BPV-20-sun,BFPV-21-mpsun}.

\subsection{Phase diagram and critical behaviors of
  lattice SU($N_c$) gauge Higgs models}
\label{su2higgs}

We now outline the main features of the phase diagram of the LNAH
models (\ref{hgaugesun}), discussing the nature of their phase
transitions and critical behaviors.  We mainly focus on the LNAH
models satisfying the unit-length constraint
(\ref{flpotentialsun}). Qualitatively similar phase diagrams are
expected when relaxing the unit-length constraint, i.e. for generic
SU($N_c$) gauge LNAH models with Hamiltonian
(\ref{hgaugesun}).

\subsubsection{General considerations on the
  low-temperature Higgs phases}
\label{higgsphases}

For $N_f=1$, the phase diagram of the SU($N_c$) gauge LNAH models is
trivial: there is a single thermodynamic phase for any
$N_c$~\cite{FS-79, DRS-80,FMS-81}, without transitions. In this case
we may only distinguish two different regimes: a confined regime
for small values of $\beta J$ and $\beta\gamma$, and a Higgs regime
for large $\beta J$ and $\beta\gamma$, which are analytically
connected.  For $\gamma=0$ it is easy to realize that no transitions
are possible.  Indeed, using the gauge and global symmetry, we can set
(unitary gauge) $\Phi = (1,0,\ldots)$, obtaining a model of decoupled
gauge fields.  The same result holds for finite $\gamma$~\cite{FS-79}.

For $N_f\ge 2$, as we have discussed in Sec.~\ref{sec:NAHSUfasi},
there are various Higgs phases characterized by different spontaneous
breakings of the global symmetry~\cite{BFPV-21-mpsun}.  For $\beta \to
\infty$, the symmetry of the ordered Higgs phases can be determined by
looking at the minima of the scalar potential. Thus, results analogous
to those presented in Sec.~\ref{sec:NAHSUfasi} are obtained. For
$N_f\ge N_c$ there are two Higgs phases, one corresponding to negative
values of $v$ for large $\beta$, and a second one corresponding to
positive values of $v$. For $N_f < N_c$, there is instead a single
ordered phase in the negative-$v$ region. Indeed, for large values of $\beta$
the system is always disordered for $v>0$.

The transitions between the disordered phase and the negative-$v$
Higgs phase are characterized by the symmetry-breaking pattern
$\mathrm{U}(N_f) \to {\mathrm U}(1) \oplus {\mathrm U}(N_f-1)$,
cf. Eq.~(\ref{gsbp-v-less-0}). On the basis of the RG flow of the
NAHFT close to four dimensions, see Sec.~\ref{epsexpsuNAHFT}, these
transitions are not expected to be charged continuous transitions,
because of the absence of a corresponding stable charged FP. However,
LGW transitions are still possible.  Since the U(1) symmetry is not
broken, the relevant symmetry-breaking pattern is $\mathrm{SU}(N_f)
\to {\mathrm U}(N_f-1)$, cf.  Eq.~(\ref{eq:gsbp-v-less-0}), which is
the symmetry-breaking pattern of the CP$^{N_f-1}$ transitions
discussed in Sec.~\ref{cpnmodels}. This is not unexpected, as the
order parameter of the transition is the operator $P_{\bm x}^{ab}$,
defined in Eq.~(\ref{pdefsun}), which is a bilinear Hermitian
traceless field as the CP$^{N_f-1}$ order parameter defined in
Eq.~(\ref{qdefcpn}). Using the results presented in
Sec.~\ref{cpnmodels}, we thus conclude that continuous transitions
only occur for $N_f=2$---they belong to the O(3) universality class.
First-order transitions are expected for larger values of $N_f$.

For $N_f \ge N_c$ a second Higgs phase occurs, since the system orders
for $v > 0$ when $\beta$ is increased.  The nature of this phase
depends on $N_f$ and $N_c$.  For $N_f = N_c$ the global symmetry
breaking pattern is $\mathrm{U}(N_f)\rightarrow \mathrm{SU}(N_f)$, cf.
Eq.~(\ref{glsymbrpatNfNc}). In this case the SU($N_f$) symmetry is not
broken in the Higgs phase, so the transition is only related to the
breaking of the U(1) symmetry, the order parameter being
$\mathrm{det}\, \Phi$, as discussed in Sec.~\ref{sec:NAHSUfasi}.
Thus, the system can undergo continuous LGW transitions belonging to
the XY universality class. No charged continuous transitions described
by the corresponding NAHFT are expected, since charged FPs exist only
for $N_f\gg N_c$ (unless some new features emerge in three dimensions,
as it occurs for the one-component LAH model with noncompact gauge
variables; see the discussion at the end of
Sec.~\ref{cboncLAHM}). Finally, for $N_f>N_c$ and $v>0$ we have
transitions with symmetry-breaking pattern $\mathrm{U}(N_f)\rightarrow
\mathrm{SU}(N_c)\otimes\mathrm{U}(N_f-N_c)$,
cf. Eq.~(\ref{glsymbrpat}).  In this case, charged continuous
transitions corresponding to the stable FP of the corresponding NAHFT
can be observed for $N_f > N_f^*(3)$, where $N_f^*(3)$ is the
($N_c$-dependent) critical flavor number defined in
Sec.~\ref{epsexpsuNAHFT}.\footnote{It is worth noting that a LGW
field-theoretical description in terms of a gauge-invariant
order-parameter field does not exist, since the Higgs phase symmetry
$\mathrm{SU}(N_c)\otimes\mathrm{U}(N_f-N_c)$ depends on the number of
colors. Therefore, any effective description should somehow take into
account some specific properties of the gauge interaction. In
principle, this color-dependent symmetry breaking may be realized in
LGW$^{\times}$ transitions with a gauge-dependent order-parameter
field. However, the latter scenario is quite unlikely in 3D systems,
because gauge fluctuations are expected to be generally relevant at
the uncharged FP of theories with continuous gauge groups, see
Sec.~\ref{epsexpsuNAHFT}.}

\subsubsection{Lattice SU(2) gauge Higgs models}
\label{phdianc2}

\begin{figure}
\begin{center}
    \includegraphics[width=0.43\columnwidth]{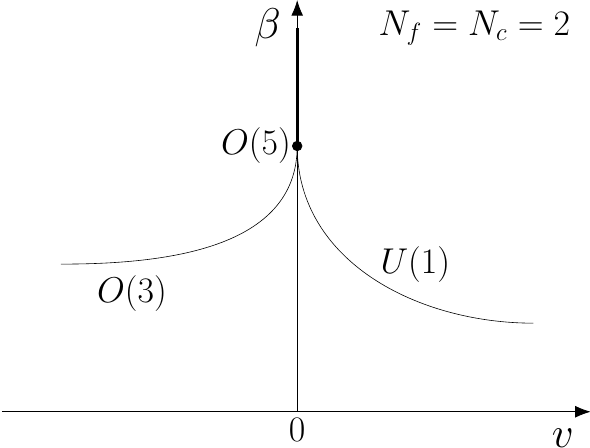}
    \hspace*{5mm}
    \includegraphics[width=0.43\columnwidth]{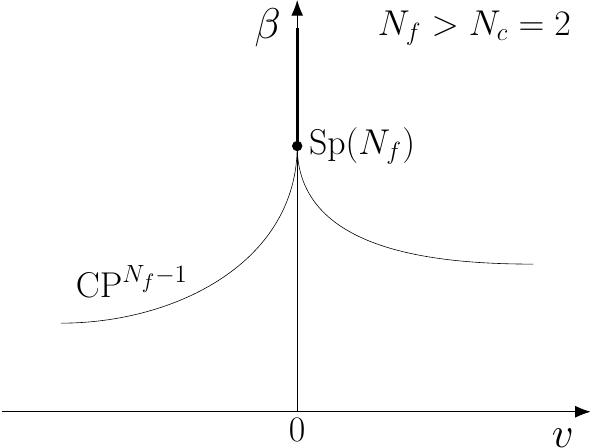}
    \caption{Sketch of the phase diagram of 3D SU(2) gauge LNAH models
      with $N_f$ scalar flavors in the fundamental representation, for
      $N_f= N_c=2$ (left) and $N_f>N_c=2$ (right) in the $\beta$-$v$
      plane for finite values of $\gamma\ge 0$ (the qualitative
      behavior should not depend on $\gamma$).  Left ($N_f=N_c=2$):
      there are two transition lines for $v\not=0$, intersecting at a
      MCP located at $(v =0, \beta = \beta_{\rm mc})$ with O(5)
      symmetry; continuous transitions belong to the O(3) and
      $\hbox{U(1)}=\hbox{SO(2)}$ vector universality classes for $v <
      0$ and $v > 0$, respectively; first-order transitions are
      expected on the line $(v =0, \beta > \beta_{\rm mc})$.  Right
      ($N_f>N_c=2$): for $v < 0$ transitions are of first order (as in
      the CP$^{N_f-1}$ model); for $v > 0$ the nature of the
      transitions might depend on $\gamma$ for sufficiently large
      values of $N_f$; in particular, for large values of $N_f$ the
      system is expected to undergo charged continuous transitions
      associated with the stable charged FP of the corresponding
      NAHFT; for $v=0$ we have a first-order line ending at the
      intersection of the transition lines (the transitions are
      expected to be of first order close to the intersection).  }
\label{phdianfnc2}
  \end{center}
\end{figure}

We now focus on LNAH models with SU(2) gauge
symmetry, which present some peculiar features that are not present
for $N_c \ge 3$.  The phase diagrams of multicomponent LNAH models
with $N_c=2$ are sketched in Fig.~\ref{phdianfnc2} for a generic value
of $\gamma$.  For any $N_f\ge 2$ there are two low-temperature Higgs
phases separated by a first-order transition line.  Since the LNAH
model (\ref{hgaugesun}) with $v=0$ is invariant under a larger group
of transformations, the Sp($N_f$)$/\mathbb{Z}_2$ group~\cite{BPV-19,
  BPV-20-sun}, the first-order line always lies on the line $v=0$.  At
the intersection point of the three transition lines, the behavior can
be critical only for $N_f=2$, with a MCP belonging to the
$\mathrm{Sp(}N_2\mathrm{)}/\mathbb{Z}_2=\mathrm{O(5)}$ universality
class (see Sec.~\ref{sec:MCPNc2NAH} for a more extensive discussion).
For not too large values of $N_f$ satisfying 
$N_f \ge 3$,\footnote{We should mention that
the field-theory model with $v=0$ admits a charged FP for large values
of $N_f$. Thus, for $N_f\gg 1$ a charged multicritical behavior can also
appear.}   RG arguments based on a multicritical LGW theory
predict the intersection point to correspond to a  first-order
transition~\cite{BPV-19,BPV-20-sun}.

The behavior along the transition lines that separate the disordered
phase from the ordered Higgs phases has been discussed in
Sec.~\ref{higgsphases}.  For $v < 0$, continuous transitions are only
possible for $N_f=2$. They belong to the CP$^1$, i.e., vector O(3),
universality class (this has been numerically checked for $\gamma=0$
in Ref.~\cite{BFPV-21-mpsun}). For larger values of $N_f$, transitions
are always of first order. For $v > 0$, XY transitions are expected
for $N_f=2$ (numerical results confirm this
prediction~\cite{BFPV-21-mpsun}). For larger values of $N_f$, we may
have charged continuous transitions associated with the stable charged
FPs of the NAHFT, see Sec.~\ref{epsexpsuNAHFT}.  Numerical
studies~\cite{BFPV-21-mpsun,BPSV-24} for relatively large values of
$N_f$ confirm this scenario. Indeed, for a sufficiently large number
$N_f$ of flavors the SU(2) gauge LNAH models undergo continuous
transitions for finite (sufficiently large) values of $\gamma$,
with a critical behavior consistent with that predicted by the 3D 
SU(2) gauge Higgs field theory. This is confirmed by the comparison of 
numerical lattice results and 
large-$N_f$ field-theoretical predictions,
see also Sec.~\ref{rNASFT}.

\subsubsection{O(5) multicritical behavior of
  the two-flavor SU(2) gauge Higgs model}
\label{sec:MCPNc2NAH}

We now focus on a particular region of the phase diagram of SU(2)
gauge LNAH models with $N_f$ flavors. We discuss the nature of the 
transition point that lies on the $v=0$ axis, where the transition lines 
intersect, see Fig.~\ref{phdianfnc2} (we
discuss the behavior at fixed $\gamma$; in the full phase diagram, we
have a $\gamma$ dependent line of such intersection points). 
Note that the global symmetry of the model enlarges
to Sp($N_f$)/$\mathbb{Z}_2$ for $v=0$.

The behavior at the intersection point depends on $N_f$.  For $N_f > 2$
and $N_f$ not too large,
the intersection point is
expected to be a first-order transition point, as predicted by an
analysis based on the corresponding multicritical LGW
theory~\cite{BPV-19,BPV-20-sun}.
Therefore, we expect the three transition lines to be of
first order close to the intersection. Note that the arguments 
presented in Sec.~\ref{higgsphases} imply that, for $N_f > N_f^*(3)$,
charged continuous transitions are
possible for $v > 0$ (for $v < 0$ only first-order transitions are possible). 
If the intersection point is a first-order transition,
we expect first-order transitions for sufficiently small positive
values of $v$ also in this case. Finally, note that 
the field-theory model with $v=0$ admits a charged FP for large values
of $N_f$. Thus, for $N_f\gg 1$ a charged multicritical behavior can also
appear.

For $N_f=2$, see the left panel of Fig.~\ref{phdianfnc2}, it is
possible to observe a multicritical behavior on the $v=0$ line,
arising from the competition of the O(3) order parameter, driving the
negative-$v$ transitions, and the U(1), i.e., O(2), order parameter,
associated with the positive-$v$ transitions. Because of the exact
symmetry under Sp(2)/${\mathbb Z}_2$=O(5) transformations of the model
with $v=0$, the multicritical behavior belongs to the multicritical
O(5) vector universality class.  Correspondingly, at the MCP the two
operators $P_{\bm x}$ and $Y_{\bm x}$ defined in
Sec.~\ref{modelsubfu} can be combined~\cite{BPV-19,BPV-20-sun} into
a five-component real order parameter---the magnetization---for O(5)
vector systems~\cite{BPV-19,BPV-20-sun}. Note that all transitions are
uncharged---gauge modes do not play any role---and thus the
multicritical behavior can be effectively described by the LGW
multicritical theory with O(2)$\oplus$O(3) symmetry discussed in
Sec.~\ref{sec:multi}.

As discussed in \ref{fssmcp}, the multicritical behavior is controlled
by two relevant RG perturbations, represented by two scaling fields
$u_1$ and $u_2$ that are functions of the model parameters and have
positive RG dimensions $y_1$ and $y_2$.  In the case of the
O(5)-symmetric MCP we can identify $u_1 \approx v$ and $u_2\approx
1-\beta/\beta_{\rm mc}$, whose RG dimensions are determined by the
O(5) vector universality class~\cite{CPV-03,HPV-05}. They are given by
$y_1 = y_{2,2} = 1.832(8)$~\cite{CPV-03} ($y_{2,2} = 1.838(10)$
from Monte Carlo simulations \cite{BC-25-O5}), where $y_{2,2}$ is the RG
scaling dimension of the coupling associated to the spin-two
perturbation of the O(5)-vector FP, and $y_2 =1/\nu_{{\rm O}(5)} =
1.2818(10)$ (using the best available estimate~\cite{Hasenbusch-22}
$\nu_{{\rm O}(5)}= 0.7802(6)$ for the length-scale exponent of the
O(5) vector universality class).

It is important to remark that MCPs arising from the competition of
O(3) and U(1) order parameters do not generally belong to the O(5)
vector universality class~\cite{HPV-05,CPV-03}. Indeed, the analysis
of the RG flow of the O(3)$\oplus$O(2) LGW theory (\ref{on1on2}) shows
that the multicritical O(5) FP is unstable against the quartic spin-4
perturbation, whose coupling has a positive RG scaling
dimension~\cite{HPV-05}, given by $y_{4,4} = 0.23(2)$.\footnote{The
only stable FP in the O(3)$\oplus$O(2) LGW theory is the decoupled FP
\cite{CPV-03, Aharony-02, PV-05} Therefore, if the system is in its
attraction domain, a decoupled critical behavior should be observed at
the MCP, with a tetracritical phase diagram, see the 
central panel of
Fig.~\ref{multicri}.}  Therefore, the observation of an O(5)
multicritical behavior generally requires an additional  tuning of the
model parameters to decouple the spin-4 perturbation of the O(5) FP.
In the case at hand, the O(5) symmetry enlargement is guaranteed by
the fact that the LNAH model for $v=0$ is exactly invariant under the
larger group Sp(2)/${\mathbb Z}_2=$SO(5). The presence of the SU(2)
gauge invariance thus provides a robust mechanism---no fine tuning is
needed--- to obtain an O(5) multicritical behavior in O(3)$\oplus$U(1)
systems, effectively canceling the spin-4 relevant perturbation of the
O(5) FP~\cite{CPV-03,HPV-05}.\footnote{It is worth mentioning that an
effective O(5) symmetry enlargement at multicritical transitions driven by
the competition of O(3) and U(1) critical modes has been suggested, 
and apparently observed, in various physical contexts, e.g.,
in systems with deconfined criticality~\cite{DES-24, NCSOS-15,
  NSCOS-15, TH-05, SF-06, TS-20, CS-23, ZHZH-23, DLGL-24} and in high-$T_c$
superconductors~\cite{Zhang-97, ZHAHA-99, AH-99, AH-00, DHZ-04,
  HPV-05}.}  It is also interesting to note that, by fixing
$\beta=\beta_{\rm mc}$ ($\beta_{\rm mc}$ corresponds to the location
of the MCP) and by varying $v$ from $v<0$ to $v>0$, the system goes
from an O(3) ordered phase to a U(1) ordered phase across a continuous
transition, the O(5) MCP.

\subsubsection{Lattice SU($N_c$) gauge
  Higgs models
  with $N_c\ge 3$}
\label{phdiancl2}

\begin{figure}
\begin{center}
    \includegraphics[width=0.3\columnwidth]{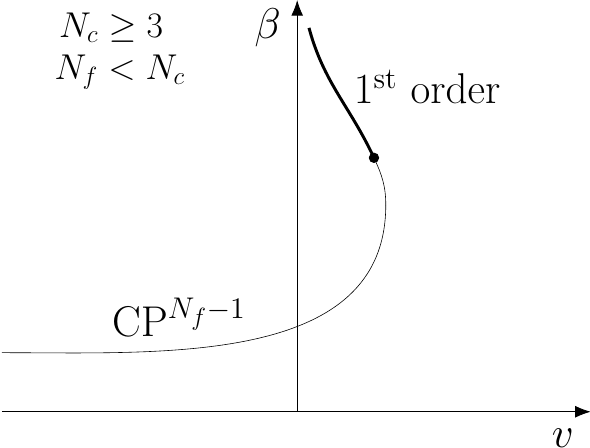}
    \hspace*{3mm}
    \includegraphics[width=0.3\columnwidth]{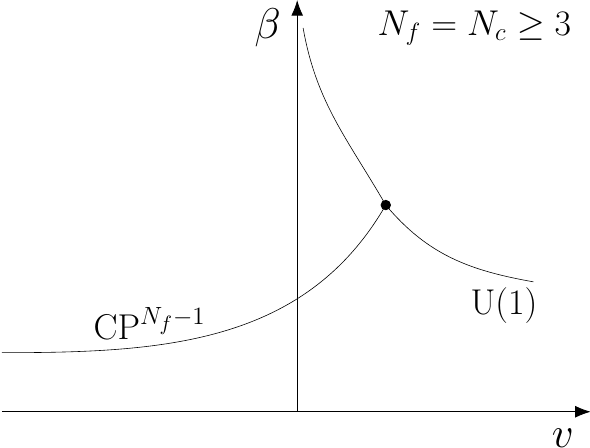}
    \hspace*{3mm}
    \includegraphics[width=0.3\columnwidth]{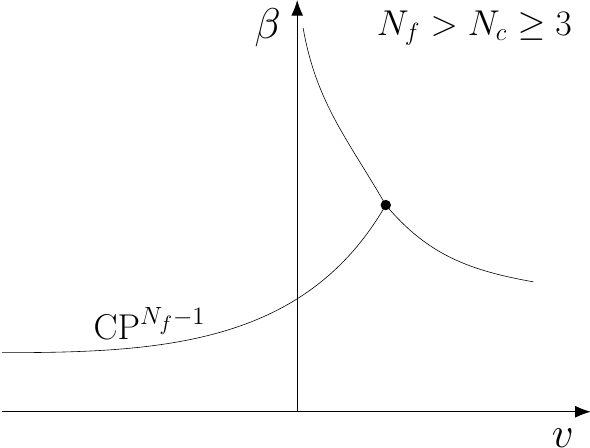}    
    \caption{Sketches of the phase diagrams of 3D SU($N_c$) gauge LNAH
      models with $N_f$ scalar fields in the fundamental
      representation, at fixed $\gamma\ge 0$, for $N_c\ge 3$,
      and $N_f<N_c$ (left), $N_f=N_c$ (middle), $N_f>N_c$ (right).
  \label{phdianfncge3}
}
\end{center}
\end{figure}

We now discuss the possible phase diagrams of the SU($N_c$) gauge LNAH
model for $N_c\ge 3$, see Fig.~\ref{phdianfncge3}, distinguishing
three cases, $N_f<N_c$, $N_f=N_c$, and $N_f>N_c$.

For $N_f<N_c$, see the left panel of Fig.~\ref{phdianfncge3}, as
discussed in Sec.~\ref{higgsphases}, there is a single negative-$v$
Higgs phase.  Along the transition line, the system behaves as the
CP$^{N_f-1}$ model, so continuous transitions are only possible for
$N_f=2$, see Sec.~\ref{cpnmodels}.  For $N_f\ge 3$ the boundary of the
Higgs phase is a first-order transition line. For $N_f=2$, the
behavior along the boundary is the one reported in the left panel of
Fig.~\ref{phdianfncge3}, see Refs.~\cite{BPV-19,BPV-20-sun}.  For
large $\beta$ the boundary corresponds to a first-order transition
line, then it turns into a continuous O(3) transition line, with the
presence of a tricritical point. Note that for $N_c\ge 3$, the line
$v=0$ does not play any role, and therefore the boundary does not lie
on the line $v=0$. It starts at $v=0$ for $\beta\to\infty$, then it
lies in the $v > 0$ plane before bending and moving in the $v < 0$
half plane.

Possible phase diagrams for $N_f=N_c$ and $N_f>N_c$ are shown in the
central and right panels of Fig.~\ref{phdianfncge3}, respectively.
The behavior is qualitatively similar to that observed for $N_c =2$.
The only difference is the absence of an enlarged symmetry for $v=0$,
so that the $v=0$ axis does not play any particular role at finite
$\beta$. Therefore, the intersection point of the transition lines
should be located in a generic point.  Analogously, the
first-order transition line that separates the two low-temperature
Higgs phases will be a generic line in the $\beta-v$ plane for each
value of $\gamma$. Concerning the nature of transition lines, we may
apply considerations similar to those reported in Sec.~\ref{phdianc2}
for $N_c=2$, so we do not repeat them. See Ref.~\cite{BFPV-21-mpsun}
for a more detailed description.

\subsection{Critical behaviors associated with the 
charged fixed point of the SU($N_c$) gauge Higgs field theory}
\label{rNASFT}

\begin{figure}
\begin{center}
    \includegraphics[width=0.43\columnwidth]{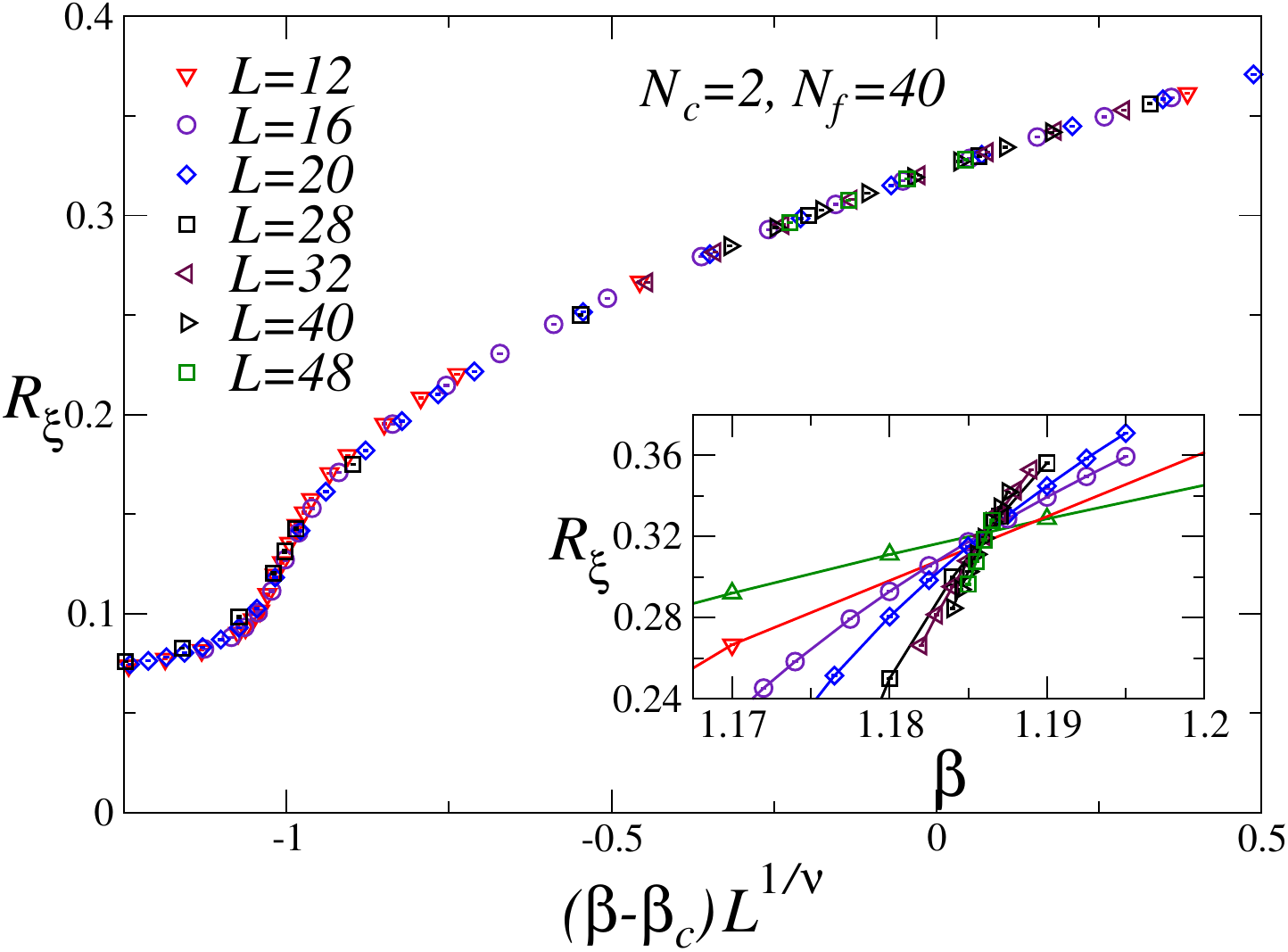}
    \hspace*{3mm}
    \includegraphics[width=0.43\columnwidth]{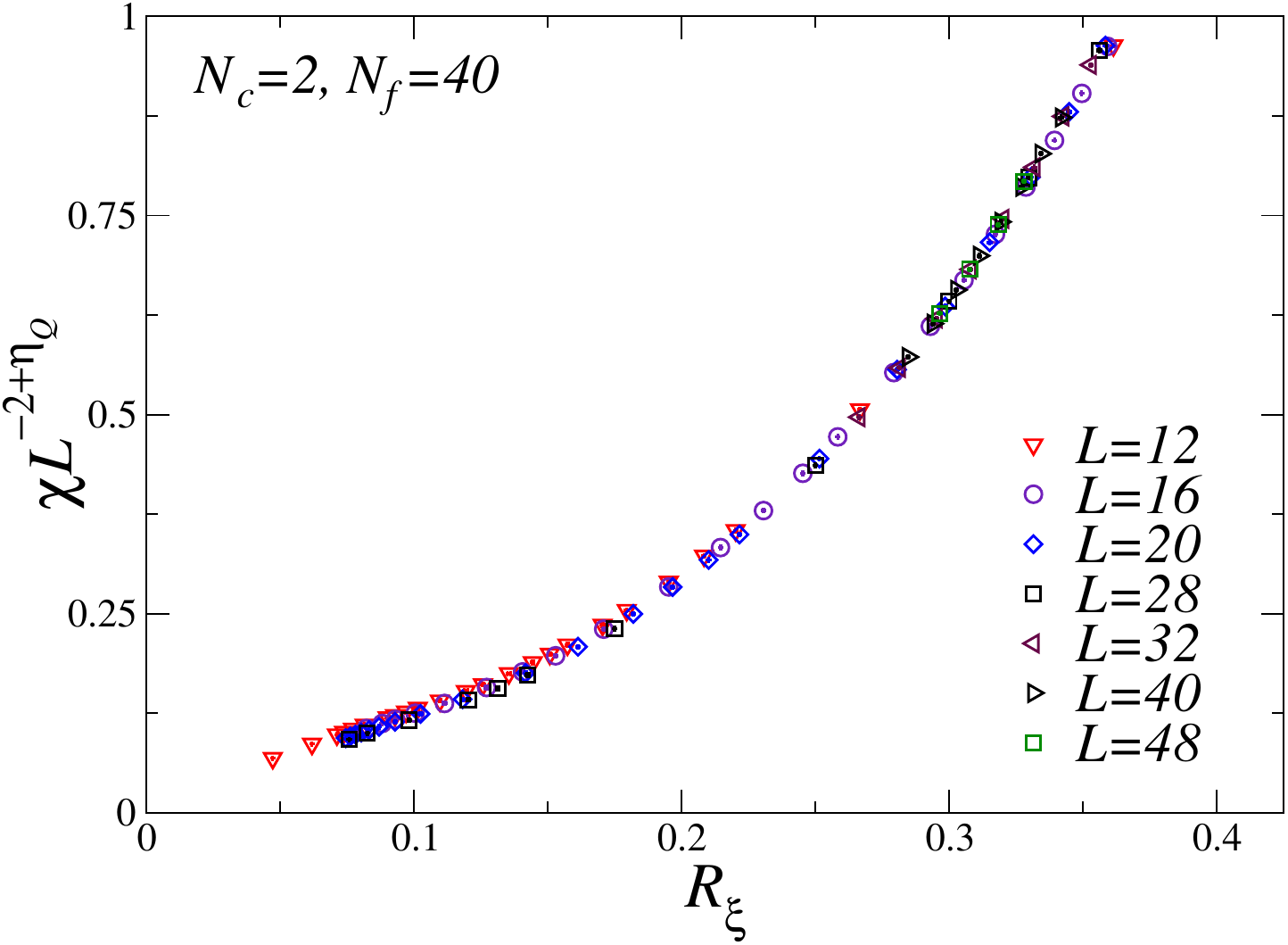}
    \caption{FSS behavior of the 3D SU(2) gauge LNAH model along the
      disorder-Higgs transition line. Results for $N_f=40$, $v=1$, and
      $\gamma=1$, varying the inverse temperature $\beta$.  The left
      panel shows data for $\xi_P/L$, where $\xi_P$ is the
      second-moment correlation length computed from the correlation
      function $G_P$ defined in Eq.~(\ref{pdefsun}), versus $X =
      (\beta - \beta_c) L^{1/\nu}$, using the best estimates $\beta_c
      = 1.18625$ and $\nu = 0.745$. The excellent collapse of the data
      confirms that the transition is continuous.  The inset shows the
      same data versus $\beta$.  The right panel shows the 
      rescaled susceptibility $\chi_P$ of the operator $P_{\bm x}$, which
      asymptotically behaves as $\chi_P \approx L^{2-\eta_P} {\cal
        X}_P(\xi_P/L)$, see Eq.~(\ref{chisclaw2}), with $\eta_P =
      0.87(1)$.
\label{numressu2}
}
\end{center}
\end{figure}

An important issue is the relation between the statistical lattice
gauge model (\ref{hgaugesun}) and the corresponding NAHFT
(\ref{cogausun}), which is the field theory with the same field
content and the same local and global symmetries. In particular, one
would like to identify the continuous transitions that can be
described by the stable charged FP of the RG flow of the continuum
SU($N_c$) gauge Higgs field theory, see Sec.~\ref{epsexpsuNAHFT}. The
analogous problem for charged transitions in AH models has been
discussed in Secs.~\ref{LAHMnc} and \ref{LAHMc}.

As we have discussed in Sec.~\ref{epsexpsuNAHFT}, the RG flow of the
NAHFT (\ref{cogausun}) has a stable charged FP in the domain
$v>0$---therefore, the symmetry-breaking pattern is given in
Eq.~(\ref{glsymbrpat})---for $N_f>N^*_f(d)$. The presence of a stable
FP indicates that lattice models can undergo charged continuous
transitions for sufficiently large values of $N_f$.  Phase transitions
with symmetry-breaking pattern $\mathrm{U}(N_f)\rightarrow
\mathrm{SU}(N_c)\otimes\mathrm{U}(N_f-N_c)$ are expected to occur
along the surface separating the disordered phase from the
positive-$v$ Higgs phase (see the right panel of
Fig.~\ref{phdianfnc2}).  Moreover, since charged continuous
transitions are characterized by the presence of critical gauge
fluctuations, they are expected to occur for sufficiently large values
of $\gamma$ (they are not expected for small values of $\gamma$, given
that for $\gamma = 0$ gauge fields can be integrated out).

Refs.~\cite{BPSV-24,BFPV-21-mpsun} report a numerical study of the
SU(2) gauge model ($N_c=2$) aiming at identifying possible charged
transitions for large values of $N_f$.  Continuous transitions are
observed for $N_f=30,40$, and 60, keeping $v=1$ and $\gamma=1$ fixed
and varying $\beta$. Some numerical FSS results are shown in
Fig.~\ref{numressu2} for $N_f=40$; they clearly show that the
transition is continuous. We also mention that the observed FSS
behavior for $N_f=40$ and
$\gamma=1$ differs from that observed
for the same value of $N_f$ ($N_f=40$) in the limit
$\gamma\to\infty$~\cite{BFPV-21-mpsun}, confirming that gauge
fluctuations are relevant along the large-$\gamma$ charged transition
line.

\begin{figure}[tbp]
\begin{center}
  \includegraphics[width=0.45\columnwidth]{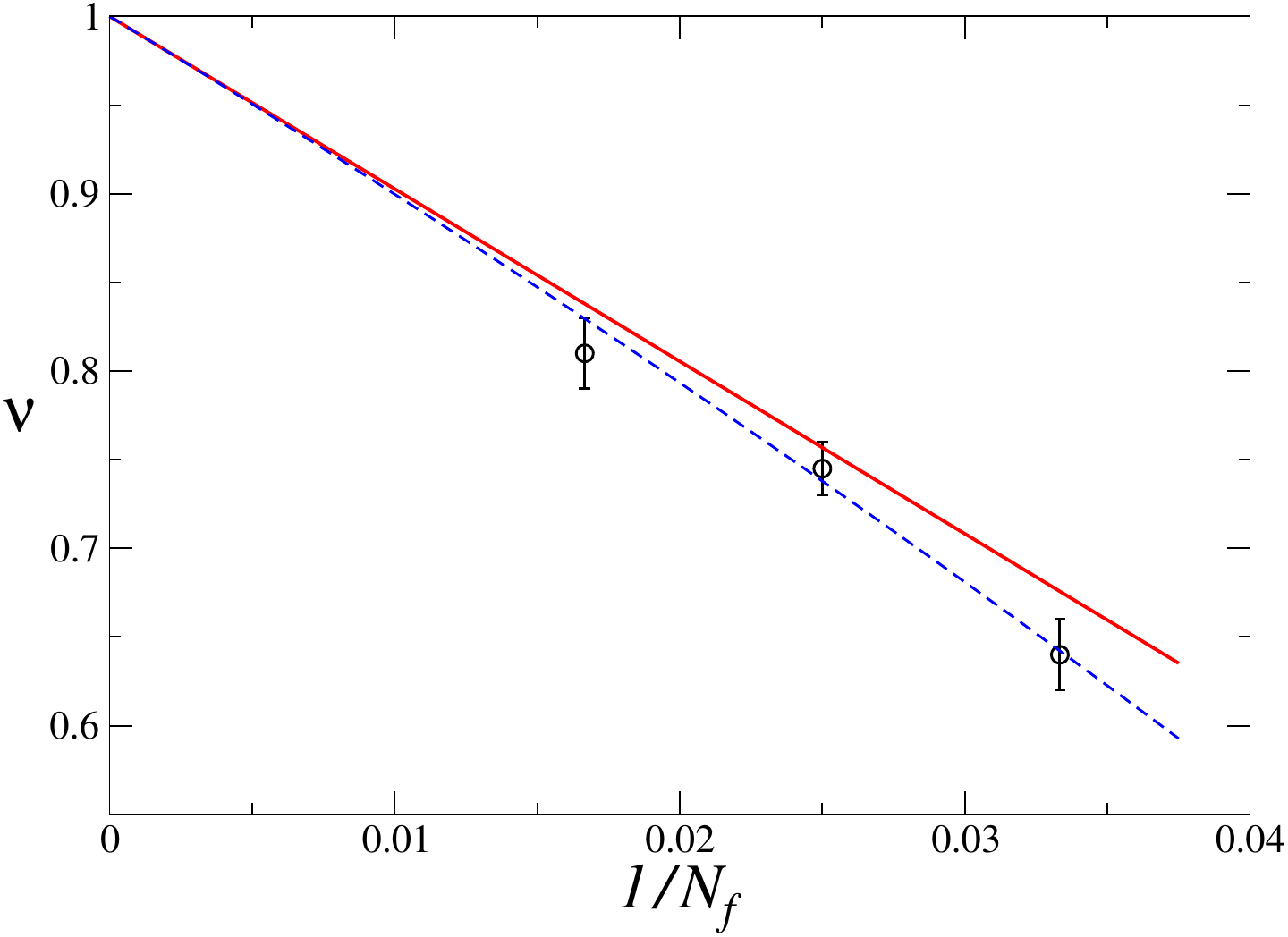}
  \caption{(Adapted from Ref.~\cite{BPSV-24}) Numerical estimates of the
    critical exponent $\nu$ versus $1/N_f$ for the SU(2) gauge LNAH
    model (\ref{hgaugesun}).  For comparison we also report the
    $O(1/N_f)$ theoretical prediction, Eq.~(\ref{nulargensun}) (solid
    line) and a next-to-leading interpolation including a $1/N_f^2$
    term, i.e.  $\nu=1+a_1/N_f+a_2/N_f^2$ (dashed line), with
    $a_2\approx -30$.
  \label{su2_crit_exp}}
  \end{center}
\end{figure}

To confirm that these large-$N_f$ transitions have a critical behavior
associated with the stable charged FP of the NAHFT, one may compare
the numerical estimates of the correlation-length exponent $\nu$ with
the field-theoretical 1/$N_f$ result reported in
Eq.~(\ref{nulargensun}). Such a comparison is reported in
Fig.~\ref{su2_crit_exp}, which shows that numerical data and
large-$N_f$ field-theoretical predictions are in nice agreement.  In
particular, the estimate $\nu=0.745(10)$ for $N_f=40$ appears in
reasonable agreement (the small difference can be interpreted as a
1/$N^2_f$ correction, see Ref.~\cite{BPSV-24}) with the value
$\nu\approx 0.757$ obtained from the $O(N_f^{-1})$ formula reported in
Eq.~(\ref{nulargensun}).  The results of
Refs.~\cite{BPSV-24,BFPV-21-mpsun} allow one to estimate $N^*_f$ in
three dimensions for $N_c=2$. Since only first-order transitions are
observed for $N_f=20$ for different values of $v$ and $\gamma$, and
continuous transitions are found for $N_f=30$, one may conclude that
$N_f^*=25(4)$~\cite{BPSV-24}. Note that this value is significantly
smaller than $N_f^*\approx 375$, the result in four dimensions. This
difference is not surprising, since a similar difference was observed
for the AHFT, see Sec.~\ref{epsexpAHFT}.

The agreement between the NAHFT results and the lattice numerical
estimates nicely supports the conjecture that the continuous
transitions observed in the lattice model can be associated with the
stable charged FP of the RG flow of the SU(2) gauge NAHFT formally
defined by the Lagrangian (\ref{cogausun}) and the corresponding
functional path integral. We expect that analogous results hold for
$N_c\ge 3$, although larger values of $N_f$ are likely required to
observe charged continuous transitions.  We finally remark that the
existence of these 3D charged universality classes implies the
existence of a well-defined nonperturbative continuum limit of the
SU($N_c$) gauge NAHFTs, for a sufficiently large number of scalar
components.

\subsection{Lattice SU($N_c$) gauge Higgs models with
 scalar fields in the adjoint representation}
\label{modelsuncadj}

We now consider SU($N_c$) gauge LNAH models with scalar fields
transforming in the adjoint representation of the gauge group.  We
discuss their phase diagram and the nature of their Higgs phases,
which qualitatively differ from what is observed in models in which
the scalar fields transform in the fundamental representation.

\subsubsection{Lattice Hamiltonian}
\label{modelsubadj}

To construct representatives of this class of models, we may consider
lattice gauge models defined on cubic lattices that are invariant under
local SU($N_c$) and global O($N_f$) transformations, with scalar
fields that transform under the adjoint representation of SU($N_c$)
and under the fundamental representation of the O($N_f$) group.

The fundamental variables are real matrices $\Phi^{af}_{\bm x}$, with
$a=1,...,N_c^2-1$ ({\em color} index) and $f=1,...,N_f$ ({\em flavor}
index), defined on the lattice sites, and gauge fields $U_{{\bm
    x},\mu} \in {\rm SU}(N_c)$ associated with the lattice
links~\cite{Wilson-74}.  The Hamiltonian and the corresponding partition
function
are~\cite{SPSS-20,BFPV-21-adj,Nadkarni-90,HPST-97,KLRS-97,HP-00, GGR-22}
\begin{eqnarray}
  H  = H_K(\Phi,U) + H_V(\Phi) + H_G(U),
  \qquad Z = \sum_{\{\Phi,U\}} e^{-\beta H},  
\label{hgaugeadj}
\end{eqnarray}
where the lattice Hamiltonian $H$ is the sum of the scalar kinetic
term $H_K$, of the local scalar potential $H_V$, and of the pure-gauge
Hamiltonian $H_G$.  As usual, we set the lattice spacing equal to one,
so all lengths are measured in units of the lattice spacing. The
scalar kinetic term $H_K$ is given by
\begin{eqnarray}
  H_K(\Phi,U) =   - {1\over 2} J
  N_f \sum_{{\bm x},\mu} {\rm Tr} \,\Phi_{\bm x}^t \,
  \widetilde{U}_{{\bm x},\mu} \, \Phi_{{\bm x}+\hat{\mu}}^{\phantom t},
  \label{Kintermadj}
\end{eqnarray}
where the matrix $\widetilde{U}^{ab}_{{\bm x},\mu}$ is the adjoint
representation of the original link variable $U_{{\bm x},\mu}$.  The
matrix $\widetilde{U}^{ab}_{{\bm x},\mu}$ is explicitly given by
$\widetilde{U}^{ab}_{{\bm x},\mu} = 2 \, {\rm Tr}(\,U^{\dagger}_{{\bm
    x},\mu} T^a U_{{\bm x},\mu} T^b\,)$, where $a,b=1,...,N_c^2-1$,
and $T^a$ are the $N_c^2-1$ generators in the fundamental
representation, with normalization ${\rm Tr} \, T^a T^b =
\frac{1}{2}\delta^{ab}$. In the following we fix $J=1$, so t
energies are measured in units of $J$.  The scalar potential term
$H_V$ is written as
\begin{eqnarray}
  H_V(\Phi) = \sum_{\bm x} V(\Phi_{\bm x}),\qquad
  V(\Phi)={r\over 2} \, {\rm Tr}\,\Phi^t\Phi + {u\over 4} \, \left( {\rm
    Tr}\,\Phi^t\Phi\right)^2 + {v\over 4} \, {\rm Tr}\,(\Phi^t\Phi)^2,
  \label{potentialsunadj}
\end{eqnarray}  
which is the most general quartic potential invariant under
O($N_f$)$\otimes$O($N_c^2-1$) transformations.  For $v=0$, the
symmetry group of $H_V(\Phi)$ is larger, namely, the O($M$) group with
$M=N_f(N_c^2-1)$. Finally, the pure-gauge plaquette term reads
\begin{eqnarray}
H_G(U) =
  - {\gamma\over N_c} \sum_{{\bm x},\mu>\nu}  {\rm Re} \,
  {\rm Tr}\, \Pi_{{\bm x},\mu\nu},\qquad
\Pi_{{\bm x},\mu\nu}=
      U_{{\bm x},\mu} \,U_{{\bm x}+\hat{\mu},\nu} \,U_{{\bm
          x}+\hat{\nu},\mu}^\dagger \,U_{{\bm x},\nu}^\dagger,
\label{plaquetteadj}
\end{eqnarray}
where the parameter $\gamma$ plays the role of inverse gauge coupling.

The model is invariant under global O($N_f$) transformations,
$\Phi_{\bm x}^{af} \to \sum_g O^{fg} \Phi_{\bm x}^{ag}$, and under
local SU($N_c$) transformations
\begin{equation}
U_{{\bm x},\mu} \to V_{\bm x} U_{{\bm x},\mu} V_{{\bm x} + \hat\mu}^\dagger, 
\qquad
\Phi^{af}_{\bm x} \to \sum_b \widetilde{V}_{\bm x}^{ab} \Phi^{bf}_{\bm x},
\end{equation}
where $V_{\bm x}$ is a SU($N_c$) matrix and $\widetilde{V}_{\bm x}$
is the corresponding matrix in the adjoint representation.  The
Hamiltonian is also invariant under the transformations $U_{{\bm
    x},\mu}\to z(x_{\mu})U_{{\bm x},\mu}$, where $z(x_{\mu})$ is an
element of the gauge-group center ${\mathbb Z}_{N_c}$ that depends
only on $x_{\mu}$ (the $\mu$-th component of the position
vector). When this symmetry is not spontaneously broken, Wilson loops
obey the area law and color charges transforming in the fundamental
representation are confined.

One may again consider a simplified model, requiring the scalar fields
to obey the fixed-length constraint ${\rm Tr}\, \Phi_{\bm x}^t
\Phi_{\bm x} = 2$.  Formally, this model can be obtained by taking the
limit $u,r\to\infty$ keeping the ratio $r/u=-2$ fixed. The
corresponding lattice Hamiltonian reads
\begin{eqnarray}
H = - {N_f\over 2} \sum_{{\bm x},\mu} {\rm Tr} \,\Phi_{\bm x}^t \,
\widetilde{U}_{{\bm x},\mu} \, \Phi_{{\bm x}+\hat{\mu}}^{\phantom t}
+ {v\over 4} \sum_{\bm x}  {\rm Tr}\,(\Phi_{\bm x}^t\Phi_{\bm x})^2
- {\gamma\over N_c} \sum_{{\bm x},\mu>\nu}  {\rm Re} \, {\rm Tr}\,
\Pi_{{\bm x},\mu\nu},\qquad {\rm Tr}\, \Phi_{\bm x}^t \Phi_{\bm x} = 2.
\label{hfixedlengthadj} 
\end{eqnarray}
Models with generic values of $r$ and $u$ are expected to have the
same qualitative behavior as this simplified model.

The critical properties of the scalar fields can be monitored by using
the correlation function $G_P$ of the gauge-invariant bilinear
operator
\begin{equation}
  P_{\bm x}^{fg} = {1\over 2} \sum_a \Phi_{\bm x}^{af} \Phi_{\bm x}^{ag}
  - {1\over N_f} \delta^{fg},
  \qquad
    G_P({\bm x},{\bm y}) \equiv \langle {\rm Tr}\, P_{\bm x} P_{\bm y} \rangle,
\label{pdefsunadj}
\end{equation}
which satisfies ${\rm Tr} \, P_{\bm x}=0$ due to the fixed-length
constraint.  The bilinear scalar operator $P_{\bm x}$ provides the
natural order parameter for the breaking of the global O($N_f$)
symmetry.

\subsubsection{Higgs phases and phase transitions}
\label{higgsphasesadj}

SU($N_c$) gauge models with a single ($N_f=1$) adjoint scalar field
have been the subject of several studies, see, e.g.,
Refs.~\cite{Nadkarni-90, DHKR-02, NRSW-22}, since they are the lattice
analogues of the Georgi-Glashow model, in which magnetic monopoles are
responsible for
confinement~\cite{GG-72,tHooft-74,Polyakov-74,Polyakov-77}. The
single-flavor adjoint model has also been considered to describe some
features of electron-doped cuprates~\cite{SSST-19}. Its phase diagram
is trivial, as it presents a single thermodynamic phase, with two
continuously connected regimes, a high-temperature disordered regime
and a low-temperature Higgs-like
regime~\cite{DHKR-02,SSST-19}. Indeed, the existence of a distinct
low-temperature Higgs phase requires the breaking of a global
symmetry, which is only possible for $N_f\ge 2$.  In the following we
mostly focus on the multiflavor models, which present a complex phase
diagram, with several different phases and transition lines.

For $N_f\ge 2$ the lattice models show different Higgs phases
associated with different symmetry-breaking patterns.  The symmetries
of the Higgs phases can be inferred by means of a mean-field
analysis~\cite{SSST-19,SPSS-20,BFPV-21-adj}---critical fluctuations
are only relevant along the transition lines. The symmetries of the
Higgs phases are indeed the symmetries of the minima of the local
scalar potential $V(\Phi)={v\over 4} \,{\rm Tr}\,(\Phi^t\Phi)^2$ in
the fixed-length limit ${\rm Tr}\, \Phi^t \Phi = 2$.  In the following
we summarize the main properties of these phases, which depend on the
number of colors $N_c$, of flavors $N_f$, and on the parameter
$v$~\cite{SSST-19,SPSS-20,BFPV-21-adj}.  Moreover, their nature also
depends on the behavior of the gauge modes, which may undergo
topological transitions due to the fluctuations of variables
associated with the gauge-group center ${\mathbb Z}_{N_c}$.  Numerical
studies for $N_c=2$ and $N_c=3$ have been reported in
Refs.~\cite{SPSS-20,BFPV-21-adj,CHHJYLRS-24}.

Two qualitatively different phase diagrams are found, depending on the
number of colors $N_c$ and of flavors $N_f$.  For $N_f\le N_c^2-1$,
independently of the specific value of $N_c$, there is a single Higgs
phase that lies in the half-space $v < 0$ for large values of $\beta$.
Indeed, the mean-field analysis shows that for $\beta\to \infty$ the
model orders only for $v < 0$.  For $v > 0$, the system is disordered
for $\beta\to\infty$.

For any $N_f$ and $N_c$ satisfying $N_f>N_c^2-1$, instead, there are  two
different low-temperature Higgs
phases~\cite{SSST-19,SPSS-20,BFPV-21-adj}.  For $\beta\to \infty$,
they correspond to the minima of the lattice potential for $v<0$ and
$v>0$, respectively, and thus we again refer to these two phases as
the negative-$v$ and positive-$v$ phases, respectively, although, this
characterization does not hold for finite $\beta$. The two Higgs
phases have different global symmetries and also different residual
gauge symmetries, as defined in Sec.~\ref{sec.ZQhiggs}.  Some details
can be found in Refs.~\cite{SSST-19,BFPV-21-adj}, and in
Sec.~\ref{higgsphasesadjNc2Nf4} for the specific case $N_c=2$ and
$N_f=4$.  In particular, the positive-$v$ Higgs phase has a
residual ${\mathbb Z}_{N_c}$ gauge symmetry, so 
topological ${\mathbb Z}_{N_c}$ gauge
transitions can occur when varying the inverse gauge coupling $\gamma$, 
as discussed in Sec.~\ref{sec.ZQhiggs}.

The possible existence of topological transitions associated with the
gauge-group center ${\mathbb Z}_{N_c}$ can also be inferred by looking
at the behavior of the model in the limit $\beta\to\infty$ keeping
$\kappa\equiv \beta\gamma$ fixed, see Ref.~\cite{BFPV-21-adj} for
details.  For $v > 0$ the relevant configurations, i.e., those that
minimize the scalar potential and the scalar kinetic energy $H_K$, are
characterized by $\widetilde{U}_{{\bm x},\mu}=1$, so $U_{{\bm x},\mu}
= \lambda_{{\bm x},\mu}\in {\mathbb Z}_{N_c}$.  Therefore, in this
limit the model (\ref{hfixedlengthadj}) reduces to the lattice
${\mathbb Z}_Q$ gauge theory with $Q=N_c$ discussed in
Sec.~\ref{sec.ZQ}.  In three dimensions, this lattice discrete gauge
model undergoes a transition at a finite value of the gauge coupling
$K$. For example,the lattice ${\mathbb
  Z}_{2}$ gauge theory~\cite{Wegner-71} presents a small-$K$
confined phase and a large-$K$ deconfined phase (which may carry
topological order at the quantum level~\cite{Sachdev-19}), separated
by a critical point at $K_c =
0.761413292(11)$~\cite{BPV-20-hcAH,FXL-18}. If the ${\mathbb Z}_{N_c}$
gauge transition persists for finite values of $\beta$, then, when
varying $\gamma$ or $\kappa$, one may have different low-temperature
Higgs phases, which have the same global symmetry but differ in the
large-scale behavior of the ${\mathbb Z}_{N_c}$ modes.

 These arguments indicate that there are two different positive-$v$
 Higgs phases, which differ for the behavior of the topological modes
 associated with the gauge-group center ${\mathbb
   Z}_{N_c}$~\cite{SSST-19,SPSS-20}. Some numerical evidence of the
 existence of topological ${\mathbb Z}_2$ transitions in the SU(2)
 gauge LNAH models with four adjoint scalar fields is presented in
 Ref.~\cite{BFPV-21-adj}.

\subsubsection{The adjoint SU(2) gauge Higgs model with four flavors}
\label{higgsphasesadjNc2Nf4}

Let us now focus on the model with $N_c=2$ and $N_f=4$, which is
expected to describe some aspects of the optimal doping criticality in
cuprate high-$T_c$ superconductors~\cite{SSST-19}.  Since
$N_f>N_c^2-1$, there are several Higgs phases characterized by
different global and gauge symmetry-breaking patterns.  For finite
$\beta$ the two Higgs phases are divided by a transition line that
ends at $v = 0$, $\beta = \infty$, which is expected to be generically
of first order, as it is the boundary of two different ordered phases.

In the negative-$v$ region, the global symmetry-breaking pattern is
$\hbox{O(4)} \to \hbox{O(3)} \oplus {\mathbb Z}_2$ and the gauge
symmetry-breaking pattern is $\hbox{SU(2)} \to \hbox{U(1)}$.  Since
the remnant gauge-invariance group of the Higgs phase is U(1), and the
three dimensional U(1) lattice gauge theory does not undergo phase
transitions~\cite{Polyakov-75,Polyakov-77},\footnote{Confinement is
present in the 3D U(1) gauge model also at weak coupling, see
Refs.~\cite{Polyakov-75,Polyakov-77}, on the contrary of what happens
in four-dimensional models, see Ref.~\cite{Guth-79}.} no topological
transition is expected, so the gauge coupling should be
irrelevant. Therefore, this suggests that there is a single
negative-$v$ Higgs phase, irrespective of the value of $\gamma$ (of
course we cannot exclude the existence of first-order transitions due
to the greater complexity of the adjoint LNAH model).  The transition
line separating the negative-$v$ Higgs phase from the disordered phase
is expected to have the same nature as the transition in the 3D
RP$^{3}$ model (for $v < 0$ one can establish a correspondence between
the behavior of the SU($N_c$) gauge model and the 3D RP$^{N_f-1}$
model~\cite{BFPV-21-adj2d}), which undergoes a first-order transition,
see Sec.~\ref{RPN}.

The structure of the positive-$v$ Higgs phase is more interesting. The
global symmetry-breaking pattern of the positive-$v$ Higgs phase is
O(4)$\to$O(3).  Thus, uncharged continuous transitions between the
positive-$v$ Higgs phase and the disordered phase are expected to
belong to the O(4) vector universality class, provided that gauge
modes are irrelevant.  The gauge symmetry-breaking pattern along the
disorder-Higgs transition line is $\hbox{SU(2)} \to {\mathbb Z}_2$, so
the Higgs phase is invariant under the center of the gauge group,
opening the possibility of ${\mathbb Z}_2$ topological transitions
controlled by the gauge coupling $\gamma$, as discussed in
Sec.~\ref{sec.ZQhiggs}.

\begin{figure}
\begin{center}
\includegraphics[width=0.43\columnwidth, clip]{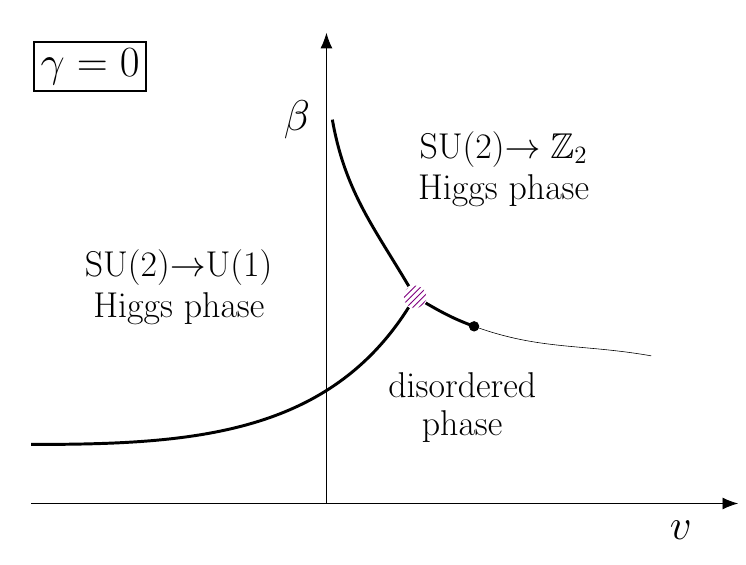}
\hspace*{5mm}
\includegraphics[width=0.43\columnwidth,clip]{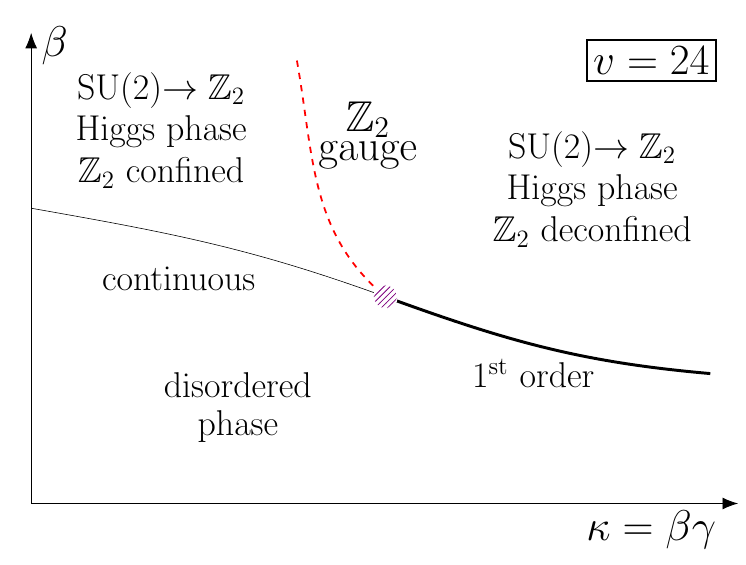}
\caption{ Sketches of the phase diagram of 3D SU(2) gauge LNAH models
  with four scalar fields in the adjoint
  representation~\cite{BFPV-21-adj}, at fixed $\gamma = 0$ (left) and
  at fixed $v=24$ (right). Thick lines denote first-order transitions,
  while thin lines correspond to continuous transitions.  Left panel:
  the shaded point is a first-order intersection point; the filled
  black point that separates first-order from continuous transitions
  occurs at $v=v^*$ with $6 < v^* < 12$.  Right panel: the ${\mathbb
    Z}_2$-gauge transition line (dashed line) starts at $\kappa_c
  \approx 0.761$, $\beta = \infty$ and ends at a multicritical point
  (shaded point) at $\kappa = \kappa_{mc}$ with $1 < \kappa_{mc} < 2$.
  In each Higgs phase we also report the gauge-symmetry-breaking
  pattern associated with the transition that divides the Higgs phase
  from the disordered phase.}
\label{phdianc2nf4}
  \end{center}
\end{figure}

In the left panel of Fig.~\ref{phdianc2nf4} we report a sketch of the
phase diagram for $\gamma = 0$, based on the theoretical expectations
just discussed and the numerical simulations reported in
Ref.~\cite{BFPV-21-adj}.  Note that the negative-$v$ phase extends in
the positive-$v$ region for intermediate values of $\beta$. The
transitions between the two low-temperature phases and between the
negative-$v$ and the disordered phase are of first order, as expected.
The nature of the transitions between the positive-$v$ and the
disordered phase depends instead on $v$, and for large values of $v$,
they become apparently continuous.

For larger values of $\gamma$ it is possible to have topological
transitions in the positive-$v$ region.  Indeed, one may have two
different positive-$v$ Higgs phases, which differ for the behavior of
the topological modes associated with the gauge-group center ${\mathbb
  Z}_2$~\cite{SSST-19,SPSS-20}.  Numerical evidence of the
existence of such topological transitions in the Higgs phase has been
reported in Ref.~\cite{BFPV-21-adj}, looking at the scaling behavior
of the energy cumulants, as described in~\ref{fssencum}.

To monitor the role played by the gauge fluctuations for $v > 0$, one
may focus on the phase diagram for a specific positive value of $v>0$,
as a function of $\kappa = \beta \gamma$ and $\beta$, see the right
panel of Fig.~\ref{phdianc2nf4}, where a sketch of the phase diagram
at fixed $v=24$ is shown.  The nature of the transition changes
significantly with increasing $\kappa$. While a continuous transition
occurs for $\kappa\lesssim 1$, for $\kappa\ge 2$ transitions are of
first order, their strength decreasing with increasing $\kappa$. A
natural hypothesis is that this abrupt change is due to the different
nature of the Higgs phase: Up to $\kappa \simeq 1$ the low-temperature
phase is characterized by confined ${\mathbb Z}_{2}$ gauge
excitations, while for $\kappa \ge 2$ the ${\mathbb Z}_{2}$ gauge
modes are deconfined. This requires the existence of the ${\mathbb
  Z}_2$ gauge transition line.

To summarize, the phase diagram shown in the right panel of
Fig.~\ref{phdianc2nf4} appears to be characterized by three phases, a
small-$\beta$ disordered phase, and two large-$\beta$ Higgs phases,
which are distinguished by the behavior of gauge-group center
modes. These phases are separated by three transition lines: (i) a
disordered-Higgs transition line for small $\kappa$, which appears to
be continuous; (ii) a disordered-Higgs transition line for large
$\kappa$, which is of first order; (iii) a continuous ${\mathbb
  Z}_{2}$ gauge transition line, which separates the two
low-temperature Higgs phases.

We also mention the numerical study of Ref.~\cite{CHHJYLRS-24} on the
confinement/deconfinement properties of the different Higgs phases,
looking at the behavior of the Polyakov loops.  The disordered
symmetric phase is confining, so Wilson loops obey the area law, due
to the fact that the adjoint scalar field cannot screen the gauge
fluctuations.  On the other hand, in the Higgs phase, in which the
topological ${\mathbb Z}_{2}$ gauge excitations deconfine, Polyakov
loops take a nonvanishing value.

As a final remark, we note that the identification of continuous
transitions associated with the stable charged FP of the RG flow of
the adjoint NAHFT, see Sec.\ref{suNAHFTadj}, remains an open issue.
We believe that further
numerical investigations of 
the phase diagram and of the critical behaviors of SU($N_c$)
gauge LNAH models with adjoint scalar fields is needed
to achieve a satisfactory characterization
of their main features.

\subsection{Lattice SO($N_c$) gauge models with multiflavor scalar fields}
\label{modelon}

A lattice SO($N_c$) gauge model with $N_f$ degenerate scalar fields
in the fundamental representation can be derived by gauging the most
general O($N_c$)$\otimes$O($N_f$)-symmetric $\varphi^4$ model defined
on a cubic lattice. We associate $N_c\times N_f$ real matrix
variables $\varphi^{af}_{\bm x}$ with each site ${\bm x}$ 
and SO($N_c$) group elements $O_{{\bm x},{\mu}}$ with each link. The 
lattice Hamiltonian~\cite{BPV-20-on} is 
\begin{eqnarray}
&&H =  - J\,N_f \sum_{{\bm x},\mu} 
{\rm Tr} \left[ \varphi_{\bm x}^t \, O_{{\bm x},{\mu}}
  \, \varphi_{{\bm x}+\hat{\mu}}\right]
+ \sum_{\bm x}  V(\varphi_{\bm x}) 
- {\gamma\over N_c} \sum_{{\bm x},\mu>\nu} {\rm Tr}\,
\left[ O_{{\bm x},{\mu}} \,O_{{\bm x}+\hat{\mu},{\nu}} 
\,O_{{\bm x}+\hat{\nu},{\mu}}^t  
\,O_{{\bm x},{\nu}}^t\right],
\label{hgaugeson}\\
&&V(\varphi_{\bm x}) = r\,{\rm Tr} \,\varphi_{\bm x}^t \,\varphi_{\bm x}
+ u \, ({\rm Tr} \,\varphi_{\bm x}^t \,\varphi_{\bm x})^2
+ 
v \, \Bigl[ {\rm Tr} \,\varphi_{\bm x}^t \,\varphi_{\bm x}
\,\varphi_{\bm x}^t \,\varphi_{\bm x}
- ({\rm Tr}\,\varphi_{\bm x}^t \,\varphi_{\bm x})^2\Bigr].
\label{potentialon}
\end{eqnarray}
For any
value of $N_c$ and $N_f$, the lattice gauge theory is invariant under
the local gauge transformation $\varphi_{\bm x}\to G_{\bm x}
\varphi_{\bm x}$ and $O_{\bm x,{\mu}}\to G_{\bm x} O_{\bm x,{\mu}}
G_{\bm{x}+\hat{\mu}}^t$ with $G_{\bm x}\in $ SO($N_c$), and under the
global transformation $\varphi_{\bm x}\to \varphi_{\bm x} W$ and
$O_{\bm x,{\mu}}\to O_{\bm x,{\mu}}$ with $W\in $ O($N_f$).

One may again consider the unit-length limit of the scalar field,
formally obtained by taking the limit $r,u\to\infty$ keeping 
$r/u = - 1$ fixed, leading to the constraint ${\rm Tr}\,\varphi_{\bm x}^t
\varphi_{\bm x} = 1$.  In the one-flavor case, $N_f=1$, the phase
diagram of the lattice SO($N_c$) gauge model (\ref{hgaugeson}) is
expected to show only one phase. This can be easily verified for
$\gamma=0$, where the model becomes trivial in the unitary gauge, and
cannot have any phase transition. For $N_c=2$ the gauge group is
equivalent to the Abelian group U(1), and the model is equivalent to
LAHMs with compact gauge variables discussed in Sec.~\ref{LAHMc}. New
critical behaviors are expected only in the non-Abelian case, $N_c\ge 3$.
The natural order parameter for the breaking of the SO($N_f$) global
symmetry is the gauge-invariant real traceless and symmetric
bilinear operator
\begin{equation}
R^{fg}_{\bm x} = \sum_a \varphi^{af}_{\bm x} \varphi^{ag}_{\bm x} - 
     {\delta^{fg}\over N_f}\, ,
\label{rdefon}
\end{equation}
which is a rank-2 operator with respect to the global O($N_f$)
symmetry group.

The field-theoretical $\varepsilon$-expansion analysis of the RG flow of the 
NAHFT (\ref{cogausonc}) predicts the existence of charged universality
classes for a sufficiently large number $N_f$ of components,
controlled by the stable charged FP appearing for positive values of
$v$. No stable FPs emerge for $v<0$, at least close to four
dimensions.  Beside charged transitions 
statistical systems may also develop other
critical behaviors described by appropriate LGW field theories, in
which the gauge correlations are not critical, as is the case 
in the strong-coupling ($\gamma\to 0$) gauge limit.

An analysis of the phase diagram of the model with 
$V(\varphi_{\bm x}) = 0$ ($v = 0$ in the fixed-length limit)
was reported in
Ref.~\cite{BPV-20-on}.  For $N_f\ge 2$ the phase diagram is
characterized by two phases: a low-temperature phase in which the
order parameter $R_{\bm x}^{fg}$ defined in Eq.~(\ref{rdefon})
condenses, and a high-temperature disordered phase.  The two phases
are separated by a transition line, where the SO$(N_f)$ symmetry is
broken.  The gauge parameter $\gamma$, corresponding to the inverse
gauge coupling, is not expected to play any particular role, at least
for sufficiently small values.  This scenario has been checked 
numerically for several values of
$\gamma$.  Along the disorder-order transition line only the
correlations of the gauge-invariant operator $R_{\bm x}^{fg}$ are
critical, while gauge modes are not critical. 
This is consistent with 
the predictions of the continuum gauge model (\ref{cogausonc}): since
stable FPs exist only for $N_f\gg 1$, one expects only 
uncharged transitions for small $N_f$.
Since the gauge modes do
not show critical behaviors, one may predict the critical behavior by
using an effective O($N_f$)-invariant LGW theory based on a
gauge-invariant order parameter associated with the rank-two symmetric
real traceless tensor $R_{\bm x}^{ab}$.  The relevant model 
is therefore the RP$^{N_f-1}$ model discussed
in Sec.~\ref{RPN}. One predicts first-order
transitions for $N_f\ge3$ 
and continuous transitions for $N_f=2$, which
belong to the XY universality class for any $N_c\ge 3$.

Note that charged continuous transitions are possible for large
values of $N_f$ according to the RG flow of the SO($N_c$) gauge NAHFT,
see Sec.~\ref{soNAHFT}, because of the presence of a stable
large-$N_f$ charged FP~\cite{Hikami-80,BPV-20-on,PRV-01}.  As it occurs in 
SU($N_c$) gauge LNAH models with scalar fields  in the fundamental
representation, continuous charged transitions 
are expected to appear
for positive values of $v$, sufficiently large values of $\gamma$, and
a sufficiently large number $N_f$ of components.

%% file: twomodels.tex
\section{Two-dimensional lattice Abelian and non-Abelian gauge Higgs models}
\label{2dmod}

In the previous sections we discussed  3D Abelian and
non-Abelian Higgs models, to investigate the role of the gauge and
global symmetries in determining their main thermodynamic properties,
such as the Higgs phases, the nature of the phase transitions, and the
continuum limits associated with their critical behaviors.  Analogous
issues can be addressed in lower-dimensional models.

In two dimensions the phase diagram of lattice models with short-range
interactions (such as the nearest-neighbor models considered in the
previous sections) and continuous global symmetries are generally
characterized by a single disordered phase. Indeed, according to the
Mermin-Wagner theorem~\cite{MW-66,Mermin-67}, they cannot have
magnetized phases characterized by the condensation of an order
parameter, and therefore they do not undergo phase transitions driven
by the spontaneous breaking of a continuous global symmetry.

However, in the zero-temperature limit they can develop an asymptotic
critical behavior, whose universal features are determined by the
symmetries of the system and are generally associated with 2D
nonlinear $\sigma$ models defined on symmetric spaces, see, e.g.,
Refs.~\cite{BHZ-80,ZJ-book,PV-02}.  The O($N$) $\sigma$ model with
$N\ge 3$ and the CP$^{N-1}$ model with $N\ge 2$ are paradigmatic
models that show this type of behavior.  Indeed, the lattice O($N$)
model behaves as the nonlinear $\sigma$ model defined on the
O($N$)/O($N-1$) symmetric space, while the CP$^{N-1}$ lattice model is
related to the nonlinear $\sigma$ model defined on the
U($N$)/[U(1)$\times$U($N-1$)] symmetric space, see, e.g.,
Ref.~\cite{ZJ-book}.  In this type of models, the correlation
functions in the thermodynamic limit are characterized by a length
scale $\xi$ that diverges as $T^p e^{c/T}$ for $T\to 0$, where $p$ and
$c$ are universal coefficients that only depend on the universality
class and that can be determined in perturbation theory (the
coefficient $c$ also depends on the normalization of $T$, which can be
uniquely fixed perturbatively), see \ref{asyfree} for details.

Systems with an Abelian O(2) global symmetry are peculiar in this
respect, since they may undergo a finite-temperature topological
Berezinskii-Kosterlitz-Thouless (BKT)
transition~\cite{KT-73,Berezinskii-70,Kosterlitz-74}, which separates
the high-$T$ disordered phase from the low-temperature nonmagnetized
spin-wave phase characterized by correlation functions that decay
algebraically.

In this section we discuss how the interplay of global and gauge
symmetries determines the effective low-energy field theory realized in
the zero-temperature continuum limit (critical behavior) of 2D lattice
gauge models using the standard Wilson formulation of gauge
variables~\cite{Wilson-74,MM-book}.

\subsection{General conjecture for the zero-temperature continuum limit}
\label{conjecture}

In the case of models characterized by both global and gauge
symmetries, the asymptotic zero-temperature critical behavior
(equivalently, continuum limit)
is expected to arise from the interplay
between the two different symmetries. For the purpose of understanding
which features are relevant and which continuum limits are effectively
realized, several 2D lattice Abelian and non-Abelian gauge models have
been investigated~\cite{BPV-20-2dAH,BPV-20-sqcd2d,BFPV-20-rpn,
  BFPV-20-on2d,BFPV-21-adj2d,BFPV-21-BKT,BF-22}.
The results of these analyses have led to the following general
conjecture:
\begin{quote}
The zero-temperature critical behavior, and therefore the continuum
limit, of 2D lattice gauge models with scalar fields belongs to the
universality class associated with a 2D theory defined on a symmetric
space~\cite{BHZ-80,ZJ-book} with the same global symmetry.
\end{quote}

2D field-theoretical models defined on symmetric spaces have been
largely investigated, see, e.g., Ref.~\cite{ZJ-book}, and several
high-order perturbative computations, in particular of their $\beta$
functions, have been reported in the
literature~\cite{Polyakov-75a,BZ-76,BZL-76,BLS-76,DVL-78,Witten-79,
  Hikami-79,BHZ-80,Hikami-80,BW-86,FT-86,CHN-89,CR-92,CRV-92,CR-93,CP-94,
  CP-95,ACPP-99,Polyakov-book,ZJ-book,PV-02}.  The O($N$) $\sigma$ and
the CP$^{N-1}$ models are the simplest examples.  Other examples of
lattice gauge models with critical behavior corresponding to a
symmetric-space field theory will be discussed in the following
of this section.

\subsection{Continuum limits of multiflavor lattice gauge theories}
\label{partcases}

We now summarize the main results obtained for several 2D multiflavor
lattice Abelian and non-Abelian gauge
models~\cite{BPV-20-2dAH,BPV-20-sqcd2d,BFPV-20-on2d,BFPV-21-adj2d,
  BFPV-21-BKT,BF-22}. They support the conjecture on the nature
of their continuum limit presented in Sec.~\ref{conjecture}. 
The analysis of the
zero-temperature behavior requires the determination of the relevant
low-energy degrees of freedom, which correspond to the minimum
configurations of the Hamiltonian, and of the residual global
symmetry, which essentially determines the appropriate corresponding
field-theoretical $\sigma$ model.  These analyses are then
supplemented by nonperturbative FSS analyses of MC data.

\subsubsection{Lattice U(1)  gauge models}
\label{2dabmod}

We first consider 2D lattice multiflavor Abelian U(1) gauge models
with compact gauge variables and $N$-component scalar fields. 
On a square lattice the Hamiltonian reads
\begin{eqnarray} 
  H = - N \sum_{{\bm x}, \mu} \left( \bar{\bm{z}}_{\bm x} \cdot
  \lambda_{{\bm x},\mu}\, {\bm z}_{{\bm x}+\hat\mu} + {\rm
    c.c.}\right) - \gamma \sum_{{\bm x},\mu>\nu} \left( \lambda_{{\bm
      x},{\mu}} \,\lambda_{{\bm x}+\hat{\mu},{\nu}}
  \,\bar{\lambda}_{{\bm x}+\hat{\nu},{\mu}} \,\bar{\lambda}_{{\bm
      x},{\nu}} + {\rm c.c.}\right),
\label{clahmH2d}
\end{eqnarray}
where ${\bm z}_{\bm x}$ is a unit-length $N$-component complex vector,
$\lambda_{{\bm x},\mu}$ are U(1) link variables, $\gamma$ plays the
role of inverse gauge coupling, and $\hat\mu=\hat{1},\hat{2}$ are unit
vectors along the lattice directions. As usual, the partition function
is defined as $Z = \sum_{\{{\bm z},\lambda\}} e^{-\beta H}$ with
$\beta = 1/T$.

For some particular values of $\gamma$ the model (\ref{clahmH2d})
becomes equivalent to simpler well-known models.  In the limit
$\gamma\to\infty$, the gauge link variables are all equal to one
modulo gauge transformations, so model (\ref{clahmH2d}) becomes
equivalent to the standard 2D O($2N$)-symmetric vector model, whose
properties are reviewed in \ref{asyfree}.  For $\gamma=0$ the AH model
(\ref{clahmH2d}) is instead a particular lattice formulation of the CP$^{N-1}$
model~\cite{ZJ-book,RS-81,DHMNP-81,BL-81,CRV-92,CR-93}.\footnote{2D
CP$^{N-1}$ models have been much studied in the literature, both
analytically and numerically, because they provide a theoretical
laboratory to understand some of the mechanisms of quantum field
theories of fundamental interactions. In particular, they share some
notable features with QCD, the theory of the hadronic strong
interactions, such as the asymptotic freedom and the so-called
$\theta$ dependence related to topology, see, e.g.,
Refs.~\cite{DVL-78,DVL-79,Witten-79,CR-92,CRV-92-b,VP-09}.}  For both
$\gamma=0$ and $\gamma=\infty$, the lattice model shows a universal
critical behavior for $\beta\to\infty$. In both cases, the correlation
length $\xi$ increases exponentially, $\xi\sim \beta^{-p} e^{c\beta}$,
with $p$ and $c$ taking different values, as in the 
CP$^{N-1}$ and O($2N$) $\sigma$ models, see \ref{asyfree}. 

As in the 3D LAH models discussed in Sec.~\ref{LAHMc}, the
zero-temperature critical behavior can be investigated by analyzing
the correlations of the gauge-invariant bilinear $Q_{\bm
  x}^{ab}=\bar{z}_{\bm x}^a z_{\bm x}^b -\delta^{ab}/N$.  The FSS
analyses of the MC results reported in Ref.~\cite{BPV-20-2dAH} show
that the asymptotic behavior is independent of $\gamma$, and
corresponds to that found for $\gamma=0$.  Therefore, the
zero-temperature critical behavior of the 2D lattice U(1) model with
an $N$-component complex scalar field belongs to the universality
class of the 2D CP$^{N-1}$ model, thus corresponding to the symmetric
space U($N$)/[U(1)$\times$U($N-1$)].  The independence of the critical
behavior on $\gamma$ is related to the subleading behavior of the
gauge modes.

\subsubsection{Lattice SU($N_c$) gauge models}
\label{2dsun}

We now consider 2D non-Abelian SU($N_c$) models defined on square
lattices, with multiflavor matter fields in the fundamental and
adjoint representations, analogous to those defined in
Secs.~\ref{modelsubfu} and \ref{modelsubadj},
respectively.

The zero-temperature critical behavior of the SU($N_c$) gauge model
with $N_f$ scalar fields in the fundamental representation and 
in the unit-length limit was analyzed in
Ref.~\cite{BPV-20-sqcd2d} for the particular value $v=0$ of the
Hamiltonian parameter, see Eq.~(\ref{flpotentialsun}). In this more
symmetric case, the numerical results show that the asymptotic
low-temperature behavior belongs to the universality class of the 2D
CP$^{N_f-1}$ field theory when $N_c\ge 3$, and to that of the 2D
Sp($N_f$) field theory for $N_c=2$. This suggests that their RG flows
are associated with the coset $S^M$/SU($N_c$), where $S^M$ is the
$M$-dimensional sphere and $M=2 N_f N_c$, thus they are asymptotically
controlled by the 2D SFTs associated with the symmetric spaces that
have the same global symmetry, i.e., U($N_f$) for $N_c\ge 3$ and
Sp($N_f$) for $N_c=2$.

\begin{figure}[tbp]
\begin{center}
  \includegraphics[width=0.45\columnwidth]{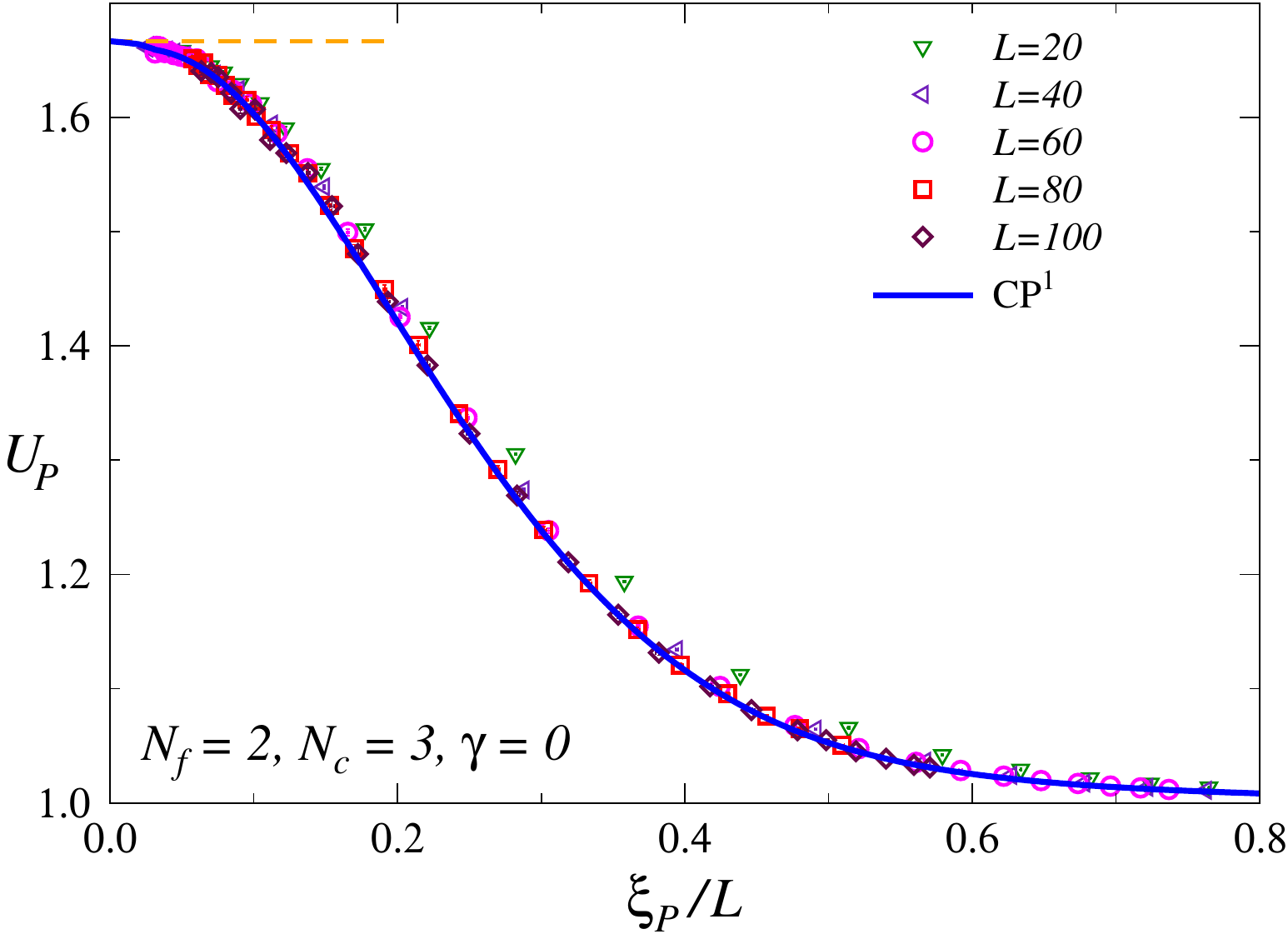}
  \hspace{1cm}
  \includegraphics[width=0.45\columnwidth]{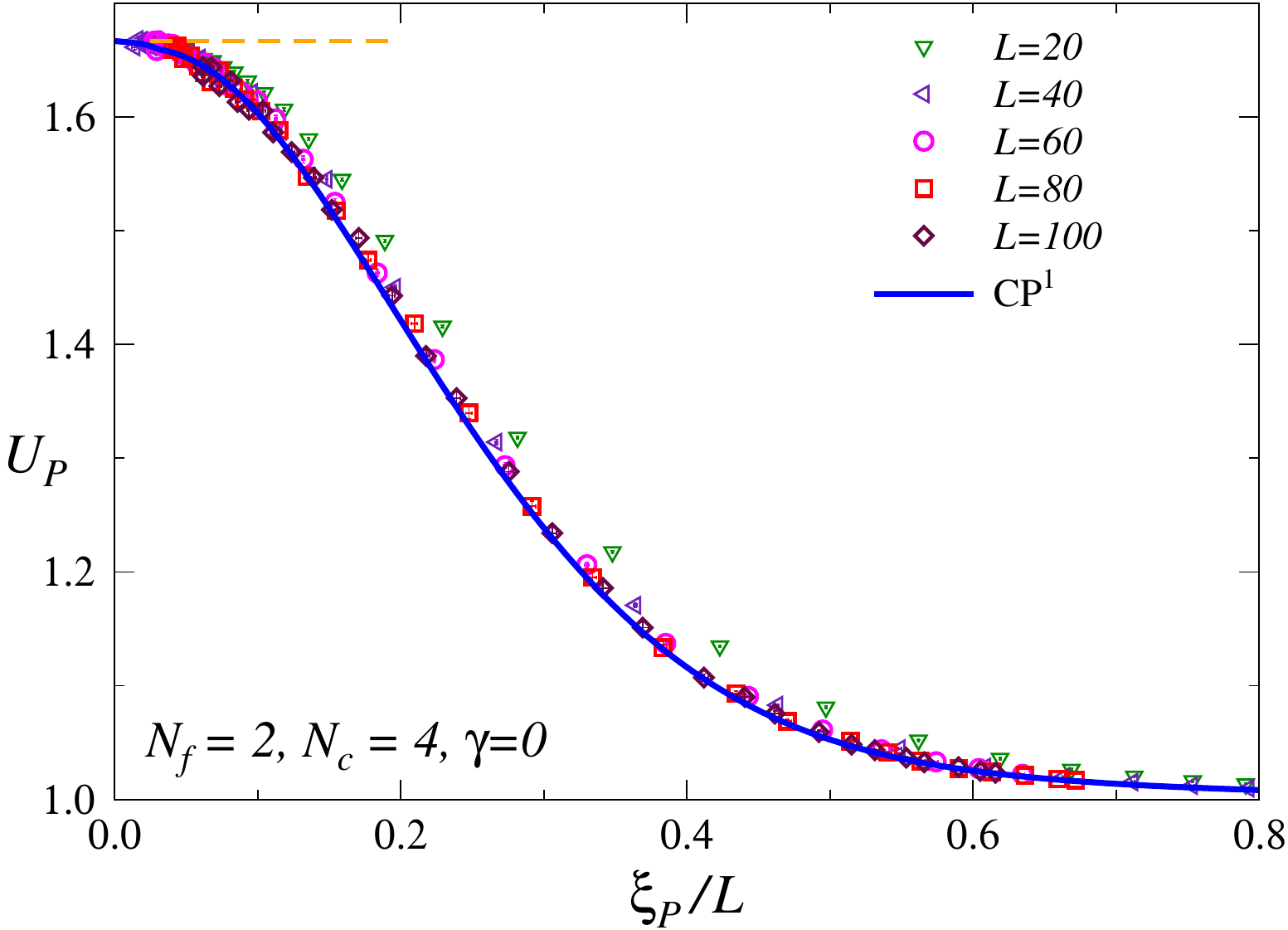}
  \caption{(Adapted from Ref.~\cite{BPV-20-sqcd2d}) We show numerical
    results for the 2D SU($N_c$) gauge LNAH models with $N_c=3$ (left)
    and $N_c=4$ (right); $N_f=2$ and $\gamma=0$ in both cases.  We
    plot the Binder parameter $U_P$ associated with the operator
    $P_{\bm x}^{fg}$, defined in Eq.~\eqref{pdefsun}, versus
    $\xi_P/L$, where $\xi_P$ is the second-moment correlation length
    associated with the correlation function $G_P$, defined as in
    Eq.~\ref{xidefpb}. In both panels, with increasing $L$, data
    approach the solid line representing the universal scaling curve
    of the 2D CP$^1$ model (see Ref.~\cite{BPV-20-2dAH} for a
    parameterization), associated with the symmetric space that has
    the same global symmetry, thus supporting the conjecture reported
    in Sec.~\ref{conjecture}. As discussed in Sec.~\ref{fssrginv}),
    this comparison does not involve any nonuniversal normalization.
  \label{fig_2dsqd}}
  \end{center}
\end{figure}

SU($N_c$) gauge models with $N_f$ scalar fields in the adjoint
representation were investigated in Ref.~\cite{BFPV-21-adj2d},
focusing on systems with $N_f\ge 3$.  For these models the
zero-temperature critical behavior, and therefore their continuum
limit, depends on the sign of the parameter $v$ appearing in the
scalar potential, see Eq.~(\ref{potentialsunadj}).  For $v\le 0$ the
lattice gauge model has the same continuum limit as the RP$^{N_f-1}$
model~\cite{BFPV-20-rpn,CHHR-98,Hasenbusch-96,NWS-96,CPS-94,CEPS-93},
for any value of $N_c$.  For positive $v$ instead, the critical
behavior depends on both $N_c$ and $N_f$.  For $N_f\le N_c^2-1$, there
is no continuum limit: correlation functions are always short ranged.
On the other hand, for $N_f>N_c^2-1$ the system develops long-range
correlations for $T\to 0$. The corresponding continuum limit is the
same as that of the $\sigma$ model defined on the symmetric space
O($N_f$)/O($q$)$\otimes$O($N_f-q$) with $q=N_c^2-1$. Indeed, the
correlation functions of the bilinear operator $P_{\bm x}^{ab}$ in the
two models have the same critical behavior in the zero-temperature
continuum limit.  In particular, for $N_f=N_c^2$, the gauge model is
equivalent to the O($N_f$) vector $\sigma$ model.  Numerical data
support these predictions. Some of them are shown in
Fig.~\ref{fig_2dsqd}.  Again, these results support the conjecture
outlined in Sec.~\ref{conjecture}.

\subsubsection{Lattice SO($N_c$) gauge models}
\label{2dson}

We finally consider 2D square-lattice models with non-Abelian
SO($N_c$) gauge symmetry and O($N_f$) global symmetry, which are
defined by the same Hamiltonian reported in Eq.~(\ref{hgaugeson}) in
the unit-length limit.  Again for $N_c=2$, these lattice gauge models
are equivalent to the 2D Abelian U(1) gauge models discussed in
Sec.~\ref{2dabmod}. Therefore, here we assume $N_c\ge 3$. Analytical
and numerical studies of their critical behavior are reported in
Refs.~\cite{BF-22,BFPV-20-on2d,BFPV-21-BKT}.

For $N_f\ge 3$ the models show a zero-temperature critical behavior.
By studying the minimum-energy configurations, which are the relevant
ones in the zero-temperature limit, and the structure of their
fluctuations, two different low-temperature regimes are identified. If
the Hamiltonian parameter $v$ is negative, see
Eq.~(\ref{potentialon}), then the model shares the same
low-temperature critical behavior as that of the 2D RP$^{N_f-1}$
models~\cite{BFPV-20-rpn,CHHR-98,Hasenbusch-96,NWS-96,CPS-94,CEPS-93}.
Gauge degrees of freedom do not play any active role in the critical
domain, and indeed the low-temperature effective theory is independent
of the number of colors $N_c$. For positive values of $v$, the nature
of the low-temperature regime depends on the number of colors and
flavors. If $N_f\le N_c$ the configurations minimizing the Hamiltonian
do not show any residual global symmetry, and no diverging correlation
length and critical behavior is present. If instead $N_f>N_c$, the
minimizing configurations maintain a residual nontrivial global
symmetry, and the low-temperature behavior is expected to be described
by the non-linear $\sigma$ model defined on the real Grassmannian
manifold SO($N_f$)/[SO($N_c$)$\times$SO($N_f-N_c$)].  This
identification was also supported by numerical results reported in
Refs.~\cite{BF-22,BFPV-20-on2d} for several values of $N_f$ and $N_c$.
They show the emergence of a color-reflection symmetry in the critical
domain, since the results obtained using SO($N_c$) and SO($N_f-N_c$)
groups, keeping fixed $N_f$, approach the same asymptotic curve in the
FSS limit~\cite{BF-22}.

We finally consider the model with $N_f=2$, in which the global
symmetry group is the Abelian O(2) group. Two-flavor SO($N_c$)-gauge
models behave differently, in that they undergo a finite-temperature
topological BKT
transition~\cite{KT-73,Berezinskii-70,Kosterlitz-74,JKKN-77,
  HMP-94,HP-97,Balog-01,Hasenbusch-05,PV-13-BKT}, between the
high-temperature disordered phase and a low-temperature spin-wave
phase characterized by quasi-long range order with vanishing
magnetization.\footnote{We recall that BKT transitions are
characterized by a finite critical temperature with 
an exponentially divergent correlation length $\xi$:
$\xi$ behaves as $\xi\sim
\exp(c/\sqrt{T-T_c})$ approaching the BKT critical temperature $T_c$
from the high-temperature phase.} The FSS analyses of the MC results
reported in Ref.~\cite{BFPV-21-BKT} confirm the existence of 
finite-temperature BKT transitions.

%% file: gaugebreaking.tex
\section{Effects of perturbations breaking the local gauge symmetry}
\label{gaugebreaking}

Gauge symmetries can emerge as low-energy effective symmetries of
many-body
systems~\cite{MSF-01,Sachdev-19,SJTS-22,Senthil-23}. However, because
of the presence of microscopic gauge-symmetry violations, it is
crucial to understand the effects of perturbations that explicitly
break gauge invariance. One would like to understand whether, and
when, gauge-symmetry breaking (GSB) perturbations destabilize the
emergent gauge model--- consequently, a gauge-invariant dynamics would
be observed only if an appropriate tuning of the model parameters is
performed---or whether, and when, GSB perturbations do not
effectively change the low-energy dynamics---in this case the
gauge-symmetric theory would describe the asymptotic dynamics even in
the presence of some (possibly small) violations. This issue is also
crucial in the context of analog quantum simulations, for example when
controllable atomic systems are engineered to effectively reproduce
the dynamics of gauge-symmetric theoretical models.  Indeed, in the
proposed artificial gauge-symmetry
realizations~\cite{ZCR-15,Banuls-etal-20,BC-20}, the gauge symmetry in
not exact and it is only expected to effectively emerge in the
low-energy dynamics, see, e.g., Refs~\cite{Martinez-etal-16,
  Bernien-etal-17, Klco-etal-18, Schweizer-etal-19, Gorg-etal-19,
  Mil-etal-20} for some experimental results.

In this section we discuss GSB effects in LAH models with noncompact
and compact gauge fields.  The gauge symmetry can be broken by adding
different nongauge invariant terms to the gauge-invariant LAH
Hamiltonian.  A first peculiar class of GSB terms is that of gauge
fixings: they leave gauge-invariant correlations invariant, without
changing the physical gauge-invariant behavior of the model, although
they may allow one to observe gauge degrees of freedom that are
relevant in the RG description of the critical behavior, see, e.g.,
Secs.~\ref{LAHMnc} and \ref{LAHMc}. Here we discuss GSB perturbations
that also affect gauge-invariant correlations. Specifically, we
consider operators analogous to photon-mass terms, which break the
gauge invariance of LAH models introduced in Secs.~\ref{LAHMnc} and
\ref{LAHMc}, respectively.  However, we expect the main features of
the results reviewed here to be valid also in the presence of more
general GSB terms.

\subsection{Gauge-symmetry breaking
  in lattice Abelian Higgs models with noncompact gauge fields}

\begin{figure*}
\includegraphics*[width=0.50\columnwidth]{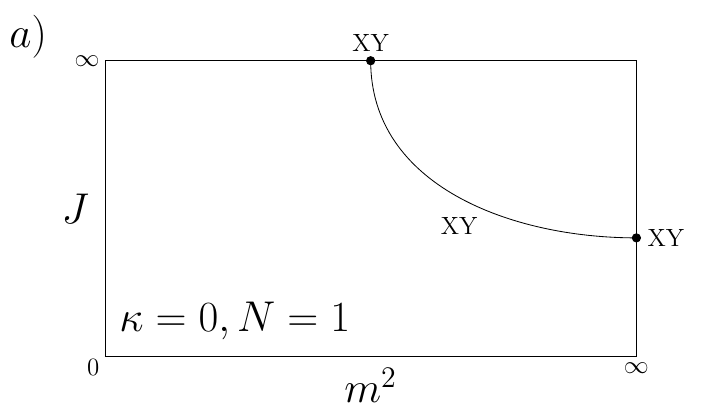}
\includegraphics*[width=0.50\columnwidth]{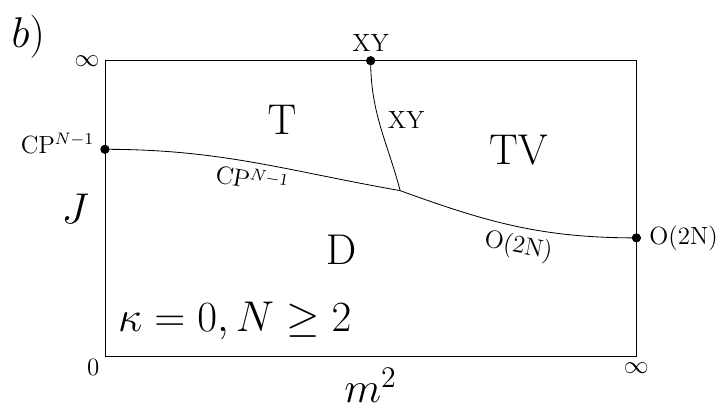}
\\ \includegraphics*[width=0.50\columnwidth]{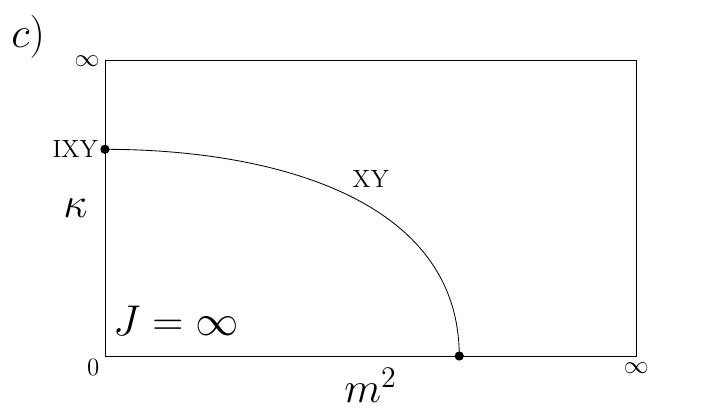}
\includegraphics*[width=0.50\columnwidth]{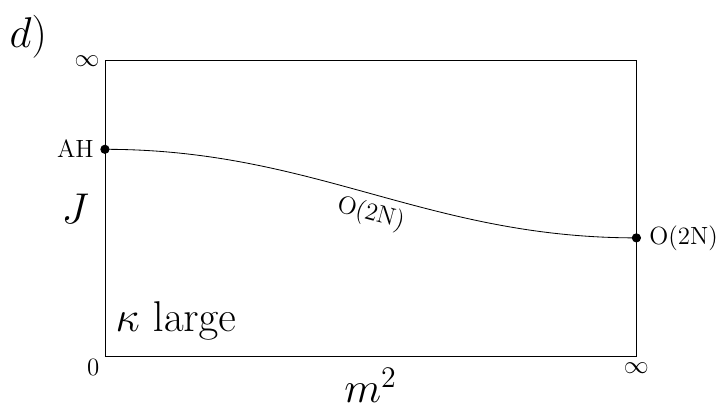}
\\ \includegraphics*[width=0.50\columnwidth]{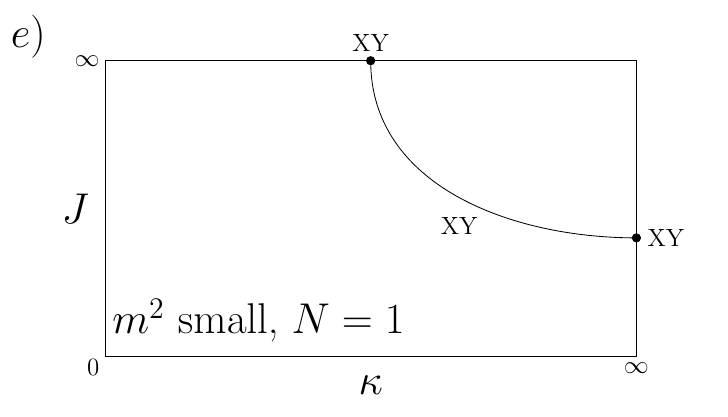}
\includegraphics*[width=0.50\columnwidth]{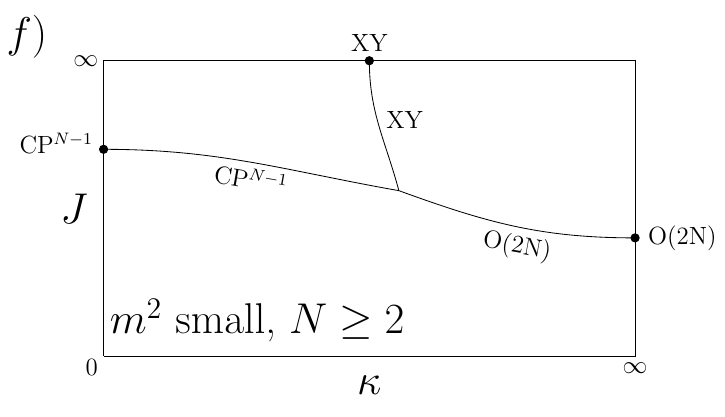}
\caption{Sketch of the phase diagrams for the noncompact Higgs model
  with GSB perturbation (\ref{GSB-massa-fotone}).  (Top panels) Phase
  diagram for $\kappa = 0$ (it also holds for the compact model, with
  $w$ replacing $m^2$): (a) $N = 1$ and (b) $N\ge 2$; scalar
  correlations are disordered in the D phase, tensor correlations are
  ordered in the T and TV phase, while vector correlations are ordered
  only in the TV phase. (Central panels) Phase diagram (c) for
  $J=\infty$ as a function of $\kappa$ and $ m^2$, and (d) for
  $\kappa$ large as a function of $J$ and $m^2$.  (Bottom panels)
  Phase diagrams for small $m^2$: (e) $N=1$, (f) $N\ge 2$.  Along each
  transition line we report the expected corresponding universality
  class.  
  }
\label{phd_ncah_gb_fig}
\end{figure*}

Let us consider the LAH model with noncompact gauge variables
introduced in Sec.~\ref{ncLAHMsec}, in the presence of a GSB
perturbation obtained by adding a photon-mass term
\begin{equation}
H_m = {m^2\over2} \sum_{{\bm x},\mu} A_{{\bm x},\mu}^2
\label{GSB-massa-fotone}
\end{equation}
to the Hamiltonian (\ref{ncLAHM}). To understand the phase diagram of
the combined model as a function of the parameters $J$, $\kappa$, and
$m^2$, it is convenient to first discuss some specific subcases.

For $\kappa = 0$, the global Hamiltonian can be written as
\begin{equation}
  H = - 2 N J\sum_{{\bm x},\mu} \hbox{Re} \left( e^{i A_{{\bm x},\mu}}\,
        \bar{\bm z}_{\bm x} \cdot {\bm z}_{{\bm x} +\hat{\mu}} \right)
  +  {m^2\over2} \sum_{{\bm x},\mu} A_{{\bm x},\mu}^2.
\end{equation}
For $m^2\to \infty$, one has $A_{{\bm x},\mu} = 0$, so one obtains
a simple scalar Hamiltonian with enlarged O$(2N)$ symmetry.  Indeed,
if we set $z_{\bm x}^a = \sigma^a_{\bm x} + i \sigma^{a+N}_{\bm x}$,
where $\sigma^a_{\bm x}$ is a real $2N$-component vector, we obtain
\begin{equation}
H = -J N  \sum_{{\bm x},\mu}  \sigma_{\bm x} \cdot \sigma_{{\bm x} + \hat\mu}.
\label{O2N-model}
\end{equation}
This is the $2N$-vector model defined in Eq.~(\ref{Nvectormod})
(rescaling the coupling $J$ by a factor $N$), which undergoes a
continuous transition at a finite value of $J$ in the O$(2N)$
universality class, see Sec.~\ref{sec:ON}.  Note that the presence of
an enlarged symmetry at the transition is a general property of
U($N$) invariant scalar models, as discussed in Sec.~\ref{sec:UN}.

For $\kappa=0$ and $J\to \infty$, the gauge fields are constrained to
be multiples of $2\pi$ modulo gauge transformations, see
Sec.~\ref{cboncLAHM}, i.e.,
\begin{equation}
 A_{{\bm x},\mu} = 2 \pi n_{{\bm x},\mu} + \nabla_\mu \phi_{\bm x},
\label{A-largeJ}
\end{equation}
where $\phi_{\bm x}$ is a real scalar field and $n_{{\bm x},\mu}$ is an
integer-valued link variable. The GSB Hamiltonian
(\ref{GSB-massa-fotone}) becomes
\begin{equation}
H_m = {m^2\over2} \sum_{{\bm x},\mu} 
    (\nabla_\mu \phi_{\bm x} + 2 \pi n_{{\bm x},\mu})^2,
\label{GSB-Villain}
\end{equation}
where $\phi_{\bm x}$ and $n_{{\bm x},\mu}$ play the role of independent
dynamic variables. This is the Villain formulation of the XY model, so
there must be  an XY transition at a finite value of $m^2$.

Taking into account the fact that for $\kappa=0$ and $m^2=0$ the LAH
model reduces itself to a lattice CP$^{N-1}$ model, see
Sec.~\ref{cpnmodels}, and therefore there is a corresponding
finite-temperature transition for $N \ge 2$, and no transition for
$N=1$, one arrives at the phase diagrams reported in the two upper
panels of Fig.~\ref{phd_ncah_gb_fig}. For $N=1$ [phase diagram (a)]
there is a single transition line, while for $N\ge 2$ [phase diagram
  (b)], there are three different transition lines intersecting in a single
point.  For small values of $m^2$ the GSB perturbation is not strong
enough to induce long-range order in the vector correlations of ${\bm
  z}_{\bm x}$, while the gauge invariant spin-2 tensor operator in
Eq.~\eqref{QdefncAH} is ordered, as for $m^2=0$.  We expect the
CP$^{N-1}$ behavior to extend along the D-T transition line, which
starts at $m^2 = 0$, as shown in Ref.~\cite{BPV-21-bgi-lattice} for a
general GSB perturbation of the compact model, and verified
numerically in Ref.~\cite{BPV-21-bgi2}. In particular, for $N=2$,
continuous transitions should belong to the O(3) universality
class. The irrelevance of the GSB perturbation 
for small $m^2$ is not unexpected,
since gauge modes do not play any active role at the CP$^{N-1}$
transition.  For large $m^2$ values, in the low-temperature phase
also vector correlations become long ranged.  A natural
hypothesis is that the critical behavior observed along the T-TV and
D-TV transition lines, for finite values of $J$ and $m^2$, is the same
as that observed for $J\to\infty$ and $m^2\to\infty$,
respectively. Thus, continuous transitions should belong to the XY and
O($2N$) universality classes, respectively. 
The phase diagrams reported in the two
upper panels of Fig.~\ref{phd_ncah_gb_fig} are supposed to hold also
for small values of $\kappa$.  On the other hand, for large values of
$\kappa$, we expect a different behavior.

Before discussing the large-$\kappa$
behavior, let us consider the limit $J\to \infty$ for generic $\kappa$
and $m^2$. Since in this limit the gauge fields take the form
(\ref{A-largeJ}), the effective Hamiltonian is equivalent to an IXY
model (discussed in Sec.~\ref{sec.IXY}) coupled with the Villain XY
model (\ref{GSB-Villain}). In panel (c) of Fig.~\ref{phd_ncah_gb_fig}
we show the expected phase diagram, showing a transition line
connecting the Villain XY model transition for $\kappa = 0$ with the
transition of the IXY model on the $m^2=0$ axis.  Continuous
transitions should belong to the XY vector universality class for
$m^2\not=0$.  While there is an XY transition if $m^2$ is varied for
fixed small $\kappa$, no transition occurs when $\kappa$ is fixed to a
large value.  Thus, for large $\kappa$ we expect a single transition
line, see panel (d) of Fig.~\ref{phd_ncah_gb_fig}, connecting the
$m^2=0$ transition to the O($2N$) transition occurring for
$m^2\to\infty$. It is natural to expect that continuous transitions
along the whole line belong to the O($2N$) universality class for $m^2
> 0$. In particular, the charged transitions that occur for $m^2 = 0$
and $N>N^*$ should be unstable with respect to the addition of the GSB
term.

We finally consider the expected phase diagrams for fixed small values
of $m^2$, see panels (e) and (f) of Fig.~\ref{phd_ncah_gb_fig}.  They
are qualitatively similar to the phase diagrams occurring for $m^2=0$,
see Fig.~\ref{phadiancLAH}.  However, the nature of the phases, and
therefore of the transitions, is different. The Coulomb phase is
replaced by a disordered phase, in which both scalar and gauge modes
are gapped, while in the low-temperature phase both scalar and gauge
fields are ordered. Note that, in the latter phase, because of the
absence of gauge invariance, both the gauge-invariant tensor
correlations and the non-gauge-invariant vector correlations are long
ranged for $N\ge 2$. In the molecular phase (which is present only for
$N\ge 2$), the addition of the GSB term changes the nature of the
gauge excitations---the Coulomb phase is replaced by a gapped
gauge-field phase---but not that of the scalar modes: tensor
correlations display long range order and vector correlations are
disordered, as in the gauge-invariant model. As for the transition
lines, the small-$\kappa$ transition line (which is only present for
$N\ge 2$) corresponds to CP$^{N-1}$ transitions as for $m^2=0$: the
GSB term is an irrelevant perturbation. On the other hand, the nature
of the other two transitions changes.  The topological IXY transitions
are now replaced by standard XY vector transitions, while the
continuous charged transitions that occur for large values of $N$
become O($2N$) vector transitions.

\subsection{Gauge-symmetry breaking in lattice Abelian Higgs
  models with compact gauge fields}

The effect of adding a GSB perturbation to the unit-charge ($Q=1$) LAH model
with compact gauge variables, see Sec.~\ref{coLAJHM}, is discussed in
Refs.~\cite{BPV-21-bgi2,BPV-21-bgi-lattice}.  We consider a GSB term
that is analogous to the photon-mass term in the small gauge limit,
i.e.,
\begin{equation}
    H_m = -w \sum_{{\bm x},\mu} \hbox{Re } \lambda_{{\bm x},\mu}.
\label{GSB-co}
\end{equation}
Since for $Q=1$ the plaquette term proportional to $\kappa$,
cf. Eq.~(\ref{clahmH}), is irrelevant---the critical behavior is the
same for any finite $\kappa$, see Sec.~\ref{q1phdia}---it is enough to
discuss the behavior for $\kappa = 0$, considering the simplified
lattice model with Hamiltonian
\begin{eqnarray}
H = - 2 N J  \sum_{{\bm x}, \mu} {\rm Re}\,(
\lambda_{{\bm x},\mu}\,\bar{\bm{z}}_{\bm x} \cdot
       {\bm z}_{{\bm x}+\hat\mu}) -w \sum_{{\bm x},\mu}
       \hbox{Re } \lambda_{{\bm x},\mu}.
\label{clahmHgsb}
\end{eqnarray}
The behavior of this combined model is analogous to the one presented
in the noncompact case. Indeed, for $w \to \infty$, we can set
$\lambda_{{\bm x}\mu} = 1$, and thus we obtain again the O($2N$)
vector model (\ref{O2N-model}). On the other hand, for $J\to \infty$,
$\lambda_{{\bm x},\mu}=1$ modulo gauge transformations, i.e.,
$\lambda_{{\bm x},\mu} = e^{i (\phi_{\bm x} - \phi_{{\bm
      x}+\hat{\mu}})}$. Substituting this expression 
in the GSB term (\ref{GSB-co}),
we obtain the effective Hamiltonian
\begin{equation}
    H_m = -w \sum_{{\bm x},\mu} 
    \cos(\phi_{\bm x} - \phi_{{\bm x} + \hat\mu}),
\end{equation}
which is the standard XY Hamiltonian. Thus, also in this case we
obtain the phase diagrams reported in the upper panels of
Fig.~\ref{phd_ncah_gb_fig}, with $m^2$ replaced by $w$.

We now present an argument---it rephrases a similar argument of
Ref.~\cite{FS-79}---that shows that the phase behavior discussed above
generally holds for
generic GSB perturbations at transitions where gauge fields are not
critical. Indeed, let us consider the partition function of a generic
model with a GSB perturbation that only depends on the gauge fields:
\begin{equation}
  Z = \int [dU_{{\bm x},\mu}]\,[d\Phi_{\bm x}]\,
  \exp[-\beta H(U_{{\bm x},\mu},\Phi_{\bm x})],
  \qquad H(U_{{\bm x},\mu},\Phi_{\bm x}) =
  H_{\rm GI}(U_{{\bm x},\mu},\Phi_{\bm x}) + H_{\rm GSB}(U_{{\bm x},\mu}),
\end{equation}
where $H_{\rm GI}$ is the gauge-invariant Hamiltonian, and $H_{\rm GSB}$ is
a generic Hamiltonian term breaking the gauge symmetry, depending only
on the link gauge variables $U_{{\bm x},\mu}$.  We now perform a
change of variables on the scalar and gauge fields that corresponds to
a gauge transformation---thus $Z$, or any expectation value of
gauge-invariant operators, does not change.  In particular, we
redefine $U_{{\bm x},\mu} \to V_{{\bm x}} U_{{\bm x},\mu}
V^\dagger_{{\bm x} + \mu}$.  As $H_{\rm GI}$ is gauge invariant, the
partition function becomes
\begin{equation}
  Z = \int [dU_{{\bm x},\mu}][d\Phi_{\bm x}]\,
  \exp\left[-\beta H_{\rm GI}(U_{{\bm x},\mu},\Phi_{\bm x}) -
    \beta H_{\rm GSB}(V_{{\bm x}} U_{{\bm x},\mu} V^\dagger_{{\bm x} + \mu})
 \right].
\end{equation}
The partition function does not depend on the set of variables
$V_{{\bm x}}$ and thus we can integrate over them without changing the
partition function.  We define a new Hamiltonian $\widehat{H}$ so that
\begin{equation}
  \widehat{H} = H_{\rm GI} + \widehat{H}_{\rm GSB},
\qquad
  \widehat{H}_{\rm GSB}(U_{{\bm x},\mu}) =  - \ln
   \int [dV_{\bm x}]\,  \exp[-
     \beta H_{\rm GSB}(V_{{\bm x}} U_{{\bm x},\mu} V^\dagger_{{\bm x} + \mu})].
\end{equation}
The new Hamiltonian $\widehat{H}$ is clearly gauge invariant, and
equivalent to the original one, if we consider the partition function
and, more generally, any gauge-invariant correlation function.  The
Hamiltonian $\widehat{H}$ contains interactions between the gauge
variables $U_{{\bm x},\mu}$ and $U_{{\bm y},\nu}$ at any distance
$|{\bm x}-{\bm y}|$. However, if these interactions are exponentially
suppressed for $|{\bm x}-{\bm y}| \to \infty$, then $\widehat{H}$ represents a
gauge-invariant model with effective short-range interactions. This is
the general scenario that we expect to emerge when the gauge fields are not
critical.  This can be verified by performing a strong-coupling
expansion assuming $\beta w$ small. For example, in the case of the
GSB perturbation (\ref{GSB-co}), $\widehat{H}$ can be written as a sum
of lattice loops. In the strong-coupling expansion, a lattice loop of
length $L$ is weighted by a factor that behaves as $(\beta w)^L$ for
$\beta w \to 0$.  For instance, the leading term is the plaquette,
with a weight of order $(\beta w)^4$, which renormalizes the value of
$\gamma$.  The next term corresponds to the $2\times 1$ plaquette,
with a coefficient proportional to $(\beta w)^6$, and so on.
Couplings therefore scale as $\exp(-a |{\bm x}-{\bm y}|)$ with $a
\sim -\ln(\beta w)$, showing the short-range nature of the
interactions.

This argument proves that, for sufficiently small, but finite, values
of the GSB parameter $w$, the partition function and the gauge-invariant
correlation functions  can be computed in an equivalent gauge-invariant theory,
without GSB terms, with short-range interactions. Finally, to conclude
the argument, let us note that we are considering a model in which
gauge fields do not play any role, i.e., the critical behavior is
independent of the gauge-field interactions, and therefore it is the same as in
the original model with $w = 0$. We conclude that the
phase structure is independent of $w$ for sufficiently small values.
Note that the argument does not rely on the Abelian nature of the
theory and thus is should also hold in non-Abelian gauge models.

We finally mention that the same arguments also apply to a different
model without gauge fields. Indeed, for $\kappa = 0$, one can
integrate out the link variables $\lambda$, obtaining an effective
Hamiltonian
\begin{equation}
H_{{\rm eff}} = -\sum_{{\bm x},\mu}
\ln I_0\left(2 J N |\widehat{w} + \bar{\bm z}_{\bm x} \cdot
{\bm z}_{{\bm x}+\hat\mu}|\right),
\label{hlaeff2}
\end{equation}
with $\widehat{w}=w/(2JN)$. As it occurs in the CP$^{N-1}$ model, we
expect the Hamiltonian $H_{{\rm eff}}$ to have the same critical
behavior as the Hamiltonian obtained by replacing the Bessel function
$I_0(x)$ with its small-$x$ expansion.  Since
\begin{equation}
|\widehat{w} + \bar{\bm z}_{\bm x} \cdot {\bm z}_{{\bm x}+\hat\mu}|^2
 = \widehat{w}^2  +
2 \widehat{w}\, \hbox{Re}\, (\bar{\bm z}_{\bm x} \cdot {\bm z}_{{\bm x}+\hat\mu}) +
|\bar{\bm z}_{\bm x} \cdot {\bm z}_{{\bm x}+\hat\mu}|^2,
\end{equation}
an equivalent compact model is 
\begin{equation}
 {H} =
- N^2 \widetilde{J} \sum_{{\bm x},\mu}
|\bar{\bm z}_{\bm x} \cdot {\bm z}_{{\bm x}+\hat\mu}|^2 +
\widetilde{w} \sum_{{\bm x},\mu} 
\hbox {Re }\bar{\bm z}_{\bm x} \cdot {\bm z}_{{\bm x}+\hat\mu} .
\label{hlaeffexpmix}
\end{equation}
The term proportional to $\widetilde{J}$ is the standard formulation of
the U($N$)-invariant CP$^{N-1}$ model without gauge fields, see
Eq.~(\ref{hcpn}), while the second term is the O($2N$) invariant
vector model, which represents here the GSB perturbation. The phase
diagram of this model is similar to the one presented in panel (b) of
Fig.~\ref{phd_ncah_gb_fig}, with three different phases characterized
by the different behavior of the scalar fields.  The only qualitative
difference is the shape of the D-TV transition line that ends on the
$J=0$ axis at the critical point $w_c$ of the O($2N$) vector model.

%% file: concluding.tex
\section{Conclusions  and outlook}
\label{conclu}

We have reviewed the statistical properties of lattice Abelian and
non-Abelian Higgs theories, in which multicomponent scalar fields
are coupled with Abelian and non-Abelian gauge fields, mostly in three
dimensions.  As remarked several times, the interplay between local
and global symmetries determines the nature of the Higgs phases
and of the phase transitions of these lattice gauge systems. Their
nature also crucially depends on the role played by gauge modes at
criticality, whether they are critical at the transition or not. In
the latter case, their role is that of selecting the critical scalar
modes, so that only critical correlations associated with
gauge-invariant scalar modes can be observed.  The topological
properties of the gauge modes can also give rise to topological
transitions between Higgs phases with the same scalar global
symmetry. These transitions are related to the breaking pattern of the
gauge symmetry in the Higgs phase. We discuss examples with a residual
${\mathbb Z}_Q$ gauge symmetry, leading to topological ${\mathbb Z}_Q$
gauge transitions when varying the gauge coupling.

The results reviewed here show that the possible critical behaviors
observed in 3D lattice gauge Higgs systems are quite varied, showing
also features that cannot be explained by the standard LGW paradigm.
The continuous transitions of the statistical Abelian and non-Abelian
gauge systems considered in this review can be classified as follows
(see Sec.~\ref{difftype}): (i) LGW transitions with gauge-invariant
order parameter and noncritical gauge modes; (ii) LGW$^{\times}$
transitions, where gauge modes are also not critical, but where the
effective order-parameter field is gauge dependent; (iii) GFT
transitions, where gauge modes are critical and that require an
effective description that includes the gauge fields; (iv) topological
transitions driven only by topological gauge modes, without any local
gauge-invariant scalar order parameter.

We have focused on the equilibrium thermodynamic behavior of classical
statistical systems, whose partition function is defined as a
path-integral functional, by integrating over continuum or lattice
configurations. However, the results for classical statistical systems
in ($d+1$) dimensions also apply to quantum transitions in $d$
dimensions, using the quantum-to-classical mapping.  Indeed, they are
relevant for the $d$-dimensional quantum transitions that can be
related with classical thermal transitions in $(d+1)$-dimensional
isotropic systems.  In particular, this requires that the critical
exponent $z$, associated with the vanishing gap $\Delta \sim \xi^{-z}$
at the quantum critical point, satisfies $z=1$. However, further
interesting developments may come from a thorough analysis of quantum
transitions in the presence of emerging gauge symmetries with $z\neq
1$~\cite{Sachdev-book,RV-21}. This issue definitely calls for further
studies.

Another interesting issue is the critical dynamics of statistical
systems with gauge symmetries. On one side, one would like to
understand if the presence of gauge symmetries gives rise to
distinctive new features for the different types of dynamics that are
usually considered in standard statistical
systems~\cite{HH-77,Ma-book,FM-06}.  A systematic study of the
critical dynamics in the presence of gauge symmetries may provide
further interesting characterizations of the different types of
transitions that we have outlined in Sec.~\ref{difftype}. On the other
side, it would be interesting to understand whether it is possible to
define dynamics that are not considered in the standard classification
of systems without gauge symmetries. As an example, one may consider
the purely relaxational dynamics with only dissipative couplings and
no conservation laws, usually called dynamic model
A~\cite{HH-77}. This dynamics can be realized by considering a
relaxational stochastic Langevin dynamics, which is also at the basis
of the stochastic quantization of gauge
theories~\cite{Zwanziger-81,FI-82,NNOO-83,ZJ-86,ZZ-88,Parisi-book}.
It would be interesting to understand how the expected critical
slowing down, signalled by the critical divergence of the relaxation
time $\tau$ of the critical modes (at criticality, $\tau\sim \xi^z$,
where $z$ is a dynamic critical exponent), depends on type of
transition, and how the nonperturbative results compare with the
field-theoretical predictions that are typically obtained in the
presence of a gauge fixing.  Some studies of the critical dynamics
have been already reported, see, e.g.,
Refs.~\cite{BKKLS-90,LMG-91,RW-93,WJ-97,LWWGY-98,AG-01,SBZ-02,
  LVF-04,DFM-07,XCMCS-18, BPV-25-crit, BPV-25-out}, most of them
addressing the dynamics at superconducting phase transitions, and 
the topological transition in 
the 3D lattice ${\mathbb Z}_2$ gauge model. However, a
systematic analysis of dynamic phenomena is still missing.

From a methodological point of view, most of the results presented
here have been obtained by using numerical approaches. We have presented the
results of the analyses of numerical data obtained in MC simulations of lattice
systems, field-theoretical perturbative determinations (in the
$\varepsilon$-expansion framework) of the RG flow, and nonperturbative
predictions obtained in the limit of a large number of scalar components.
However, other methods may also provide interesting insights in the future. We
mention, e.g., the tensor renormalization-group method (see, e.g.,
Ref.~\cite{MSU-22} and references therein), and the bootstrap
method~\cite{PRV-19-rev,RS-24}, which has been mostly applied to systems
without gauge symmetries, see results reported in the tables of \ref{univclass}
(some results for QED$_3$ with fermions are reviewed in Ref.~\cite{RS-24}).
Another interesting possibility consists in the use of optical lattices and
trapped ions to simulate lattice Abelian and non-Abelian gauge theories, see,
e.g., Refs.~\cite{BY-06,ZR-11,ZCR-12,BBDRSWZ-13,ZCR-13} and the reviews,
Refs.~\cite{Banuls-etal-20,BC-20}.

We finally mention that new scenarios may arise from the inclusion of
fermionic fields in LAH and LNAH systems. This topic is relevant in
condensed-matter physics as many field-theoretical models with
emerging gauge symmetries include both fermionic and scalar
excitations, see, e.g.,
Refs.~\cite{SSST-19,Sachdev-19,BK-19,BS-19,BLS-20,STS-21}.  The study
of the corresponding lattice gauge systems may extend the
phenomenology of phase transitions in the presence of gauge
symmetries. As for LAH and LNAH systems with only scalar matter, one
key issue would be that of identifying the 3D transitions where both
matter and gauge fields are critical, allowing one to define the
continuum limit of the corresponding SFTs. The same issue is also
important in lattice gauge models with fermionic matter only.  Some
studies of the phase diagram and critical behavior of 3D models
including fermionic fields can be found in
Refs.~\cite{ANW-88,DK-90,SS-91,Gracey-93-qedgn,Gracey-93-qedgn2,
  Gracey-94-qed,VZS-00,HKSS-04,HSF-05,SK-08,NK-08,KS-08,NGPNG-09,
  AKS-10,AHKLS-11,BGJR-14,CP-16,JH-17,AB-18,BRM-19,WDXYM-19,ZBMGM-19,
  XQZAXM-19,JWSMX-20,ZBMGM-20, LP-21, AELPX-22, HRSV-24}.

\section*{Acknowledgements}

We thank Alessio Franchi, Niccol\`o Francini, Ivan Soler Calero, Giacomo Bracci
Testasecca for collaborating with us on some of the topics considered in this
review, and Alessandro Vichi for useful discussions.

%% file: appa.tex
\section{Renormalization-group theory of critical phenomena}
\label{rgtheory}

In this appendix we report an overview of the RG theory, which 
provides a general framework to explain the observed scaling
behavior at continuous phase transitions~\cite{Fisher-71, WK-74,
  Fisher-74, Ma-book, BGZ-76, Wegner-76, Aharony-76, Wilson-83,
  Abraham-86, CJ-86, Cardy-editor, Privman-90, PHA-91, Fisher-98,
  ZJ-book, ID-book1, ID-book2, Cardy-book, AM-book, PV-02}.
We present the RG scaling relations in the thermodynamic
(infinite-volume) limit and in the FSS limit and, in particular, we
discuss the leading scaling behavior and the scaling corrections
characterizing the deviations from the asymptotic relations.  We only
consider classical transitions, generally driven by thermal
fluctuations; the extension to quantum transitions driven by quantum
fluctuations is discussed in
Refs.~\cite{SGCS-97,Sachdev-book,CPV-14,RV-21}.  FSS behaviors also
emerge at first-order transitions~\cite{Binder-87,PV-24}.  These
phenomena for classical and quantum transitions are discussed in
Refs.~\cite{PV-24, BK-90, Binder-87, CNPV-14}.  In the following we do
not discuss the critical dynamics at thermal and quantum
transitions. Reviews can be found in Refs.~\cite{HH-77,FM-06,RV-21}.

\subsection{Universal power laws at the critical point}
\label{univcritexp}

To fix the ideas, we consider a prototypical $d$-dimensional
statistical system undergoing a continuous transition. The behavior of
the system is controlled by two relevant Hamiltonian parameters $r$
and $h$, which can be defined so as to vanish at the critical point.
Thus, the critical point is located at $r = h = 0$.  We also assume
the presence of a ${\mathbb Z}_2$-symmetry, as it occurs, e.g., in
Ising transitions, which is unbroken in the phase with $r
> 0$ (paramagnetic phase in magnetic systems) 
and spontaneously broken in the (ferromagnetic) phase with $r < 0$.
The parameter $r$ is associated with a RG perturbation that is
invariant under the global symmetry and is usually identified with the
reduced temperature $r\sim T/T_c-1$. The parameter $h$ is assumed to
be odd with respect to the ${\mathbb Z}_2$-symmetry, thus it can be
identified with an external homogeneous field coupled with the order
parameter driving the symmetry breaking.

In the thermodynamic limit, when approaching the critical point from
the disordered phase keeping $h=0$, the length scale $\xi$ of the
critical modes diverges as $\xi\sim r^{-\nu}$, where $\nu$ is a
universal length-scale critical exponent. Another universal critical
exponent $\eta$ is traditionally introduced to characterize the space
dependence at criticality of the two-point function of the
order parameter: $G(\bm{x}_1,\bm{x}_2) \sim |\bm{x_1} -
\bm{x_2}|^{-(d-2+\eta)}$ for $r=h=0$.  The RG dimensions of the
perturbations associated with $r$ and $h$ are related with the critical
exponents $\nu$ and $\eta$, as~\cite{Wegner-76,PV-02}
\begin{equation}
  y_r = 1/\nu, \qquad y_h = {d+2-\eta\over 2}.
  \label{ymuh}
\end{equation}
Using scaling and hyperscaling arguments (see, e.g.,
Refs.~\cite{Ma-book, ZJ-book, PV-02}), the exponents associated with
the scaling behavior of other observables, like the magnetization or
the critical equation of state, can be expressed in terms of the two
independent exponents $\nu$ and $\eta$. The corrections to the
asymptotic power laws behave as $\xi^{-\omega}$ for $\xi\to\infty$,
where $\omega>0$ is another universal exponent characterizing the
critical behavior~\cite{ZJ-book, PV-02}, which is generally associated
with the leading irrelevant RG perturbation.

\subsection{Scaling laws in the thermodynamic limit}
\label{scatherbeh}

We consider 
the free-energy density of a statistical system,
\begin{equation}
  F = - {T\over V} \ln Z,\qquad Z = \sum_{\{\varphi\}} e^{-\beta H},
  \qquad \beta=1/T,\qquad V=L^d,
\label{freenZ}
\end{equation}
where the summation is over all configurations $\{\varphi\}$ of the
system. According to the RG theory, $F$ obeys a general scaling law.
Indeed, we can write $F$ in terms of the nonlinear scaling fields
associated with the RG eigenoperators at the FP of the RG
flow~\cite{Wegner-76, PV-02}. Therefore, close to a continuous
transition, the free-energy density in the 
infinite-volume limit can be written in terms of scaling
fields~\cite{Wegner-76}, as
\begin{equation}
  F(r,h) = F_{\rm reg}(r,h^2) + F_{\rm sing}(u_r,u_h,\{v_i\}).
  \label{Gsing-RG-1}
\end{equation}
Here $F_{\rm reg}$ is a nonuniversal function, which is analytic at
the critical point and must be even with respect to the parameter $h$
related with the odd perturbation (for a general symmetry group it
should depend on a combination of $h$ that is invariant under the
symmetry transformations). The singular contribution, $F_{\rm sing}$,
bears the nonanalyticity of the critical behavior and its universal
features. The arguments of $F_{\rm sing}$ are the so-called nonlinear
scaling fields~\cite{Wegner-76}: they are analytic nonlinear functions
of the model parameters, related with the eigenoperators that
diagonalize the RG flow close to the FP, and transform
multiplicatively under the RG flow.  We have singled out the relevant
scaling fields $u_r$ and $u_h$ that are associated with the model
parameters $r$ and $h$, and have positive RG dimensions $y_r$ amd
$y_h$.  The scaling fields $\{v_i\}$ (there is an infinite number of
them) are RG irrelevant, i.e., their RG dimensions $y_i$ are negative.
They give rise to scaling corrections to the asymptotic critical
behavior in the infinite-volume limit.  Assuming that they are ordered
so that $|y_1| \le |y_2|,\le \ldots$, the RG dimension $y_1<0$
generally determines the leading scaling corrections, thus we identify
$\omega=-y_1$.

The nonlinear scaling fields depend on the control
parameters $r$ and $h$. Taking into account the assumed ${\mathbb
  Z}_2$ symmetry and the respectively even and odd properties of $r$
and $h$, close to the critical point the relevant scaling fields $u_r$
and $u_h$ can be generally expanded as
\begin{equation}
u_r = r + c_r r^2 +
O(r^3,h^2r),\qquad u_h = h + c_h r h + O(h^3,r^2 h),
\label{uruh}
\end{equation}
where $c_r$ and $c_h$ are nonuniversal constants.  As for the
irrelevant scaling fields $v_i$, they are generally nonvanishing at
the critical point.

The singular part of the free energy~\eqref{Gsing-RG-1} is expected to
satisfy the homogeneous scaling law
\begin{equation}
  F_{\rm sing} \big(u_r,u_h,\{v_i\} \big) = b^{-d} \,
  F_{\rm sing}(b^{y_r} u_r , b^{y_h} u_h,\{b^{y_i} v_i\}),
  \label{Fsing-scaling}
\end{equation}
where $b$ is an arbitrary scale factor
and $y_r$ and $y_h$ are given in
Eq.~(\ref{ymuh}). To obtain more explicit scaling relations, one can
fix the arbitrariness of the scale parameter $b$, setting $b
= u_r^{-1/y_r}=u_r^{-\nu}$, which diverges in the critical limit
$r,h\to 0$.  The following asymptotic behavior of the free-energy
density emerges:\footnote{Note that the expansion~\eqref{scalsingiv}
is only possible below the upper critical dimension~\cite{Fisher-74}.
Above it, the failure of this expansion causes a breakdown of the
hyperscaling relations, allowing one to recover the mean-field
exponents.}
\begin{equation}
  F_{\rm sing} = u_r^{d\nu} \, {\cal
    F}(u_h/u_r^{y_h\nu},\{{v_i/u_r^{y_i\nu}}\}) = u_r^{d\nu} \,\Bigl[
    {\cal F}_0(u_h/u_r^{y_h\nu}) + u_r^{\omega\nu} {\cal
      F}_\omega(u_h/u_r^{y_h\nu}) + \cdots \Bigr],
\label{scalsingiv}
\end{equation}
where ${\cal F}_0$ and ${\cal F}_\omega$ are universal scaling
functions, and we kept only the leading correction to scaling, which
vanishes as $u_r^{\omega\nu}$ with
$\omega=-y_1$ approaching the critical point. 
Note that, since the scaling fields have
arbitrary normalization, the universality of the scaling functions
${\cal F}_0$ and ${\cal F}_\omega$ holds apart from a normalization of
each argument and an overall constant. By differentiating the
free-energy density, one can straightforwardly derive analogous
scaling relations for other observables, such as the energy density
and the magnetization.  To eventually obtain the scaling relations in
terms of the external parameters $r$ and $h$ controlling the approach
to the critical point, one needs to expand the scaling fields as in
Eq.~(\ref{uruh}). The subleading terms in these expansions give rise
to analytic scaling corrections.

Analogous scaling relations can be obtained for the correlation
functions of local operators. For example, the two-point
correlation function of a local operator $O({\bm x})$, whose
RG dimension is $y_o$, is expected to asymptotically behave as
\begin{equation}
  G_O({\bm x},{\bm y}) = \langle O({\bm x}) O({\bm y}) \rangle 
  \approx \xi^{-2y_o} {\cal G}_O({\bm x}/\xi, {\bm y}/\xi,
    h \,\xi^{y_h}),\qquad \xi\sim r^{-\nu},
  \label{GOsca}
\end{equation}
when approaching the critical point from the disordered phase
  $r>0$.

\subsection{Finite-size scaling}
\label{fssapp}

The RG scaling relations reported in \ref{scatherbeh} hold in
the thermodynamic limit, which is well defined for any
$r\neq 0$ or $h\neq 0$, since the correlation length $\xi$ is
finite.  Nonetheless, for practical purposes, both experimentally and
numerically, one typically has to face with systems of finite
size $L$. Also in this case, for large values of $L$, 
 it is possible to observe a universal scaling behavior.

Finite-size effects in critical phenomena have been the object of many
theoretical studies~\cite{FB-72, Barber-83, Privman-90, PHA-91,
  Cardy-editor, PV-02}. The FSS theory describes the critical behavior
around a critical point, when the correlation length $\xi$ of the
critical modes becomes comparable with the size $L$ of the system.
The FSS approach is one of the most effective techniques to determine
the universal features of continuous phase transitions, such as 
the critical exponents.  While infinite-volume methods require the
condition $1 \ll \xi \ll L$ to be satisfied, FSS methods apply in the less
demanding regime $\xi\sim L\gg 1$.  The FSS theory provides the asymptotic
scaling behavior when $L,\xi\to\infty$, keeping their ratio
$\xi/L$ fixed. This regime presents universal features, shared by all
systems whose transition belongs to the same universality class.

\subsubsection{Free-energy density}
\label{fssfreen}

In the FSS framework, the finite size of the system is taken into
account by adding the length scale $L$ in the scaling
laws~\cite{Wegner-76, Diehl-86, Privman-90, PHA-91, SGCS-97, SS-00,
  PV-02, CPV-14,RV-21}. In the following we focus for concreteness on
finite-size systems without boundaries, such as systems with periodic
or antiperiodic boundary conditions. In these cases the scaling laws,
and in particular the corrections to the asymptotic scaling behavior,
are simpler.  The corrections to the bulk scaling laws due to the
presence of boundaries are discussed in, e.g.,
Refs.~\cite{Privman-90,PHA-91,SS-00,CPV-14,RV-21}.  In the FSS limit
the free-energy density can be written as
\begin{equation}
  F(L,r,h) = F_{\rm reg}(r,h^2) + F_{\rm sing}\big(L,u_r,u_h,\{v_i\}),
  \label{Gsing-RG-1fss}
\end{equation}
where $F_{\rm reg}$ is again a nonuniversal analytic function, which
is assumed to be independent of $L$, or, more plausibly, to depend on
$L$ only through exponentially small
corrections~\cite{Privman-90,PHA-91,SS-00,CPV-14}.  According to the
RG theory, the singular part $F_{\rm
  sing}$ satisfies  the homogeneous scaling law
\begin{equation}
  F_{\rm sing}(L,u_r,u_h,\{v_i\}) = b^{-d} \,
  F_{\rm sing}(L/b, u_r\,b^{y_r}, u_h\,b^{y_h},
  \{v_i\,b^{y_i}\}),
  \label{Fsing-scaling-fss}
\end{equation}
where $b$ is an arbitrary constant. To derive the FSS behavior, we fix
$b = L$, thus obtaining
\begin{equation}
  F_{\rm sing} =  
  L^{-d} {\cal F}\big(L^{y_r} u_r, L^{y_h}u_h, 
  \{L^{y_i} v_i\} \big).
  \label{scalsing}
\end{equation}
The arguments $\{ L^{y_i} v_i \}$, corresponding to the irrelevant
scaling fields with $y_i<0$, vanish for $L\to\infty$.  Thus, provided
that $F_{\rm sing}$ is finite and nonvanishing in this limit, and
neglecting the analytic corrections arising from the higher-order
terms of the polynomial expansion of the scaling fields $u_r$ and
$u_h$, see Eq.~(\ref{uruh}), we can expand the singular part of the
free energy as
\begin{equation} 
  F_{\rm sing} \approx
  L^{-d} \, \left[ {\cal F}_0(W,Z)
  +v_1 L^{-\omega} \,
  {\cal F}_\omega(W,Z)
  + \ldots\right],\qquad W= L^{y_r}r, \quad Z=L^{y_h}h,
  \label{Fsing-RG}
\end{equation}
where we only retain the contribution of the dominant (least)
irrelevant scaling field of RG dimension $y_1=-\omega$.  The scaling
behavior in the thermodynamic limit can be recovered from the FSS
homogenous law (\ref{Fsing-scaling-fss}), by choosing $b=u_r^{-\nu}$
and taking the limit $L/b \to \infty$.  The scaling laws of several
interesting quantities can be obtained by taking derivatives of the
free energy with respect of $r$ and $h$, such as the energy density,
the specific heat, the magnetization, the magnetic susceptibility,
etc...

The above scaling laws hold in generic systems.  However, in some
cases the scaling behavior is more complex, due to the appearance of
logarithmic terms~\cite{Wegner-76}. They may be induced by the
presence of marginal RG perturbations, as it happens for the 
Berezinskii-Kosterlitz-Thouless (BKT)
transitions that occur in 2D U(1)-symmetric systems~\cite{Berezinskii-72, KT-73,
  AGG-80, Hasenbusch-05, PV-13-BKT}, or by resonances between the RG
eigenvalues, as it occurs in 2D Ising transitions~\cite{Wegner-76,
  CHPV-02}, or in 3D O$(N)$-vector models in the large-$N$
limit~\cite{Diehl-etal-12, PV-98, PV-99}.

The FSS limit, which corresponds to taking $r\to 0$, thus
$\xi\to\infty$, and $L\to\infty$ at fixed $\xi/L$, definitely differs
from the critical limit in the thermodynamic limit, which is generally
obtained by first taking the large-$L$ limit keeping $r$ or $\xi$
fixed, and then the critical limit $r\to 0$ or, equivalently
$\xi\to\infty$.  However, assuming that the FSS and infinite-volume
behavior asymptotically match, one may straightforwardly derive the
infinite-volume scaling behavior from the FSS relations, by taking the
limit $\xi/L\to 0$.

We finally remark that, in numerical and experimental investigations
of finite-size systems, the knowledge of the leading asymptotic
behavior may not be enough to estimate the critical parameters,
because data are generally available only for limited ranges of parameter
values and system sizes, which are often relatively small.  Under
these circumstances, the asymptotic FSS predictions may be affected by
sizable scaling corrections. Accurate estimates of the critical
parameters thus require a robust control of the corrections to the
asymptotic scaling behavior. This is also important for a conclusive
identification of the universality class of a continuous transition,
when there are no solid theoretical arguments to predict it.
Moreover, an understanding of the finite-size effects is relevant for
experiments, when relatively small systems are considered (see,
e.g., Ref.~\cite{GKMD-08}), or in particle systems trapped by external
(usually harmonic) forces, as in cold-atom setups (see, e.g.,
Refs.~\cite{CW-02, Ketterle-02, Donner-etal-07, BDZ-08, CV-09,
  CV-10-tr}).

\subsubsection{Energy cumulants}
\label{fssencum}

As discussed in this review, statistical models with gauge symmetries
also undergo topological transitions characterized by the absence of a
local order parameter, such as those belonging to the 3D ${\mathbb
  Z}_Q$ gauge and IXY universality classes (see Sec.~\ref{toptra}),
which are also found in LAH and LNAH models (see Secs.~\ref{LAHMnc}
and \ref{LAHMc}). The numerical study of the critical properties at
these topological transitions can be performed by analyzing the
scaling properties of the energy cumulants, which are
always gauge invariant and well defined, see, e.g.,
Refs.~\cite{SSNHS-03,BPV-20-hcAH,BPV-22-z2g,BPV-24-cAH}.

The gauge-invariant energy cumulants $C_k$ are intensive quantities
that can be defined by taking inverse-temperature derivatives of the
free-energy density, more precisely
\begin{equation}
C_k=\frac{1}{V}\left(\frac{\partial}{\partial\beta}\right)^k
\log Z(\beta),
\label{Ckdef}
\end{equation}
where $Z$ is the partition function defined in
Eq.~(\ref{freenZ}).\footnote{The cumulants $C_k$ can be related to
the energy central moments $M_k = \langle \, (H-\langle H\rangle)^k \,
\rangle$, by $C_1=-\langle H\rangle/V$, $C_2=M_2/V$, $C_3 = -L^{-3}
M_3$, $C_4 = (M_4 - 3 M_2^2)/V$, etc.} Note that $C_1=-E$ where
$E=\langle H \rangle/V$ is the energy density, and $C_2$ is
proportional to the specific heat.  

At continuous transitions, the cumulant $C_k$ is expected to
show the FSS behavior (we assume $h=0$ in the following)
\begin{eqnarray} 
C_k(r,L) \approx  L^{k/\nu-d} 
   \Bigl[ {\cal U}_k(W) + O(L^{-\omega)})\Bigr]  
+  B_k(r),\qquad W = r L^{y_r},
\label{cksca}
\end{eqnarray}
where $B_k(r)$ is a regular function---the so-called analytic
background~\cite{PV-02,BPV-20-hcAH}.  Note that the relation $C_{k+1}
= \partial C_k/\partial\beta$ implies ${\cal U}_{k+1}(W) \sim
- \partial_W {\cal U}_k(W)$. Further properties of the scaling functions
${\cal U}_{k+1}$ are derived in Ref.~\cite{BPV-24-cAH}.

As one can easily check, the first cumulant $C_1=-E$ is dominated by
the analytic background contribution $B_1$, due to the relation
$\nu>1/d>0$, which is generally satisfied at continuous
transitions~\cite{PV-02}. This makes the scaling behavior of the
energy density subleading, and therefore rather difficult to observe
(this issue has been addressed in Refs.~\cite{PV-24-e,RV-24}). Also
the second cumulant $C_2$, which is proportional to the specific heat,
is dominated by the analytic background when the specific-heat
exponent $\alpha = 2\nu - d$ is negative.  The singular scaling part
is always the dominant term in the higher energy cumulants $C_k$,
i.e., for $k\ge 3$. Taking into account that their numerical
computation becomes harder and harder with increasing $k$, due to
significant cancellations which make their relative error quite large
and increasing with $k$, optimal results are obtained by focusing on
$C_3$ and $C_4$.

\subsubsection{The order-parameter correlation function}
\label{fsscorrfunc}

For vanishing magnetic field, $h=0$, the leading scaling behavior of
the two-point correlation function of the order-parameter field
$\phi({\bm x})$ is given by
\begin{equation}
  G({\bm x},L,r) \equiv \langle \phi({\bm
    x}_1)\phi({\bm x}_2)\rangle \approx L^{-2y_\phi} \left[ {\cal
      G}({\bm x}/L, W) + O(L^{-\omega})\right],\qquad
  y_\phi = {d-2+\eta\over 2}, \quad W=L^{y_r}r,
\label{gxysca}
\end{equation}
where $y_\phi$ is the RG dimension of the field $\phi({\bm x})$, ${\bm x}
\equiv {\bm x}_1-{\bm x}_2$, and we assumed translation invariance (i.e., the
absence the boundaries).  The integral of $G({\bm x},L,r)$
over the whole finite volume $L^d$ 
provides the magnetic susceptibility
\begin{equation}
  \chi(L,r) = L^{2-\eta} \big[ {\cal X}(W) +
    O(L^{-\omega}) \big] + B_\chi(r) .
  \label{chisclaw}
\end{equation}
The scaling function ${\cal X}$ is universal, apart from a
multiplicative factor and a normalization of the argument. The
function $B_\chi$ is an analytic background term, see,
e.g., Ref.~\cite{PV-02}.

The length scale associated with the critical modes can be obtained
from the correlation function $G$.  For this purpose different
definitions can be considered. In the infinite-volume limit,
or if at least one of the spatial dimensions is infinite,
one may
define a correlation length $\xi_e$ from the large-distance
exponential decay of the two-point correlation
function~\eqref{gxysca}, i.e., $G({\bm x})\sim \exp(-|{\bm
  x}|/\xi_e)$, provided the system is not at a critical point.
An
alternative definition, particularly useful for the analysis of finite
systems, relies on the second moment of the
order-parameter correlation function. More precisely, on a lattice one 
defines
\begin{equation}
  \xi^2 \equiv {1\over 4 \sin^2 (p_{\rm min}/2)} {\widetilde{G}({\bm
      0}) - \widetilde{G}({\bm p})\over \widetilde{G}({\bm p})},
     \label{xidefpb}
\end{equation}
where $p_{\rm min} \equiv 2 \pi/L$, ${\bm p}$ is a vector with
only one nonvanishing component equal to $p_{\rm min}$, and
$\widetilde{G}({\bm p})$ is the Fourier transform of $G({\bm x})$.
This definition is particularly convenient when using periodic
boundary conditions, but it can also be used in other cases, at least
as far as $2\pi/L$ is a legitimate lattice momentum.

\subsubsection{Renormalization-group
  invariant quantities}
\label{fssrginv}

Dimensionless RG invariant quantities are particularly useful for the 
investigation of the critical behavior. Examples of such quantities are the
ratio
\begin{equation}
  R_\xi\equiv \xi/L,
  \label{rxidef}
\end{equation}
where $\xi$ is any length scale related to the critical modes, for
example the one defined in Eq.~\eqref{xidefpb}, and the so-called
Binder parameter associated with the order-parameter field, defined as
\begin{equation}
U = \frac{\langle \mu_2^2\rangle}{\langle \mu_2 \rangle^2}, \qquad
\mu_2 = \frac{1}{V^2} \sum_{{\bm x},{\bm y}} \phi_{\bm x}
\phi_{\bm y}.
\label{binderdef}
\end{equation}
In the FSS limit, at zero external field $h$, RG invariant
quantities (we denote them generically with $R$)
 behave as 
\begin{equation}
  R(L,r) \approx {\cal R}(W) + L^{-\omega} \,{\cal
    R}_\omega(W) + \ldots .
  \label{rscaf}
\end{equation}
The scaling function ${\cal R}(W)$ is universal, apart from a
normalization of the argument.  In particular, the limit
\begin{equation}
  \lim_{L\to\infty} R(L,0) =  {\cal R}(0)
  \label{rzero}
\end{equation}
is universal. It depends on the universality class but not on the 
 microscopic details of the model. It also
depends on the shape of the finite lattice and on the boundary
conditions.

The FSS behavior of the RG-invariant quantities $R$ can be exploited
to determine the critical point~\cite{Binder-83, PV-02}. Indeed, when
the following inequalities hold
\begin{equation}
  \lim_{r\to0^-}\lim_{L\to\infty}R(L,r) > \lim_{L\to\infty}R(L,0) >
  \lim_{r\to0^+}\lim_{L\to\infty}R(L,r)
  \label{monbeh}
\end{equation}
(or the analogous ones with $<$ replacing  $>$), one can define
$r_{\rm cross}$ by requiring
\begin{equation}
  R(L,r_{\rm cross}) = R(2L,r_{\rm cross}).
  \label{rcross}
\end{equation}
The crossing point $r_{\rm cross}$ converges to $r=0$, with
corrections that typically vanish  as $L^{-1/\nu-\omega}$ as 
$L\to\infty$.

We finally note that, if a RG invariant quantity $R_m$
is monotonic with respect to the relevant parameter $r$, as is
generally the case for $R_\xi=\xi/L$, one may write for a second generic RG
invariant quantity $R_i$ the scaling relation
\begin{equation}
  \label{uvsrxi}
  R_i(L,r) = \widehat{\cal R}_i(R_m) + O(L^{-\omega}),
\end{equation}
where $\widehat{\cal R}_i$ depends on the universality class only,
without any nonuniversal normalization of the argument. This is true
once the boundary conditions and the shape of the lattice have been
fixed.  The scaling relation~\eqref{uvsrxi} is particularly convenient to test
universality-class predictions, since it allows one to compare the results for 
different models without the need of tuning 
nonuniversal parameters.  One may write the FSS behavior of any other quantity
in terms of $R_\xi$. For example, the asymptotic FSS behavior of the
susceptibility $\chi$ defined in Eq.~(\ref{chisclaw}) can be rewritten as
\begin{equation}
  \chi(L,r) \approx L^{2-\eta} \widehat{\cal X}(R_\xi),
  \label{chisclaw2}
\end{equation}
where the FSS function $\widehat{\cal X}$ is now expected to be universal
apart from a multiplicative factor only.

\subsection{Scaling behavior at a multicritical point}
\label{fssmcp}

We now briefly outline the FSS behavior at a multicritical point
(MCP), see Fig.~\ref{multicri}, which is generally characterized by
the presence of two relevant RG perturbations, which are both 
invariant under the global symmetry of the model and which 
 must be both tuned to approach the MCP. 
We indicate with $r_1$ and $r_2$ the corresponding Hamiltonian parameters,
normalized so that $r_1=r_2=0$ at the MCP. 
At a MCP, the singular part
of the free-energy density can be written
as~\cite{FN-74,NKF-74,KNF-76,PV-02,CPV-03,BPV-22-z2g}
\begin{equation}
F_{\rm sing}(r_1,r_2,L) \approx L^{-d} {\cal F}(u_1 L^{y_1},u_2 L^{y_2}),
\label{freeenmcp}
\end{equation} 
where $u_1$ and $u_2$ are the nonlinear scaling fields associated with
the two relevant parameters $r_1$ and $r_2$ (they are analytic
functions of $r_1$ and $r_2$, that satisfy $u_1\approx r_1$ and
$u_2\approx r_2$), $y_1>0$ and $y_2>0$ are the corresponding RG
dimensions (we assume $y_1>y_2$), and we neglected corrections to the
multicritical behavior due to the irrelevant scaling fields.

In the infinite-volume limit and neglecting subleading corrections, we
can rewrite the singular part of the free-energy density as
\begin{eqnarray}
  F_{\rm sing}(r_1,r_2) = |u_2|^{d/y_2} {\cal F}_\pm (X),
  \qquad
  X \equiv
  u_1 |u_2|^{-\phi}\,, \qquad \phi\equiv {y_1/y_2}>1\,,
  \label{freeenmcp2}
\end{eqnarray} 
where the functions ${\cal F}_\pm(X)$ apply to the parameter regions
in which $\pm u_2 > 0$, respectively, and $\phi$ is the so-called
crossover exponent associated with the MCP.  Close to the MCP, the
transition lines follow the equation $X = u_1 |u_2|^{-\phi} = {\rm
  const}$ with a different constant for each transition line.  Since
$\phi > 1$, they are tangent to the line $u_1 = 0$.

\subsection{Two-dimensional spin models with continuous symmetry}
\label{asyfree}

According to the Mermin-Wagner theorem~\cite{MW-66,Mermin-67}, 2D
statistical models with short-range interactions and continuous
symmetries cannot have low-temperature magnetized phases
characterized by the condensation of an order parameter, and therefore
they do not undergo phase transitions driven by the spontaneous
breaking of the global symmetry.  Examples of such 2D models are the
$N$-vector models defined by the Hamiltonian (\ref{Nvectormod}) on a
square lattice (setting the external magnetic field ${\bm h}$ to
zero), the 2D CP$^{N-1}$ and RP$^{N-1}$ models, defined 
Sec.~\ref{cpnmodels} and Sec.~\ref{RPN}, respectively.  Their phase
diagrams are generally characterized by a single disordered phase
with the notable exception of systems with Abelian O(2) symmetry.
They show a critical behavior only in the
zero-temperature limit, with universal features that are 
determined by the symmetries of the system. An effective description 
is provided by the 2D nonlinear $\sigma$ models  defined on
symmetric spaces, see, e.g., Refs.~\cite{ZJ-book,PV-02}.

These statistical systems are asymptotically free with a
nonperturbatively generated mass gap. Such a property is also present
in QCD, the theory of strong interactions, and thus these 2D
theories have often been used as toy models to understand some of the
nonperturbative mechanisms characterizing 
QCD, see, e.g., Refs.~\cite{Polyakov-book,ZJ-book,PV-02,VP-09}.  We
mention that the 2D $N$-vector model is also important in
condensed-matter physics. For $N=3$ it describes the zero-temperature 
behavior of the 2D spin-$S$ Heisenberg quantum
antiferromagnet~\cite{CHN-89}.  Indeed, at finite temperature $T$ this
quantum spin system is described by a (2+1)-dimensional O(3) classical
theory in which the Euclidean time direction has a finite extent
$1/T$. In the critical limit the relation $1/T\ll \xi$ is satisfied,
therefore the system becomes effectively two-dimensional, and thus its
critical behavior is described by the 2D O(3)-vector model.

The long-distance critical behavior in the zero-temperature limit can
be predicted by performing  perturbative computations in powers of $T$, 
see,
e.g., Refs.~\cite{Polyakov-75a,BZ-76,BZL-76,BLS-76,DVL-78,Witten-79,
  Hikami-79,BW-86,FT-86,CHN-89,CR-92,CRV-92,CR-93,CP-94,
  CP-95,ACPP-99,Polyakov-book,ZJ-book,PV-02}. In particular, the
$T$-dependence of the correlation length $\xi$ can be inferred from the 
perturbative expansion of the lattice $\beta$ function, defined in terms of the 
derivative of the correlation length $\xi$ with respect to $T$,
\begin{eqnarray}
  \beta(T) = \left( {1\over \xi}  {d \xi\over d T}\right)^{-1} = 
     - T^2 \sum_{n=0} \beta_n T^n.
  \label{betaf}
\end{eqnarray}
Indeed, one obtains, see, e.g., Refs.~\cite{ZJ-book,PV-02},
\begin{eqnarray}
\xi(T) = \xi_0 \left({\beta_0 T}\right)^{\beta_1/\beta_0^2}
\exp\left( {1\over \beta_0 T}\right)\, 
\exp\left[\int_0^T dt\left({1\over \beta(t)} +
{1\over \beta_0t^2} -
{\beta_1\over \beta_0^2 t}\right)\right],
\label{xi-2d}
\end{eqnarray}
where $\xi_0$ is a nonperturbative model-dependent constant (depending
also on the actual definition of $\xi$).  More specifically, using the
first known terms of the $\beta$ functions, the asymptotic behavior of
the length scale in the zero-temperature limit 
is (see, e.g., Refs.~\cite{ZJ-book,PV-02})
\begin{equation}
  \xi(T) \sim B^{-{1\over N-2}} \, e^B,\qquad B = {2\pi\over (N-2)T},
  \label{xion}
\end{equation}
for $N$-vector models (with $N\ge 3$),
and~\cite{CRV-92}
\begin{equation}
\xi(T) \sim B^{-{2\over N}} e^B,\qquad B = {2\pi\over T},
\label{xicpn}
\end{equation}
for lattice CP$^{N-1}$ models ($N\ge 2$). 

The asymptotic behaviors (\ref{xion}) and (\ref{xicpn}) are quite
difficult to observe because the neglected corrections decay as
inverse powers of $\ln \xi$ in the large-$\xi$ limit.  It is important
to observe that these logarithmic corrections are not present when
considering RG invariant dimensionless quantities $R(T)$. In this
case, scaling corrections are proportional to negative integer powers
of $\xi$, multiplied by powers of $\ln \xi$ (equivalently powers of
$T$).  In particular, for the $N$-vector and CP$^{N-1}$ models we have
\begin{equation}
  R(T) =  R^* + O\left[{T^p/\xi(T)^2}\right],
\label{scaling-2d-Nge3}
\end{equation}
where $p$ is a model-dependent power. Therefore, apart from
logarithmic corrections (due to $T\sim 1/\ln\xi$), the 
leading scaling-correction exponent is 
$\omega=2$.  Such a behavior has been verified explicitly in the large-$N$
limit, see, e.g., Ref. \cite{CCCPV-00}.

We also mention that the nature of the asymptotic low-temperature
behavior of the 2D RP$^{N-1}$ models has been long debated, see, e.g.,
Refs.~\cite{BFPV-20-rpn,CHHR-98,NWS-96,Hasenbusch-96,CPS-94,CEPS-93}.
Refs.~\cite{Hasenbusch-96,NWS-96,CHHR-98} reported arguments that
support the claim that RP$^{N-1}$ and $N$-vector models belong to the
same universality class, implying the irrelevance of the ${\mathbb
  Z}_2$ gauge symmetry. They are essentially related to the fact that
$N$-vector and RP$^{N-1}$ have the same asymptotic perturbative
behavior.  However, extensive numerical results for several variant
RP$^{N-1}$ models~\cite{BFPV-20-rpn} show substantial differences in
the nonperturbative scaling functions, which may be explained by the
nonperturbative relevance of the topological ${\mathbb Z}_2$ defects
characterizing the RP$^{N-1}$ models, giving rise to distinct
universality classes.

We finally mention that 2D systems with an Abelian O(2) global
symmetry, such as the $N=2$ XY vector model and the RP$^1$ model, are peculiar
in this respect, since they can undergo a finite-temperature topological
BKT transition~\cite{KT-73,Berezinskii-70,Kosterlitz-74}, which
separates the high-$T$ disordered phase from the low-temperature
nonmagnetized phase characterized by spin waves and quasi-long range order
with vanishing magnetization. At BKT transitions 
the correlation length $\xi$ 
scales as $\xi\sim \exp(c/\sqrt{T-T_c})$
approaching the BKT critical temperature $T_c$ from the
high-temperature phase. The scaling behavior at BKT transitions has
been extensively discussed in the literature, see, e.g.,
Refs.~\cite{KT-73,Berezinskii-70,Kosterlitz-74,JKKN-77,
  HMP-94,HP-97,Balog-01,PV-02,Hasenbusch-05,PV-13}.

\subsection{Finite-size scaling at first-order transitions}
\label{ffsfo}

The behaviors emerging at first-order phase transitions are more
complex than those observed at continuous transitions, see,
e.g., Refs.\cite{Binder-87,PV-24}. Critical phenomena at continuous
phase transitions are essentially related with the presence of critical
correlations, which decay as a power of the distance at the critical
point, and with a diverging length scale $\xi$.  When approaching the
critical point, the long-distance behavior, i.e., on distances of the
order of the scale $\xi$, shows universal features that only depend on
few global properties of the microscopic short-distance
interactions. The behavior at first-order transitions is instead more
complex. Indeed, we typically observe the coexistence of different
phases, with one or more phases that are ordered 
on all length scales, apart from some
short-range local fluctuations.  The presence of coexisting phases
gives rise to peculiar competing phenomena in the phase-coexistence
region, such as metastability, nucleation, droplet formation,
coarsening, etc.~\cite{Binder-87,Bray-94}. Moreover, the observed bulk
behavior crucially depends on the nature of the boundary conditions,
at variance with what happens at continuous transitions, where
boundary conditions only affect some finite-size properties of the
system, but are irrelevant in the thermodynamic limit.  The
sensitivity to the boundary conditions represents one of the main
qualitative differences between continuous and first-order phase
transitions~\cite{PV-24}.

First-order transitions are characterized by the discontinuity of
thermodynamic quantities. In $d$-di\-men\-sio\-nal finite systems
these discontinuities are smeared out, but they still give rise to
easily identifiable properties.  A nontrivial FSS behavior have been
also established at first-order classical and quantum
transitions~\cite{NN-75,FB-82,PF-83,BN-82,FP-85,CLB-86,Binder-87,BK-90,LK-91,
  VRSB-93,CNPV-14,PV-24}, generalizing the ideas that had been
previously developed in the context of critical transitions.  For
instance, the specific heat and the magnetic susceptibility show a
sharp maximum that diverges as a power of the volume and the same is
true for the cumulants of the order parameter in magnetic transitions.

To distinguish continuous from first-order transitions in numerical
studies of systems with limited size, one can exploit the substantial
differences of their FSS behavior.  A possible approach consists in
considering the large-$L$ behavior of the specific heat or of the
susceptibility of the order parameter, that should have a maximum that
diverges as $L^d$.~\cite{Ape-90,FMOU-90,Mc-90,Billoire-95} In
practice, one should compute the maximum of the susceptibility
$\chi_{A,\rm max}(L)$ of a local operator $A_{\bm x}$ for each value
of $L$ and then fit the results to a power law $a L^p$. If one obtains
$p\approx d$, the transition is of first order; otherwise, the
transition is continuous, $p$ being related to standard critical
exponents. If $A$ is the energy density, then $\chi_A$ is the specific
heat that diverges as $L^{\alpha/\nu}$ (if $\alpha$ is positive),
while if $A$ is the magnetization, then $\chi_A\sim L^{2-\eta}$. This
approach works nicely for strongly discontinuous transitions. In the
case of weakly discontinuous transition, however, the asymptotic
behavior $\chi_A\sim L^d$ may set in for values of $L$ that are much
larger than those considered in the simulations. Thus, data may show
an effective scaling $\chi_{A,\rm max}(L) \sim L^p$, with $p$
significantly smaller than $d$, effectively mimicking a continuous
transition (the Potts model~\cite{Wu-82,Baxter-book} in two and three
dimensions is a good example, see
Refs.~\cite{Ape-90,FMOU-90,Mc-90,Billoire-95}).

At order-disorder magnetic transitions, the Binder cumulant $U_m$
associated with the magnetization usually provides a better
indicator~\cite{VRSB-93}.  Indeed, it diverges at first-order
transitions, while it is smooth and finite at continuous
transitions. Thus, the observation that $U_{m,\rm max}(L)$ increases
with $L$ is an evidence of the discontinuous nature of the transition,
even if the maximum does not scale as $L^d$, as it should do
asymptotically.  This idea have been exploited to determine the nature
of the transitions in several models, including systems with gauge
symmetries, such as the LAH and LNAH model, see, e.g.,
Refs.~\cite{PV-19-CP,BPV-21-ncAH,BPV-19,BFPV-21-mpsun}.

An extended review of the FSS behavior at thermal and quantum first-order
transitions and of its  dependence on the boundary conditions can be
found in Ref.~\cite{PV-24}.

%% file: appb.tex
\section{Critical exponents of
  some O($N$)-vector universality classes}
\label{univclass}

In this appendix we report the presently most accurate estimates of
the critical exponents for the most common 3D $N$-vector universality
classes, the Ising ($N=1$), XY ($N=2$), and Heisenberg ($N=3$)
universality classes. An effective LGW model is the $\Phi^4$ field
theory for an $N$-component real field, while a lattice representative
is the standard $N$-vector model defined by the Hamiltonian
(\ref{Nvectormod}).  The estimates are reported in
Tables~\ref{isingres3d}, \ref{tablexy}, and \ref{tableo3} for $N=1$,
$N=2$, and $N=3$, respectively.  A more complete list of theoretical
and experimental results can be found in Ref.~\cite{PV-02} (at least
up to 2002); those reported here are the most accurate in the various
approaches.  We recall that the critical exponents controlling the
asymptotic exponents of other observables can be obtained from $\nu$
and $\eta$ by using scaling and hyperscaling relations (see, e.g.,
Refs.~\cite{ZJ-book, Ma-book}).  Accurate estimates of the critical
exponents of the O($N$) vector universality classes for $N\ge 4$ can
be found in Refs.~\cite{Hasenbusch-22,PV-02}.

\begin{table}
  \begin{center}
    \begin{tabular}{lllllllc}
      \hline\hline \multicolumn{2}{c}{3D Ising}&
      \multicolumn{1}{c}{$\nu$}& \multicolumn{1}{c}{$\eta$}&
      \multicolumn{1}{c}{$\omega$}& \multicolumn{1}{c}{Ref.}
      \\\hline
      SFT & 6-loop 3D expansion & 0.6304(13) &
      0.0335(25) & 0.799(11) & \cite{GZ-98} \\
      & 6-loop $\varepsilon$ expansion & 0.6292(5) & 0.0362(6) & 0.820(7) &
      \cite{KP-17} \\
      & Functional RG & 0.63012(16) &
      0.0361(11) & 0.832(14) & \cite{DBTW-20} \\
      & CFT bootstrap & 0.629971(4) & 0.036298(2) & 0.8297(2) &
      \cite{KPSV-16} \\
      \hline
      Lattice & High-$T$ expansion & 0.63012(16)
      & 0.0364(2) & 0.82(4) & \cite{CPRV-02} \\
      & MC &
      0.63020(12) & 0.0372(10) & 0.82(3) &\cite{DB-03}\\
      & MC & 0.63002(10) & 0.03627(10) & 0.832(6)
      &\cite{Hasenbusch-10} \\ \hline\hline
    \end{tabular}
    \caption{Estimates of the critical exponents of the 3D Ising
      universality class (characterized by a global ${\mathbb Z}_2$
      symmetry). We report the correlation-length exponent $\nu$, the
      order-parameter exponent $\eta$, and the exponent $\omega$
      associated with the leading scaling corrections.  We report: SFT
      results obtained by resumming high-order perturbative expansions
      (6-loop calculations in the fixed-dimension scheme~\cite{GZ-98,
        AS-95, LZ-77a, LZ-77b, BNGM-76, Parisi-80} and in the
      $\varepsilon$-expansion scheme~\cite{KP-17, CGLT-83, WF-72,
        tHV-72}), by using functional RG
      approach~\cite{DBTW-20}, and the conformal field theory (CFT)
      bootstrap~\cite{KPSV-16, PRV-19-rev} approach; results obtained
      in lattice models by resumming high-temperature
      expansions~\cite{CPRV-02} and in MC
      simulations~\cite{Hasenbusch-10, DB-03,Hasenbusch-01,
        Hasenbusch-99}.  We note that there is an overall agreement
      among the results obtained by the different
      approaches. Moreover, there is also a good agreement with
      experiments in physical systems that undergo continuous
      transitions in the Ising universality class (liquid-vapor
      systems, binary systems, uniaxial magnetic systems, Coulombic
      systems, etc.) see, e.g., Ref.~\cite{PV-02} for a list of
      experimental results.}
    \label{isingres3d}
  \end{center}
\end{table}

\begin{table}
  \begin{center}
    \begin{tabular}{lllllc}
      \hline\hline
      \multicolumn{2}{c}{3D XY} &
      \multicolumn{1}{c}{$\nu$} & 
      \multicolumn{1}{c}{$\eta$} & 
      \multicolumn{1}{c}{$\omega$}&
      \multicolumn{1}{c}{Ref.} 
      \\\hline
      SFT & 6-loop 3D expansion& 0.6703(15) & 0.035(3) & 0.789(11)
      & \cite{GZ-98} \\
      & 6-loop $\varepsilon$ expansion & 0.6690(10) & 0.0380(6) & 0.804(3)
      & \cite{KP-17} \\
      & Functional RG & 0.6716(6) & 0.0380(13) & 0.791(8) & 
      \cite{DBTW-20} \\
      & CFT bootstrap &
      0.67175(10) & 0.038176(44) & 0.794(8) & \cite{CLLPSSV-20} \\
      \hline
      Lattice & High-$T$+MC & 0.6717(1) & 0.0381(2) & 0.785(20) &
      \cite{CHPV-06} \\
      & MC & 0.6717(3) & & & \cite{BMPS-06} \\
      & MC & 0.67169(7) & 0.03810(8) & 0.789(4) &
      \cite{Hasenbusch-19} \\
      \hline\hline
    \end{tabular}
    \caption{ Estimates of the critical exponents of the 3D XY
      universality class (characterized by a global O(2) symmetry). We
      report the correlation-length exponent $\nu$, the
      order-parameter exponent $\eta$, and the exponent $\omega$
      associated with the leading scaling corrections.  We report: SFT
      results obtained by resumming high-order perturbative expansions
      \cite{GZ-98,KP-17} (6-loop calculations in the fixed-dimension
      and in the $\varepsilon$-expansion scheme), by using
      functional RG approach~\cite{DBTW-20} and the conformal
      field theory (CFT) bootstrap~\cite{CLLPSSV-20} approach; results
      obtained in lattice models by combining high-temperature and MC
      results~\cite{CHPV-06} and in MC
      simulations~\cite{BMPS-06,Hasenbusch-19}.  We also mention that
      experimental estimates in a microgravity environment have been
      obtained for the $^4$He superfluid
      transition~\cite{Lipa-etal-03,Lipa-etal-00, Lipa-etal-96}.  See
      Ref.~\cite{PV-02} for a more complete list of theoretical and
      experimental results.}
    \label{tablexy}
  \end{center}
\end{table}

\begin{table}
  \begin{center}
    \begin{tabular}{lllllc}
      \hline\hline \multicolumn{2}{c}{3D O(3)} &
      \multicolumn{1}{c}{$\nu$} & \multicolumn{1}{c}{$\eta$} &
      \multicolumn{1}{c}{$\omega$}& \multicolumn{1}{c}{Ref.}

      \\\hline
      SFT & 6-loop 3D
      expansion& 0.7073(35) & 0.0355(25) & 0.782(13) & \cite{GZ-98} \\

      & 6-loop $\varepsilon$ expansion & 0.7059(20) & 0.03663(12) & 0.795(7)
      & \cite{KP-17} \\

      & Functional RG & 0.7114(9) & 0.0376(13) & 0.769(11) & 
      \cite{DBTW-20} \\

      & CFT bootstrap &
      0.7117(4)
      & 0.03787(13) & & \cite{CLLPSSV-20-o3} \\\hline

      Lattice & High-$T$+MC & 0.7112(5) & 0.0375(5) &
      &\cite{CHPRV-02}\\

      & High-$T$+MC & 0.7117(5) & 0.0378(5) &
      &\cite{HV-11} \\

      & High-$T$+MC & 0.7116(2) & 0.0378(3) &
      &\cite{Hasenbusch-20} \\

      & MC & 0.71164(10) & 0.03784(5) &
      0.759(2) & \cite{Hasenbusch-20} \\ 
      \hline\hline
    \end{tabular}
    \caption{ Estimates of the critical exponents of the 3D Heisenberg
      universality class (characterized by a global O(3) symmetry). We
      report the correlation-length exponent $\nu$, the
      order-parameter exponent $\eta$, and the exponent $\omega$
      associated with the leading scaling corrections.  We report: SFT
      results obtained by resumming high-order perturbative expansions
      \cite{GZ-98,KP-17} (6-loop calculations in the fixed-dimension
      and in the $\varepsilon$-expansion scheme), by using
      functional RG approach~\cite{DBTW-20}, and the conformal
      field theory (CFT) bootstrap~\cite{CLLPSSV-20-o3} approach;
      results obtained in lattice models by combining high-temperature
      and MC results~\cite{CHPRV-02, HV-11} and in MC
      simulations~\cite{Hasenbusch-20}.  See Refs.~\cite{PV-02,
        Hasenbusch-20} for a more complete list of theoretical and
      experimental results.}
    \label{tableo3}
  \end{center}
\end{table}

%% file: appc.tex
\section{Mean-field analysis of the low-temperature Higgs phases}
\label{meanfieldHiggs}

  In this appendix we present a mean-field analysis of the
  low-temperature Higgs phases of the LNAH model with SU($N_c$) gauge
  symmetry and $N_f$ degenerate scalar flavors in the fundamental
  representation. This model has been discussed in
  Sec.~\ref{modelsubfu}.  For more details see
  Ref.~\cite{BFPV-21-mpsun}.  This analysis allows one to infer the
  symmetry properties of the low-temperature Higgs phases, which
  depend on the parameter $v$, the number of colors $N_c$ and of
  flavors $N_f$.  It is worth noting that this discussion applies to
  generic $d$-dimensional systems, therefore also to 4D space-time
  systems that may be relevant in the context of high-energy physics.
  Analogous analyses can be performed for SU($N_c$) theories with
  scalar matter in the adjoint
  representation~\cite{SSST-19,SPSS-20,BFPV-21-adj}, and SO($N_c$)
  gauge LNAH models~\cite{BPV-20-on}.

The SU($N_c$) gauge model has different Higgs phases, whose symmetry
corresponds to the symmetry of the minimum-energy configurations that
dominate the partition function for $\beta\to\infty$.  As discussed in
Ref.~\cite{BFPV-21-mpsun} they are obtained by considering separately
the minima of the scalar kinetic energy term (\ref{Kinterm}) and of
the scalar potential $V({\bm \Phi})$, defined in
Eq.~(\ref{potentialsun}).

The kinetic term (\ref{Kinterm}) is minimized by the condition
$\Phi_{\bm x} = U_{{\bm x},\mu} \Phi_{{\bm x}+\hat{\mu}}$, which
implies $V(\Phi_{\bm x}) = V(\Phi_{{\bm x}+\hat{\mu}})$. It follows
that minimum configurations are obtained by minimizing the potential
energy on each lattice site. Therefore, below we neglect the ${\bm x}$
dependence of the field. One can easily verify that the symmetry
properties of the minimum-potential configurations do not depend on
$r$ and $u$, and therefore we only report explicit results for the
unit-length limit in which ${\rm Tr}\, \Phi^\dagger
\Phi^{\phantom\dagger} = 1$.

To determine the minima of $V(\Phi)$, one may use the
singular-value decomposition that allows one to rewrite the field
$\Phi$ as $\Phi^{af} = \sum_{bg} C^{ab} W^{bg} F^{gf}$, where $C\in
{\rm U}(N_c)$ and $F\in {\rm U}(N_f)$ are unitary matrices, and $W$ is
an $N_c \times N_f$ rectangular matrix. Its nondiagonal elements
vanish ($W^{ij} = 0$ for $i\not= j$), while the diagonal elements are
real and nonnegative, $W^{ii} = w_i>0$ ($i=1,...,q$), with $q={\rm
  Min}[N_f,N_c]$.  Substituting the above expression 
in $V(\Phi)$, one obtains 
\begin{equation}
  V(\Phi) = {v\over 4} {\rm Tr}\,(\Phi^\dagger \Phi)^2=
  \frac{v}{4} \sum_{i=1}^q w_i^4,\qquad
  \hbox{Tr}\, \Phi^\dagger \Phi = \sum_{i=1}^q
w_i^2 = 1.
\end{equation}
A straightforward minimization of the potential, subject to the
unit-length constraint, gives two different solutions, depending on
the sign of $v$, which are:
\begin{eqnarray}
&& w_1 = 1\,, \quad w_{2}=...=w_q = 0, \qquad {\rm
    for}\;v<0;\label{solution1} \\
  && w_1 = \ldots = w_q = {1/\sqrt{q}},
  \qquad {\rm for}\;v>0, \qquad {\rm where}\;\; q={\rm
    Min}[N_f,N_c].
\label{solution2}
\end{eqnarray}
The solution (\ref{solution1}) for $v<0$ allows one to rewrite the
field as
\begin{equation}
  \Phi^{af} = s^a z^f,
\label{minvlt0}
\end{equation}
where ${\bm s}$ and ${\bm z}$ are unit-length complex vectors with
$N_c$ and $N_f$ components, respectively, satisfying $\bar{\bm s}\cdot
{\bm s}=1$ and $\bar{\bm z}\cdot {\bm z}=1$.  Modulo gauge
transformations, these solutions are invariant under
$\hbox{U}(1)\oplus\hbox{U}(N_f-1)$ transformations, leading to the
global-symmetry breaking pattern
$\hbox{U}(N_f)\to\hbox{U}(1)\oplus\hbox{U}(N_f-1)$. 

As discussed in Sec.~\ref{higgsphases}, the model has a global U(1) 
symmetry that is not broken, so the relevant symmetry-breaking pattern
is $\hbox{SU}(N_f)\to\hbox{U}(N_f-1)$, which  is the same as 
for the CP$^{N_f-1}$ transition.
Thus, if the gauge dynamics is not relevant at the
transition, for $v < 0$ we expect the non-Abelian gauge model with
U($N_f$) global symmetry and the CP$^{N_f-1}$ model to have the same
critical behavior, for any $N_c$.  The correspondence between the two
models can also be established by noting that the relevant order
parameter at the transition is the bilinear operator $P_{\bm x}^{fg}$
defined in Eq.~(\ref{pdefsun}), which assumes the simple form $P_{\bm
  x}^{fg} = \bar{z}_{\bm x}^f z_{\bm x}^g-\delta^{fg}/N_f$ on the
minimum configurations. It is thus equivalent to a local projector
$\bar{z}_{\bm x}^f z_{\bm x}^g$ onto a one-dimensional space.  If we
assume that the critical behavior of the gauge model is only
determined by the fluctuations that preserve the minimum-energy
structure, the effective scalar model that describes the critical
fluctuations can be identified with the CP$^{N_f-1}$ model.

To exclude the possible presence of topological transitions, 
one should determine the gauge symmetry breaking pattern, as discussed
in Sec.~\ref{sec.ZQhiggs}. If we repeatedly apply the relation
$\Phi_{\bm x} = U_{{\bm x},\mu} \Phi_{{\bm x}+\hat{\mu}}$
along a plaquette, we obtain the equation
$\Phi_{\bm x} = \Pi_{{\bm x},\mu\nu} \Phi_{\bm x}$, where $\Pi_{{\bm
    x},\mu\nu}$ is the plaquette operator defined in
Eq.~(\ref{plaquette}).
For minimum configurations of type
(\ref{solution1}), using Eq.~(\ref{minvlt0}), we have $s_{\bm x}^{a} =
\sum_b \Pi_{{\bm x},\mu\nu}^{ab} s_{\bm x}^b$, i.e., the plaquette
$\Pi_{{\bm x},\mu\nu}$ has one unit eigenvalue.  Thus, for
$\beta\to\infty$ there is still a residual SU($N_c-1$) symmetry,
independently of the flavor number $N_f$. Since SU($N_c-1$) gauge 
models are always disordered, no topological transitions are expected.

The solution (\ref{solution2}) for $v>0$ implies more complex 
symmetry-breaking patterns. 
In particular, we must distinguish three different
cases: $N_f<N_c$, $N_f=N_c$, and $N_f>N_c$. 
The minimum-potential configurations take the form:
\begin{eqnarray}
\Phi^{af} &=& {1\over \sqrt{q}}\,\delta^{af},
    \quad {\rm for}\;N_f<N_c;\qquad\qquad 
   \label{eq:minnfnc} \\
\Phi^{af} &=& {1\over \sqrt{q}}\,\delta^{af}\phi,\quad
  \phi \in {\rm U}(1),  \quad {\rm for}\;N_f=N_c.
   \label{phiagnfltltnc} \\
 \Phi^{af} &=& {1\over \sqrt{N_c}}\,F^{af}, \qquad F\in {\rm U}(N_f),
  \qquad {\rm for}\;\;N_f>N_c.
  \label{minnfltnc}
\end{eqnarray}
If we substitute these expressions in the relation 
$\Phi_{\bm x} = \Pi_{{\bm x},\mu\nu} \Phi_{\bm x}$ that follows 
from the minimization of the kinetic energy, one can verify that
the plaquette $\Pi_{{\bm x},\mu\nu}$ has $q$ unit eigenvalues.
Thus, for $N_f \ge N_c$, $\Pi_{{\bm
    x},\mu\nu} = 1$ and the gauge variables are gauge equivalent to
the trivial configuration, i.e., $U_{{\bm x},\mu} = V_{\bm x}^\dagger
V_{{\bm x}+\hat{\mu}}$ where $V_{\bm x}\in$ SU($N_c$).
These results allow us to exclude the presence of topological transitions,
related to the residual gauge symmetry in the Higgs phase, see 
Sec.~\ref{sec.ZQhiggs}. Indeed, for $N_f \ge N_c$ there is no 
residual gauge symmetry, while for $N_f< N_c$, gauge modes are controlled by 
a residual
SU($N_c-N_f$) gauge theory that does not undergo finite-temperature 
transitions.

For $N_f \le  N_c$ the SU($N_f$) symmetry cannot be broken. Indeed, if we 
compute the bilinear operator $P_{\bm x}^{fg}$ defined in 
Eq.~(\ref{pdefsun}) on the minimum configurations,
it vanishes trivially. Equivalently, any SU($N_f$)
transformation leaves the minimum-potential configurations invariant.
Thus, there is no global SU($N_f$) symmetry breaking. 
As discussed in Sec.~\ref{higgsphases}, this implies that no
transition is expected for $N_f < N_c$. For $N_f = N_c$, as discussed in 
Sec.~\ref{higgsphases}, the minimum configurations 
break the U(1) symmetry, and thus U(1) transitions are possible.

For $N_f>N_c$, the minimum-potential
configurations take the form (\ref{minnfltnc}).
In the ordered phase the relevant fluctuations are supposed to be
those that preserve this structure.  Therefore, the field
$\Phi^{af}_{\bm x}$ can be parameterized as in Eq.~(\ref{minnfltnc}),
with a site-dependent unitary matrix $F_{\bm x}$. The link
variable can be expressed as $U_{{\bm x},\mu} = V_{\bm x}^\dagger
V_{{\bm x}+\hat{\mu}}$ with $V_{\bm x}\in {\rm SU}(N_c)$
as all plaquettes are equal to the identity.
Substituting this parametrization in the kinetic term (\ref{Kinterm})
of the Hamiltonian, we obtain
\begin{equation}
H_K = - {N_f\over N_c} \sum_{{\bm x}\mu} {\rm Re}\,\hbox{Tr}\, 
  (F^\dagger_{\bm x}  \widehat{V}_{\bm x}^{\dag} Y 
   \widehat{V}_{{\bm x}+\hat{\mu}} F_{{\bm x}+\hat{\mu}} ), 
\end{equation}
where $Y = I_{N_c} \oplus 0$ is an $N_f\times N_f$ diagonal matrix in
which the first $N_c$ elements are one and the other $N_f-N_c$
elements are zero, and $\widehat{V} = V \oplus I_{N_f-N_c}$.  This
Hamiltonian is invariant under the global transformations $F_{\bm
  x}\to F_{\bm x} M$, with $M\in \mathrm{U}(N_f)$, and under the local
transformations $F_{\bm x}\to W_{\bm x}F_{\bm x}$ and
$\widehat{V}_{\bm x}\to \widehat{V}_{\bm x} G_{\bm x}$, with $W_{\bm
  x}=W^{(1)}_{\bm x}\oplus W^{(2)}_{\bm x}$ and $G_{\bm
  x}=W^{(1)}_{\bm x}\oplus I_{N_f-N_c}$, where $W^{(1)}_{\bm x}
\in\mathrm{SU}(N_c)$, $W^{(2)}_{\bm x} \in\mathrm{U}(N_f-N_c)$
($F_{\bm x}$ is unitary so that $F^{\dag}_{\bm x}F_{\bm x}=I_{N_f}$).
Therefore the global symmetry of the effective model that describes the critical
fluctuations is 
$\mathrm{SU}(N_f)/[\mathrm{SU}(N_c)\otimes\mathrm{SU}(N_f-N_c)]$,
which corresponds to the global symmetry-breaking pattern ${\rm
  U}(N_f)\rightarrow \mathrm{SU}(N_c)\otimes\mathrm{U}(N_f-N_c)$,
where we have also included the unbroken U(1) symmetry.